\documentclass[fleqn,usenatbib]{mnras}

\usepackage{newtxtext,newtxmath}

\usepackage[T1]{fontenc}
\usepackage{ae,aecompl}

\usepackage{graphicx}	
\usepackage{amsmath}	
\usepackage{amssymb}	
\usepackage{caption}

\title[SAMI -- Fornax Dwarfs Survey I]{The SAMI -- Fornax Dwarfs Survey I: Sample, observations and the specific stellar angular momentum of dwarf elliptical galaxies}

\author[N. Scott et al.]{
Nicholas Scott$^{1,2}$\thanks{E-mail: nicholas.scott@sydney.edu.au},
F. Sara Eftekhari$^{3}$,
Reynier F. Peletier$^{3}$,
Julia J. Bryant$^{1,2}$,
\newauthor
Joss Bland-Hawthorn$^{1,2}$,
Massimo Capaccioli$^{4}$,
Scott M. Croom$^{1,2}$,
\newauthor
Michael Drinkwater$^{5}$,
J\'esus Falc\'on-Barroso$^{6,7}$,
Michael Hilker$^{8}$,
Enrichetta Iodice$^{9}$,
\newauthor
Nuria F. P. Lorente$^{10}$,
Steffen Mieske$^{11}$,
Marilena Spavone$^{9}$,
Glenn van de Ven$^{12}$
\newauthor
and Aku Venhola$^{13}$
\\
$^{1}$Sydney Institute for Astronomy, School of Physics, A28, The University of Sydney, NSW, 2006, Australia\\
$^{2}$ARC Centre of Excellence for All Sky Astrophysics in 3 Dimensions (ASTRO 3D)\\
$^{3}$Kapteyn Institute, University of Groningen, Landleven 12, 9747, AD, Groningen, The Netherlands\\
$^{4}$University of Naples Federico II, C.U. Monte Sant'Angelo, Via Cinthia, 80126, Naples, Italy\\
$^{5}$School of Mathematics and Physics, The University of Queensland, St Lucia, QLD 4072, Australia\\
$^{6}$Instituto de Astrof\'isica de Canarias, Calle V\'ia L\'actea s/n, E-38205 La Laguna, Tenerife, Spain\\ 
$^{7}$Departamento de Astrof\'isica, Universidad de La Laguna (ULL), E-38206 La Laguna, Tenerife, Spain\\
$^{8}$ESO, European Southern Observatory, Karl-Schwarzschild-Str 2, D-85748 Garching bei M{\"u}nchen, Germany\\
$^{9}$INAF - Astronomical Observatory of Capodimonte, via Moiariello 16, I-80131 Napoli, Italy\\
$^{10}$Australian Astronomical Optics, AAO-MQ, Faculty of Science and Engineering, Macquarie University, NSW 2109, Australia\\
$^{11}$European Southern Observatory, Alonso de Cordova 3107, 7630355 Vitacura, Santiago, Chile\\
$^{12}$Department of Astrophysics, University Vienna, T{\"u}rkenschanzstrasse 17, A-1180 Wien, Austria\\
$^{13}$Space physics and astronomy research unit, University of Oulu, Pentti Kaiteran katu 1, FI-90014 Oulu, Finland\\
}

\date{Accepted XXX. Received YYY; in original form ZZZ}

\pubyear{2020}

\begin{document}
\label{firstpage}
\pagerange{\pageref{firstpage}--\pageref{lastpage}}
\maketitle

\begin{abstract}
 Dwarf ellipticals are the most common galaxy type in cluster environments, however the challenges associated with their observation mean their formation mechanisms are still poorly understood. To address this, we present deep integral field observations of a sample of 31 low-mass ($10^{7.5} <$ M$_\star < 10^{9.5}$ M$_\odot$) early-type galaxies in the Fornax cluster with the SAMI instrument. For 21 galaxies our observations are sufficiently deep to construct spatially resolved maps of the stellar velocity and velocity dispersion --- for the remaining galaxies we extract global velocities and dispersions from aperture spectra only. From the kinematic maps we measure the specific stellar angular momentum $\lambda_R$ of the lowest mass dE galaxies to date. Combining our observations with early-type galaxy data from the literature spanning a large range in stellar mass, we find that $\lambda_R$ decreases towards lower stellar mass, with a corresponding increase in the proportion of slowly rotating galaxies in this regime. The decrease of $\lambda_R$ with mass in our sample dE galaxies is consistent with a similar trend seen in somewhat more massive spiral galaxies from the CALIFA survey. This suggests that the degree of dynamical heating required to produce dEs from low-mass starforming progenitors may be relatively modest, and consistent with a broad range of formation mechanisms.
\end{abstract}




\section{Introduction}

The population of low-redshift galaxies shows a clear bimodality between blue, star-forming, late-type galaxies and red, passive, early-type galaxies \citep[e.g.][]{Strateva:2001}. Understanding the physical processes that transform galaxies from star-forming to passive --- so-called quenching mechanisms --- is at the heart of studies of galaxy evolution.

\citet{Peng:2010} argue that quenching can be separated into two distinct regimes; mass quenching and environment quenching. Mass quenching occurs when a galaxy's own halo prevents the formation of the dense gas necessary for star formation. Mass quenching operates on galaxies with stellar masses M$_* \gtrsim 10^{9.5}$ M$_\odot$, though only becomes efficient at shutting down star formation at M$_* > 10^{11}$ M$_\odot$. Environment quenching occurs when a galaxy is influenced by factors outside its own halo --- either nearby galaxies or its host group or cluster halo --- in such a way as to cause star formation to cease. Environment quenching has an impact on galaxies of all masses, but is only efficient in relatively dense environments and at low to intermediate galaxy masses. 

A variety of environmental mechanisms have been proposed for quenching galaxies; ram-pressure stripping \citep{Gunn:1972}, strangulation \citep{Larson:1980}, harassment and tidal interactions \citep[e.g.][]{Moore:1998,Hernquist:1989}, and merging. From studies of galaxies that are actively being quenched \citep[e.g.][]{Koch:2012,Poggianti:2017,Jaffe:2018,Owers:2019, Zabel:2019} we can identify signatures of the different quenching mechanisms in terms of differing timescales and spatial distributions for the shut-down of star formation \citep[e.g.][]{Schaefer:2019}. Most importantly for this work, we expect that ram pressure stripping and strangulation will leave the stellar kinematics of a galaxy relatively undisturbed and cold \citep[though][argue sudden stripping may heat the kinematics in very low mass galaxies]{Hammer:2019}, whereas harassment, mergers and large tidal interactions will all serve to heat the kinematic distribution of the stars. 

One limitation of this approach is that it focuses on only the small subset of galaxies that are undergoing quenching right now. From such targeted studies it has been historically difficult to determine which physical processes play the most significant role in quenching galaxies. An alternative approach, and one better suited to address the demographics of quenching, is to study the large population of passive galaxies that have already quenched. We can potentially identify how the passive population was quenched by examining the kinematics, stellar populations and gas content of quenched galaxies, a task ideally suited to integral field spectroscopy. Starting with the SAURON survey \citep{Emsellem:2007, Cappellari:2007}, large IFS surveys of giant (M$_\star \gtrsim 10^{9.5}$ M$_\odot$) passive galaxies have found two distinct kinematic classes of objects, slow rotators and fast rotators, based on their kinematic morphologies and specific stellar angular momentum. The Atlas$^\mathrm{3D}$ survey \citep{Emsellem:2011} found slow rotators to make up a significant proportion of the giant, passive galaxy population only for galaxies with M$_\star \gtrsim 10^{11}$ M$_\odot$. Below this mass the vast majority of passive galaxies are fast rotators \citep[see][for an overview]{Cappellari:2016}. This trend has been confirmed by a series of IFS surveys with increasingly large sample sizes \citep{vandeSande:2017,Graham:2018,Veale:2017,Falcon-Barroso:2019}. In the Fornax cluster, \citet{Scott:2014} examined the stellar kinematics of a subset of the most massive passive galaxies, finding only two slow rotators in the cluster, a result confirmed by the Fornax3D survey \citep{Sarzi:2018, Iodice:2019}. Note that the fast rotator/slow rotator classification scheme was developed for galaxies with M$_\star \gtrsim 10^{9.5}$ M$_\odot$ and it is not clear that it is applicable to lower mass galaxies with significantly different formation histories.

Unfortunately, for massive galaxies in dense environments, both mass quenching and environment quenching are expected to contribute to the shutdown of star formation, making the two independent processes difficult to separate. In contrast, for low-mass dwarf galaxies environmental processes are expected to dominate, with mass quenching playing a negligible role in influencing their observed properties. This makes low-mass galaxies in dense environments ideal objects to study the mechanisms of environmental quenching. The low-mass population of galaxies in galaxy clusters is dominated by galaxies with a dwarf elliptical (dE) morphology --- that is red, passive, spheroidal systems  \citep{Lisker:2007,Janz:2012,Roediger:2017, Venhola:2019}, though they may exhibit a range of stellar population ages \citep{Hamraz:2019} and internal structures \citep{Lisker:2006}. As predominantly passive objects \citep{Roediger:2017}, dE galaxies represent one end result of the physical processes of environmental quenching. 

The impact of environment on the quenching of dwarf galaxies has been widely studied using imaging and fibre spectroscopy. dEs are extremely rare in low-density environments \citep{Geha:2012, Davies:2016}, implying that the environment plays a key role in shutting down star formation in dwarf galaxies. Both observations \citep{Michielsen:2008,Wetzel:2013} and simulations \citep{Boselli:2008,Fillingham:2015,Fillingham:2016} suggest that the shutdown of star formation occurs rapidly on timescale of order 1 Gyr, with ram pressure stripping the leading mechanism \citep{Boselli:2014}. Kinematic heating due to harassment may occur on longer timescales \citep{Michielsen:2008,Benson:2015}. This scenario suggests that the progenitors of present day dEs are low mass, star forming spiral and irregular galaxies \citep[however these progenitors may not be similar to present day low-mass star forming galaxies, see][]{Lisker:2013}. An alternative formation scenario, where dEs represent the low-mass tail of the giant elliptical population is suggested by consistent scaling relations between the two classes \citep[e.g.][]{Geha:2003,Chilingarian:2009}, however the strong environmental dependence of dE number density is difficult to reconcile with this scenario.

While common in galaxy clusters, the resolved kinematics of dEs are challenging to measure. They are intrinsically faint, have small effective radii and are expected to have rotation velocities or velocity dispersions of, at most, a few 10s of km s$^{-1}$. Despite these challenges several recent studies have targeted dEs, predominantly in clusters, either with IFS \citep{Rys:2013,Adams:2014, Mentz:2016,Penny:2016} or long-slit spectroscopy \citep{Pedraz:2002,Toloba:2014,Penny:2015, Janz:2017}. These studies find a variety of kinematic morphologies for dE galaxies, with an increase in the relative proportion of slow rotators compared to intermediate mass (M$_\star \sim 10^{10}$ M$_\odot$) galaxies. However the sample size of these studies is modest compared to those of more massive galaxies --- for the IFS studies at most a dozen objects, and a few tens for the long slit studies. They are also restricted to the brighter end of the dE distribution, primarily targeting galaxies with M$_\star > 10^{9}$ M$_\odot$.

\begin{table}
    \centering
    \caption{Summary of SAMI Fornax Survey observing runs}
    \label{tab:observing_runs}
    \begin{tabular}{lccc}
         Date & Fields & Median & Exposure\\
         & observed & seeing & Time per\\
         & & & Field (hrs)\\
         \hline
         4$^\mathrm{th}$ -- 8$^\mathrm{th}$ Nov 2015 & 2 & 2.3\arcsec & 7.0\\
         26$^\mathrm{th}$ -- 30$^\mathrm{th}$ Oct 2016 & 3 & 1.9\arcsec & 6.7\\
         9$^\mathrm{th}$ -- 15$^\mathrm{th}$ Oct 2018 & 5 & 2.2\arcsec & 6.1
    \end{tabular}
\end{table}

\begin{table*}
    \centering
    \caption{SAMI -- Fornax Dwarf Survey dwarf galaxies with successful stellar kinematic measurements, ordered by decreasing M$_\star$}
    \label{tab:primary_sample}
\begin{tabular}{cccccccccccc}
FDS & FCC & RA & Dec & M$_r$ & M$_g$ & R$_e$ & $\log M_{\star}$ & Morph. & $\epsilon$ & $\lambda_{R_e}$ & Max Rad\\
ID & ID & & & & & & & Class & & & \\
\hline
& & (deg) & (deg) & (mag) & (mag) & (arcsec) & (M$_\odot$) & & & & (R$_e$) \\
\hline
& & & & & & & (1) & (2) & & (3) & (4) \\
\hline
\hline
\multicolumn{12}{c}{Dwarf ellipticals} \\
\hline
6\_D002 & 277 & 55.59492 & -35.1541 & -18.8 & -18.1 & 11.4 & 9.47 & e(s) & 0.42 & 0.27 & 0.95 \\
16\_D002 & 143 & 53.7466710 & -35.171061  & -18.6 & -18.0 & 9.8 & 9.45 & e(s) & 0.15 & 0.15 & 1.1  \\
7\_D000 & 301 & 56.2649 & -35.97267 & -18.3 & -17.7 & 7.6 & 9.36 & e(s) & 0.46 & 0.39 & 1.6  \\
11\_D279 & 182 & 54.2262696 & -35.374677  & -17.9 & -17.1 & 9.7 & 9.16 & e(s)* & 0.04 & 0.18 & 1.05  \\
16\_D159 & 136 & 53.6227783 & -35.546452 & -17.8 & -17.0 & 17.5 & 9.08 & e & 0.15 & 0.15 & 0.65  \\
11\_D235 & 202 & 54.5266666 & -35.43833 & -17.3 & -16.6 & 13.3 & 8.90 & e* & 0.41 & 0.13 & 0.85  \\
15\_D417 & 106 & 53.19867 & -34.23873 & -17.4 & -16.8 & 10.7 & 8.89 & e(s) & 0.51 & 0.14 & 1.15  \\
11\_D283 & 222 & 54.8054166 & -35.36972 & -17.0 & -16.3 & 16.1 & 8.77 & e* & 0.11 & 0.32 & 0.6  \\
10\_D189 & 203 & 54.5381730 & -34.518726 & -16.9 & -16.3 & 16.0 & 8.75 & e(s) & 0.45 & 0.33 & 0.75  \\
15\_D384 & 135 & 53.6285240 & -34.297455 & -16.8 & -16.2 & 14.7 & 8.70 & e(s) & 0.53 & 0.31 & 0.75  \\
16\_D417 & 100 & 52.94848 & -35.05139 & -17.0 & -16.2 & 19.8 & 8.70 & e* & 0.24 & 0.28 & 0.55  \\
11\_D069 & 252 & 55.20999 & -35.74846 & -16.4 & -15.7 & 11.1 & 8.58 & e* & 0.06 & 0.15 & 0.8  \\
11\_D458 & 245 & 55.14099 & -35.02289 & -16.5 & -15.8 & 14.5 & 8.57 & e* & 0.08 & 0.19 & 0.65  \\
7\_D326 & 300 & 56.24959 & -36.31975 & -16.4 & -15.7 & 20.8 & 8.55 & e* & 0.28 & 0.28 & 0.55  \\
11\_D396 & 207 & 54.5795833 & -35.1275 & -16.6 & -16.0 & 9.6 & 8.51 & e & 0.17 & 0.31 & 0.85  \\
6\_D455 & 266 & 55.42216 & -35.17027 & -16.3 & -15.7 & 6.9 & 8.49 & e* & 0.11 & 0.17 & 1.35  \\
11\_D155 & 188 & 54.26875 & -35.58861 & -16.2 & -15.5 & 12.2 & 8.42 & e* & 0.04 & 0.20 & 0.8  \\
11\_D339 & 211 & 54.58875 & -35.25833 & -16.1 & -15.5 & 6.6 & 8.33 & e* & 0.25 & 0.11 & 1.5  \\
12\_D367 & 164 & 54.0537009 & -36.166437 & -16.0 & -15.4 & 9.9 & 8.33 & e(s) & 0.45 & 0.08 & 1.15  \\
11\_D079 & 223 & 54.83125 & -35.72333 & -16.1 & -15.5 & 17.0 & 8.32 & e* & 0.11 & -- & --  \\
13\_D042 & 253 & 55.2303 & -37.83763 & -15.8 & -15.1 & 10.9 & 8.30 & e & 0.38 & 0.26 & 1.0  \\
6\_D208 & 274 & 55.5716666 & -35.53916 & -15.7 & -15.2 & 12.0 & 8.17 & e* & 0.04 & -- & --  \\
6\_D098 & 298 & 56.18507 & -35.68372 & -15.6 & -15.0 & 7.0 & 8.16 & e* & 0.29 & 0.18 & 1.5  \\
10\_D014 & 195 & 54.3471331 & -34.900098 & -15.4 & -14.8 & 12.8 & 8.13 & e & 0.46 & -- & --  \\
6\_D170 & 264 & 55.3820833 & -35.58777 & -15.5 & -14.9 & 10.3 & 8.09 & e & 0.60 & -- & --  \\
13\_D258 & 250 & 55.18497 & -37.40827 & -15.1 & -14.4 & 9.2 & 7.97 & e & 0.24 & -- & --  \\
10\_D302 & 178 & 54.2027295 & -34.280105 & -15.0 & -14.5 & 11.3 & 7.95 & e & 0.29 & -- & --  \\
15\_D232 & B904 & 53.4840889 & -34.561798 & -15.2 & -14.6 & 5.1 & 7.90 & e & 0.20 & -- & -- \\
10\_D003 & 181 & 54.2219391 & -34.938393 & -15.0 & -14.3 & 9.7 & 7.87 & e* & 0.40 & -- & -- \\
15\_D223 & 134 & 53.5904080 & -34.592517 & -14.6 & -14.1 & 6.5 & 7.63 & e & 0.43 & -- & --  \\
21\_D129 & B442 & 51.77604 & -36.63679 & -14.1 & -13.6 & 4.5 & 7.59 & e & 0.30 & -- & -- \\
\hline
\multicolumn{12}{c}{Star forming dwarfs (spirals and irregulars) with M$_r$ $>$ -19} \\
\hline
26\_D003 & 33 & 51.24324 & -37.00961 & -18.1 & -17.4 & 16.9 & 9.24 & l & 0.63 & 0.46 & 0.75 \\
11\_D519 & 235 & 55.041069 & -35.629093 & -18.6 & -18.2 & 42.3 & 9.03 & l & 0.32 & -- & -- \\
25\_D241 & 37 & 51.2893372 & -36.365185 & -18.2 & -17.8 & 33.9 & 9.02 & l & 0.32 & -- & -- \\
5\_D000 & 263 & 55.38557 & -34.88875 & -18.3 & -17.7 & 16.5 & 9.00 & l & 0.52 & 0.15 & 0.75 \\
7\_D360 & 285 & 55.7601471 & -36.273358 & -18.0 & 17.5 & 32.7 & 8.78 & l & 0.26 & -- & -- \\
15\_D17 & 113 & 53.279419 & -34.805576 & -17.0 & -16.5 & 18.9 & 8.48 & l & 0.31 & -- & -- \\
22\_D244 & 46 & 51.6043 & -37.12778 & -16.3 & -15.8 & 8.5 & 8.31 & l & 0.36 & 0.38 & 1.3 \\
7\_D310 & 306 & 56.43909 & -36.3461 & -15.9 & -15.5 & 7.2 & 7.91 & l* & 0.41 & 0.52 & 1.6 \\

\hline
\end{tabular}

Unless otherwise stated, all values are taken from the FDS dwarf catalogue \citep{Venhola:2019}. (1) Stellar masses are determined following \citet{Taylor:2011}, using the observed ($g-i$) colour and $r-$band absolute magnitude. (2) Visual morphological classes from \citet{Venhola:2019}: e = smooth early-type, l = smooth late-type, * = nucleated, (s) = structured. (3) $\lambda_R$ measured within a 1 R$_e$ aperture, or, where the kinematics do not extend to 1 R$_e$, derived from aperture-correcting $\lambda_R$ within the largest available aperture. The typical uncertainty on $\lambda_{R_e}$ is $\sim 5$ per cent. (4) Maximum extent of the stellar kinematics, as a fraction of R$_e$.
\end{table*}

In this work and a short series of following papers, we aim to address these two deficiencies in our current understanding of dEs by observing and analysing a new, large sample of low-mass galaxies in the nearby Fornax cluster. We cover a sample of dwarf galaxies representative of the general cluster dwarf population, and exploring down to magnitudes and surface brightness that have not been studied before. Our sample offers the opportunity to study the dark matter content of dwarf galaxies with a potentially significant amount of dark matter in a mass regime that bridges observations of the Local Group dSph population and existing studies of more massive dwarf galaxies \citep[e.g.][]{Rys:2014}. Observations, described in Section \ref{sec:observations}, were undertaken using the Sydney Anglo Australian Observatory Multi-Object Integral Field Spectrograph (SAMI). We take advantage of deep, wide-field photometry from the Fornax Deep Survey \citep[FDS][]{Iodice:2016,Venhola:2018} to select and characterise our targets, described in Section \ref{sec:sample}. We describe the reduction of the SAMI data in Section \ref{sec:data_reduction}, and briefly describe the stellar kinematics analysis in Section \ref{sec:stellar_kin}, with details provided by Eftekhari et al.(in prep). In Section \ref{sec:results} we present an analysis of the stellar kinematics of dE galaxies in the Fornax cluster, and conclude in Section \ref{sec:conclusions}. Analysis of the integrated stellar kinematics of this sample is presented by Eftekhari et al. (in prep), with an analysis of the stellar populations to appear in a future article.

Throughout this work we adopt a distance to the Fornax cluster of 20.0 Mpc \citep{Blakeslee:2009} and a virial radius for the cluster of 2.2$^\circ$ \citep[0.77 Mpc at our adopted distance,][]{Drinkwater:2001}.

\begin{figure*}
    \centering
    \includegraphics[width=6.75in,clip,trim = 10 10 10 10 ]{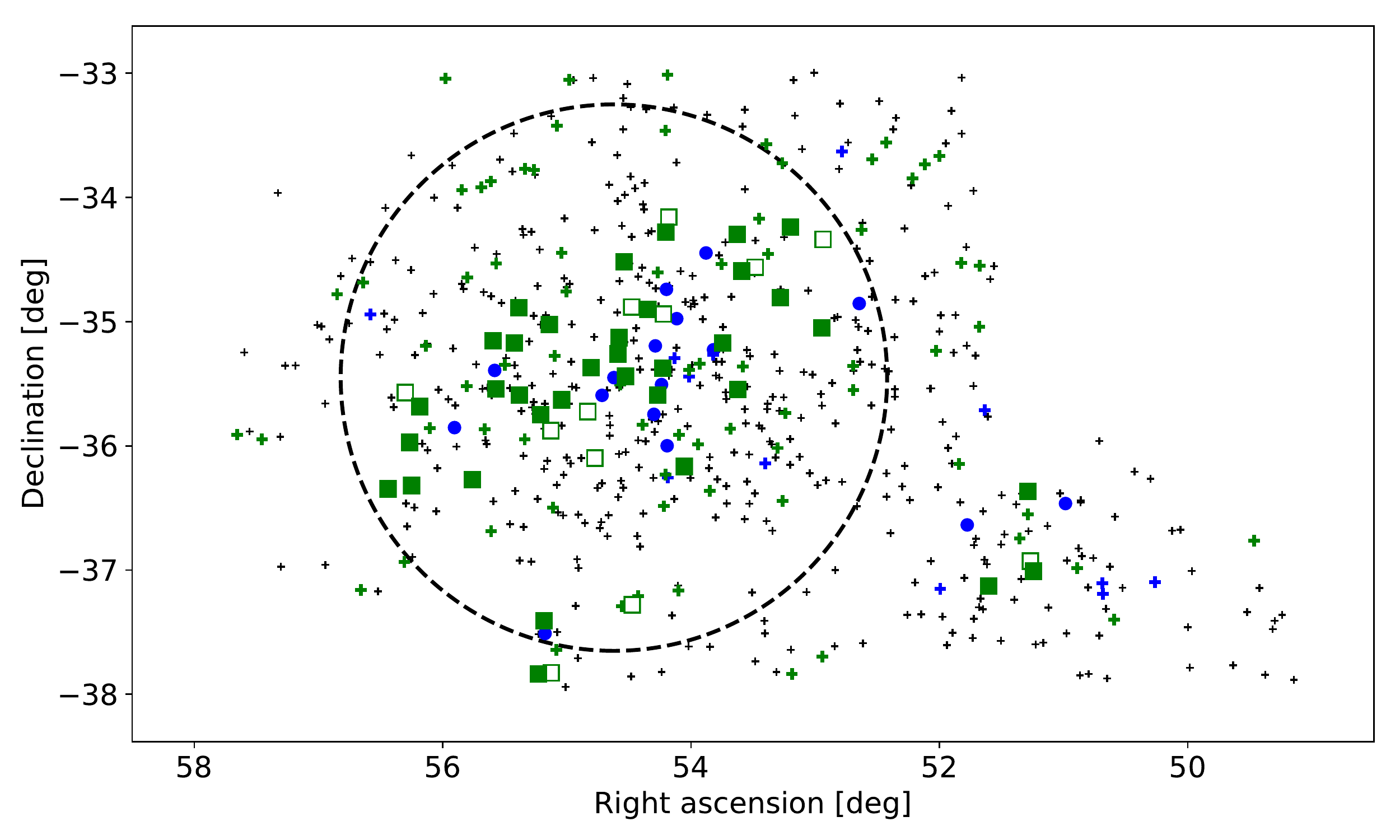}
    \caption{Map of galaxy positions within the Fornax cluster. Filled green squares indicate primary targets with successful stellar kinematic measurements, and open green squares primary targets without stellar kinematic measurements. Blue circles indicate giant galaxies with kinematic measurements. Blue and green crosses indicate respectively giant and primary galaxies that were not observed by SAMI. Small black crosses indicate low-mass dwarf galaxies. All positions for galaxies in the FDS are taken from \citet{Venhola:2018}. The dashed line indicates the cluster Virial radius of 2.2$^\circ$ \citep{Drinkwater:2001}.}
    \label{fig:cluster_map}
\end{figure*}

\section{Observations}
\label{sec:observations}

All observations were conducted with the Sydney -- Australian Astronomical Observatory (AAO) Multi-Object Integral-Field spectrograph \citep[SAMI][]{Croom:2012} on the 3.9m Anglo-Australian Telescope at Siding Spring Observatory, New South Wales. SAMI is a fibre-fed, mulit-object integral-field spectrograph that feeds the AAOmega spectrograph \citep{Sharp:2006}. 

SAMI uses the innovative ``hexabundle'' technology \citep{Bland-Hawthorn:2011,Bryant:2011,Bryant:2014}, where individual optical fibres are fused together to make integral field units (IFUs). SAMI consists of 13 such IFUs, where each IFU consists of 61 individual 1\farcs6 diameter fibres arranged in a close-packed pattern. The resulting hexabundles have an on-sky diameter of 15\arcsec which can be placed anywhere within the instrument's $1^\circ$ field-of-view. In addition to the 13 hexabundles there are 26 dedicated sky fibres that are placed at predetermined blank sky positions throughout the field-of-view to enable simultaneous observation of the night sky background.

The AAOmega spectrograph is a double-armed spectrograph covering the blue and red optical regions of the electromagnetic spectrum. AAOmega allows variable wavelength coverage and spectral resolution in each arm. For the SAMI--Fornax Dwarfs Survey  we used the 1500V grating in the blue arm and the 1000R grating in the red arm to provide high enough spectral resolution to measure velocity dispersions in low-mass galaxies. Note that this is different from the standard setup used by the SAMI Galaxy Survey \citep{Allen:2015,Green:2018,Scott:2018}, which utilised the 580V grating in the blue arm. This non-standard setup resulted in one fibre falling off the edge of the detector, leaving only 25 usable sky fibres.

For each observed field, 12 hexabundles were allocated to galaxy targets, with 1 hexabundle allocated to a secondary standard calibration star. This secondary standard star facilitates several critical steps in the data reduction (e.g. telluric correction, absolute flux calibration), as well as allowing us to assess the point spread function and transmission of each individual observation.

For each field we aimed to obtain 7 hours ($\sim 25,000$ seconds) of on-source integration time. Our observing strategy followed that of the SAMI Galaxy Survey \citep{Sharp:2015}. Individual integrations were $\sim$ 1800\ s, with dithers of 0\farcs8 (half a fibre diameter) applied between exposures, following a 7-point hexagonal dither pattern, optimised for the SAMI hexabundles. The dither pattern ensures an even distribution of S/N over a hexabundle, accounting for the gaps between fibres. This 7-point dither pattern was repeated twice for each field, yielding $2 \times 7 \times 1800 \simeq25,000$ second total exposure time per galaxy. Arc lamp calibrations and observations of primary spectophotometric standard stars from the European Southern Observatory Optical and UV Spectrophometric Standard Stars catalogue\footnote{Available at:\\ https://www.eso.org/sci/observing/tools/standards/spectra.html} were interspersed with the object exposures at regular intervals.

Observations took place over three separate observing runs between 2015 and 2018. Table  \ref{tab:observing_runs} provides an overview of the three observing runs, with the mean seeing and mean exposure time per field for each run. 

In 2015 observing was significantly affected by poor weather, resulting in only two fields being successfully observed. Four targets suffered from inaccurate catalogue coordinates, falling partially or entirely outside their allocated hexabundle field-of-view. Observations in 2016 and 2018 were not affected by inaccurate coordinates due to the improved FDS input catalogue and all targets filled their allocated bundles. The weather was somewhat improved in 2016 and 2018, allowing a higher fraction of target fields to be observed during those runs. Over the three observing runs we observed 10 complete fields, totalling 118 unique galaxies.

Our primary science targets consist of morphologically-classified dE galaxies in the Fornax cluster. Due to the on-sky density of primary targets, not all IFUs could be allocated to a dE for each observation. Spare IFUs were allocated to other dwarf galaxies, giant cluster members of any morphological type or background galaxies in that order of preference. In practice, a target catalogue was prepared for each semester, and targets were assigned priorities ranging from 5 (highest) to 1 (lowest) as described below. A tiling algorithm \citep{Bryant:2015} then assigned targets to fields in a way that maximises the number of high priority targets observed while also ensuring all IFUs have an available target.

Targets were selected from the Fornax Cluster Catalogue \citep[][hereafter FCC]{Ferguson:1990} for observations in 2015, and from the Fornax Deep Survey catalogue \citep[][hereafter FDS]{Venhola:2018} in 2016 and 2018, once this deeper, wider-field photometry became available. In the following subsections we provide details of how targets were selected for each observing run and the properties of the final sample of observed objects.

\subsection{Fornax Cluster Catalogue selection}
In 2015 we selected primary science targets from the FCC because higher fidelity catalogues with all required information were not yet available. We selected galaxies within a $3 \times 1$ degree region centred on the cluster centre, which we assume is coincident with the centre of NGC 1399, the Brightest Cluster Galaxy in Fornax. We included only galaxies with a membership class of 1 (definite member) or 2 (likely member) in the FCC. This resulted in an input catalogue of 82 galaxies.

Priorities were assigned based on the absolute {\it B-}band magnitude of the targets and the observed morphology, but were adjusted to prioritise potentially interesting objects. Galaxies with a dE or dS0 morphology and $-18 < M_B \leq -13$ were assigned the highest priority 5 (31 objects), with fainter ($-11 < M_B < -13$) dE and dS0 assigned priority 4 (31 objects). Relatively compact galaxies (effective radii less than the bundle diameter of 15'') not satisfying either of the above criteria were assigned priority 3 (8 objects), with the remaining massive early-types and late-type galaxies of all magnitudes assigned priorities 1 and 2 (12 objects). Note that these numbers refer to galaxies in our input target catalogue, the galaxies that were actually selected are described in Sections \ref{sec:primary_sample} and \ref{sec:secondary_targets}.

\section{Sample}
\label{sec:sample}

\begin{figure}
    \centering
    \includegraphics[width=3.25in]{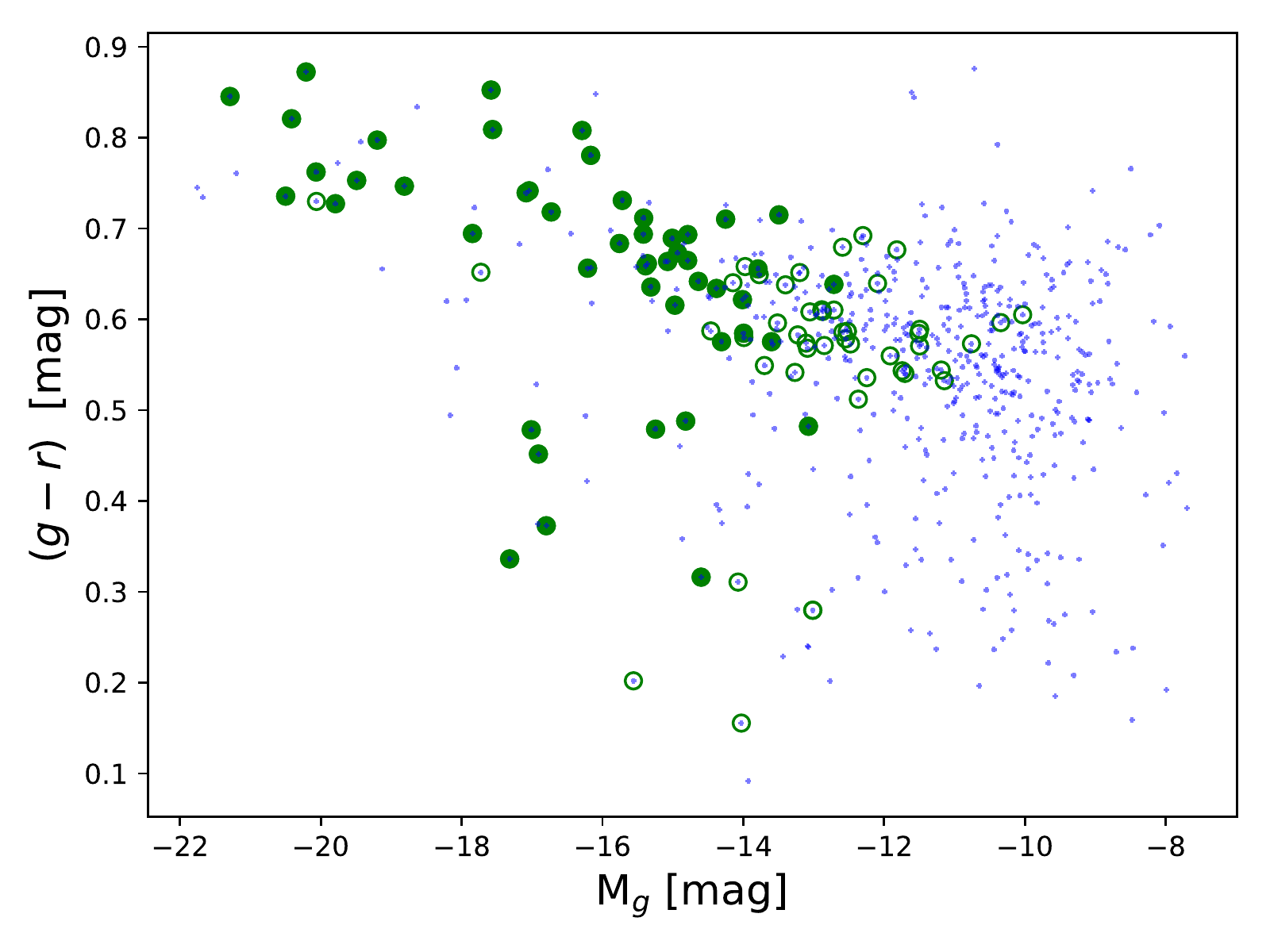}
    \caption{The ($g-r$) vs M$_g$ colour-magnitude diagram for our sample (green circles), compared to that from the full FDS sample (blue crosses). We primarily target galaxies on the red sequence, but do include some blue galaxies that satisfy our selection criteria. Filled (open) symbols indicate galaxies for which we could (could not) obtain aperture stellar kinematic measurements.}
    \label{fig:colour_mag}
\end{figure}

\begin{figure}
    \centering
    \includegraphics[width=3.25in]{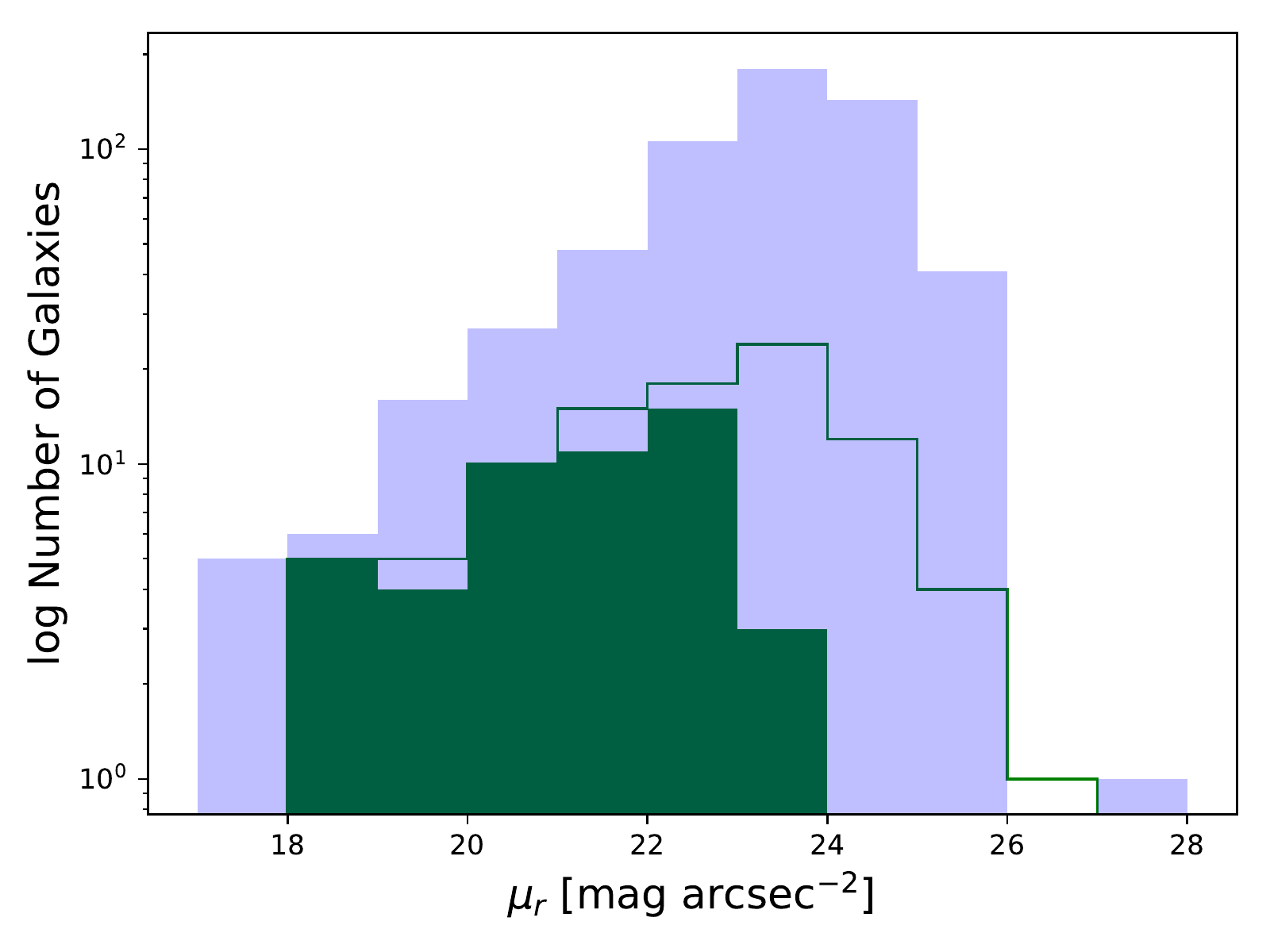}
    \caption{The logarithm of the number of galaxies as a function of effective surface brightness. The open green histogram shows the distribution for all observed SAMI-FDS galaxies, while the solid green histogram indicates only those for which we were successful in measuring stellar kinematics. The pale blue histogram indicates the effective surface brightness distribution for the full FDS sample.}
    \label{fig:surface_brightness}
\end{figure}

\begin{figure*}
    \centering
    \includegraphics[width=2.25in]{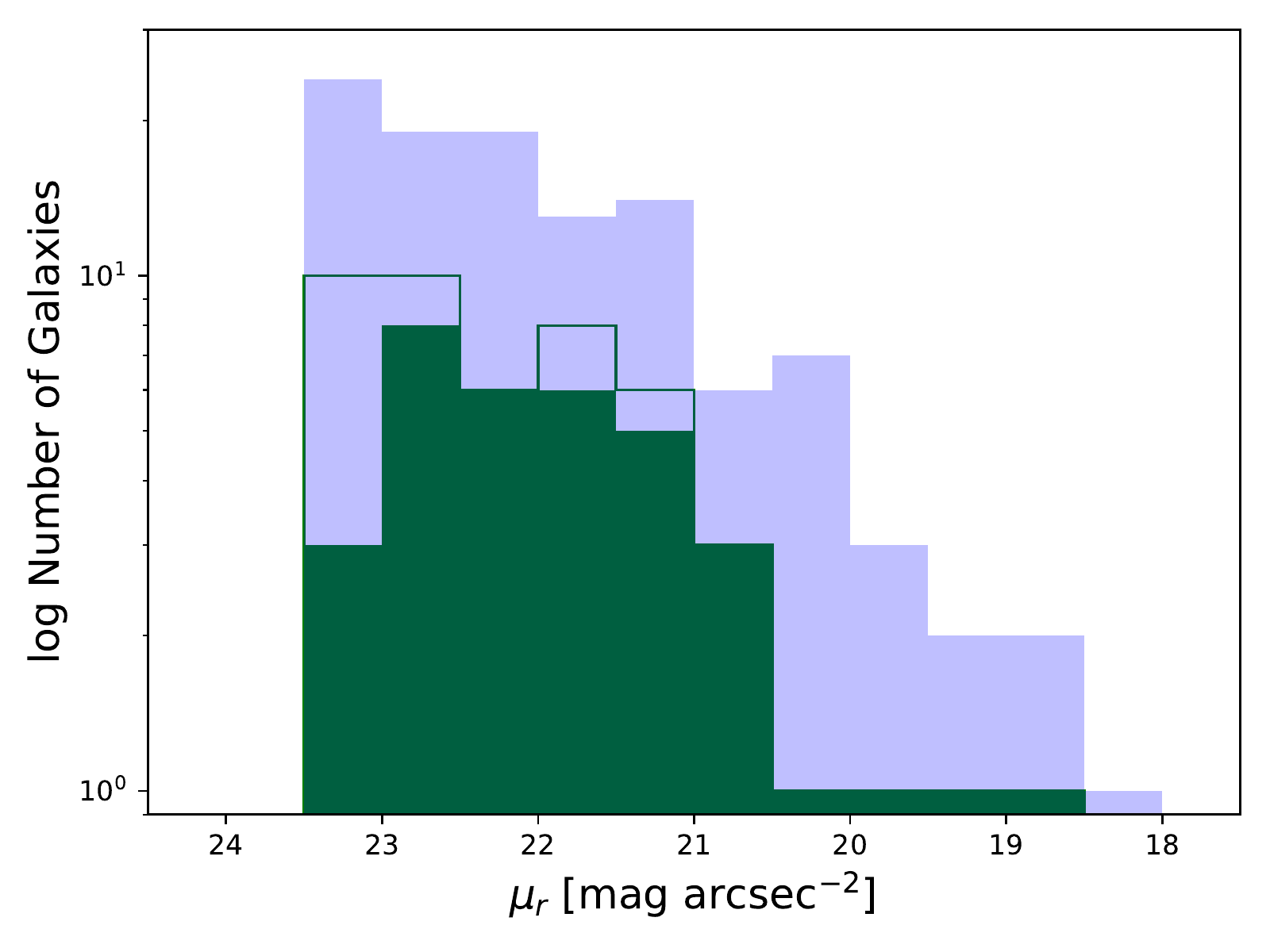}
    \includegraphics[width=2.25in]{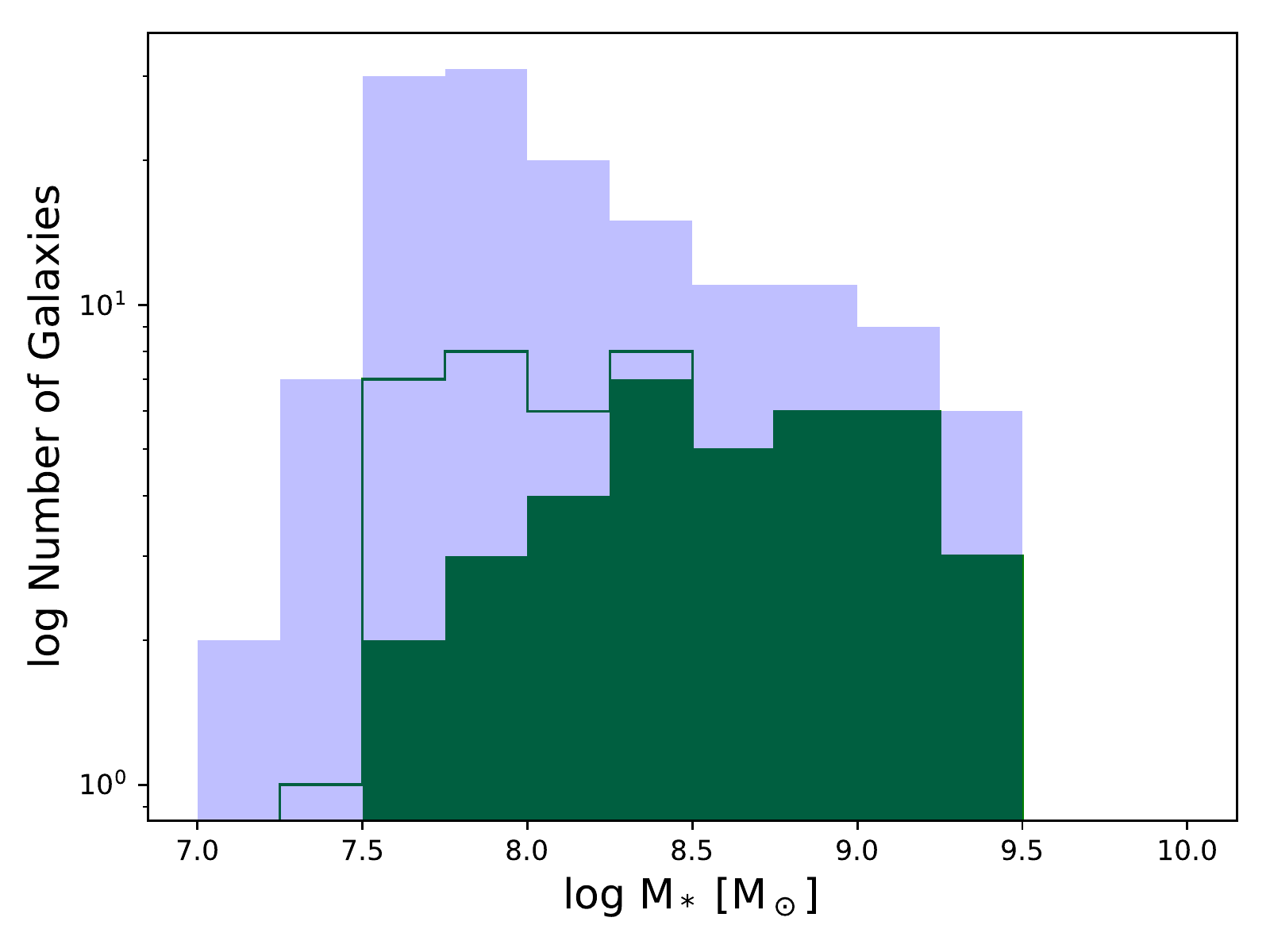}
    \includegraphics[width=2.25in]{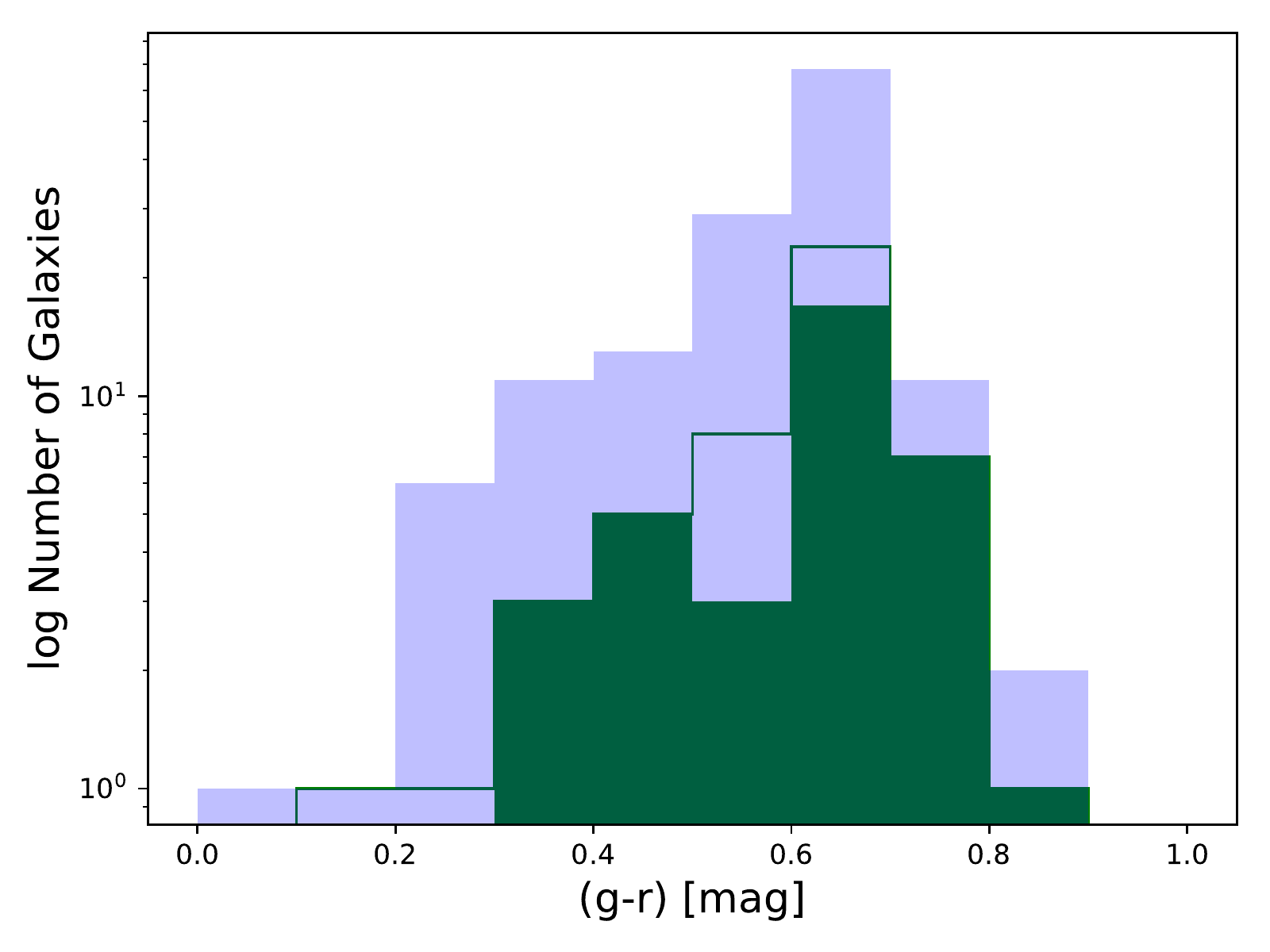}
    \includegraphics[width=2.25in]{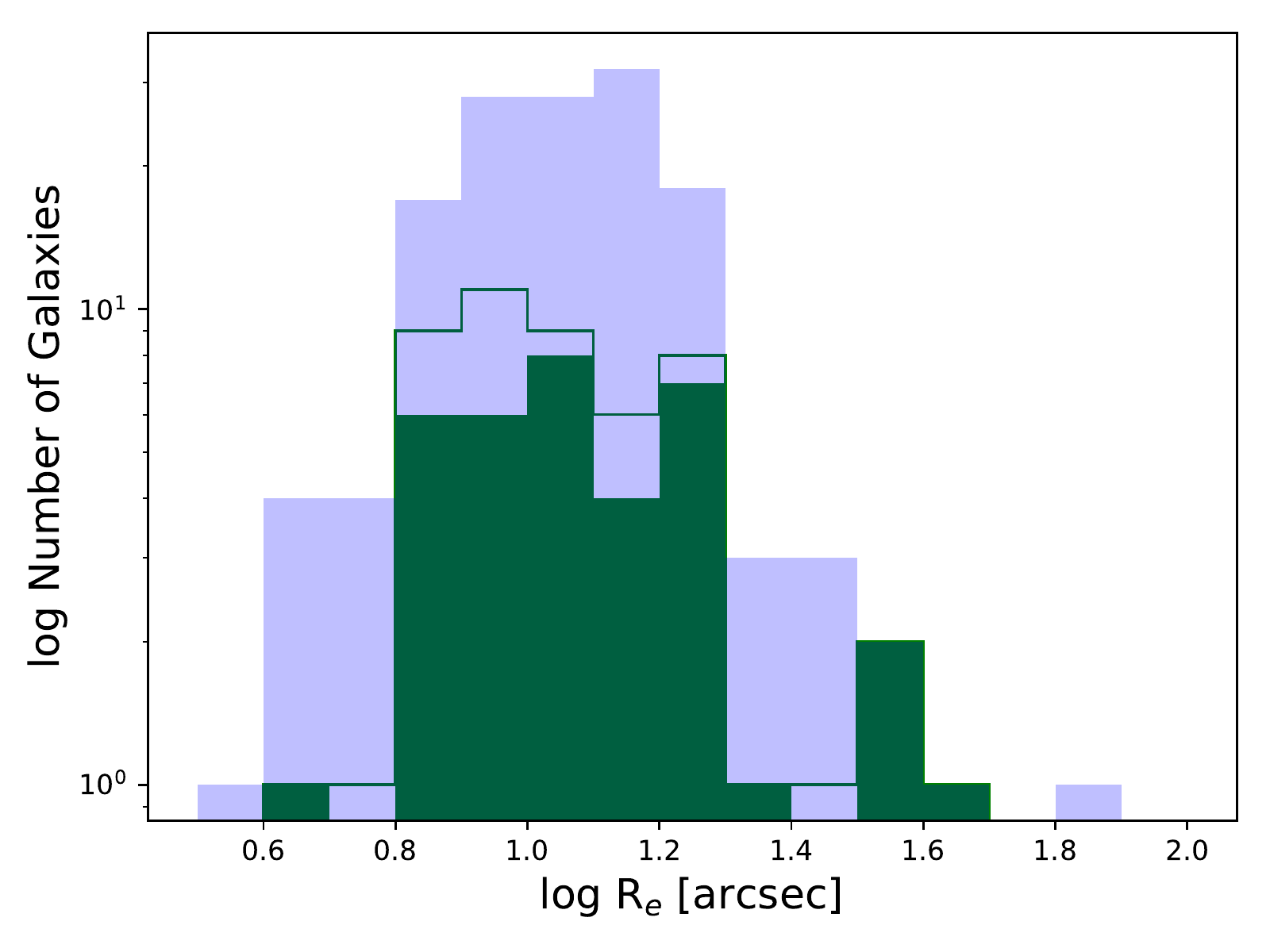}
    \includegraphics[width=2.25in]{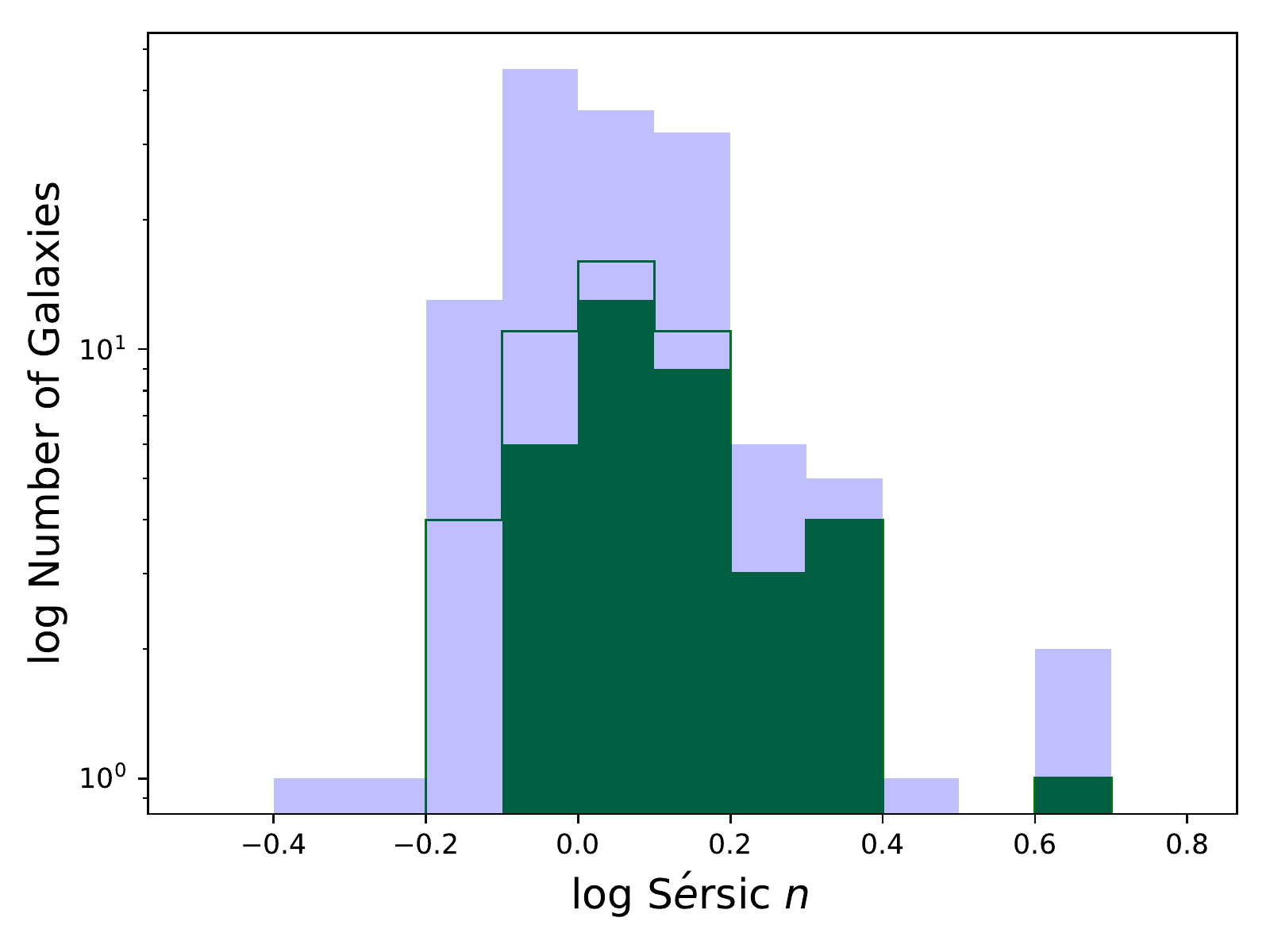}
    \includegraphics[width=2.25in]{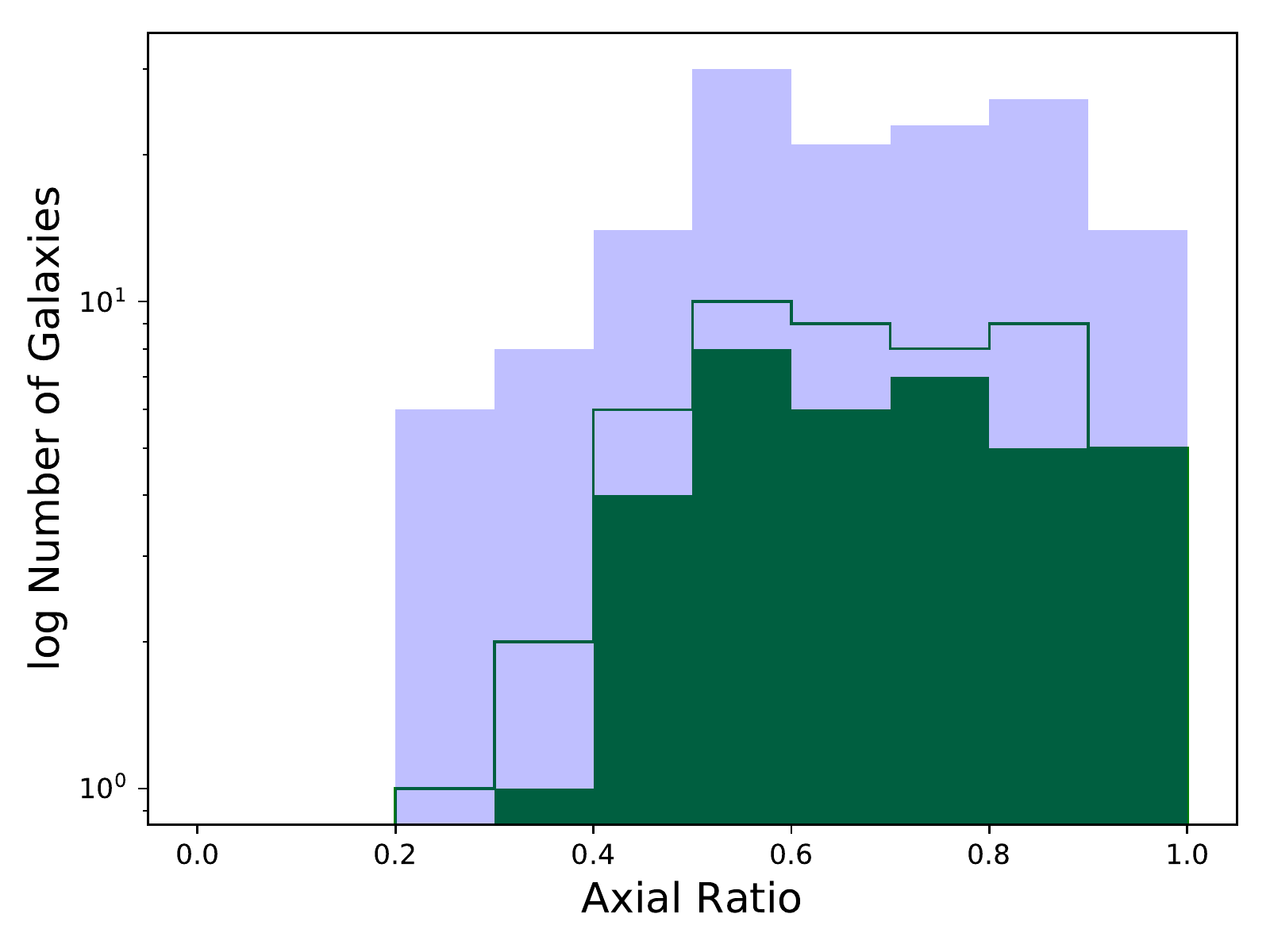}
    \caption{Distribution of observed (open green), stellar kinematics (filled green) and FDS (pale blue) galaxies that satisfy our primary sample selection criteria (see text for details) with respect to a range of galaxy properties. Top row: $r$-band effective surface brightness, stellar mass and ($g-r$) colour. Bottom row: effective radius, S\'ersic $n$ and axial ratio (b/a).}
    \label{fig:sample_properties}
\end{figure*}

\subsection{Fornax Deep Survey selection}
The selection and prioritisation of targets from the FDS was similar to that from the FCC, with modifications based on experience from the 2015 observations and to take advantage of the improved quality and quantity of the FDS catalogue compared to the FCC.

We began by selecting all galaxies in the FDS catalogue with $-14 > M_r > -19$ and $\mu_r < 23.5$ and a dE visual morphology (from the FCC). The surface brightness criterion selected against galaxies for which we were unable to obtain useful data in the 2015 observations (S/N $< 10$ integrated over the entire galaxy). Galaxies with $M_r > -14$ were not included, even as lower priority targets, again based on our experience from the 2015 observations on which targets yielded useful spectra. Secondary targets consisted of giant galaxies with $M_r < -19$, dwarf galaxies with a non-dE morphology and an additional sample of ultra-compact dwarf galaxies (UCDs) from \citep{Wittmann:2016}. We also included galaxies identified as unlikely cluster members or background galaxies as low-priority secondary targets.

In 2016, FDS coverage of the cluster was incomplete, so we selected targets from the central $\sim 1$\ degree for which the catalogue was already complete. Prior to the 2018 observations, the FDS catalogue had been extended to cover the outer regions of the cluster, allowing us to target regions further from the cluster center. In 2018, we observed regions between 1 and 3 degrees from the cluster center with sufficient galaxy density to assign the majority of hexabundles to a primary dE target.

Priorities were assigned to targets in a similar fashion as for the 2015 FCC sample. Galaxies with a dE morphology, $-14 > M_r > -19$ and $\mu_r < 23.5$ were assigned priority 5 (the highest priority). UCDs were assigned priority 4 and dwarf galaxies with a non-dE morphology were assigned priority 3. Giant galaxies were assigned priority 2, with background galaxies assigned priority 1. As before, targets were assigned to fields by the tiling algorithm to maximise the number of high-priority targets observed.

\subsection{Primary Sample}
\label{sec:primary_sample}
As noted above, our primary science targets were galaxies with a dE or dS0 morphology and $M_r > -19$. Of the 118 galaxies targeted, 59 (50 per cent) fall into this category. A complete list of low-mass, early-type galaxies with successful stellar kinematic measurements can be found in Table \ref{tab:primary_sample}. Here we describe the properties of this primary sample and compare to the full FDS catalogue to identify possible bias in our observed sample with respect to the complete cluster population. We also examine the distribution of observed galaxies within the Fornax cluster to highlight the environmental coverage of this sample, again with respect to FDS.

Figure \ref{fig:colour_mag} shows the ($g-r$) vs M$_g$ colour-magnitude diagram for our sample (green circles) and for the full FDS sample (blue crosses). Filled and open green symbols indicate galaxies where our stellar kinematic measurements were successful or unsuccessful respectively (see Section \ref{sec:stellar_kin} for details). As we are primarily interested in galaxies with an early-type morphology it is unsurprising that the majority of our targets fall on the red sequence for the cluster, however we also target a number of blue objects, where they satisfy our selection criteria.

In Figure \ref{fig:surface_brightness} we show the distribution in surface brightness for our full observed sample (open green histogram) and the subset for which we are able to measure stellar kinematic measurements (filled green histogram, see Section \ref{sec:stellar_kin} and Eftekhari et al. in prep. for details). The pale blue histogram shows the complete FDS galaxy sample from \citet{Venhola:2018}. Below an effective surface brightness of $\mu_r = 23.5$ we are unable to measure stellar kinematics, therefore in the following section we consider only galaxies with $\mu_r < 23.5$. We restrict this comparison to our primary targets only, with $-19 < M_r < -14$.

In Figure \ref{fig:sample_properties} we show the distribution in a range of galaxy properties of SAMI-FDS primary targets. With respect to the subset of the FDS sample consistent with our revised selection criteria ($\mu_r < 23.5$, $-19 < M_r < -14$), our primary sample is more massive than the full FDS sample, driven by a decrease in our completeness below M$_* = 10^{8}$ M$_\odot$. This is particularly true for the subset of galaxies for which we have successful stellar kinematic measurements, unsurprising due to the relatively high S/N required to obtain spatially resolved spectroscopy. The R$_e$ distribution of the SAMI-FDS primary sample is consistent with that of the full FDS sample. Our sample is redder and rounder and has slightly higher S\'ersic index $n$ than the full FDS sample, reflecting our primary sample selection criteria of targeting galaxies with an early-type morphology.

In summary, while the galaxies for which we successfully obtain stellar kinematic measurements are, on average, more massive, rounder and redder than FDS dwarf galaxies, our sample is representative of bright, early-type dwarf galaxies in the FDS, in the sense that the fractional completeness of the SAMI-FDS galaxies with stellar kinematics is independent of the examined galaxy properties in the mass regime M$_* > 10^8$ M$_\odot$.

\subsection{Secondary targets}
\label{sec:secondary_targets}
In addition to our primary science targets we observed 59 secondary targets. These objects consist of: faint early-type cluster members (11 objects), giant early-type cluster members (14 objects), late-type cluster members (14), Ultra Compact Dwarfs (UCDs, 5 objects) and background galaxies (15 objects). Late-type dwarf galaxies for which we could measure stellar kinematics are included in Table \ref{tab:primary_sample}, while the remaining secondary targets can be found in Table \ref{tab:secondary_sample}. These objects were not selected in any systematic fashion, but instead were drawn randomly from targets that could be allocated to an IFU while maximising the number of primary targets observed. Because of this we do not attempt to quantify their sample properties with respect to any parent distribution, but instead simply provide an overview. Secondary targets are not included in the subsequent analysis.

\section{Data Reduction}
\label{sec:data_reduction}

Our SAMI observations were reduced using the {\it sami} {\sc Python} package \citep{Allen:2015}, following the approach described in \citet{Scott:2018}, with further details provided in \citet{Sharp:2015} and \citet{Green:2018}. Here we summarise this process and provide a detailed description only where our reduction process differed from that described in the above articles.

SAMI data is reduced in two stages; the first takes the data from raw observed frames to Row-Stacked Spectra (RSS) frames, which is handled primarily by the two-degree field data reduction software package {\sc 2dfDR}\footnote{https://www.aao.gov.au/science/software/2dfdr}. The second stage takes the data from RSS frames to flux-calibrated, three-dimensional data cubes, utilising purpose-built {\sc Python} software as part of the {\it sami} package. The entire process is overseen by the {\it sami} {\sc Python} {\it manager}.

\begin{figure*}
    \centering
    \includegraphics[width=2.in,clip,trim = 20 0 70 0]{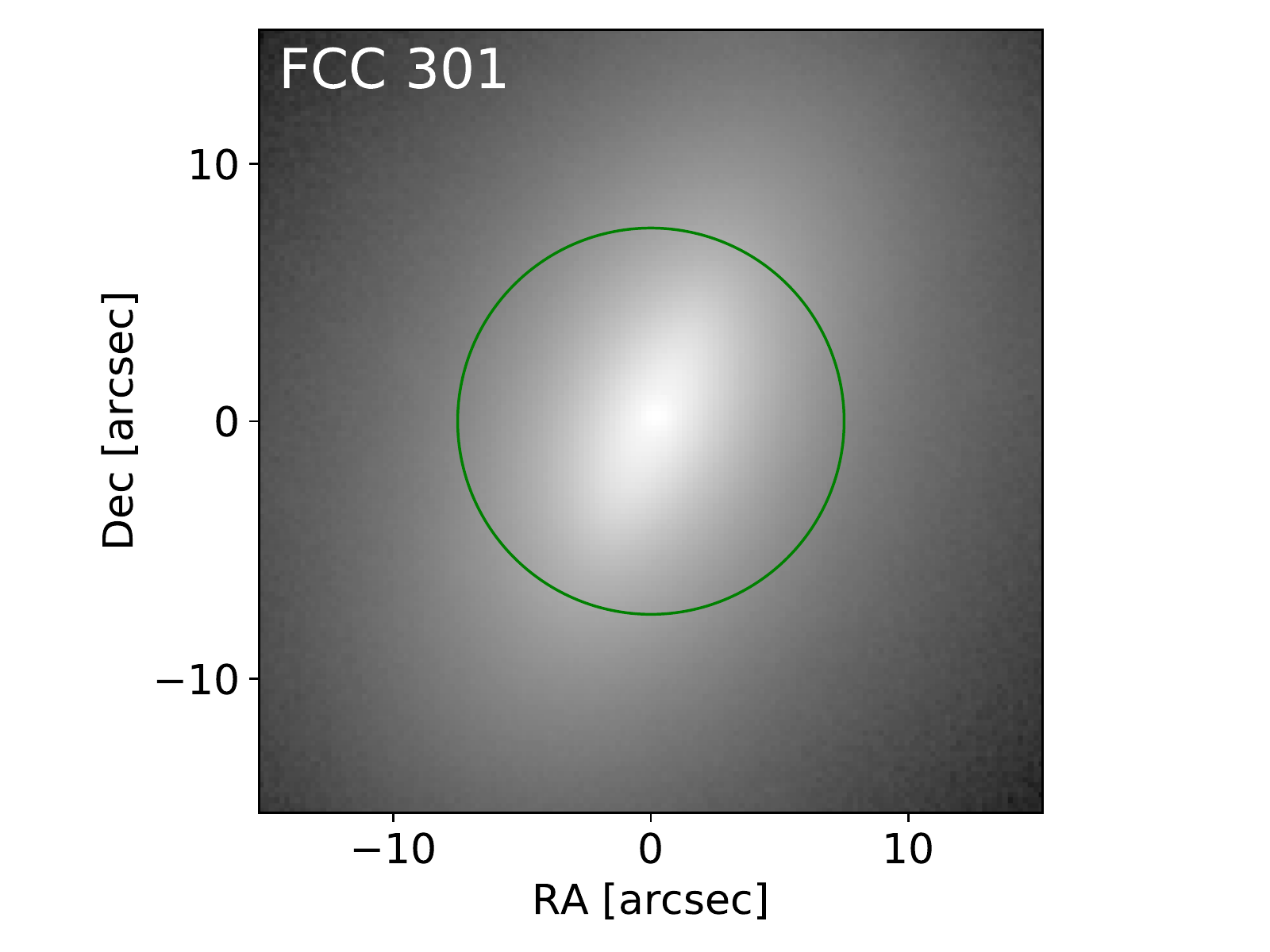}
    \includegraphics[width=2.25in,clip,trim = 20 10 30 10]{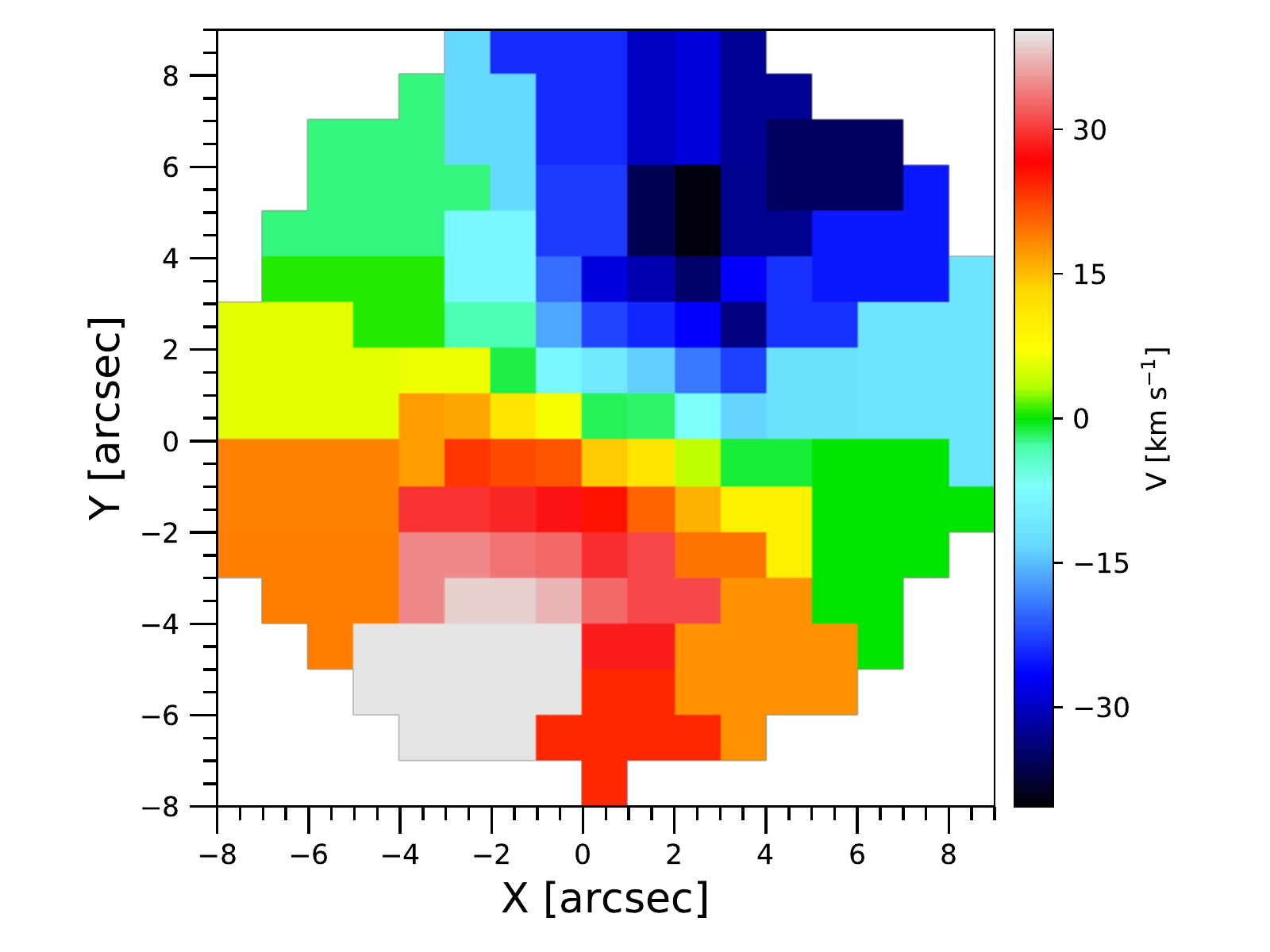}
    \includegraphics[width=2.25in,clip,trim = 20 10 30 10]{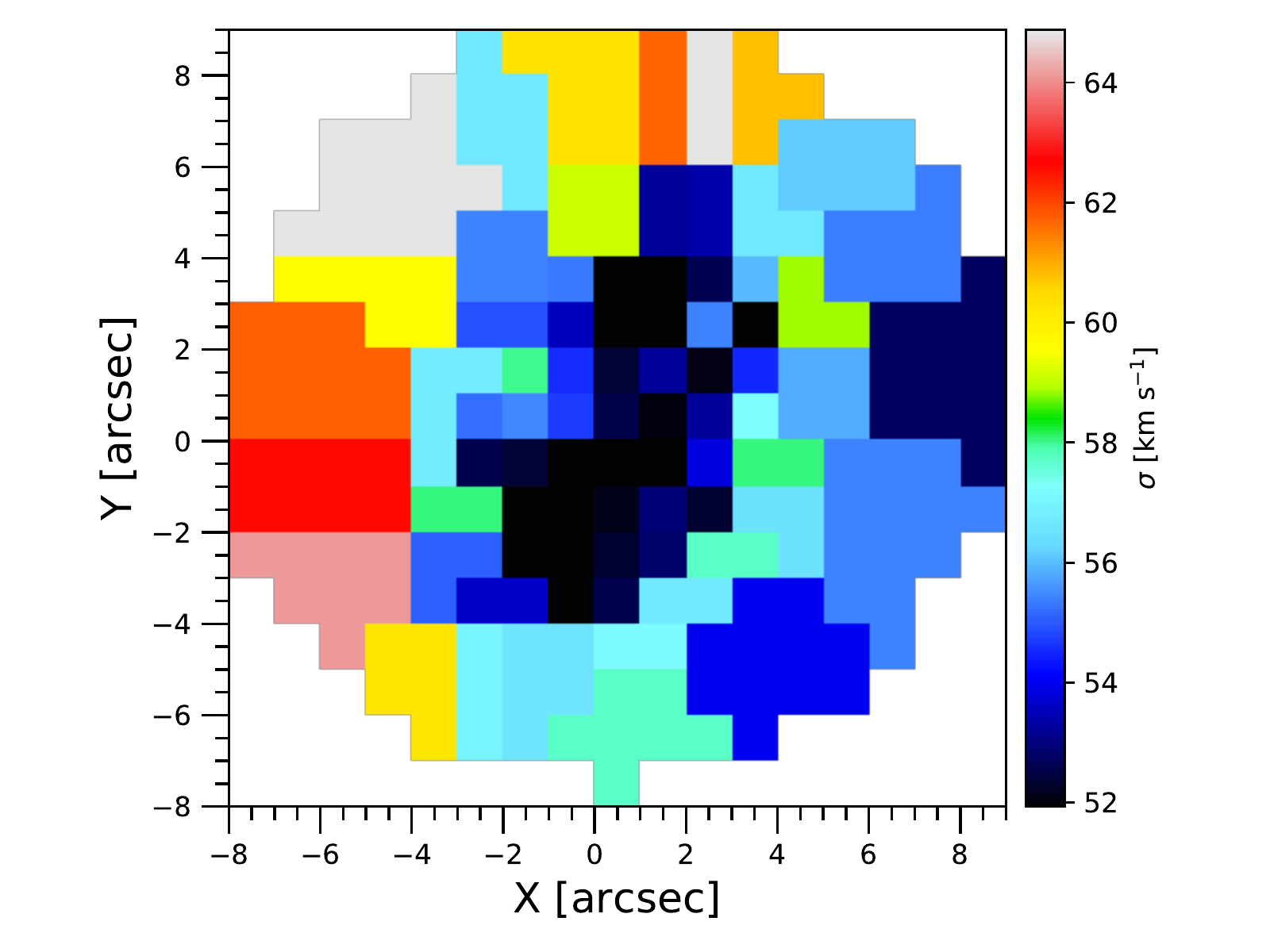}
    
    \includegraphics[width=2.in,clip,trim = 20 0 70 0]{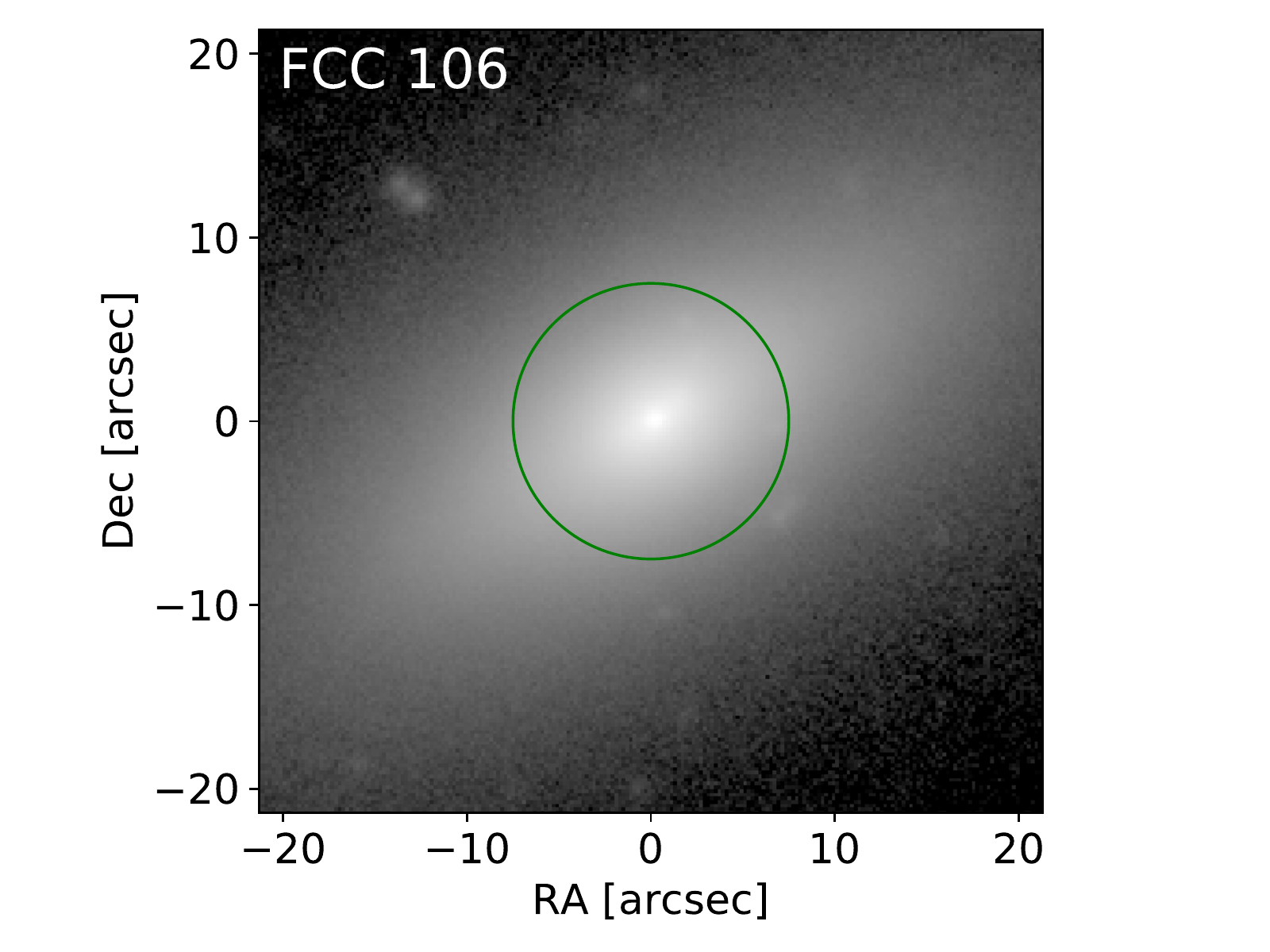}
    \includegraphics[width=2.25in,clip,trim = 20 10 20 10]{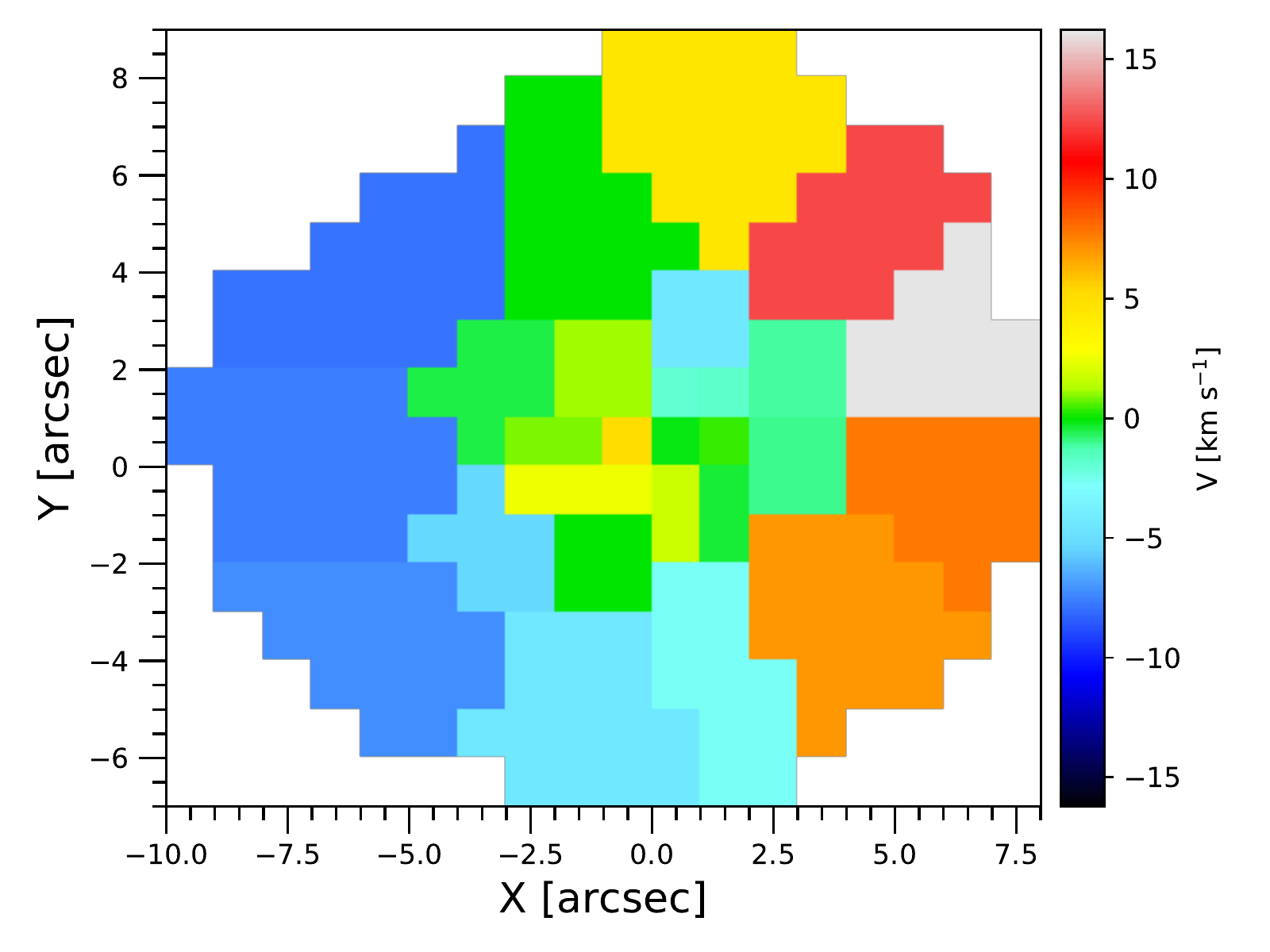}
    \includegraphics[width=2.25in,clip,trim = 20 10 20 10]{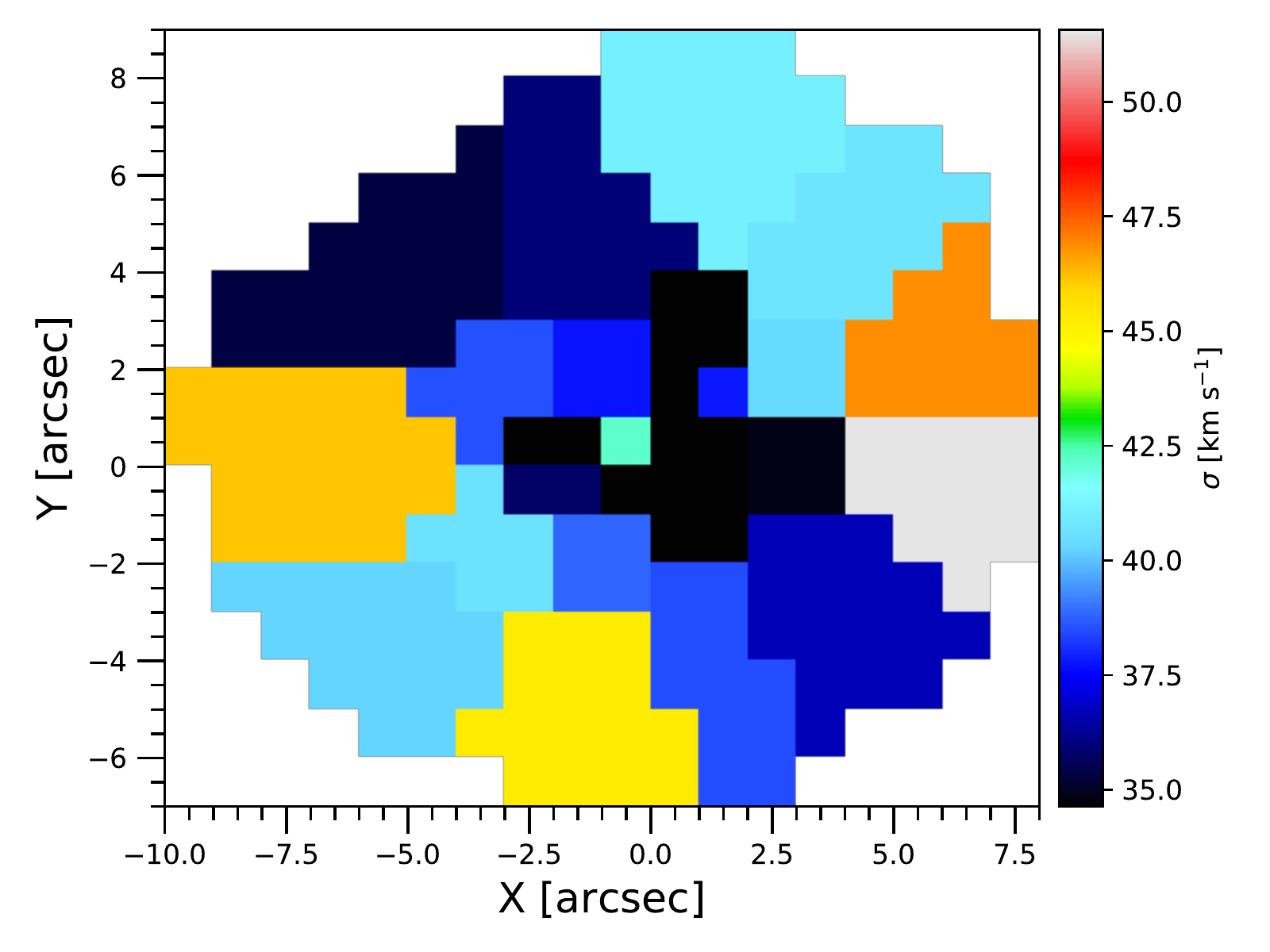}
    \caption{Stellar kinematic maps for two example dE galaxies. Upper row: FCC301, a galaxy exhibiting strong, ordered rotation. Lower row: FCC106, a dispersion dominated system with minimal rotation. Left column: {\it r}-band cutout images from FDS, with diameter 4 R$_e$. Green circle indicates the approximate position of the SAMI IFS field-of-view. Centre column: Voronoi-binned maps of the mean line-of-sight stellar velocity, $V$. Right column: Voronoi-binned maps of the mean line-of-sight stellar velocity dispersion, $\sigma$.}
    \label{fig:primary_maps}
\end{figure*}

Data reduction with {\sc 2dfDR} includes the standard steps of bias subtraction, flat-fielding, wavelength calibration and sky subtraction. In addition, spectra corresponding to individual fibres are extracted using `tramlines' fit to observations of the twilight sky. Subsequent to the fibre extraction, telluric correction and relative and absolute flux calibration steps are applied utilising the spectrophometric standard star and secondary standard star observations. Finally, the data for each individual object are extracted from the RSS frames and combined into a three-dimensional data cube using a drizzle-based algorithm.

During the 2015 observing run it was discovered that the tramlines for the 1500V grating were not well-aligned to the CCD, resulting in one or two fibre spectra partially falling off the edge of the CCD --- this primarily affected one sky fibre, but in three observations in 2016 an outer fibre from the first hexabundle was also affected. The influence on the output data quality is essentially negligible: the resultant decrease in the S/N of the sky spectrum is only 2 per cent, and the reduction in area of the first hexabundle (when it is affected) is only 1.5 per cent. This does not affect the 1000R red arm observations or observations with the more widely used 580V blue arm grating.

SAMI's circular fibre spectra are resampled onto a regular, square grid using a drizzle algorithm. There are two free parameters in this algorithm, the size of the square output spaxels and the drop factor, which effectively shrinks the size of the SAMI fibre before determining which output spaxels receive flux from a circular fibre. The SAMI Galaxy Survey adopted an output spaxel size of 0\farcs5 and a drop factor of 0.5, yielding an effective fibre diameter of 0\farcs8. Due to the lower effective surface brightness of the dwarf galaxies observed in this study we adopted an output spaxel size of 1\arcsec (which corresponds to 97 pc at the adopted Fornax distance), increasing the flux per spaxel. While this comes at the cost of reduced spatial resolution, the Fornax cluster is ten times closer than typical for the SAMI Galaxy Survey, resulting in much higher physical resolution, despite the larger spaxels.

\section{Stellar Kinematics}
\label{sec:stellar_kin}

In the following section we present stellar kinematic maps of galaxies in our primary sample with sufficiently high S/N to produce reliable maps. The stellar kinematic measurements are fully described in Eftekhari et al. (in prep.), but we briefly summarise them here. 

The data cubes were spatially binned to a median S/N of at least 10 in the blue continuum, averaged over the wavelength range 4900 -- 5100 \AA, using the Voronoi binning algorithm of \citet{Cappellari:2003}. The binned spectra were analysed using the penalized Pixel Fitting (pPXF) software of \citet{Cappellari:2004}. For spectral templates we utilised a subset of the ELODIE \citep{Prugniel:2007} stellar spectral library, as this is the empirical library best matched to the SAMI 1500V spectral resolution. Emission features were masked before extracting the moments of the mean line-of-sight velocity distribution: velocity ($V$) and velocity dispersion. The spatially resolved measurements of $V$ and $\sigma$ were used to construct the maps presented and analysed in the following section. 

\section{Results}
\label{sec:results}

\subsection{Stellar kinematic maps}

In Figure \ref{fig:primary_maps} we present the velocity and velocity dispersion maps (and {\it r}-band cut-out images from FDS) for two example galaxies in our primary sample with good quality stellar kinematics (with the full set of maps presented in Figures \ref{fig:primary_maps_app} and \ref{fig:latetype_maps}). 'Good-quality' here refers to maps for which the spectral fits to all spaxels appear good, as judged by visual inspection, and represents a very conservative selection. For kinematic measurements of fainter galaxies and a detailed examination of the measurement uncertainties see Eftekhari et al. (in prep.) Maps for secondary targets are presented in Figure \ref{fig:secondary_maps}. Galaxies in our sample show a range of kinematic morphologies from strong, ordered disk-like rotation to non-rotating and dispersion dominated. In Section \ref{sec:lambdaR} below we quantify the relative angular momentum of our sample using the $\lambda_R$ parameter.

There are a small number of objects that overlap between our primary sample and that of \citet[from the Fornax3D survey]{Iodice:2019}, with larger overlap with our secondary targets. Noting the significant differences in radial coverage and instrumental resolution between the two studies, we find excellent agreement between the velocity and velocity dispersion maps of that work and those presented here. A more detailed quantitative comparison between the two sets of measurements will be presented in Eftekhari et al. (in preparation).

\begin{figure}
    \centering
    \includegraphics[width=3.25in]{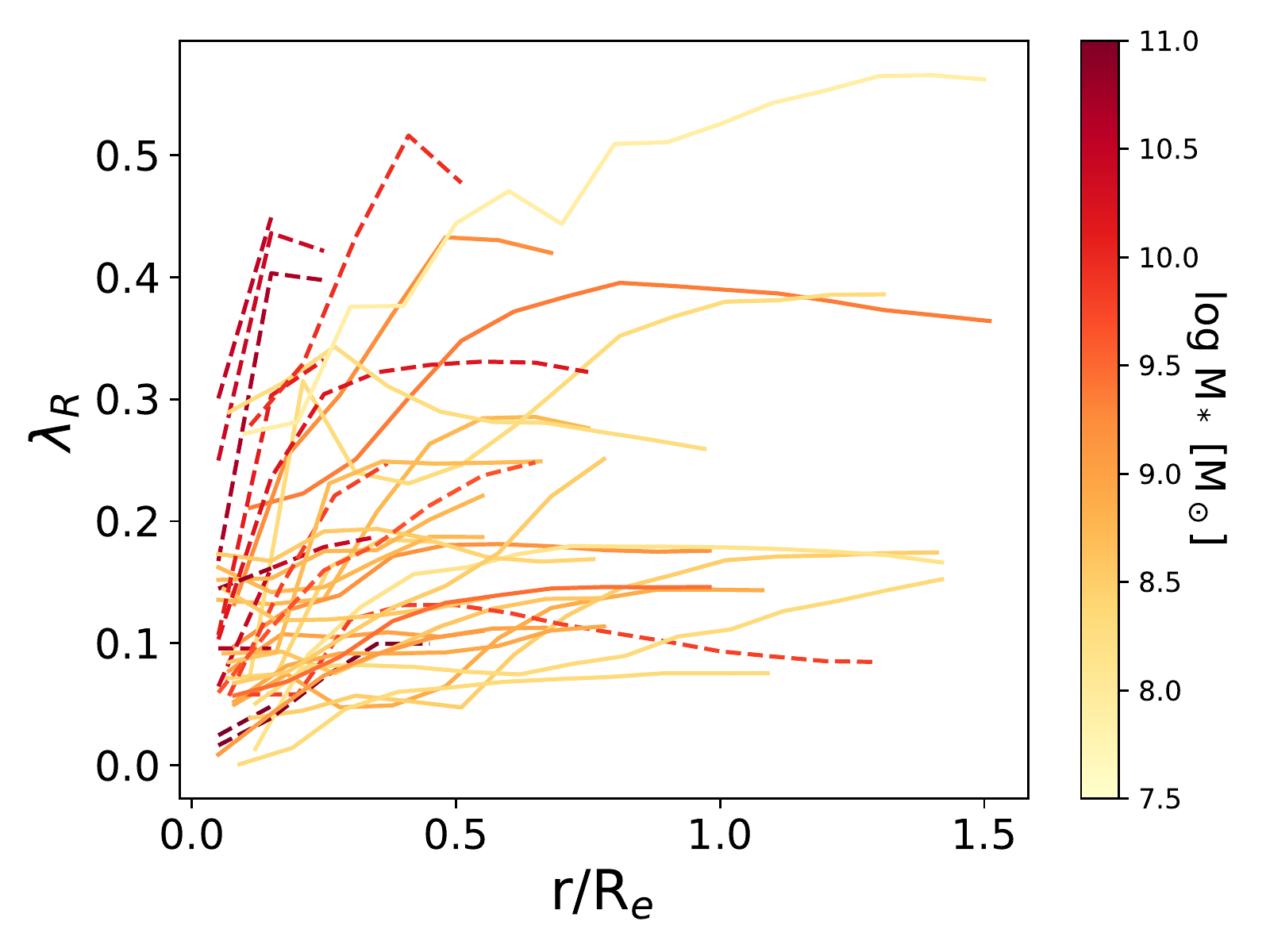}
    \caption{Profiles of $\lambda_R$ (integrated within elliptical apertures of increasing radius) versus the radius of the corresponding aperture for all galaxies for which we could derive kinematic maps. Profiles are colour coded by the stellar mass of the galaxy. Solid lines indicate primary dE targets and dwarf late-types, dashed lines indicate giant galaxies.}
    \label{fig:lambdar_profiles}
\end{figure}

\subsection{$\lambda_R$ measurements}
\label{sec:lambdaR}
\citet{Emsellem:2007} introduced the specific stellar angular momentum proxy, $\lambda_R$, as a parameter to quantify whether a galaxy is dominated by rotational or pressure support. $\lambda_R$ is similar to the common $(V/\sigma)_e$ parameter, but is more closely related to the kinematic morphology of a galaxy. $\lambda_R$ is defined as:

\begin{equation}
    \lambda_R = \frac{<R|V|>}{<R\sqrt{V^2 + \sigma^2}>} = \frac{\sum_{i=0}^{N}F_iR_i|V_i|}{\sum_{i=0}^{N}F_iR_i\sqrt(V_i^2+\sigma_i^2)},
\end{equation}
where the summation is over all $i$ spaxels under consideration. $F_i$, $V_i$ and $\sigma_i$ respectively refer to the flux, velocity and velocity dispersion of the $i$th spaxel. In this work, $R_i$ refers to the elliptical radius of the $i$th spaxel -- that is the major axis radius of the ellipse on which the $i$th spaxel lies, where the ellipticty, $\epsilon$, and the position angle of the ellipse are taken from \citet{Venhola:2018}. Following \citet{vandeSande:2017}, for each spaxel we set $V_i$ and $\sigma_i$ to the $V$ and $\sigma$ of the bin to which it belongs, whereas $F_i$ is simply the unbinned flux of each spaxel.

\begin{figure}
    \centering
    \includegraphics[width=3.25in]{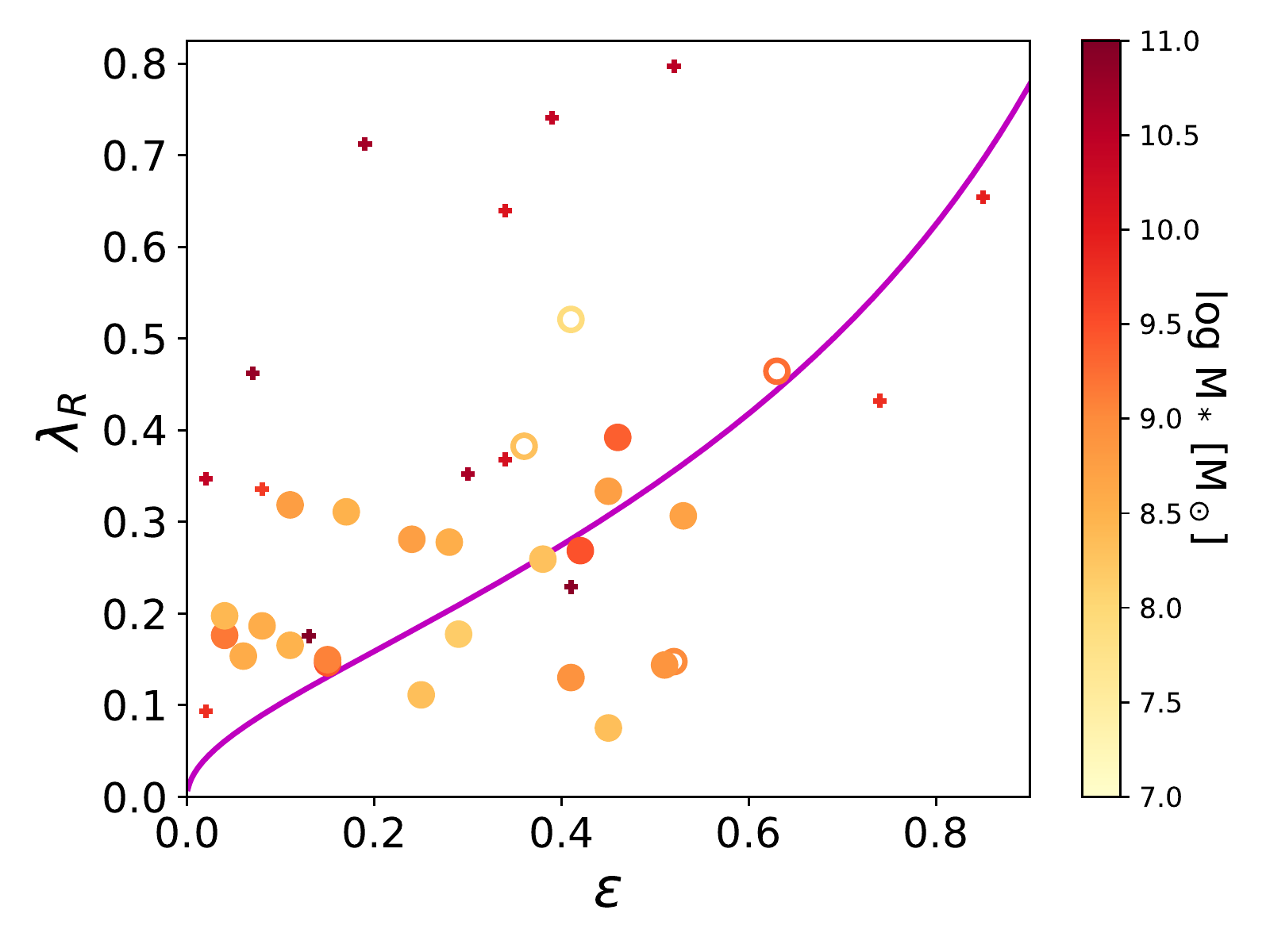}
    \caption{The $\lambda_R$ versus $\epsilon$ plane for all galaxies for which we could derive kinematic maps. Points are colour coded by the stellar mass of the galaxy, with filled circles indicating primary dE targets, open circles late-type dwarfs and small crosses giant cluster members. The magenta line indicates the model prediction for an edge-on axisymmetric galaxy from \citet{Cappellari:2007} \citep[see][for further details]{vandeSande:2017}.}
    \label{fig:lambdar_eps}
\end{figure}

$\lambda_R$ is an integrated quantity measured within an aperture, or a series of increasing apertures to create a profile. In Figure \ref{fig:lambdar_profiles} we show the $\lambda_R$ profiles as a function of the normalized radius, $r/R_e$, coloured by the stellar mass of the galaxy. Solid lines indicate primary sample galaxies, while dashed lines indicate all secondary targets for which we were able to measure kinematics (primarily giant galaxies and more massive late-type dwarfs). 

\subsubsection{Aperture correction of $\lambda_R$}

For all primary targets our $\lambda_R$ measurements sample out to at least $R_e/2$, with a median coverage of 0.85 $R_e$. Given this coverage range, we adopt $R_e$ as our canonical aperture in which we measure $\lambda_R$. For galaxies where our kinematic measurements do not reach 1 R$_e$ we apply an aperture correction from \citet[their eqns. 7 and 10]{vandeSande:2017}. We verify this aperture correction using the 9 galaxies with $\lambda_R$ measurements that extend beyond 1 R$_e$, finding  that the corrected $\lambda_R$ measurements are consistent with the measured $\lambda_{R_e}$, with a mean offset $\Delta(\lambda_R) = 0.01$ and rms scatter $\sigma (\lambda_R) = 0.04$. The median correction in $\lambda_R$ for galaxies whose measurements do not reach 1 R$_e$ is $\Delta(\lambda_R) = 0.04$.

\begin{figure*}
    \centering
    \includegraphics[width=6.75in]{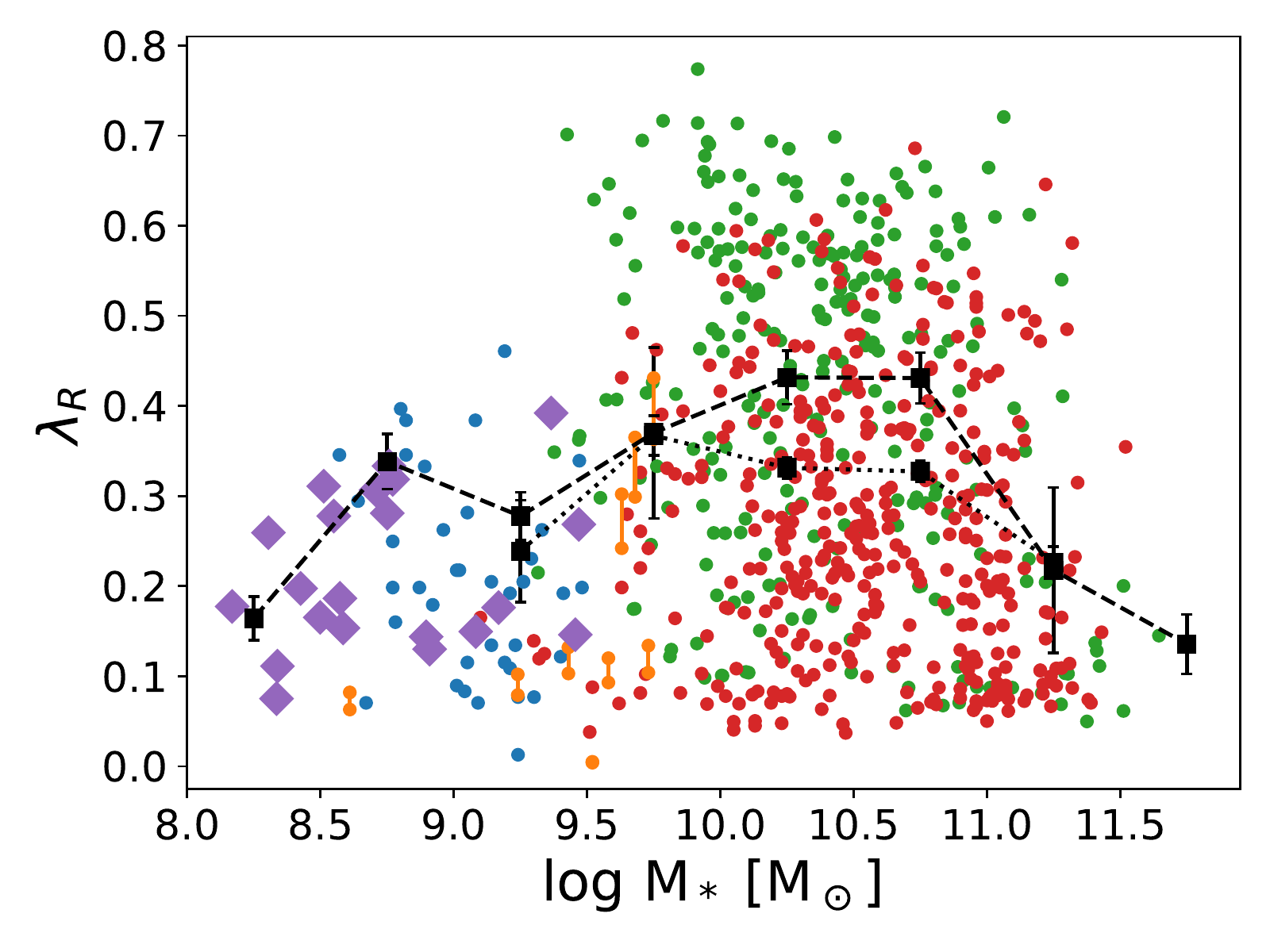}
    \caption{$\lambda_R$ as a function of log M$_\star$ for galaxies with an early-type morphology. Purple points from this work, blue points from \citet{Toloba:2014}, orange points from \citet{Janz:2017}, red points from \citet{vandeSande:2017} and green points from \citet{Emsellem:2011}. The black squares and error bars indicate the mean and error on the mean within bins of 0.5 dex in M$_\star$ for cluster (dashed) and field (dotted) subsamples respectively.}
    \label{fig:lambdar_mass}
\end{figure*}

The $\lambda_R$ -- $\epsilon$ plane is used to account for projection when comparing the specific stellar angular momentum of galaxies. It was first proposed by \citet{Emsellem:2007}, and represents a revision of the classical $V/\sigma$ -- $\epsilon$ diagram which has traditionally fulfilled this role. In Figure \ref{fig:lambdar_eps} we show $\lambda_{R_e}$ versus $\epsilon$ for our sample, coloured by the stellar mass of each galaxy, and with filled and open symbols indicating the primary sample and secondary targets respectively. We explicitly choose not to apply one of the common criteria to divide galaxies into slow rotators and fast rotators in this plane \citep{Emsellem:2007,Emsellem:2011,Cappellari:2016} as this classification was motivated by the properties of the giant galaxy population, and the utility of applying the same separation to dwarf galaxies is unclear.

\section{Discussion}

\subsection{The specific stellar angular momentum of dE galaxies}

\citet{Naab:2014} showed that for massive early-type galaxies the position of a galaxy within the $\lambda_R$ -- $\epsilon$ plane correlates with its formation history, with galaxies at low $\lambda_R$ typically having merging play a larger role in their evolution than galaxies at higher values of $\lambda_R$. For these massive galaxies, a number of studies have examined the distribution of the galaxy population in the $\lambda_R$ -- $\epsilon$ plane \citep[e.g.][]{Emsellem:2011,vandeSande:2017,Veale:2017,Graham:2018,Falcon-Barroso:2019}, noting a significant dependence on stellar mass. \citet{Rys:2013} extended this analysis to a small sample of dwarf galaxies in the Virgo cluster, finding a broader range in the degree of rotational support than in intermediate-mass early-type galaxies. Using long-slit spectroscopy, \citet{Toloba:2014} obtained a significant sample of dwarf early-type galaxies, finding that in general the rotational support in dEs is lower than in intermediate mass early-type galaxies, a result further supported by \citet{Janz:2017}. Here we extend this analysis of dE galaxies to lower masses using our sample of dEs with spatially resolved stellar kinematic measurements.

In Figure \ref{fig:lambdar_mass} we show the dependence of $\lambda_R$ on M$_\star$ for early-type galaxies spanning nearby four orders-of-magnitude in stellar mass. We combine data from this study with more massive Virgo dwarf galaxies from \citet[including a correction from long-slit to IFS-equivalent values using their Eqn. 3]{Toloba:2015} and isolated dwarfs from \citet{Janz:2017}. We include only our primary dE targets here because i) the aperture corrections for the giant early types observed in this study are large and relatively uncertain, and ii) those galaxies were selected in a way that is not necessarily representative of the giant early type population. For the \cite{Janz:2017} sample (orange points), we show two values for each galaxy, indicating the range of values given by their model fits to long-slit data. We supplement these dwarf samples with early-type giants from \citet[both Virgo and field]{Emsellem:2011} and \citet[field only]{vandeSande:2017}. All samples are corrected to a 1 R$_e$ aperture measurement following \citet{vandeSande:2017}. For the giant population, $\lambda_R$ increases with decreasing M$_\star$, reaching a peak around M$_\star \sim 10^{10.25}$. Below this mass we find $\lambda_R$ decreases with decreasing stellar mass. Low-$\lambda_R$, slowly rotating galaxies dominate both the very high (M$_\star > 10^{11.2}$ M$_\odot$) and low (M$_\star < 10^{8.5}$ M$_\odot$) mass regimes. At high masses, the transition to a slow rotator dominated population as M$_\star$ increases is quite sudden \citep[see][]{Cappellari:2016}, but at low masses the decline in $\lambda_R$ with M$\star$ is relatively smooth.  

We note that the samples used here have very different selection criteria and so interpreting the mean trend is challenging. \citet{Janz:2017} selected early types based on having a quiescent stellar population, whereas the other studies select early types based on visual morphology, perhaps accounting for the typically lower $\lambda_R$ of the \citet{Janz:2017} galaxies. More significantly, the samples used here are drawn from quite different environments. This work and \citet{Toloba:2014} consist entirely of galaxies in clusters; the sample of \citet{Emsellem:2011} includes both cluster and field galaxies and that of \citet{vandeSande:2017} and \cite{Janz:2017} include no cluster galaxies but a mixture of group and field objects. We separate the sample into cluster and non-cluster galaxies and derive the mean $\lambda_R$ as a function of mass for the two environments (dashed and dotted lines in Fig. \ref{fig:lambdar_mass}). The behaviour is consistent in both the high and low density environments, with the mean $\lambda_R$ falling at both high and low masses.

\begin{figure}
    \centering
    \includegraphics[width=3.25in]{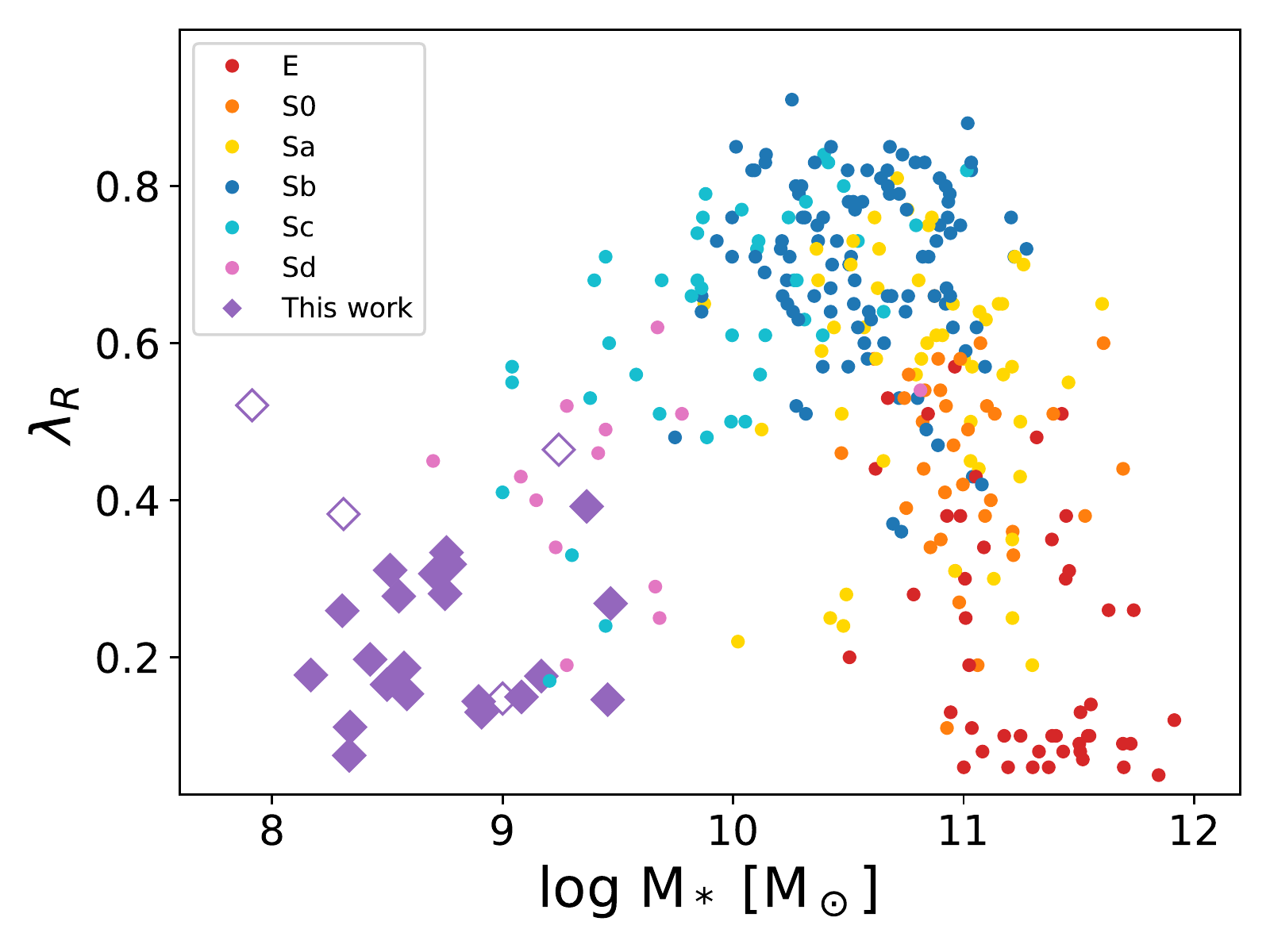}
    \caption{$\lambda_R$ vs M$_\star$ for galaxies of all morphological types. Coloured points from \citet{Falcon-Barroso:2019}, solid purple diamonds are dEs from this work, open purple diamonds are late-type dwarfs.}
    \label{fig:lambdar_morph_comp}
\end{figure}

The trend of decreasing $\lambda_R$ with decreasing M$_\star$ described above is consistent with \citet{Falcon-Barroso:2019}, who find a similar dependence of $\lambda_R$ with M$_\star$ in a sample with a mix of morphologies. That we find this trend in a pure early-type sample shows the variation in $\lambda_R$ must be driven by stellar mass, and cannot be attributed purely to the changing morphological mix of the galaxy population with stellar mass.  Although the sample of \citet{Falcon-Barroso:2019}, shown in Figure \ref{fig:lambdar_morph_comp}, only reaches masses of $\sim10^9$ M$_\odot$ and is incomplete at the lowest masses, a clear decrease is seen in $\lambda_R$ with decreasing M$_\star$ below M$_\star \sim 10^{10}$ M$_\odot$. 

All of their faint galaxies are spirals, classified by them as Sc and Sd, with the average $\lambda_R$ lower for Sds than for Scs. \citet{Falcon-Barroso:2019} find that their late-type spirals have surprisingly low $\lambda_R$ values for spiral galaxies. The systems with $\lambda_R$ between 0.35 and 0.6 are typically edge-on systems, as they say. To keep these objects thin, such galaxies would need a considerable amount of dark matter (with the enclosed dynamical mass being a factor 10 higher than the baryonic mass). 

Here we can add some more pieces to the puzzle. The objects in this paper have, in general, lower $\lambda_R$ values than the galaxies of \citet{Falcon-Barroso:2019}, as expected by the trend in mass. The objects with $\lambda_R$ $<$ 0.3, i.e. the slowly rotating dEs, are generally not irregular and not face-on, indicating that they are genuinely dynamically hot systems, suggesting that the low mass spirals of \citet{Falcon-Barroso:2019} with low $\lambda_R$ may also be dynamically hot, even though they are spirals. The dark matter fractions in our galaxies are also found to be higher than for intermediate mass galaxies (Eftekhari et al., in prep.). 

While low mass spiral and irregular galaxies show lower values of $\lambda_R$ than massive spirals \citep{Falcon-Barroso:2019}, there remains a modest offset between the $\lambda_R$ of dEs and Sc/Sd galaxies at fixed stellar mass. \citet{Rys:2014} show that the dynamics of dE galaxies in the Virgo cluster are consistent with their progenitors being comparable mass spiral galaxies that are tidally heated by the cluster environment. It is unclear whether the same argument can be extended to the lower mass dEs in this study as no comparison sample of Sc/Sd/Irr galaxies with masses M$_\star < 10^9$ M$_\odot$ exist. If the trend of decreasing $\lambda_R$ with decreasing M$_\star$ for spirals seen in \citet{Falcon-Barroso:2019} continues then the dynamical gap between spirals and dEs may close at M$_\star \sim 10^8$ M$_\odot$ and no kinematic heating may be required. For low mass galaxies this trend is consistent with the hypothesis that dEs can originate from low mass spirals and irregulars, as they enter a galaxy cluster and lose their gas due to ram pressure stripping. On a kinematic basis alone we cannot rule out different physical mechanisms causing the increase in $\lambda_R$ observed in dEs and spirals, however the similarity in other observed properties, such as their exponential surface brightness distributions, is also consistent with a common formation path \citep[see e.g.,][]{Venhola:2019}. In a future paper we will address this question in more detail. 

\begin{figure}
    \centering
    \includegraphics[width=3.25in]{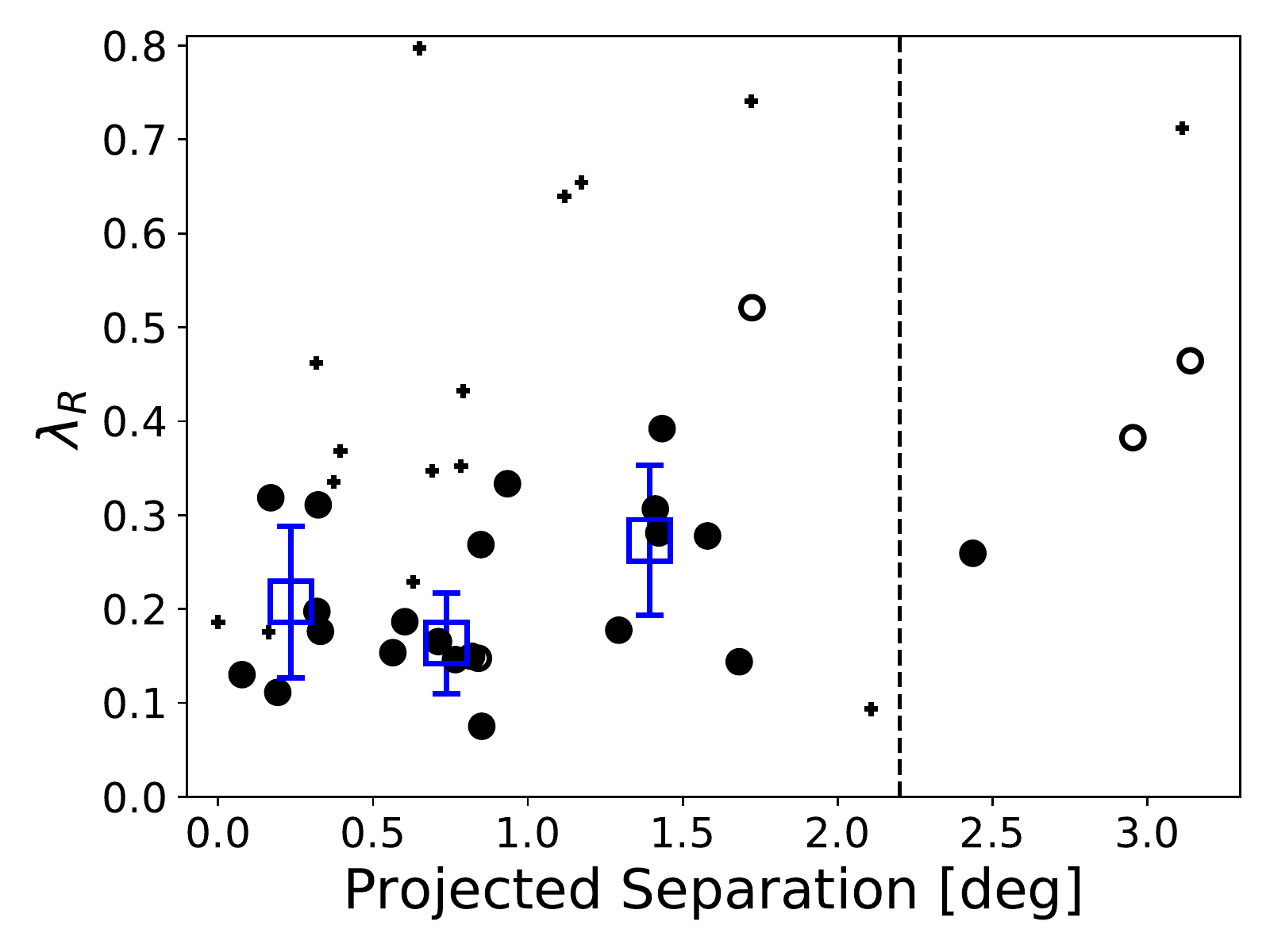}
    \caption{Specific stellar angular momentum $\lambda_R$ vs distance from the centre of the cluster. Symbols as in Fig. \ref{fig:lambdar_eps}. The dashed line indicates the position of the Virial radius of the cluster. The blue squares and errorbars indicate the mean and standard deviation for primary dE galaxies in three bins of projected separation.}
    \label{fig:lambdar_crad}
\end{figure}

An alternate scenario \citep[also discussed in][]{Rys:2014} is that the progenitors of present day dEs were low mass $z = 1 -2$ spiral galaxies that are dynamically hotter than present day spirals. Croom et al. (in prep) explore the ability of a pure disk-fading scenario to explain the properties of present-day lenticular galaxies, finding that the kinematics and structure of $z \sim 0$ S0s are consistent with passively evolved $z = 1 - 2$ spirals, though their results are only directly applicable at M$_\star > 10^{9.5}$ M$_\odot$. No sample of M$_\star \sim 10^8$ M$_\odot$ spiral or irregular galaxies with stellar kinematic measurements exists to directly compare possible progenitor kinematics.

\subsection{The environmental dependence of dE kinematics}

As noted above, there is evidence for a kinematic morphology -- density relation at fixed M$_\star$ amongst giant galaxies. For dwarf galaxies a similar relation may exist. In their sample of dE galaxies in the Virgo cluster, \citet{Toloba:2015} found that the fraction of slow rotators decreases, and the average $\lambda_R$ increases, with increasing projected distance from the cluster centre. In Figure \ref{fig:lambdar_crad} we show a consistent result for the Fornax cluster --- the average $\lambda_R$ of dwarf galaxies (indicated by the open blue squares) is lower in the central $\sim 1^\circ$ of the cluster compared to the cluster outskirts. This is consistent with the kinematic morphology -- density relation found in e.g. \citet{Graham:2019}, however, given the relatively small sample of dE galaxies with spatially resolved kinematics, we cannot directly examine the dependence with local galaxy number density. Importantly, given our modest sample of galaxies, we cannot yet disentangle the effect of environment from the observed trends with stellar mass. We also note that \cite{Drinkwater:2001b} found significant H$\alpha$ emission in some dE galaxies, particularly towards the edge of the cluster, again suggesting an environmental dependence of dE properties. We will examine the emission line properties of our sample in an upcoming work.

\section{Conclusions}
\label{sec:conclusions}

In this work we have presented integral field spectroscopy observations of a sample dwarf early-type galaxies in the Fornax cluster. These observations represent the largest sample of low-mass early-type galaxies in a cluster to date. We have demonstrated that our observed sample is fully representative of the dwarf galaxy population in the Fornax cluster above a stellar mass of $10^8$ M$_\odot$.

For a subset of 21 dwarf early-type galaxies and 4 dwarf late-types we construct maps of the spatially resolved stellar velocity and velocity dispersion. Using these maps we assign kinematic morphological classifications to these galaxies, as well as deriving their specific stellar angular momentum, $\lambda_R$. 

Combining our sample with observations of giant and massive dwarf early-type galaxies, we find that $\lambda_R$ increases from the most massive galaxies to a peak around M$_\star \sim 10^{10}$ M$_\odot$, before declining towards lower masses. This is similar to the trend for slightly more massive spiral galaxies found by \citet{Falcon-Barroso:2019}. No comparable kinematic measurements exist for spirals of a similar mass range to our sample. Without a direct comparison sample we cannot definitively confirm or rule out particular formation mechanisms, however our kinematic results are consistent with a scenario where dEs form from low mass spiral galaxies that have their gas stripped by cluster processes, with possibly some modest external heating to fully account for dynamical differences between dEs and low mass spirals, consistent with the findings of \citet{Koleva:2014}. Further observations of the stellar kinematics of star-forming dwarf galaxies are required to ascertain whether alternative formation mechanisms with increased dynamical heating are required to explain the observed kinematic differences between the two populations. We also confirm a trend first reported in \citet{Toloba:2015}, that the $\lambda_R$ of dE galaxies is consistent with a modest increase with increasing cluster-centric radius, however studies in a larger number of cluster and or massive group environments are needed to solidify this result.

This paper is the first in a short series examining the dwarf early-type galaxy population of the Fornax cluster. In Eftekhari et al. (in prep) we present the scaling relations of these galaxies and examine how the dark matter content of galaxies varies as a function of their mass. In a later paper we will present the stellar population properties of these galaxies.

\section*{Acknowledgements}
We would like to thank Dilyar Barat, Francesco D'Eugenio, Greg Goldstein and Jesse van de Sande for their role in the 2015 observing run, and Joachim Janz for providing $\lambda_R$ measurements for isolated dE galaxies. We would also like to thank the anonymous referee for their constructive comments that have greatly improved this paper. NS acknowledges support of an Australian Research Council Discovery Early Career Research Award (project number DE190100375) funded by the Australian Government and a University of Sydney Postdoctoral Research Fellowship. SE acknowledges funding support by the ESO PhD studentship programme. RFP acknowledges financial support from the European Union's Horizon 2020 research and innovation program under the Marie Sklodowska-Curie grant agreement No. 721463 to the SUNDIAL ITN network. JF-B has been supported through the RAVET project by the grant AYA2016-77237-C3-1-P from the Spanish Ministry of Science, Innovation and Universities (MCIU) and through the IAC project TRACES which is partially supported through the state budget and the regional budget of the Consejer\'\i a de Econom\'\i a, Industria, Comercio y Conocimiento of the Canary Islands Autonomous Community. GvdV acknowledges funding from the European Research Council (ERC) under the European Union's Horizon 2020 research and innovation programme under grant agreement No 724857 (Consolidator Grant ArcheoDyn). AV acknowledges financial support from the Emil Aaltonen foundation. Parts of this research were supported by the Australian Research Council Centre of Excellence for All Sky Astrophysics in 3 Dimensions (ASTRO 3D), through project number CE170100013.




\bibliographystyle{mnras}
\bibliography{fornax_dEs} 



\appendix

\section{Kinematics of primary targets}

A number of the galaxies in our primary sample exhibit peculiar kinematic features and we note and discuss those objects here:

{\it FCC 106:} This galaxy shows a central region of low velocity dispersion, with evidence for a counter-rotating kinematic component in the central $\sim 5$ arcsec.

{\it FCC 143:} This galaxy exhibits an elongated region of enhanced velocity dispersion, which \citet{Iodice:2019} show is surrounded by a region of low dispersion and low net rotation. They suggest this kinematically distinct structure may be related to an ongoing interaction with the nearby galaxy FCC 147.

{\it FCC 182:} This galaxy shows rotation that rises sharply in the central $\sim 2$ arcsec, before declining slightly towards 1 R$_e$, suggesting the presence of a additional central component.

{\it FCC 303:} This galaxy exhibits ordered rotation and a strong central dip in velocity dispersion \citep[consistent with][]{Iodice:2019}. The elongated low-dispersion structure suggest the presence of a central cold component, likely a small inner disk.

\begin{figure*}
    \centering
    \includegraphics[width=2.in,clip,trim = 20 0 70 0]{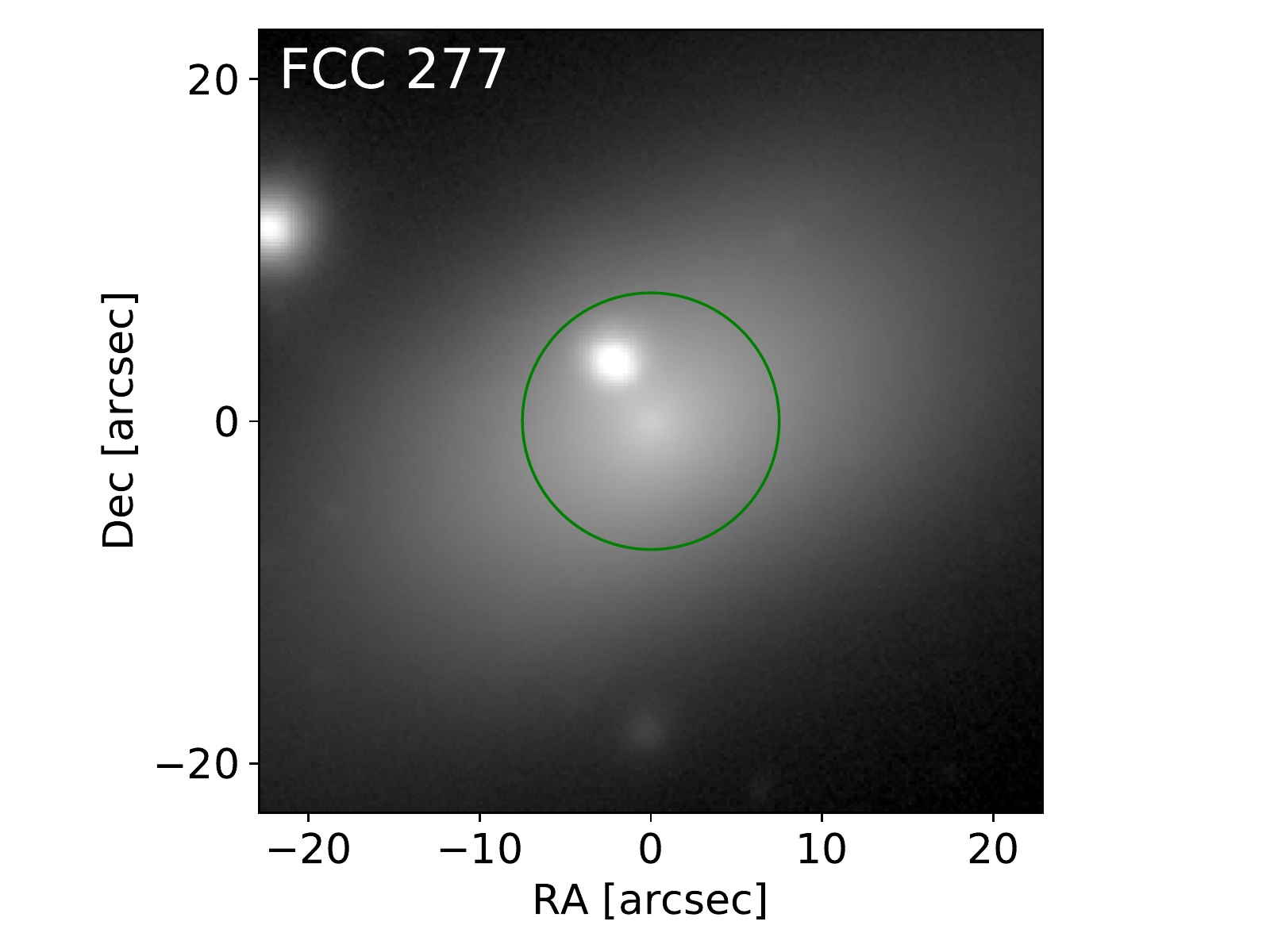}
    \includegraphics[width=2.25in,clip,trim = 20 10 30 10]{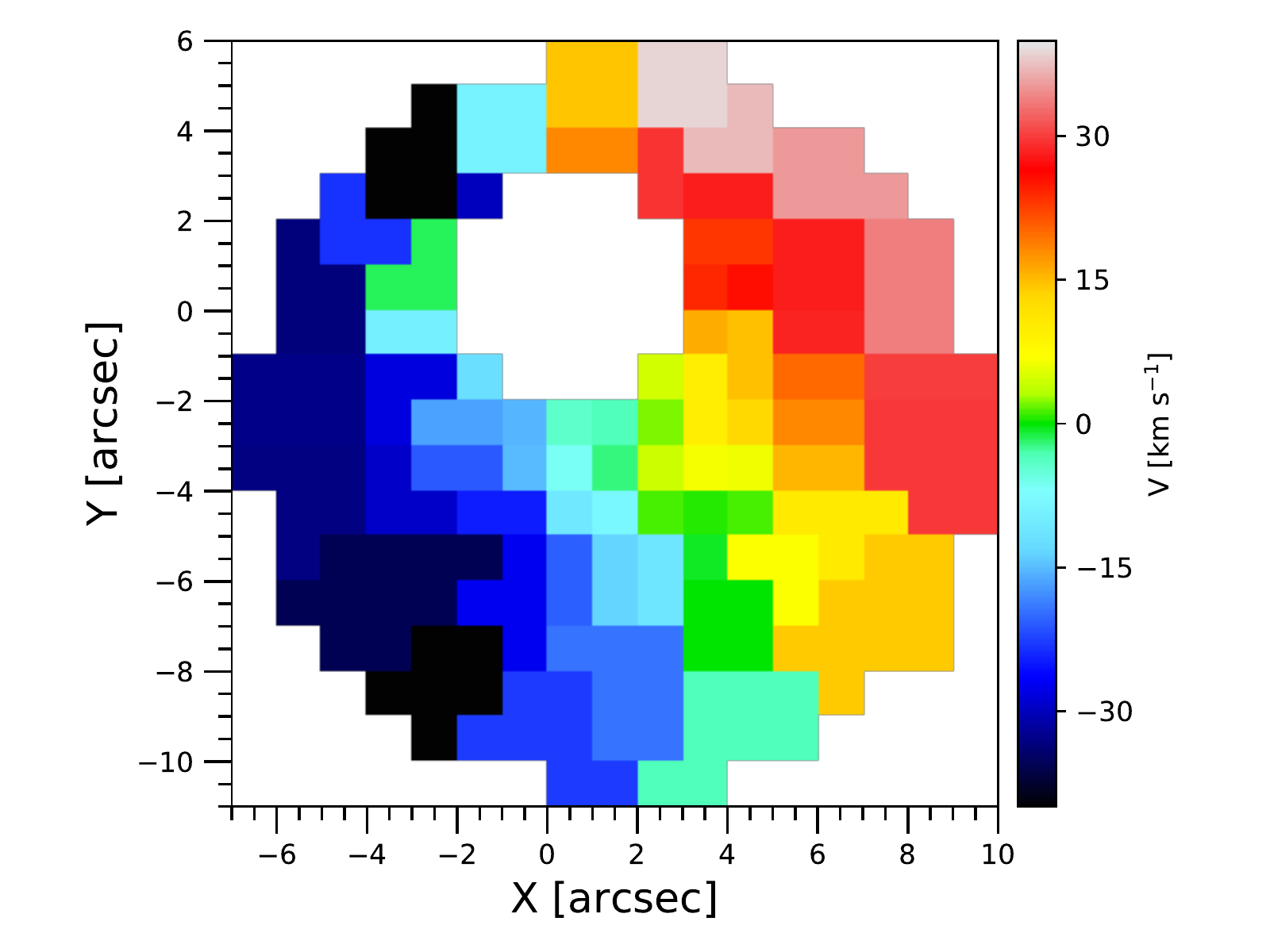}
    \includegraphics[width=2.25in,clip,trim = 20 10 30 10]{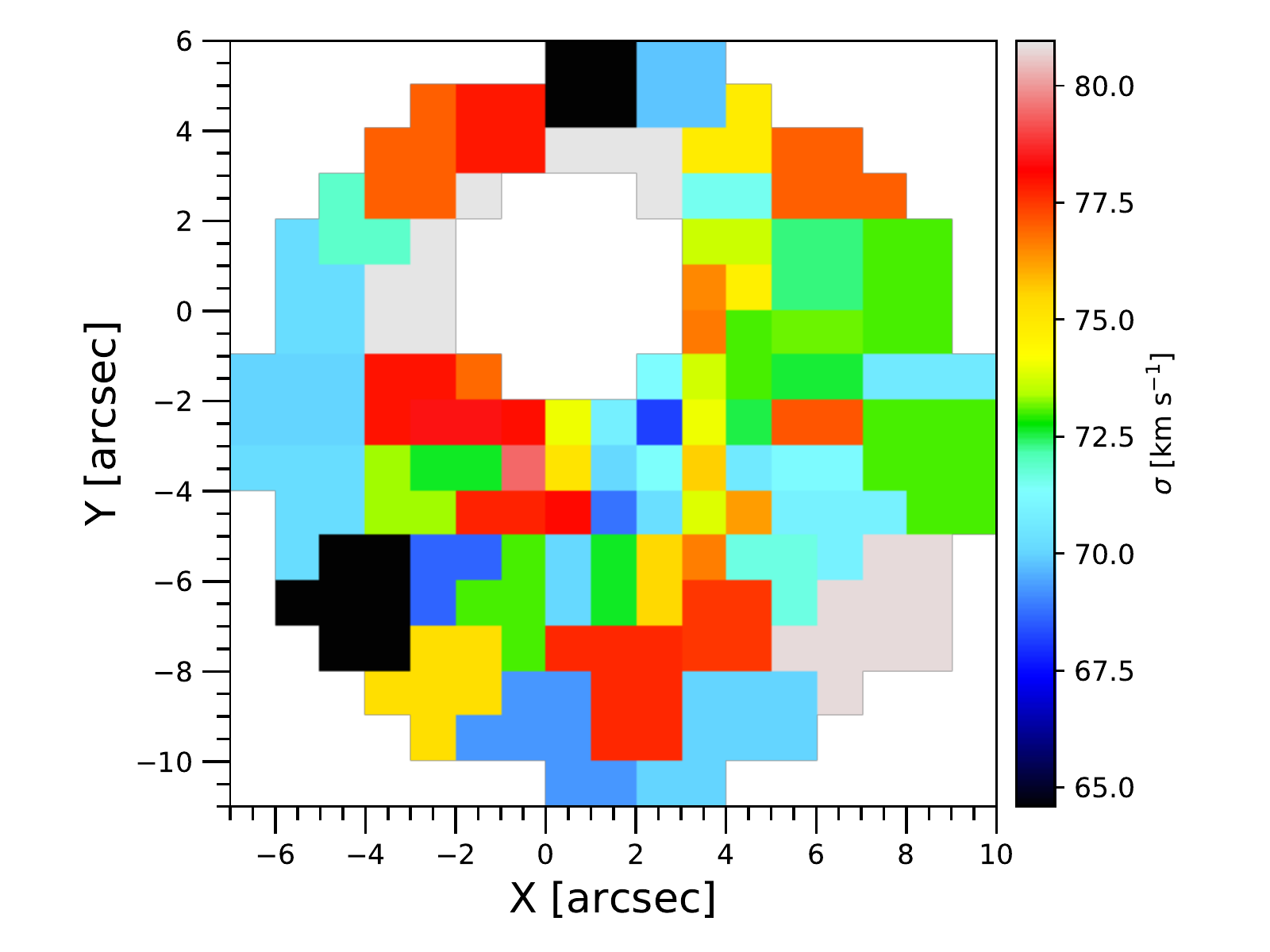}
    
    \includegraphics[width=2.in,clip,trim = 20 0 70 0]{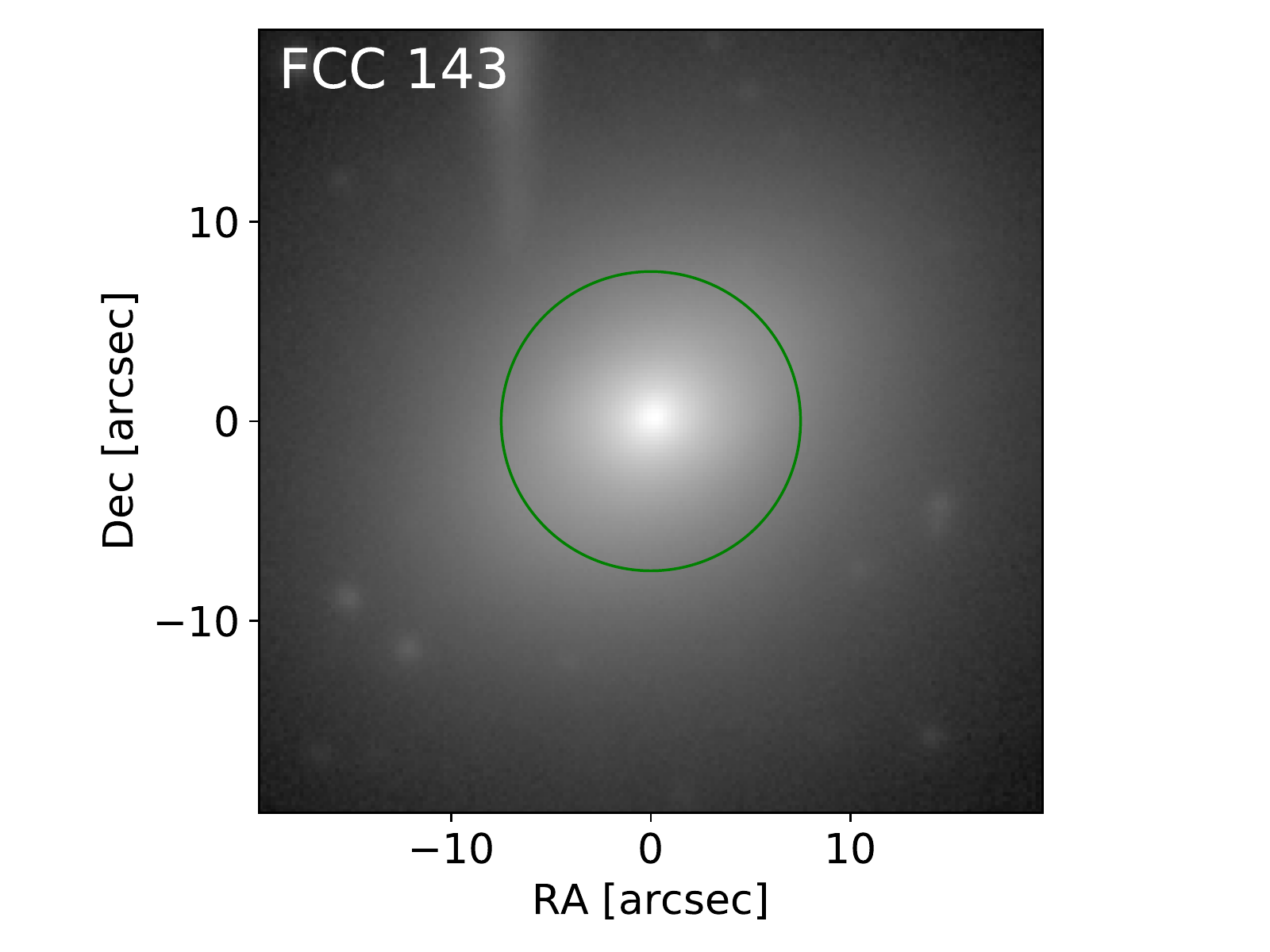}
    \includegraphics[width=2.25in,clip,trim = 20 10 30 10]{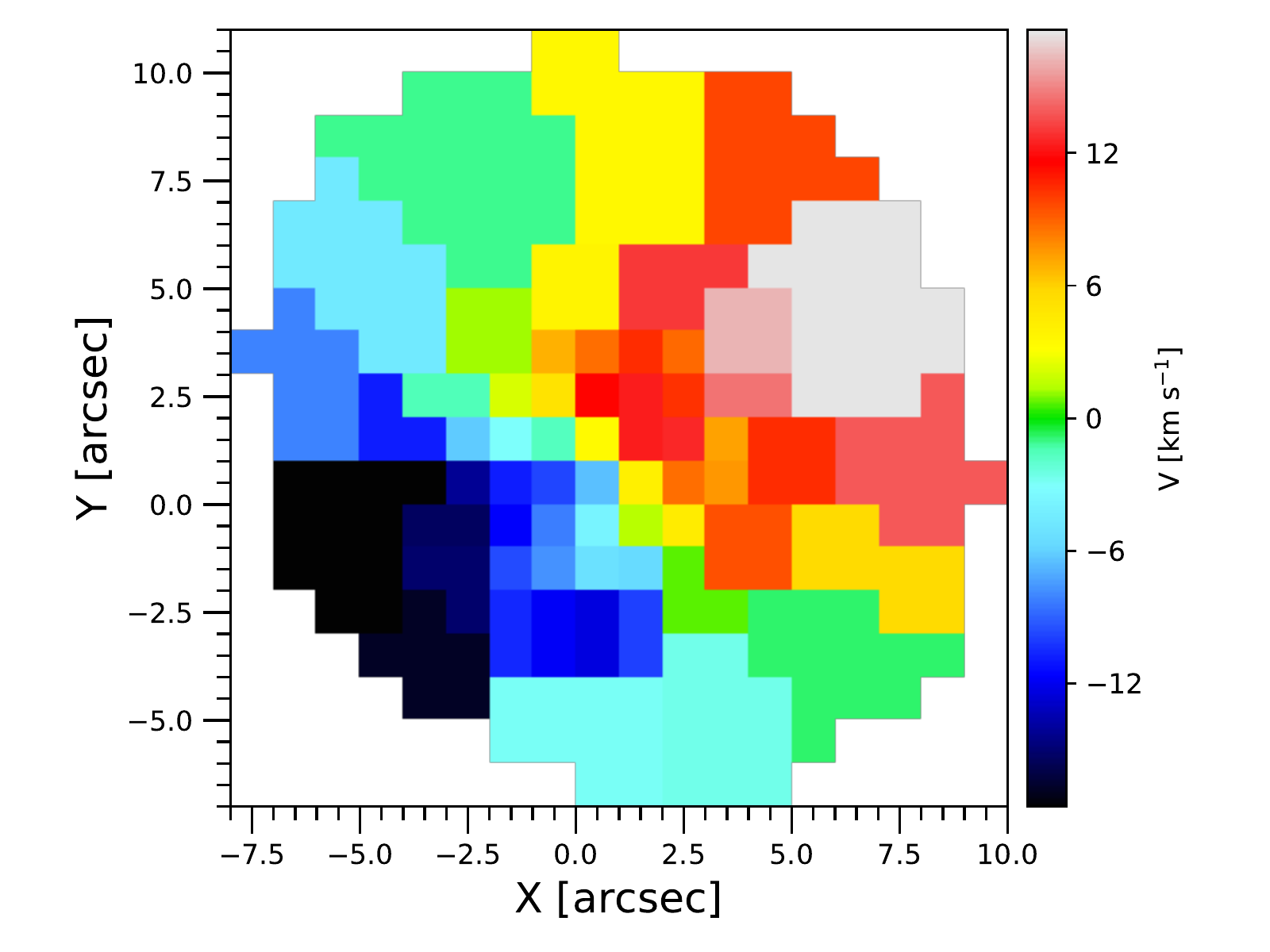}
    \includegraphics[width=2.25in,clip,trim = 20 10 30 10]{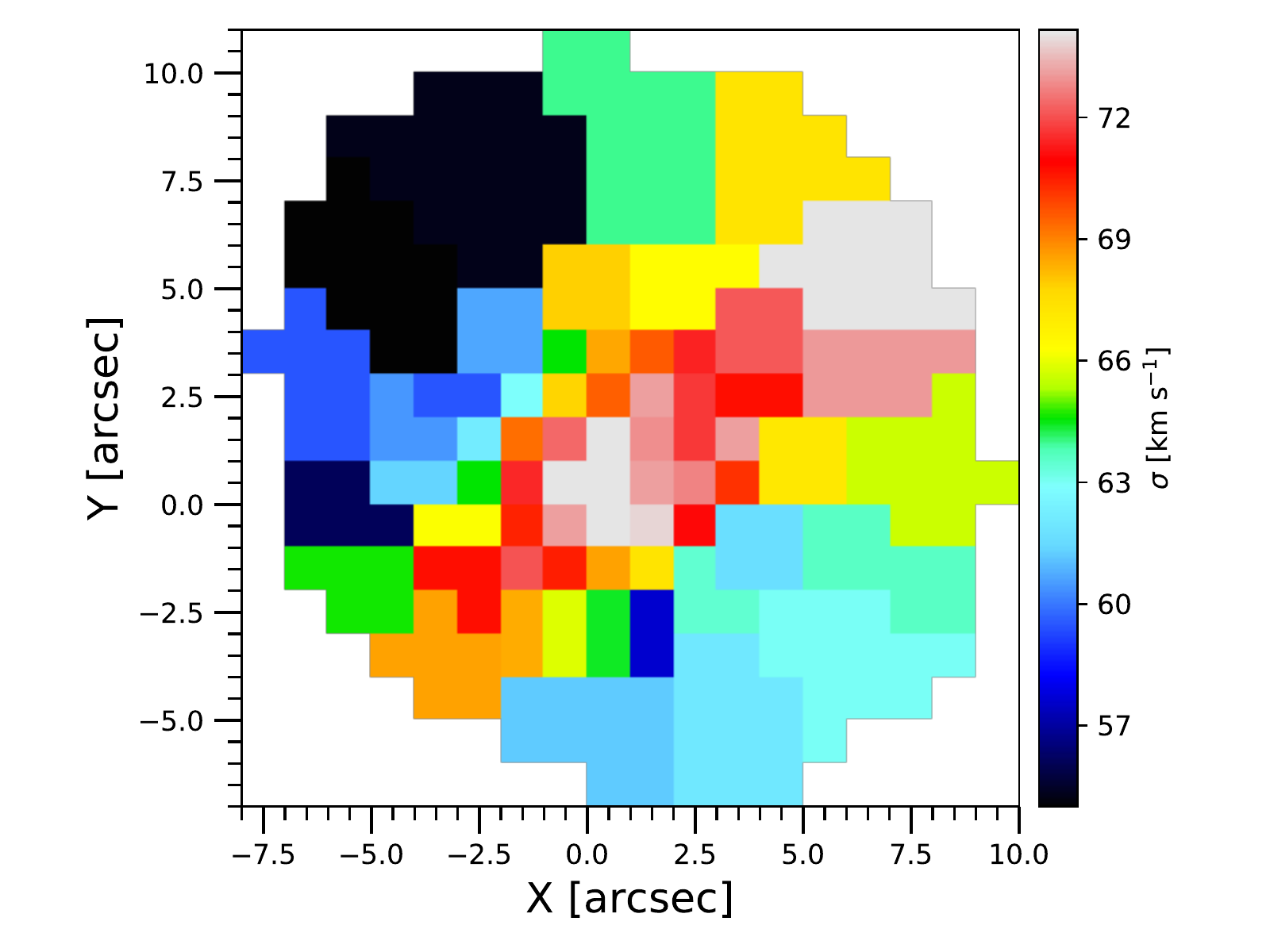}
    
    \includegraphics[width=2.in,clip,trim = 20 0 70 0]{figures/cutouts/301_cutout.pdf}
    \includegraphics[width=2.25in,clip,trim = 20 10 30 10]{figures/maps/301_vel_map.pdf}
    \includegraphics[width=2.25in,clip,trim = 20 10 30 10]{figures/maps/301_sig_map.pdf}

    \includegraphics[width=2.in,clip,trim = 20 0 70 0]{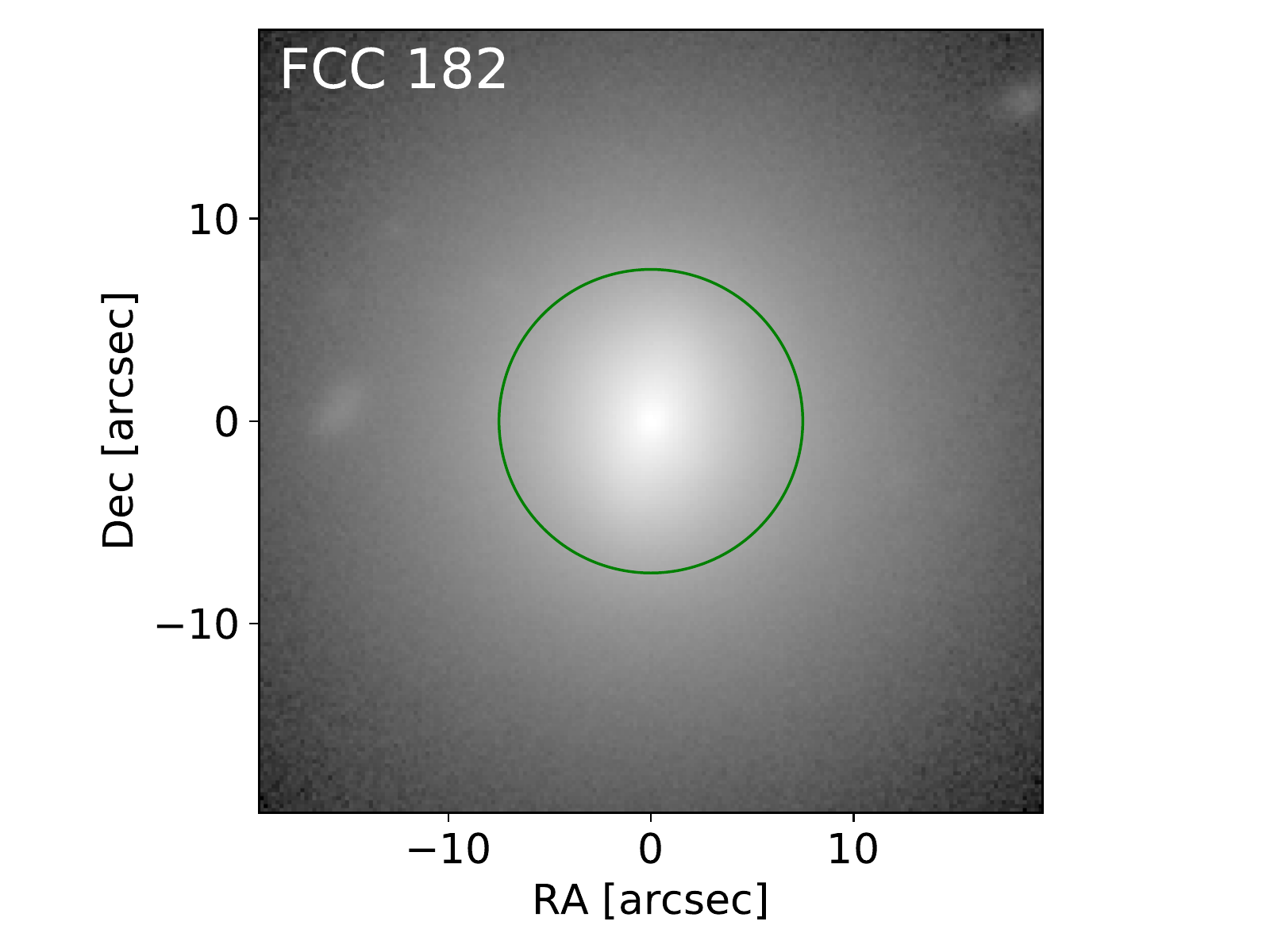}
    \includegraphics[width=2.25in,clip,trim = 20 10 40 10]{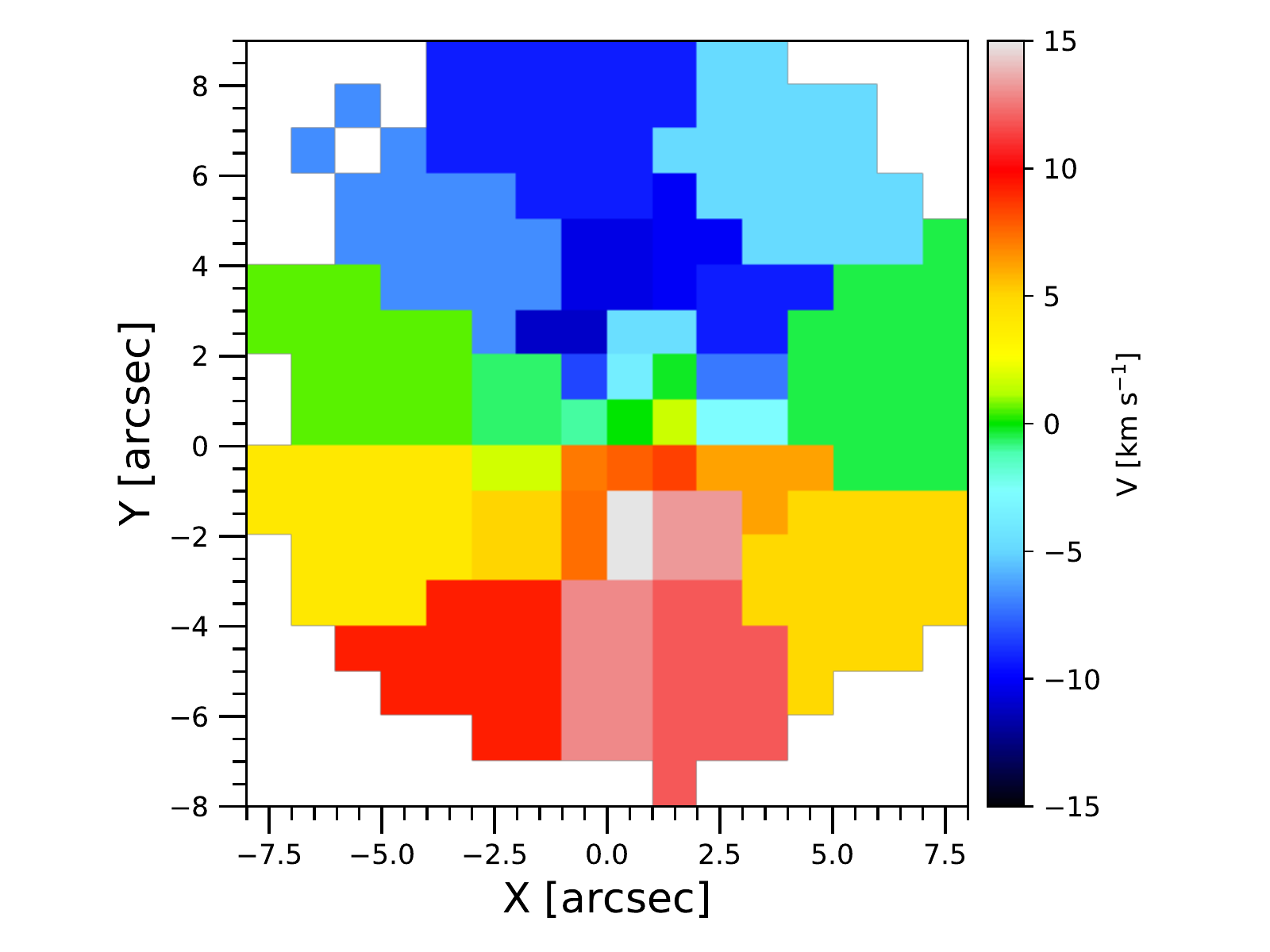}
    \includegraphics[width=2.25in,clip,trim = 20 10 40 10]{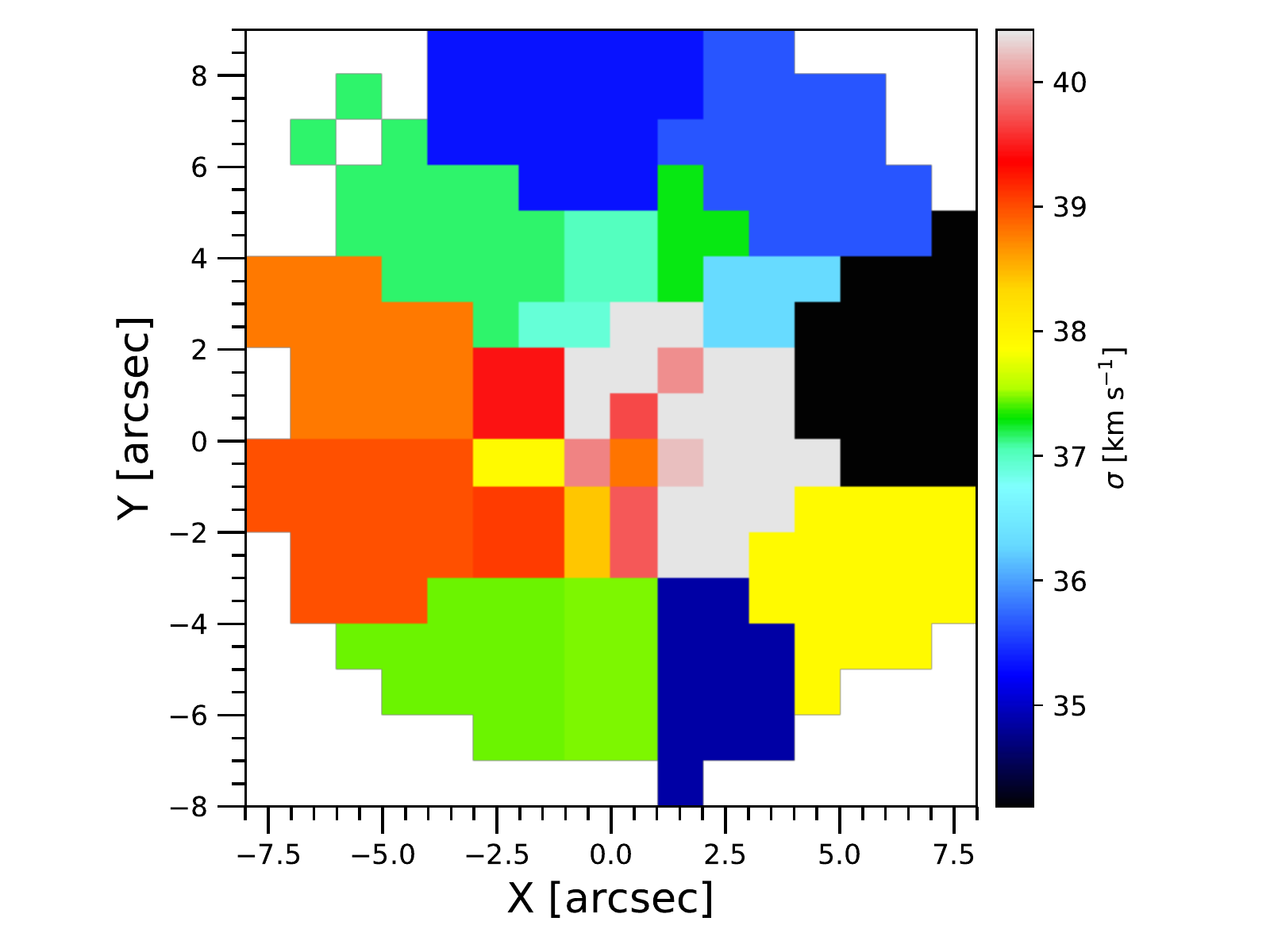}
    
    \caption{Images and stellar kinematic maps for all primary targets with good-quality kinematics, ordered by increasing M$_r$. Left column: {\it r}-band cutout images from FDS, with diameter 4 R$_e$. Green circle indicates the approximate position of the SAMI IFS field-of-view. Centre column: Voronoi-binned maps of the mean line-of-sight stellar velocity, $V$. Right column: Voronoi-binned maps of the mean line-of-sight stellar velocity dispersion, $\sigma$. For the majority of galaxies the displayed $V$ and $\sigma$ ranges are scaled to show the full range for each galaxy. For galaxies with low rotation or dispersion we impose a minimum range in $V$ and $\sigma$, reflective of the uncertainties on the measurements and the limits of the instrumental resolution.}
    \label{fig:primary_maps_app}
\end{figure*}

\begin{figure*}
    \centering
    \includegraphics[width=2.in,clip,trim = 20 0 70 0]{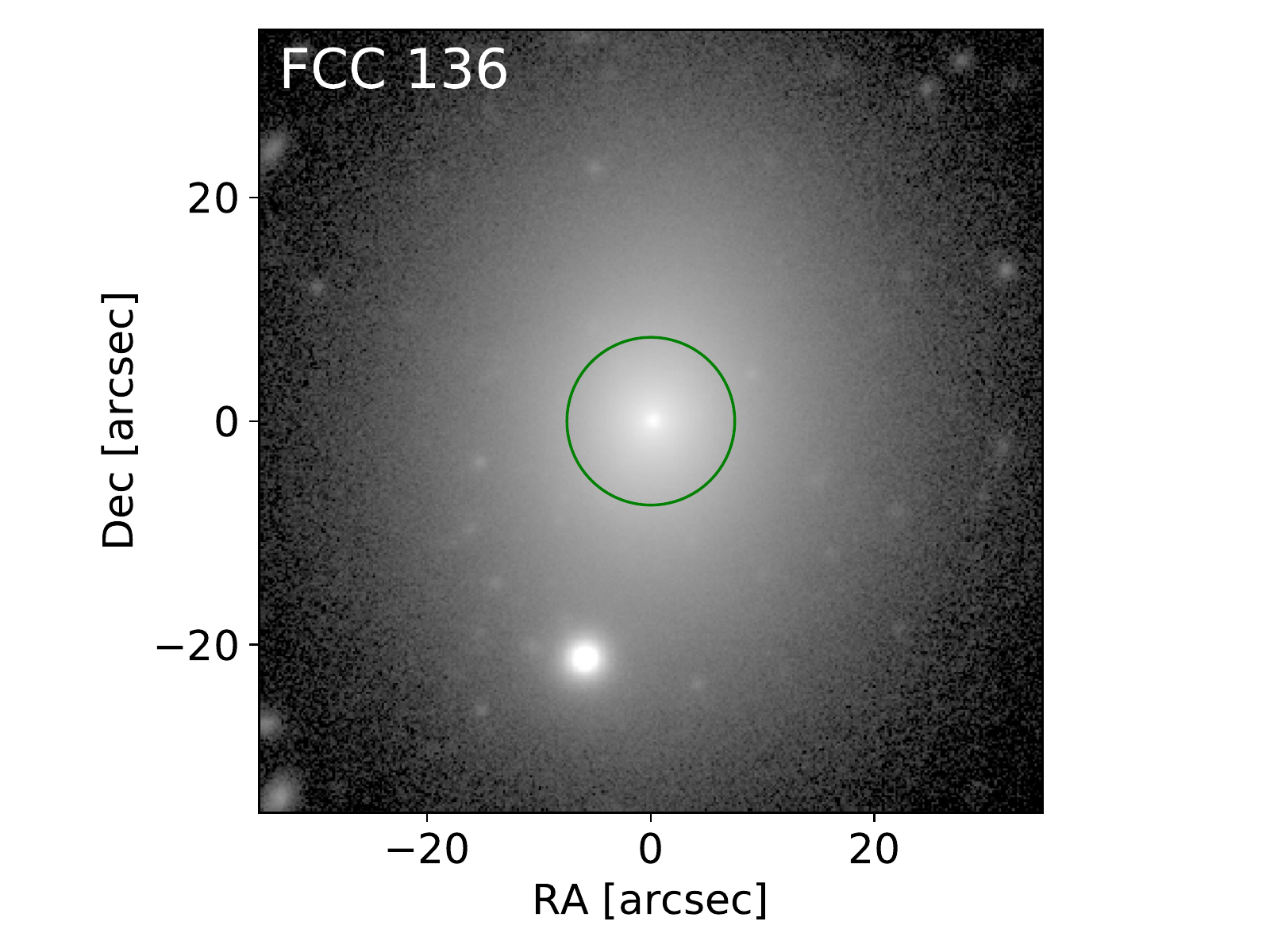}
    \includegraphics[width=2.25in,clip,trim = 20 10 40 10]{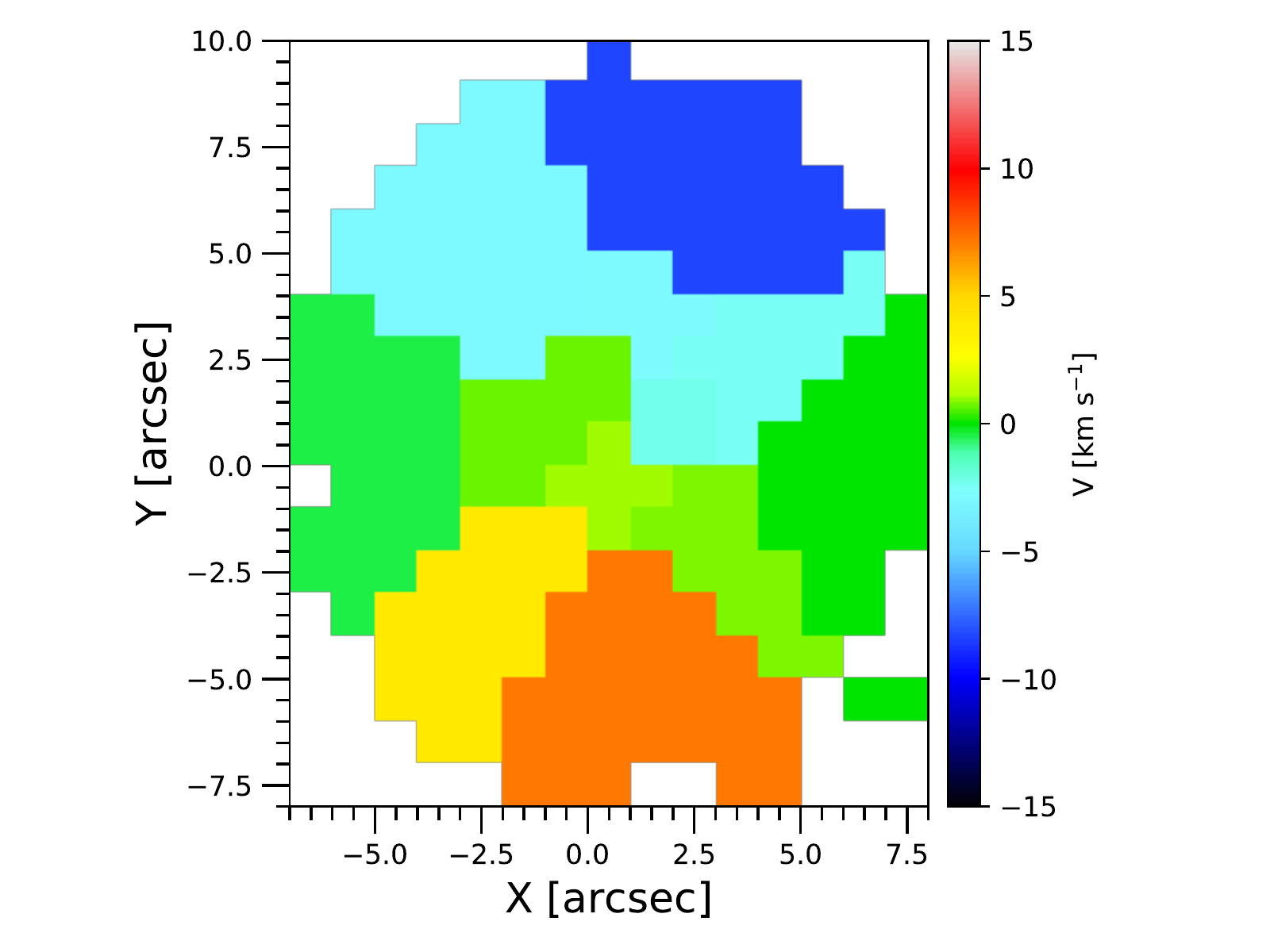}
    \includegraphics[width=2.25in,clip,trim = 20 10 40 10]{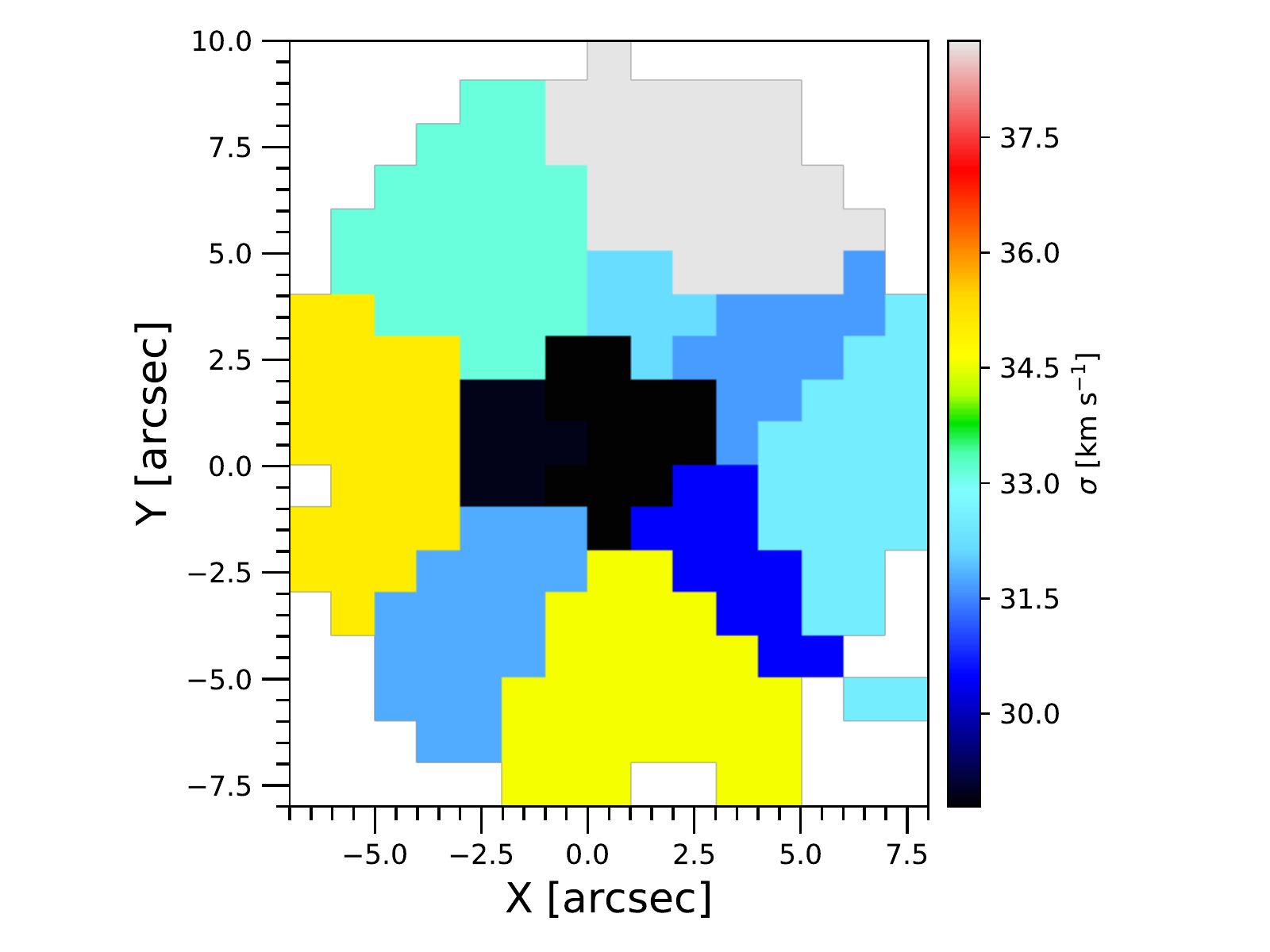}
    
    \includegraphics[width=2.in,clip,trim = 20 0 70 0]{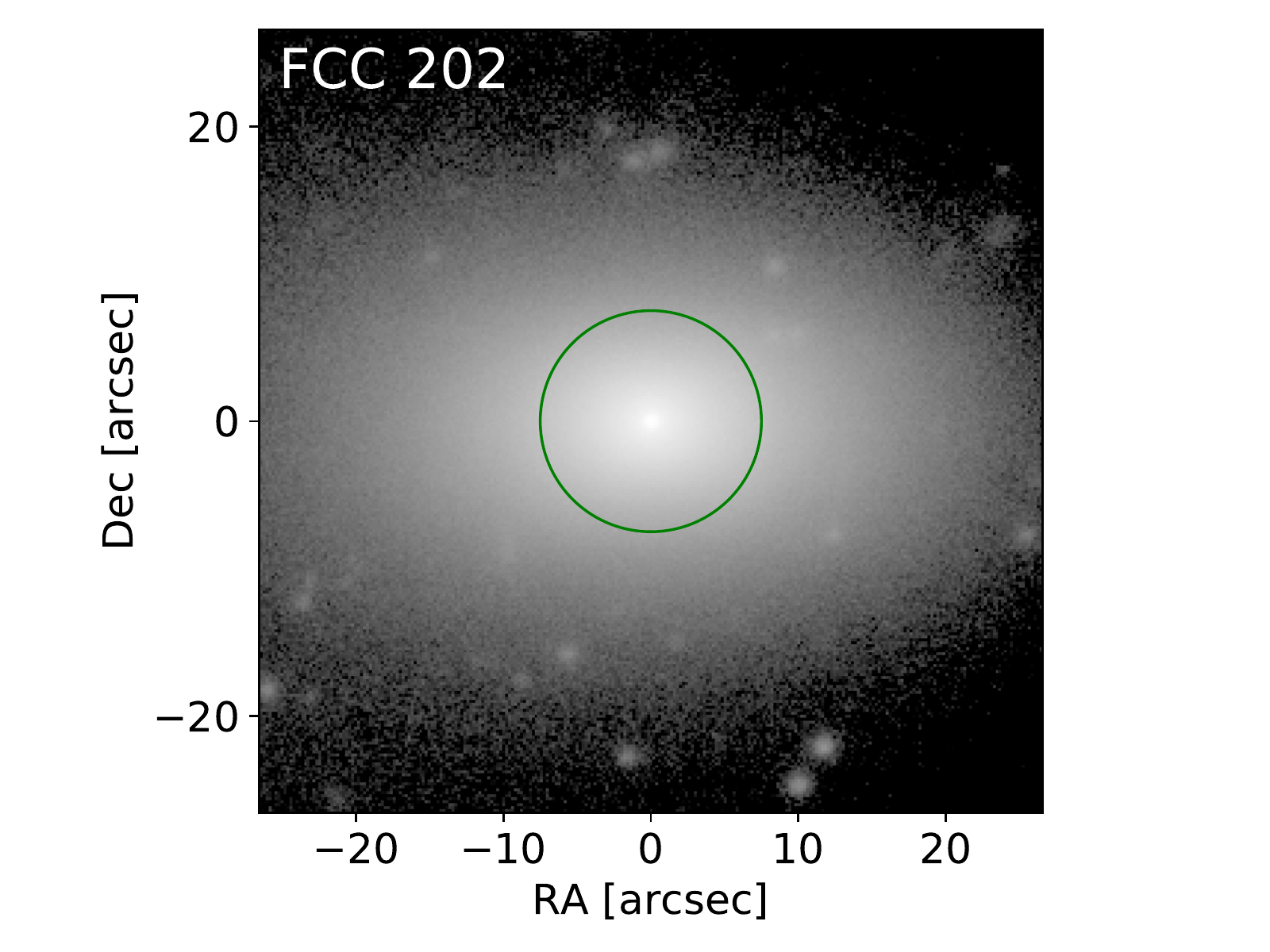}
    \includegraphics[width=2.25in,clip,trim = 20 10 40 10]{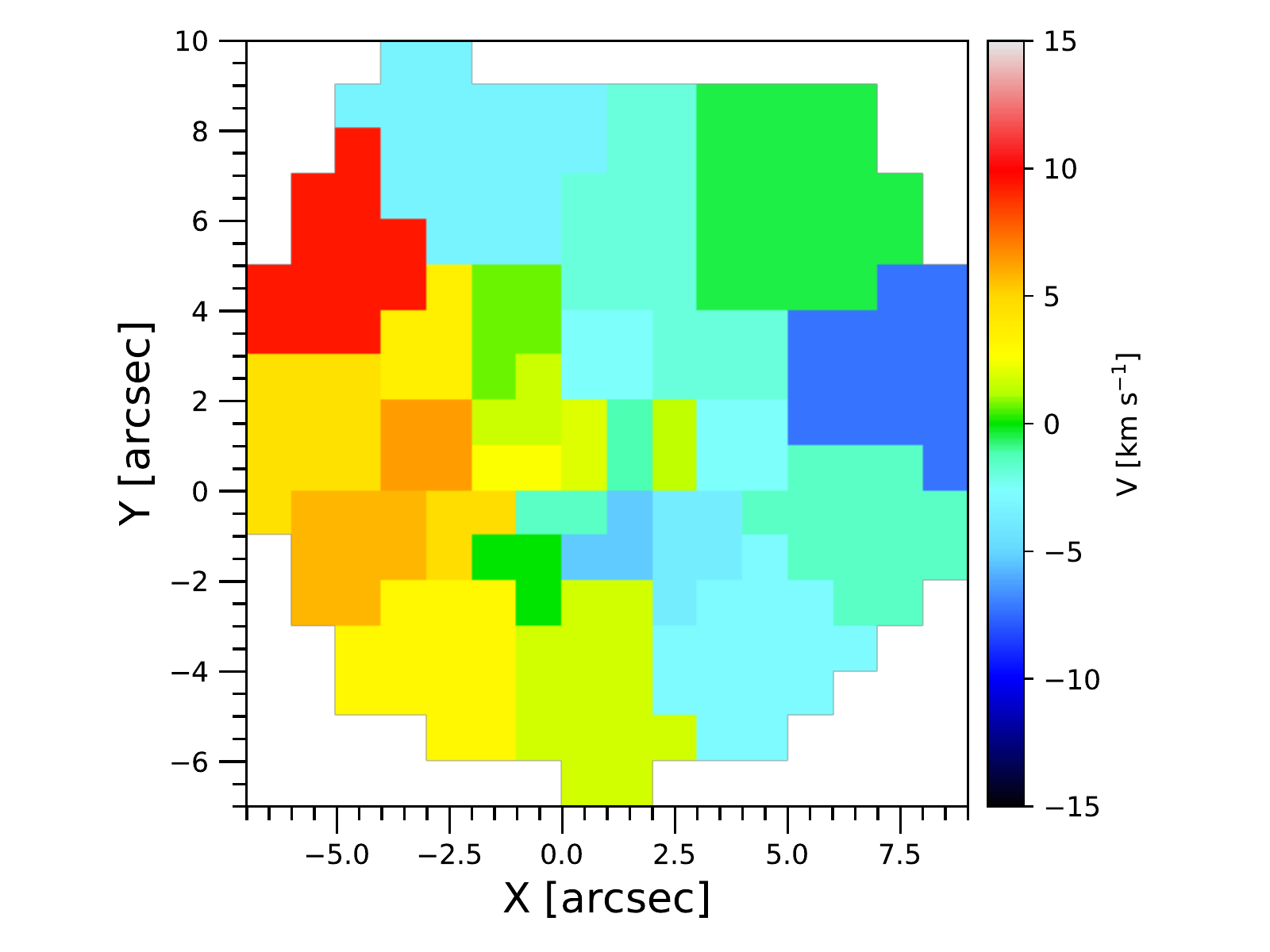}
    \includegraphics[width=2.25in,clip,trim = 20 10 40 10]{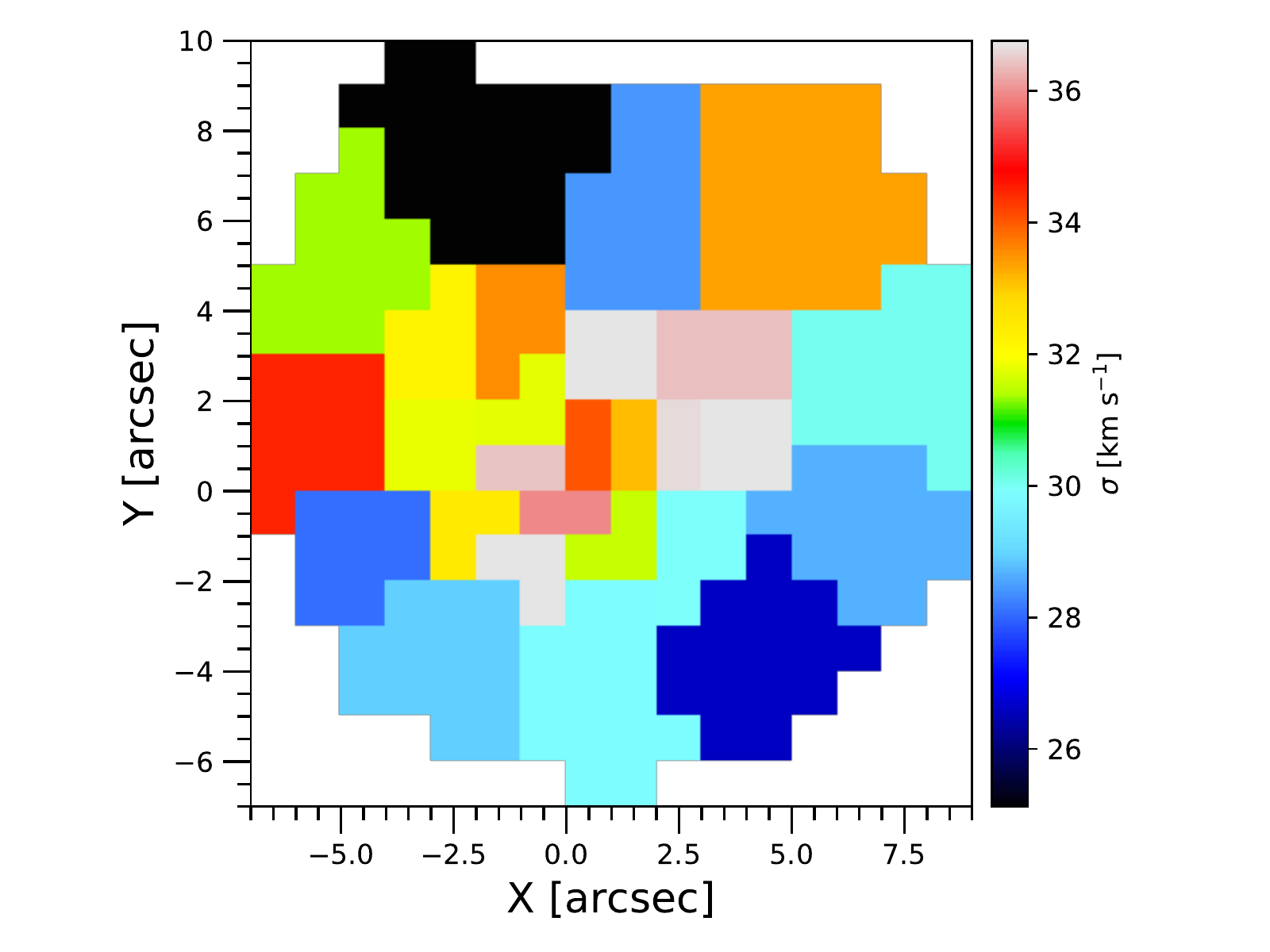}
    
    \includegraphics[width=2.in,clip,trim = 20 0 70 0]{figures/cutouts/106_cutout.pdf}
    \includegraphics[width=2.25in,clip,trim = 10 10 20 10]{figures/maps/106_vel_map.pdf}
    \includegraphics[width=2.25in,clip,trim = 20 10 20 10]{figures/maps/106_sig_map.pdf}
    
    \includegraphics[width=2.in,clip,trim = 20 0 70 0]{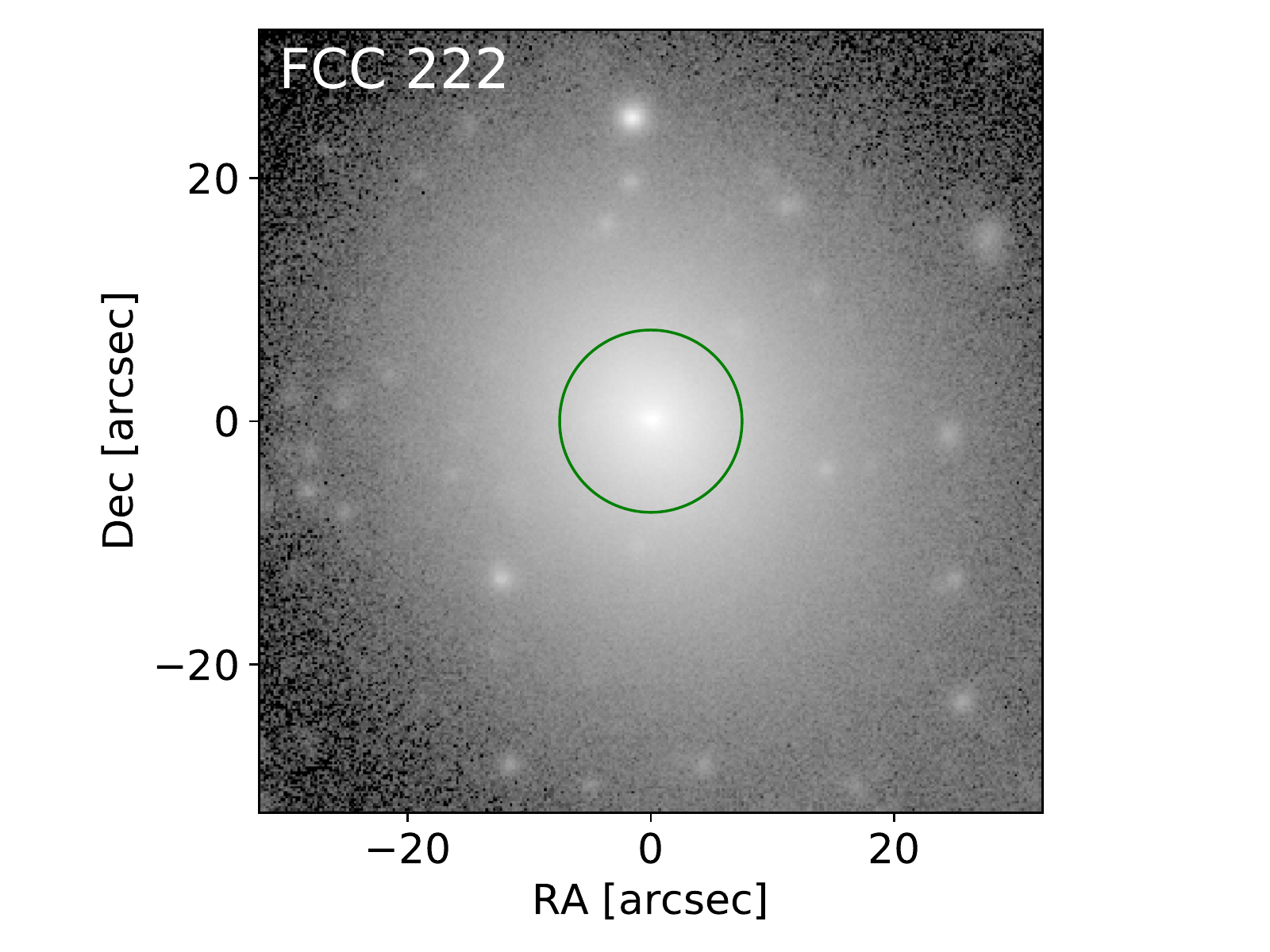}
    \includegraphics[width=2.25in,clip,trim = 20 10 30 10]{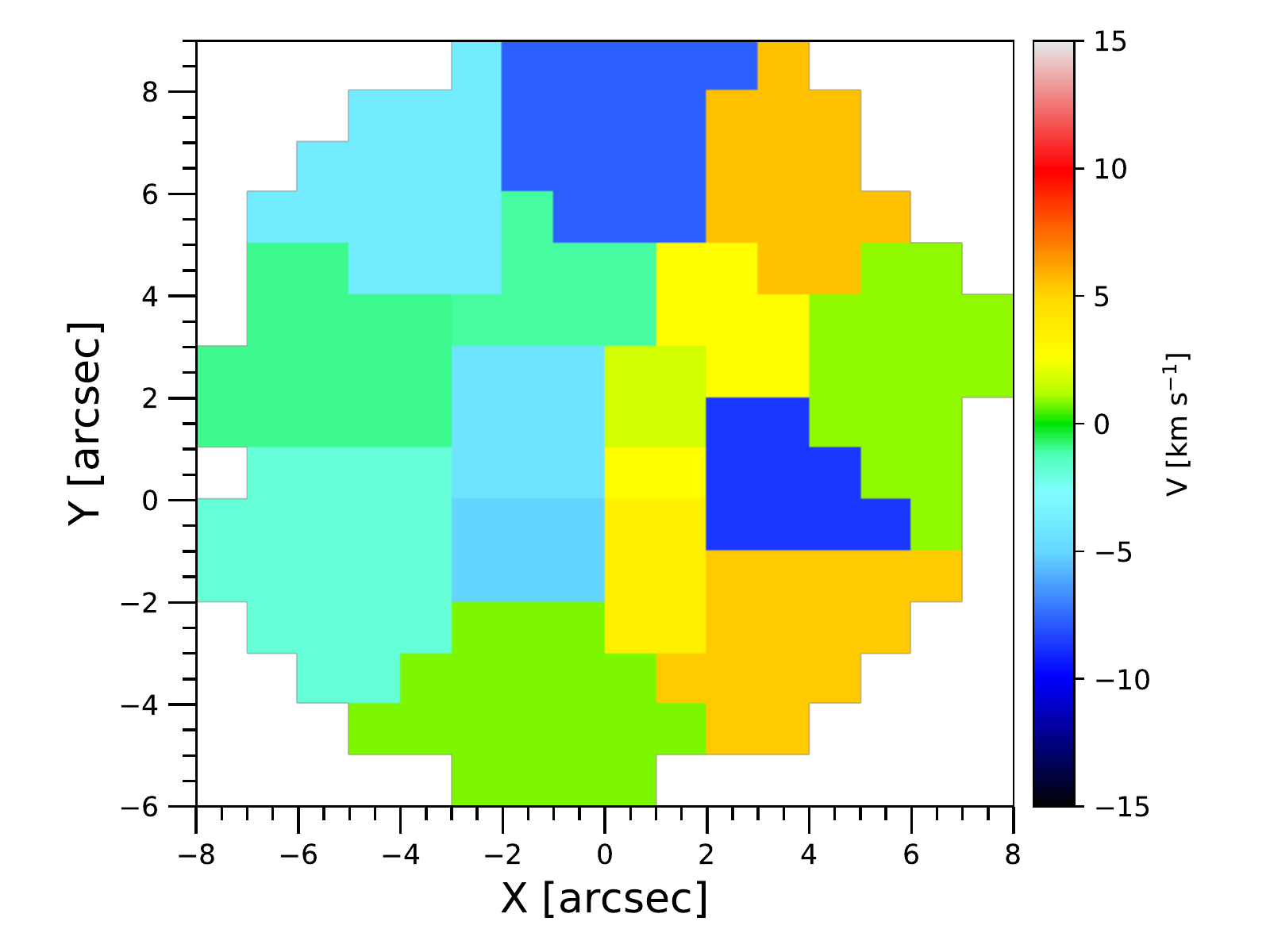}
    \includegraphics[width=2.25in,clip,trim = 20 10 30 10]{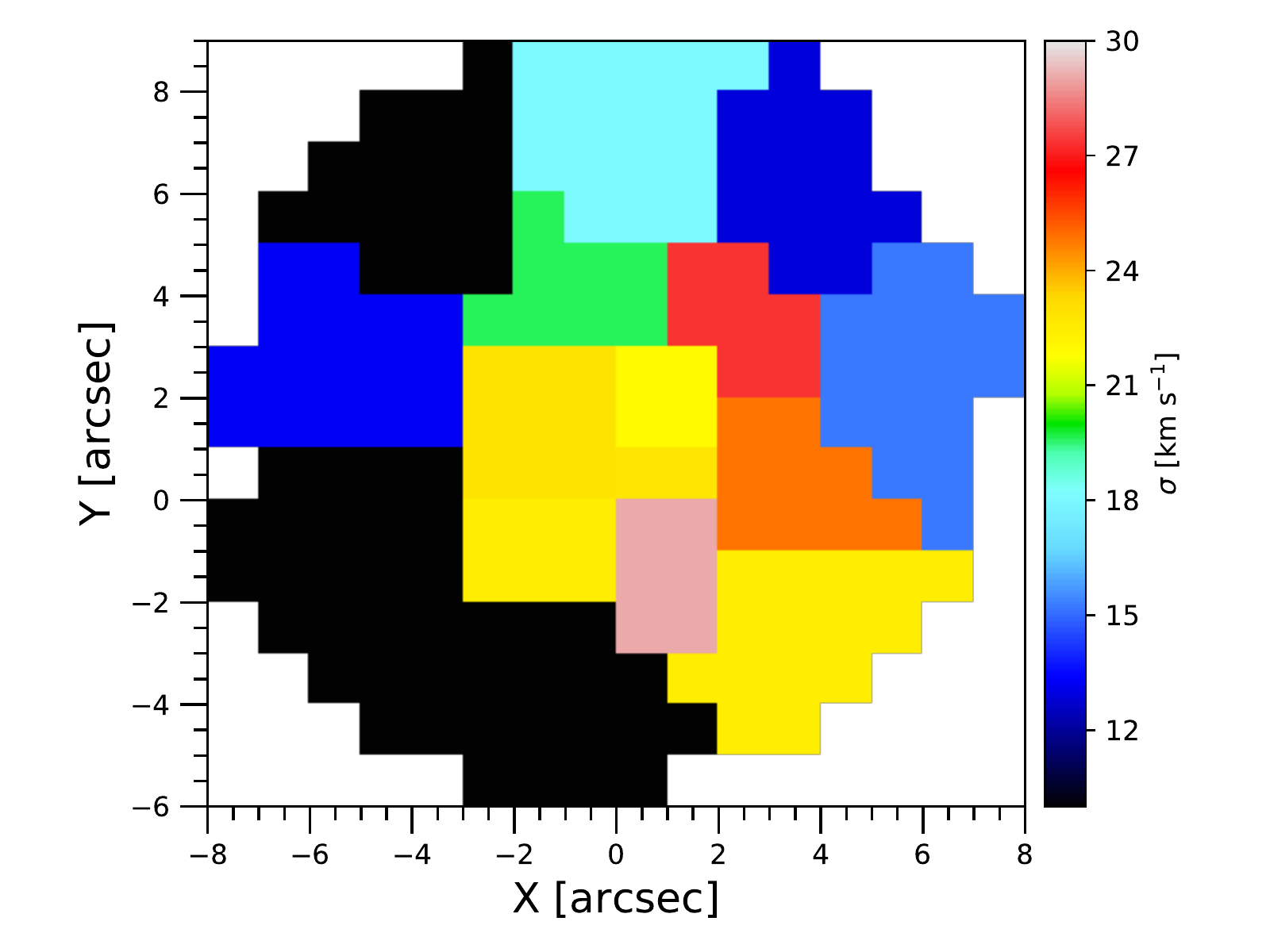}
    
     \includegraphics[width=2.in,clip,trim = 20 0 70 0]{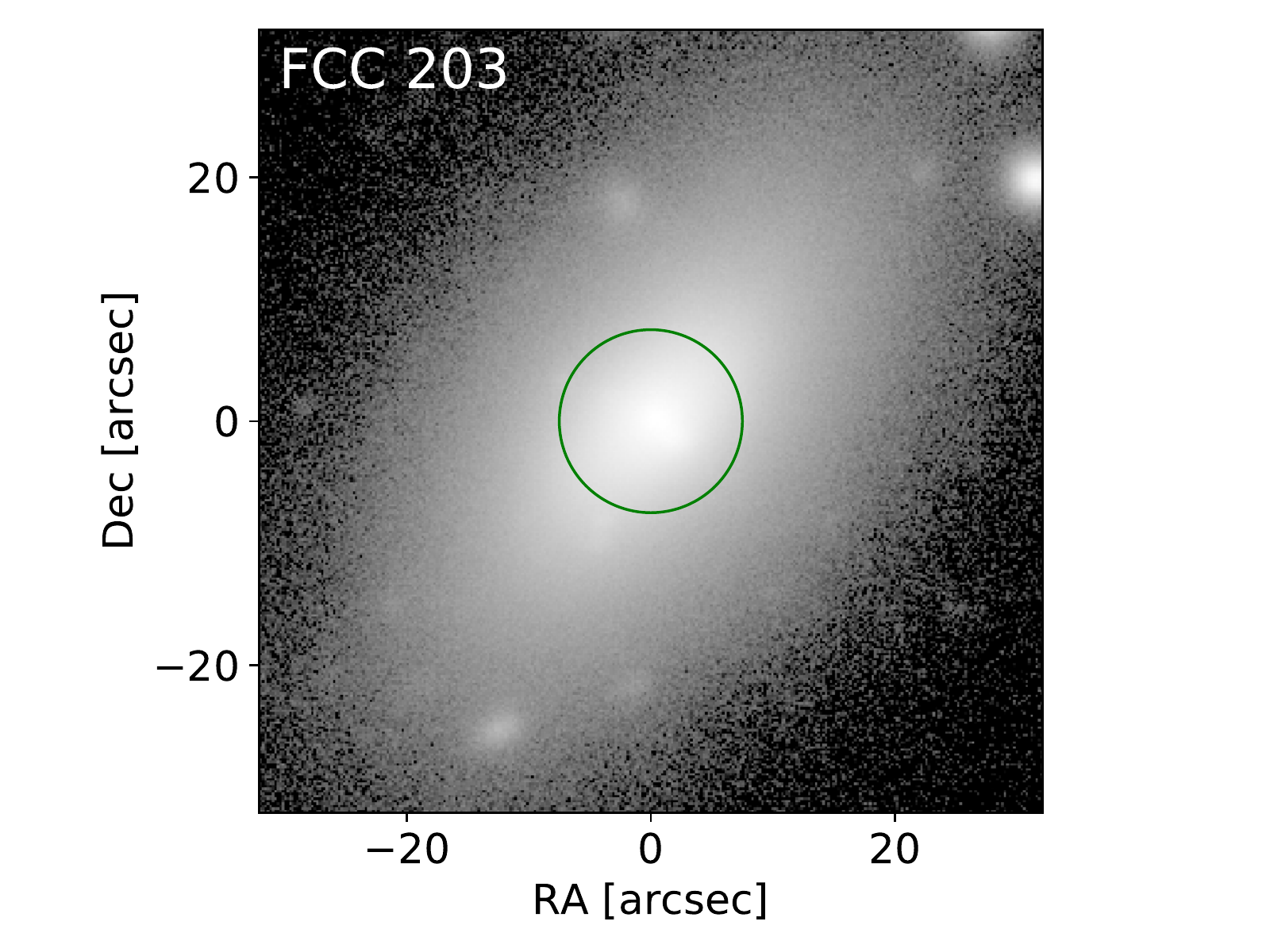}
    \includegraphics[width=2.25in,clip,trim = 20 10 30 10]{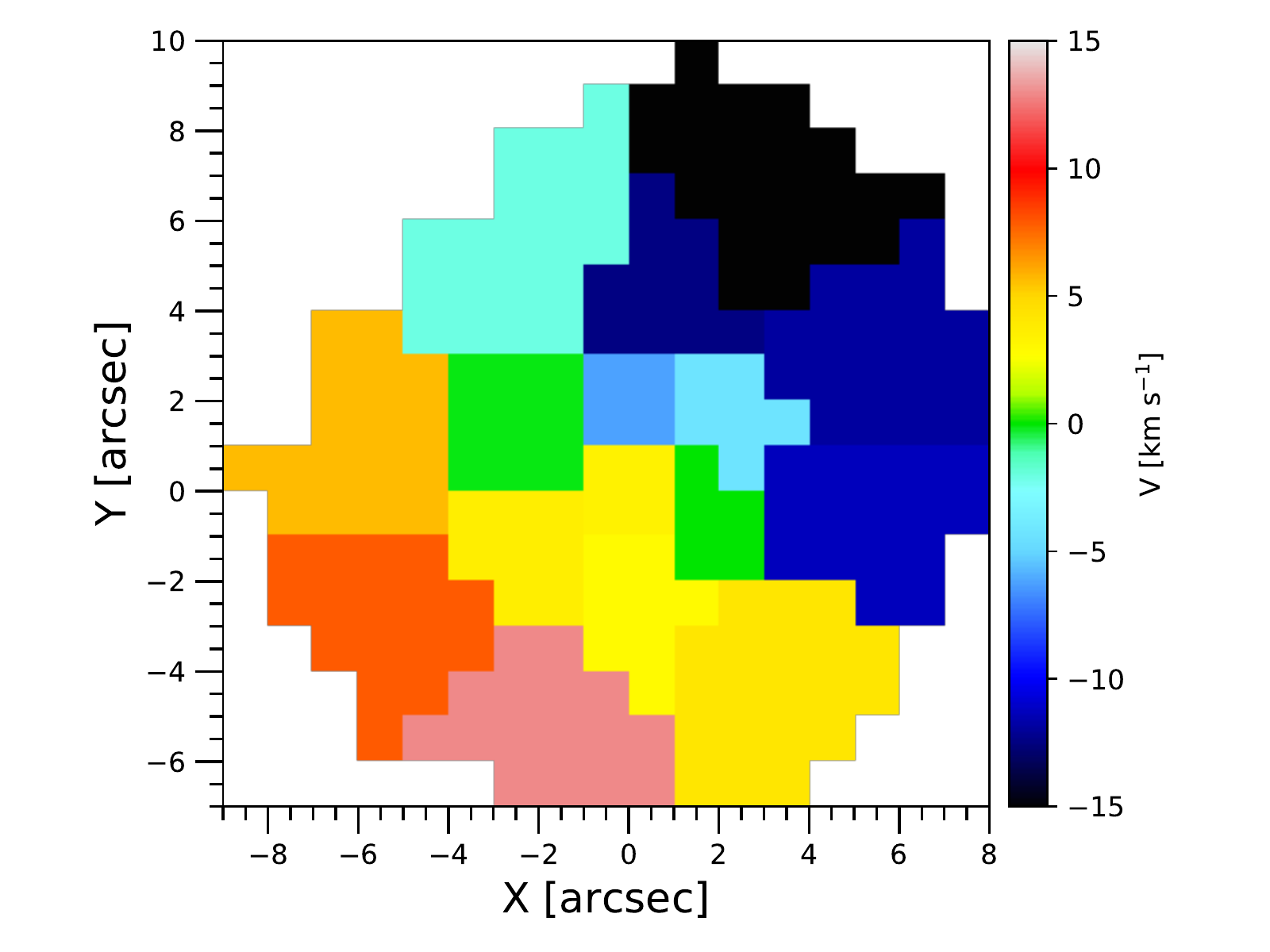}
    \includegraphics[width=2.25in,clip,trim = 20 10 30 10]{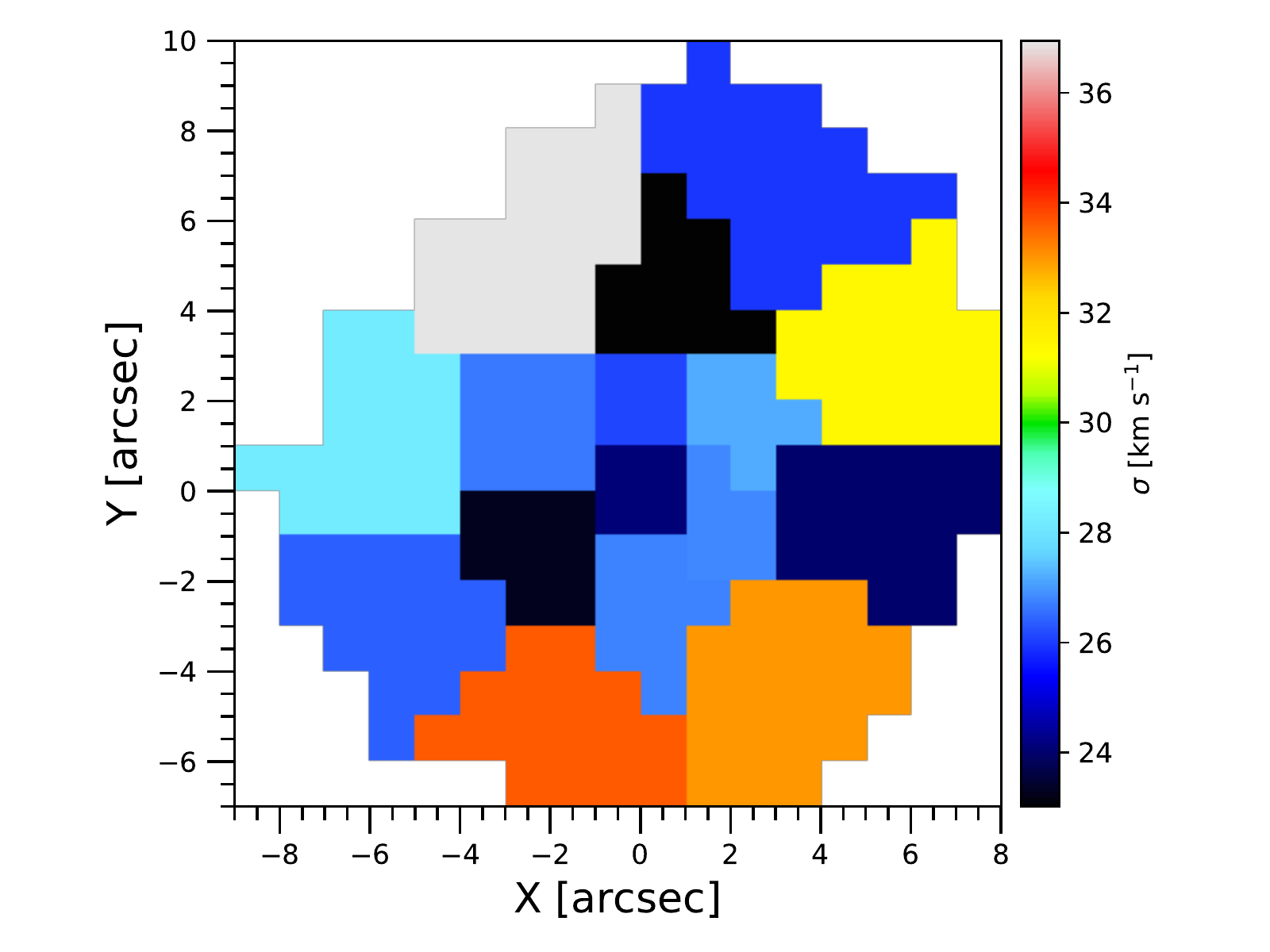}   
    \vspace{3mm}
    {\bf Figure \ref{fig:primary_maps_app}.} continued
\end{figure*}

\begin{figure*}
    \centering
    \includegraphics[width=2.in,clip,trim = 20 0 70 0]{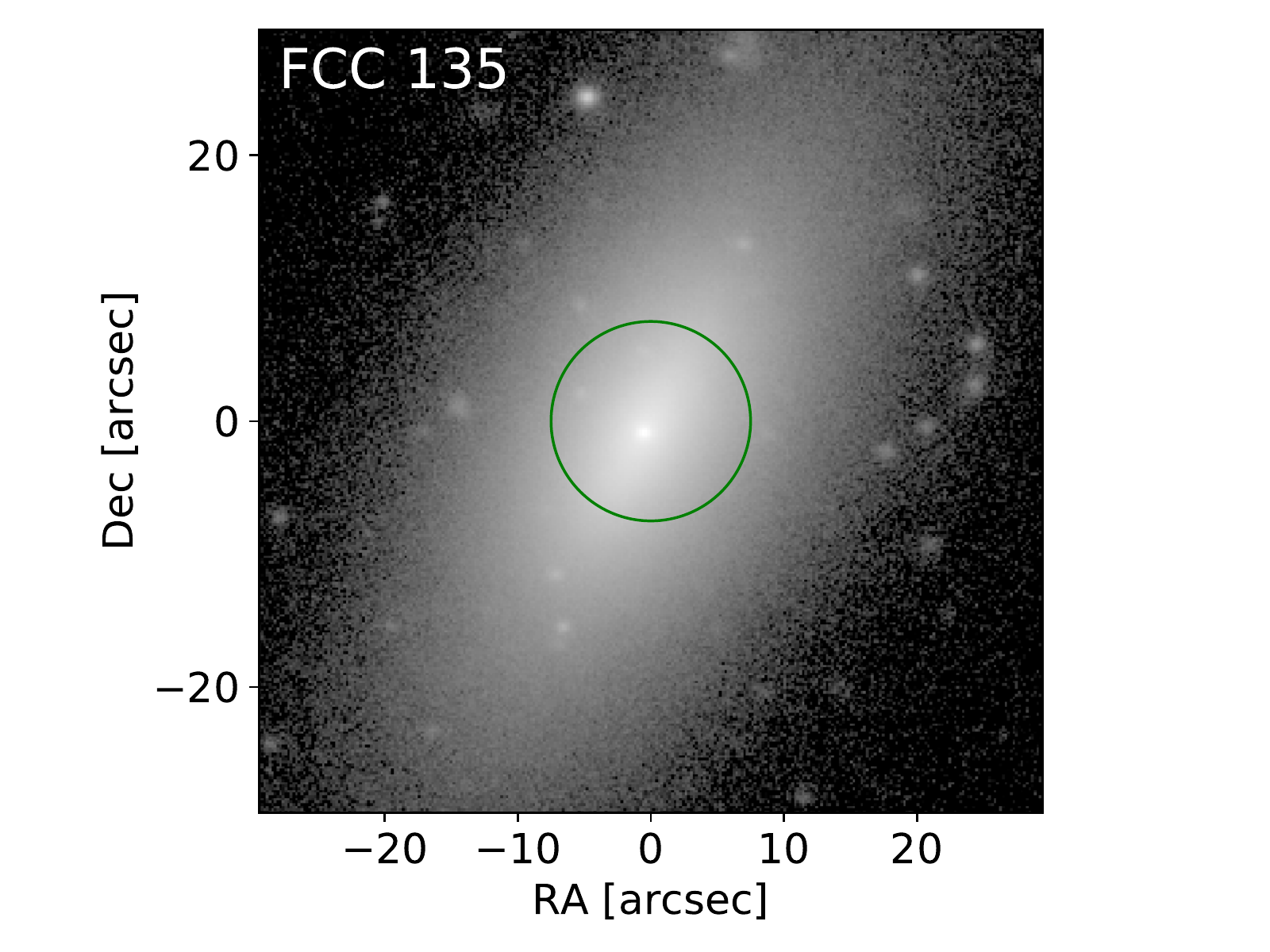}
    \includegraphics[width=2.25in,clip,trim = 20 10 40 10]{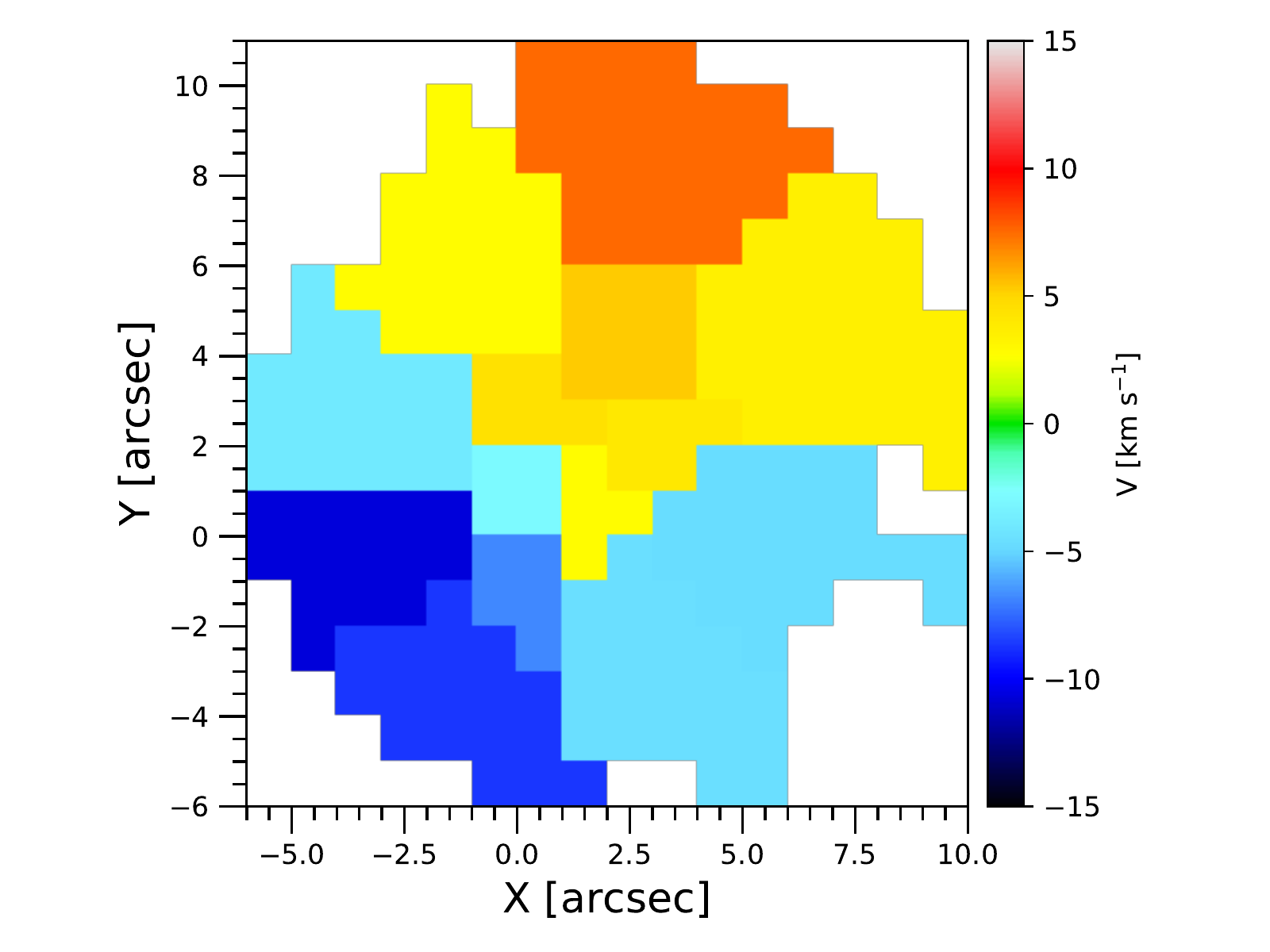}
    \includegraphics[width=2.25in,clip,trim = 20 10 40 10]{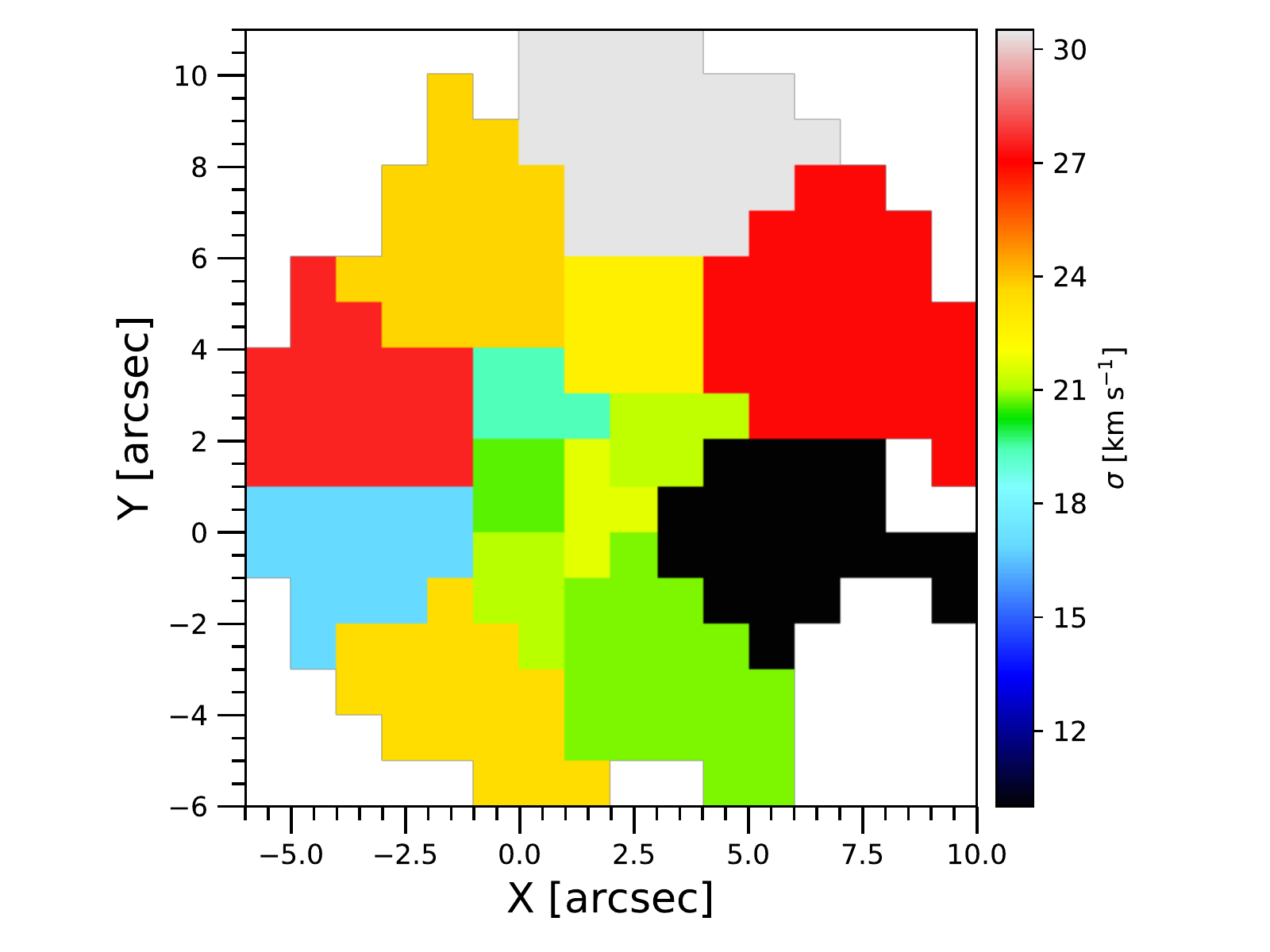}
    
    \includegraphics[width=2.in,clip,trim = 20 0 70 0]{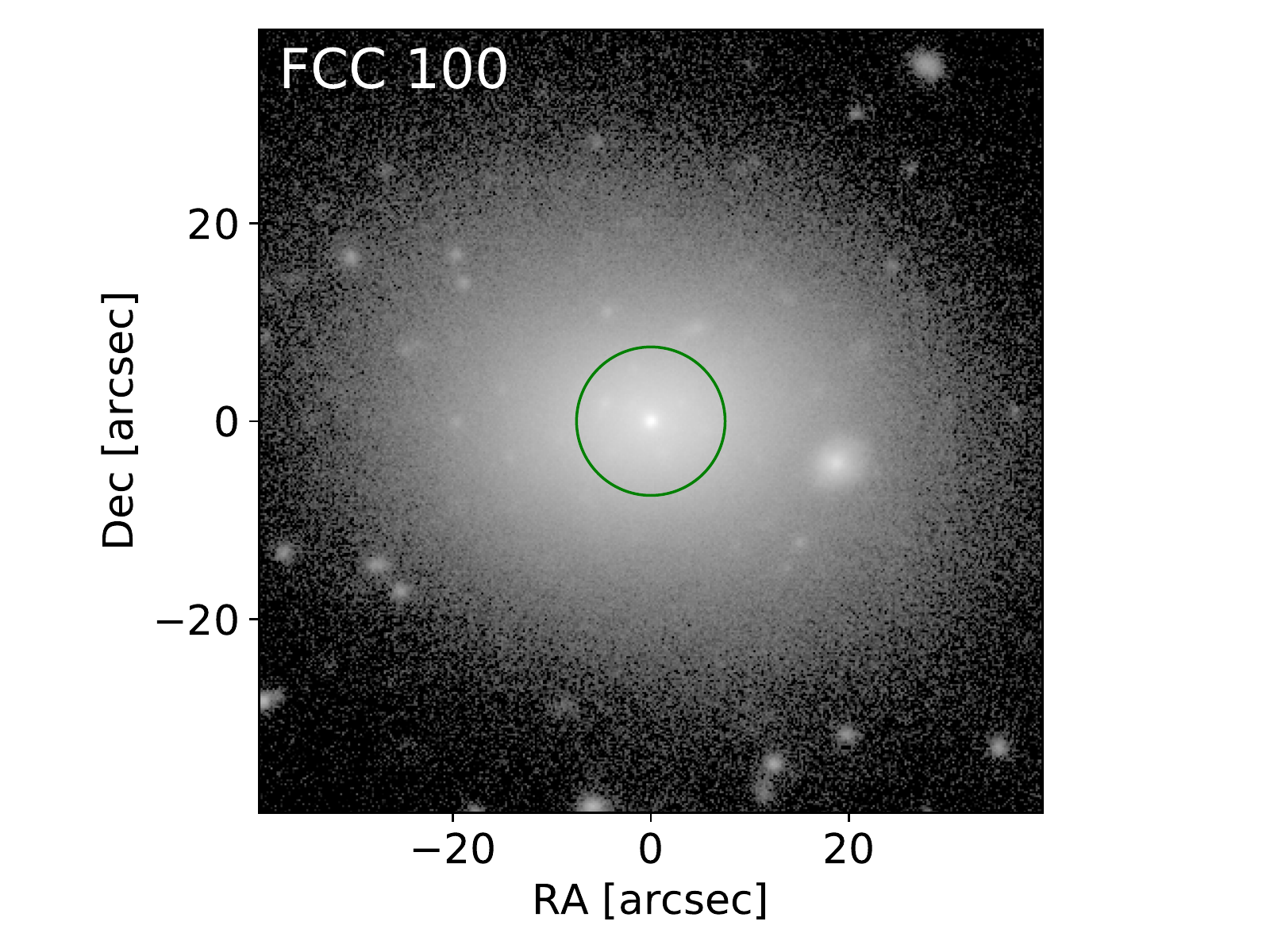}
    \includegraphics[width=2.25in,clip,trim = 10 10 10 10]{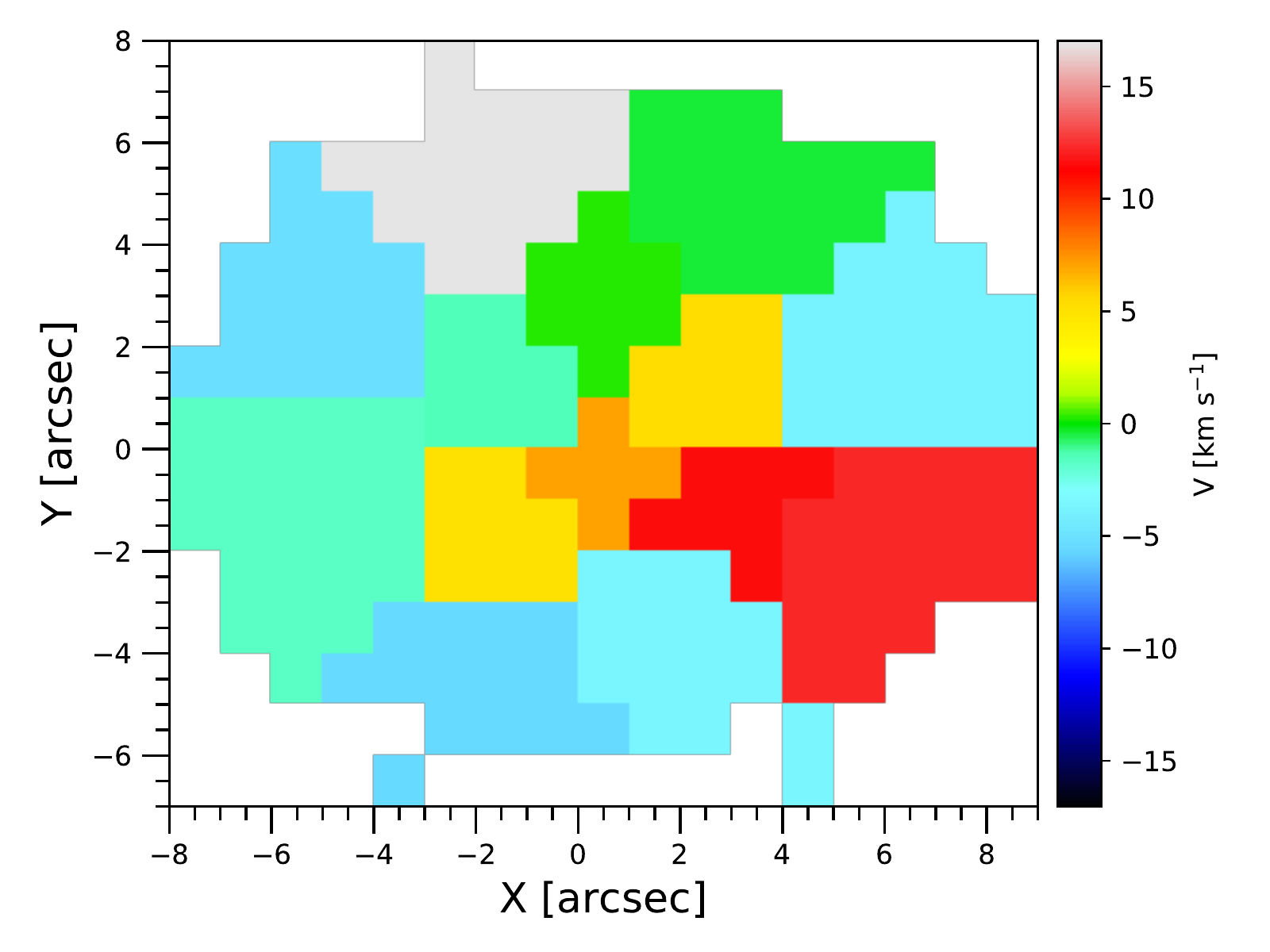}
    \includegraphics[width=2.25in,clip,trim = 20 10 10 10]{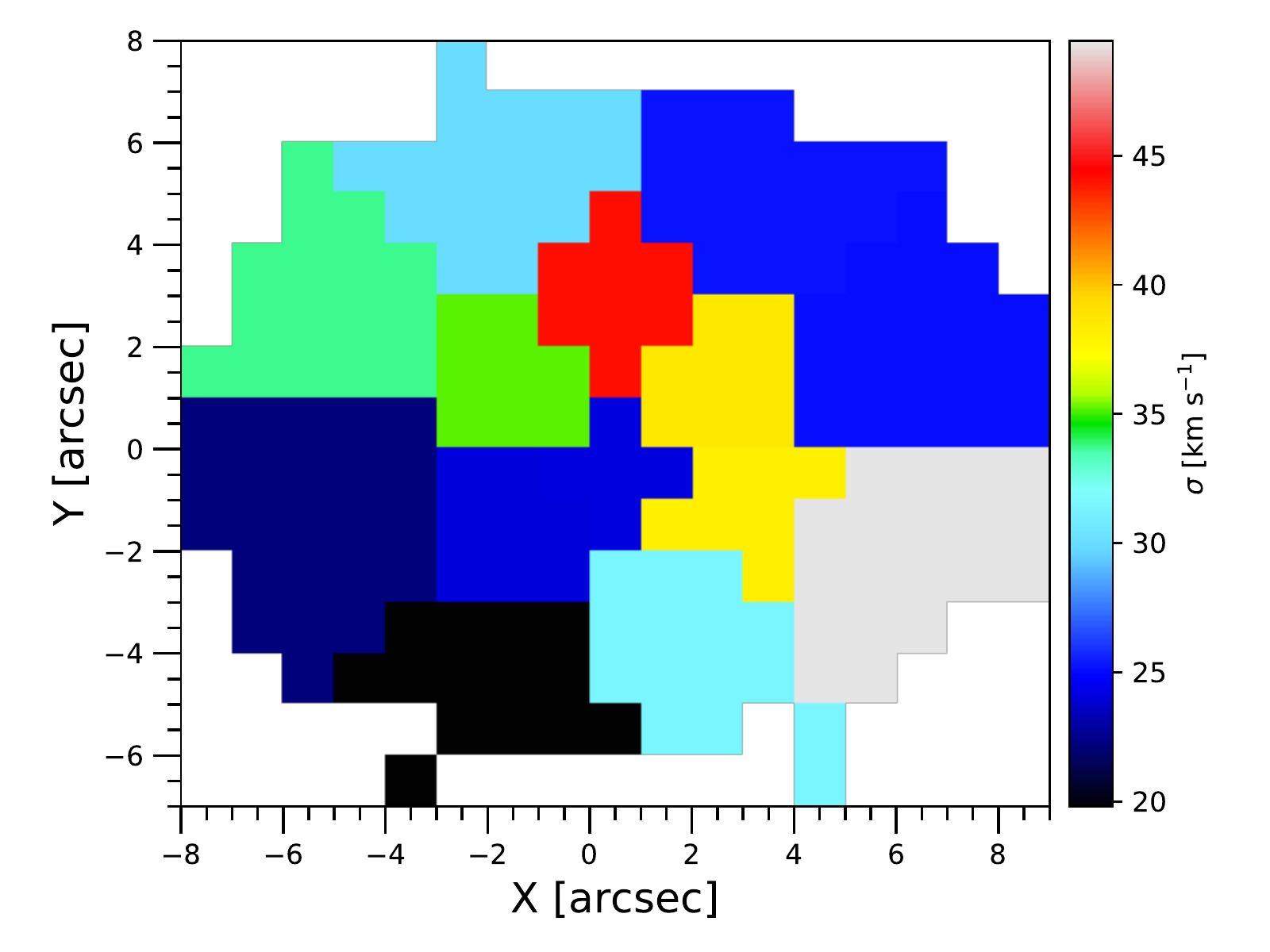}
    
    \includegraphics[width=2.in,clip,trim = 20 0 70 0]{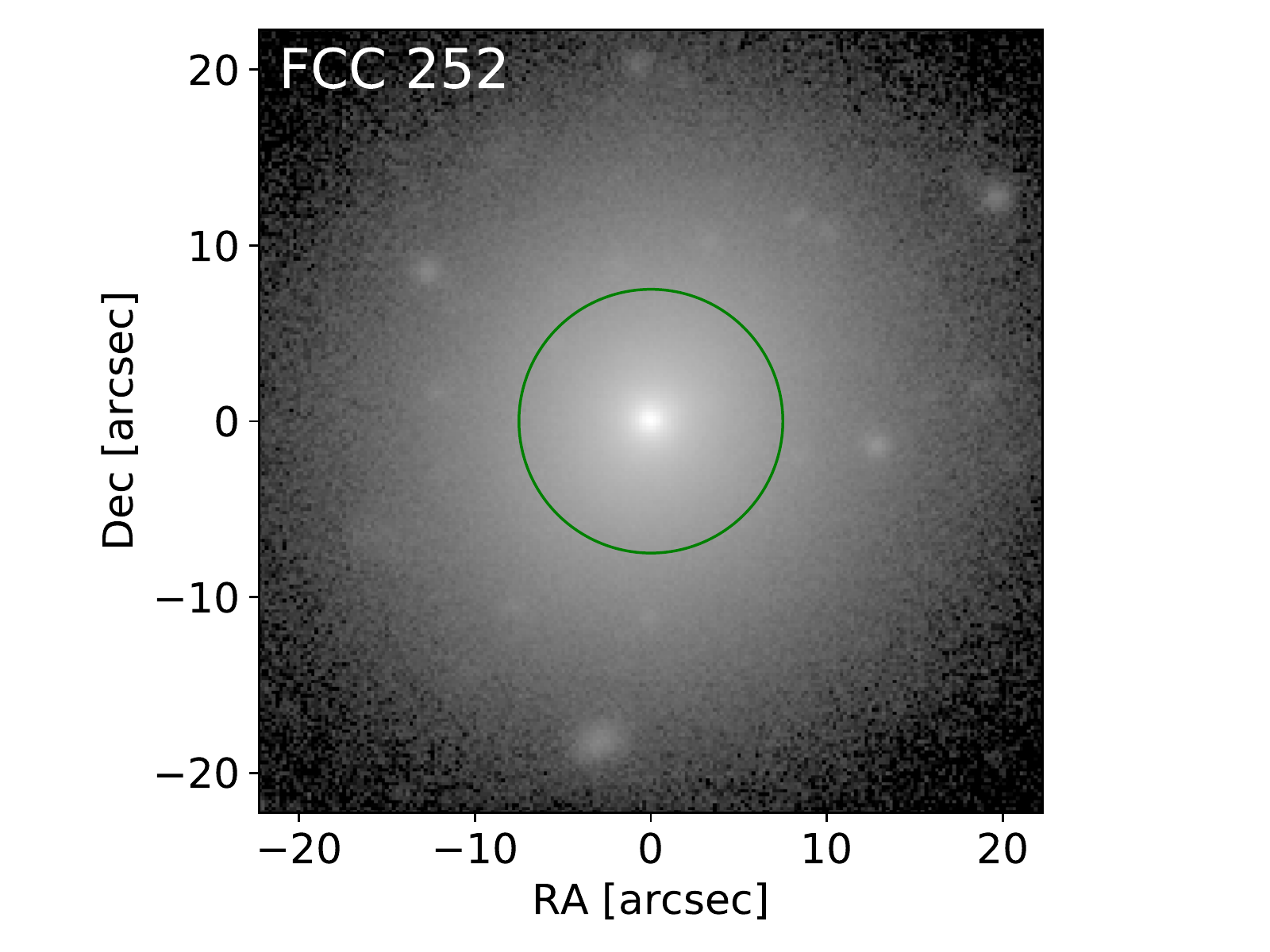}
    \includegraphics[width=2.25in,clip,trim = 20 10 30 10]{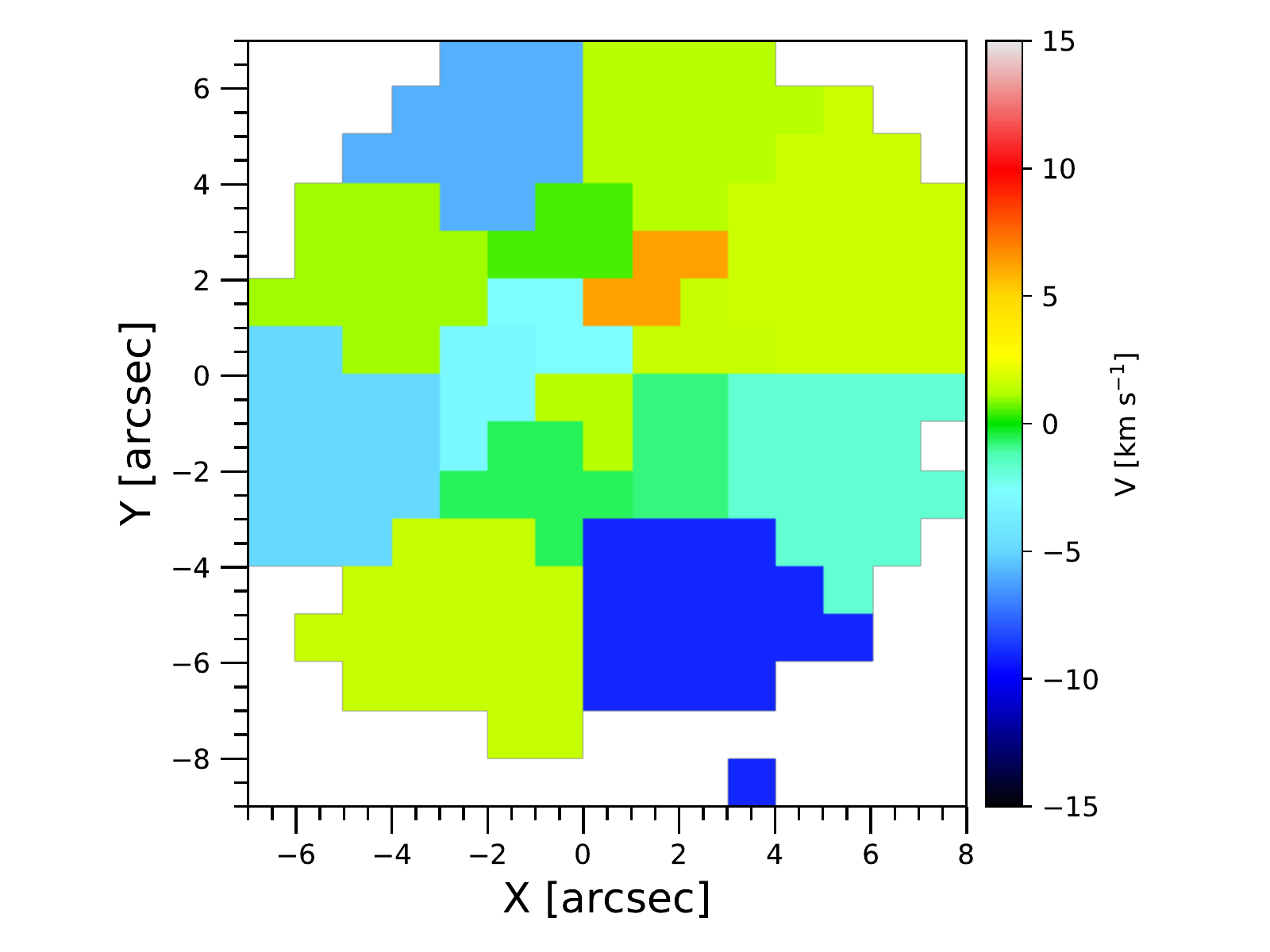}
    \includegraphics[width=2.25in,clip,trim = 20 10 30 10]{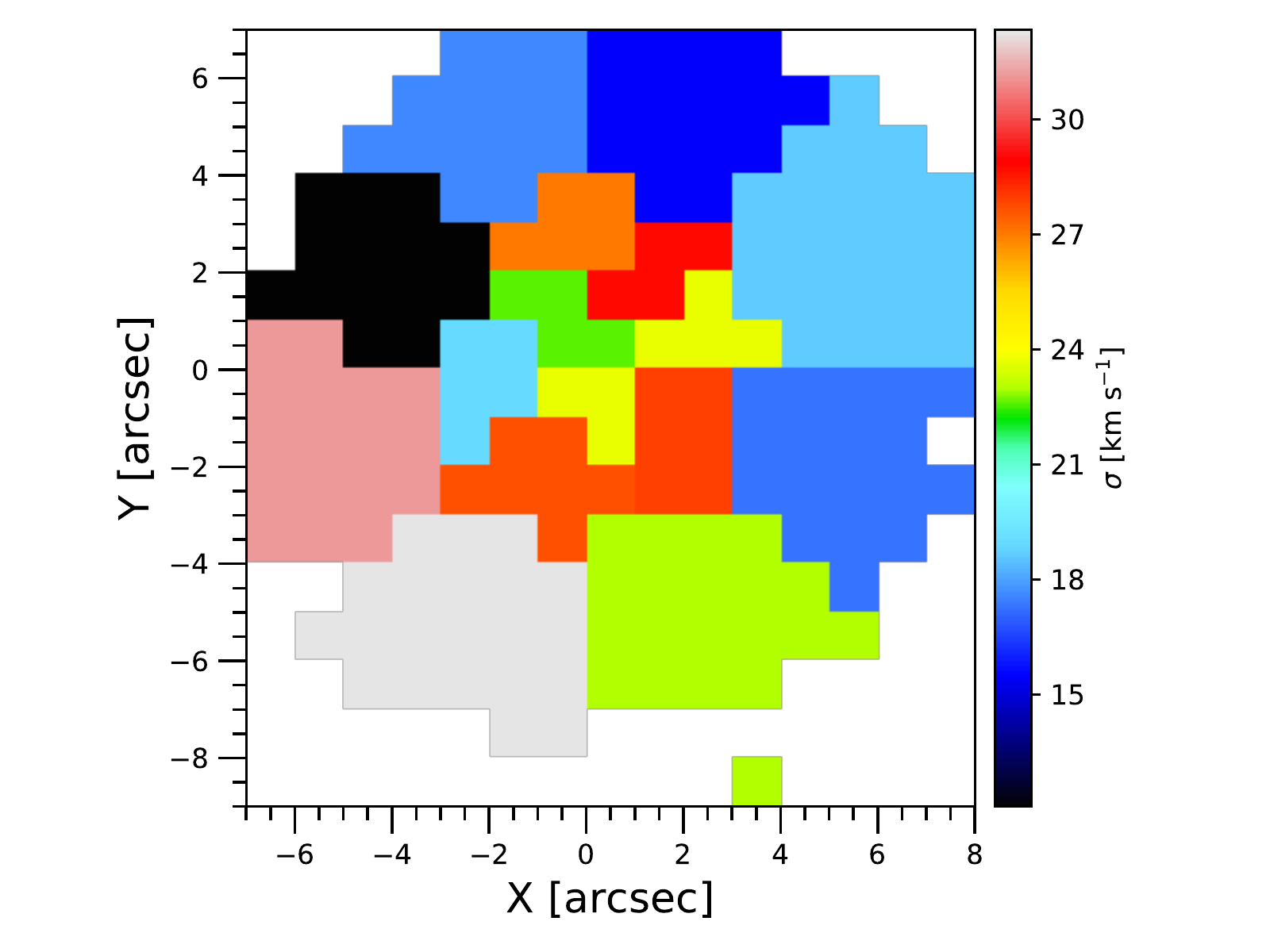}
    
    \includegraphics[width=2.in,clip,trim = 20 0 70 0]{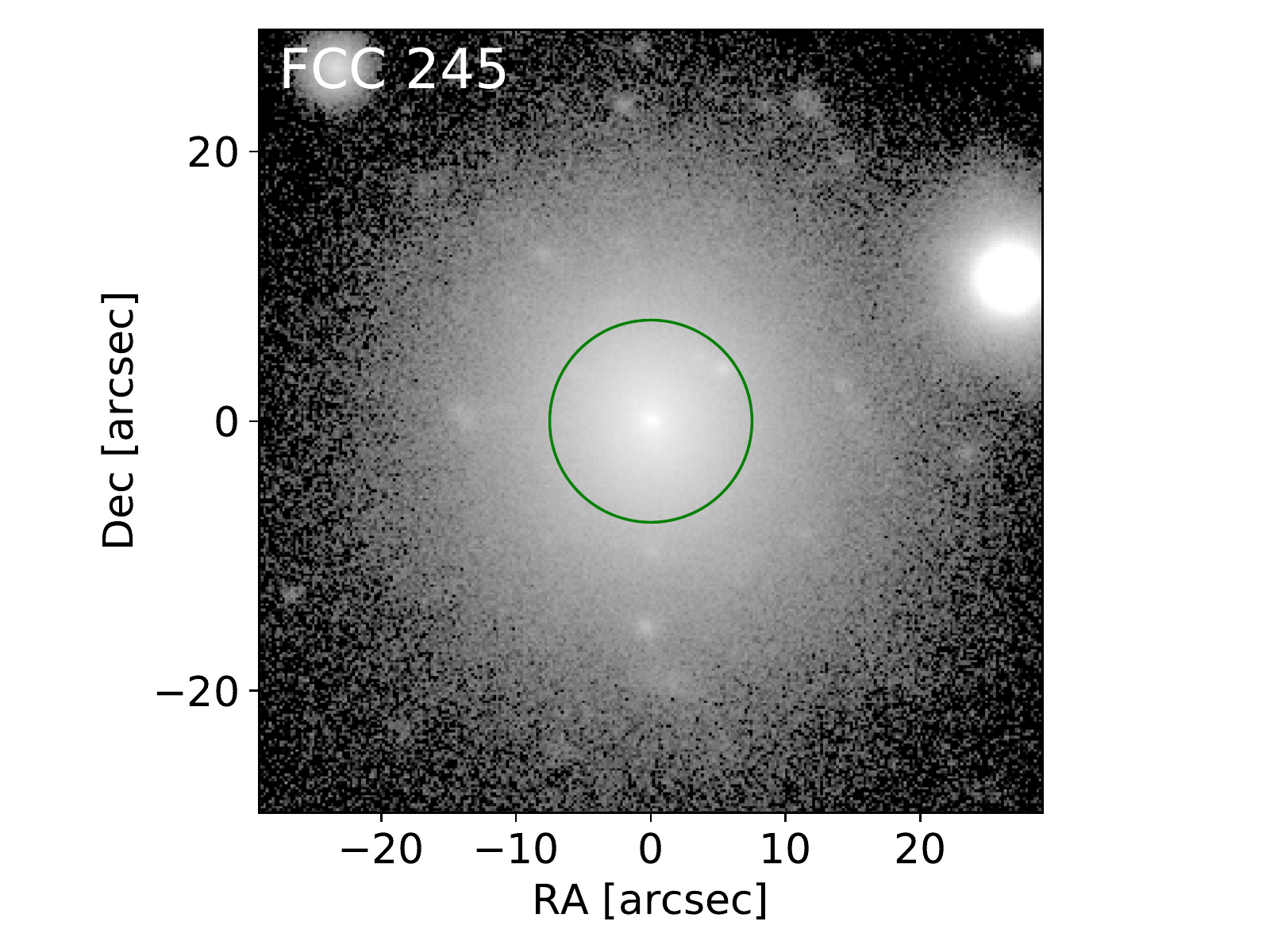}
    \includegraphics[width=2.25in,clip,trim = 20 10 40 10]{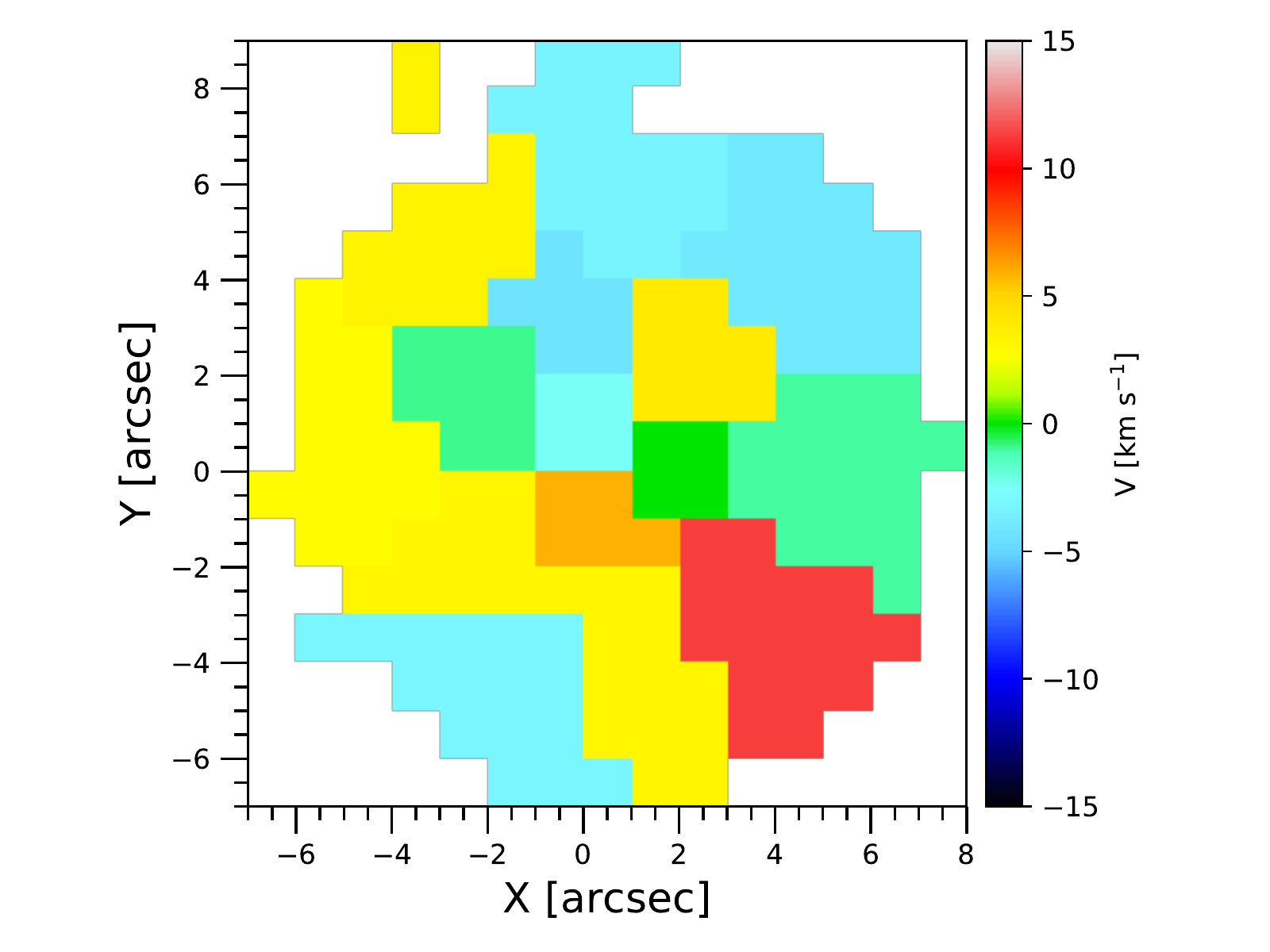}
    \includegraphics[width=2.25in,clip,trim = 20 10 40 10]{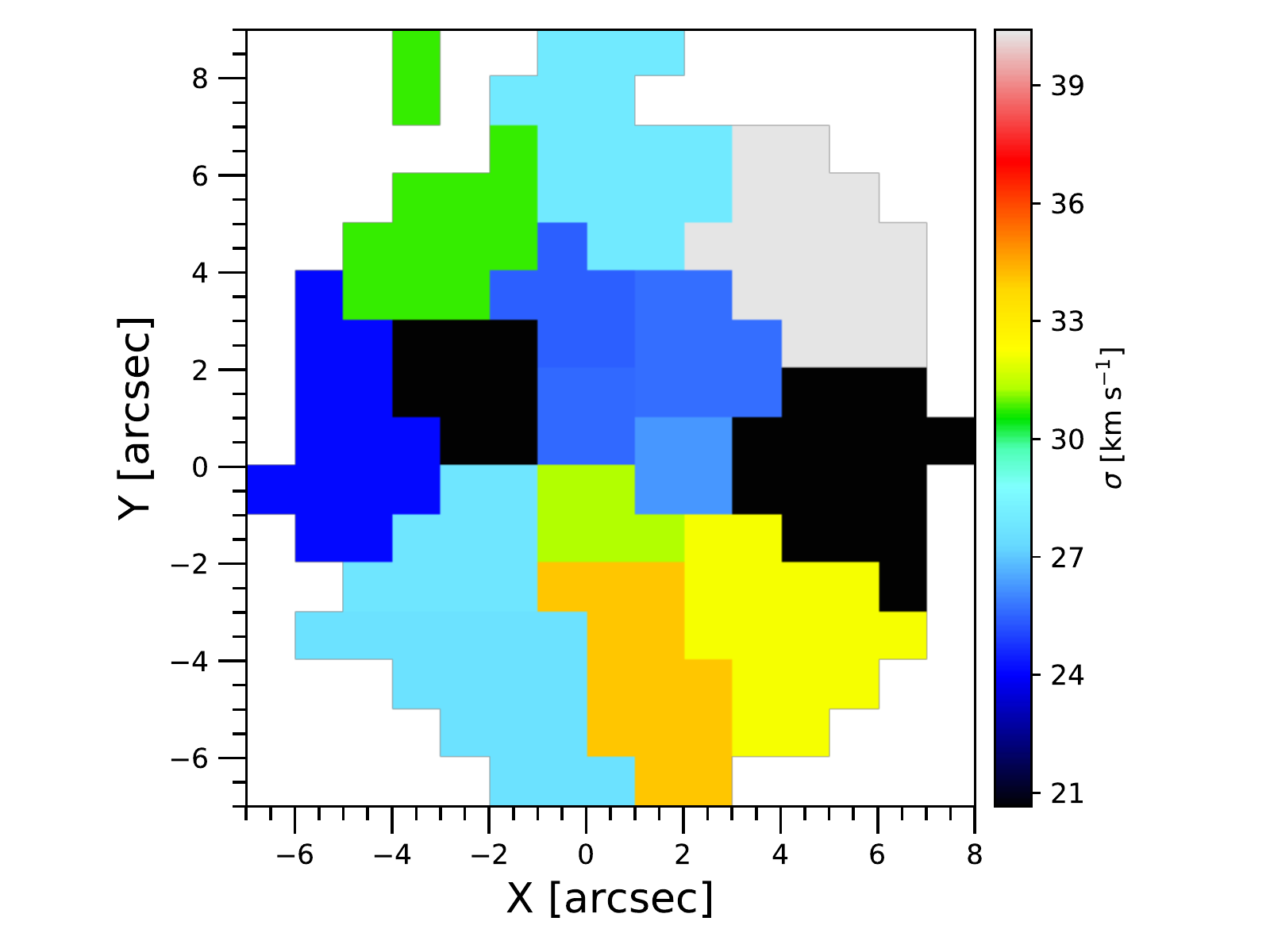}
    
    \includegraphics[width=2.in,clip,trim = 20 0 70 0]{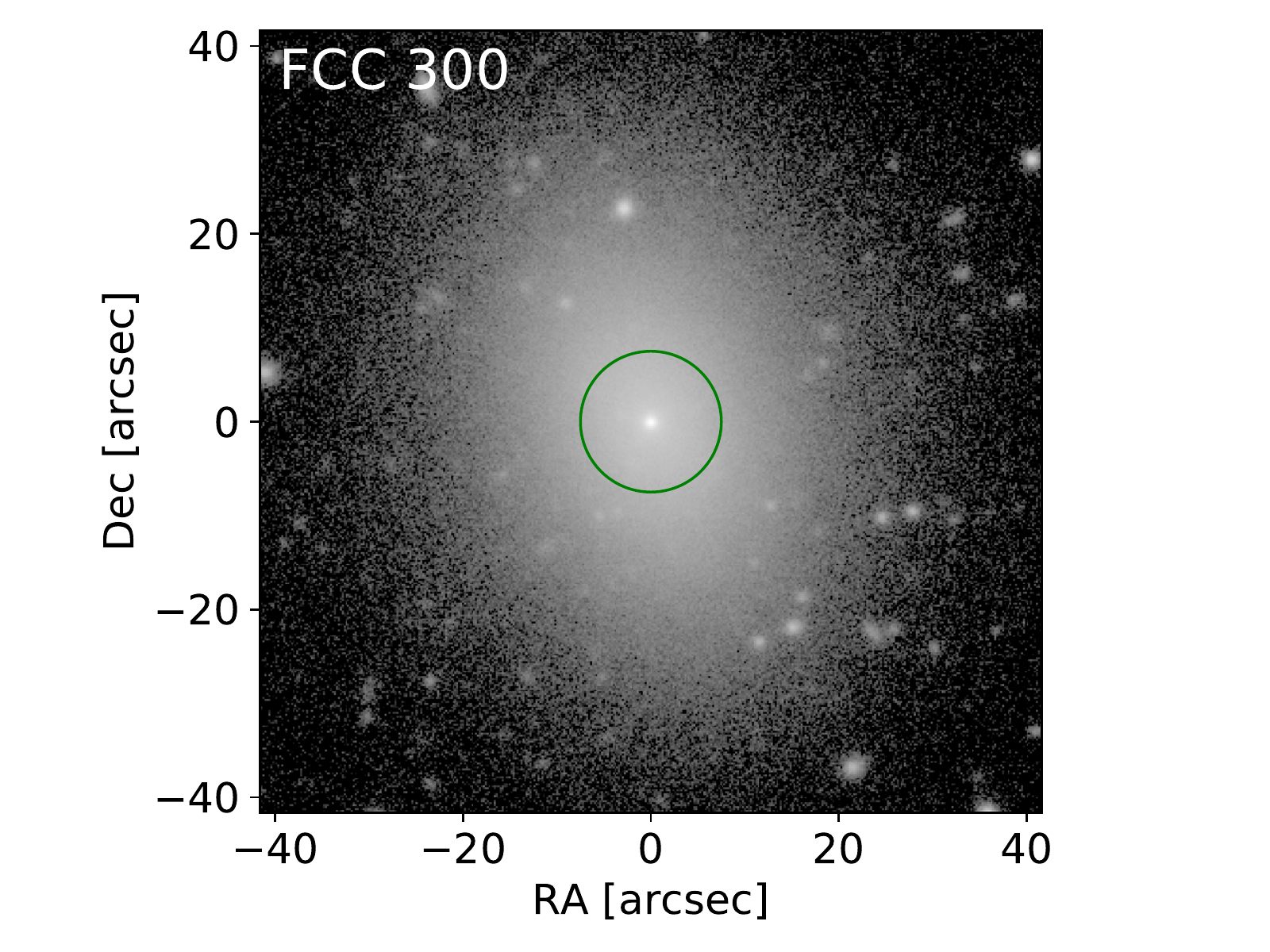}
    \includegraphics[width=2.25in,clip,trim = 30 10 30 10]{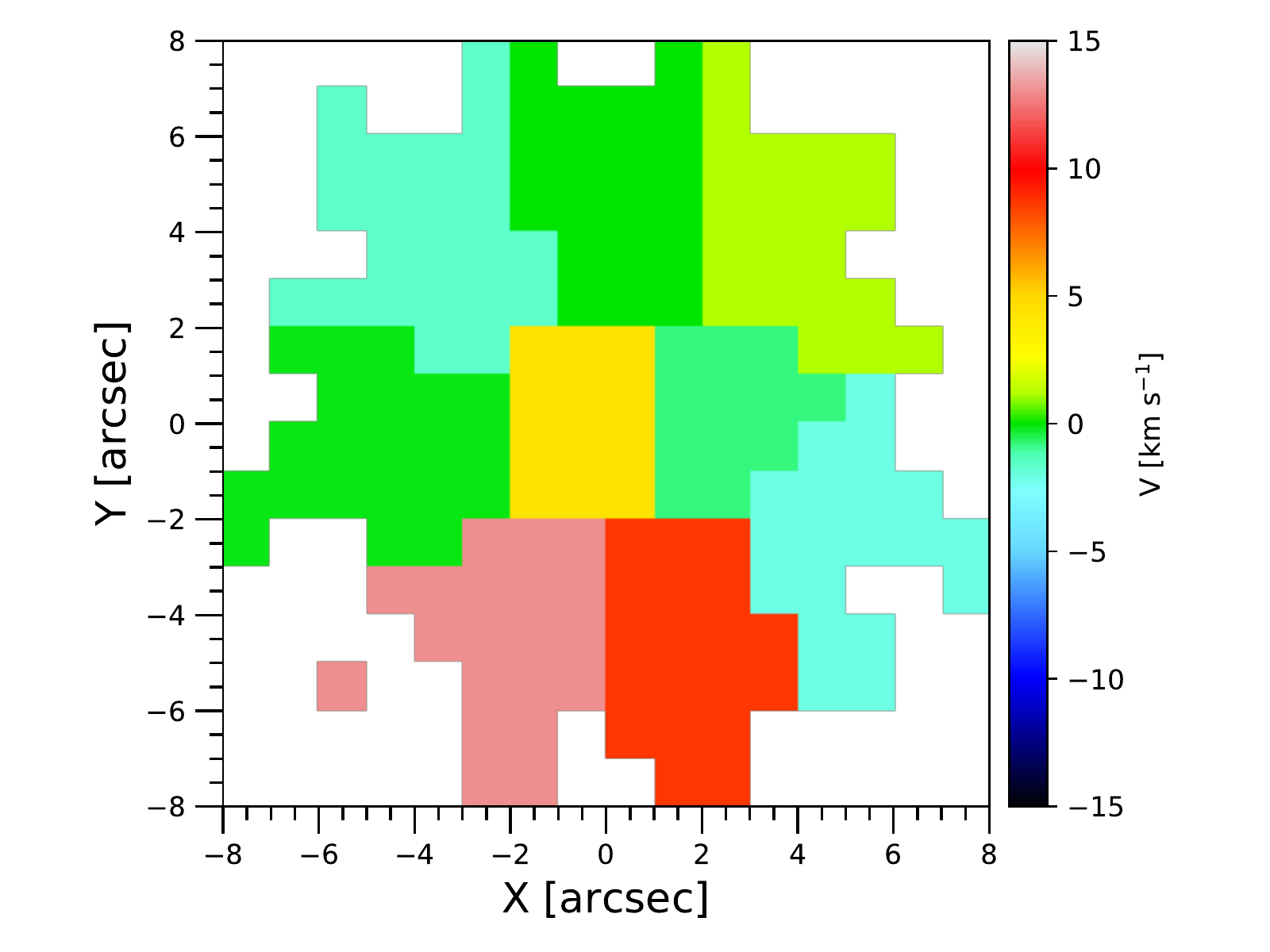}
    \includegraphics[width=2.25in,clip,trim = 30 10 30 10]{figures/maps/300_vel_map.pdf}

    \vspace{3mm}
    {\bf Figure \ref{fig:primary_maps_app}.} continued
\end{figure*}

\begin{figure*}
    \centering
    \includegraphics[width=2.in,clip,trim = 20 0 70 0]{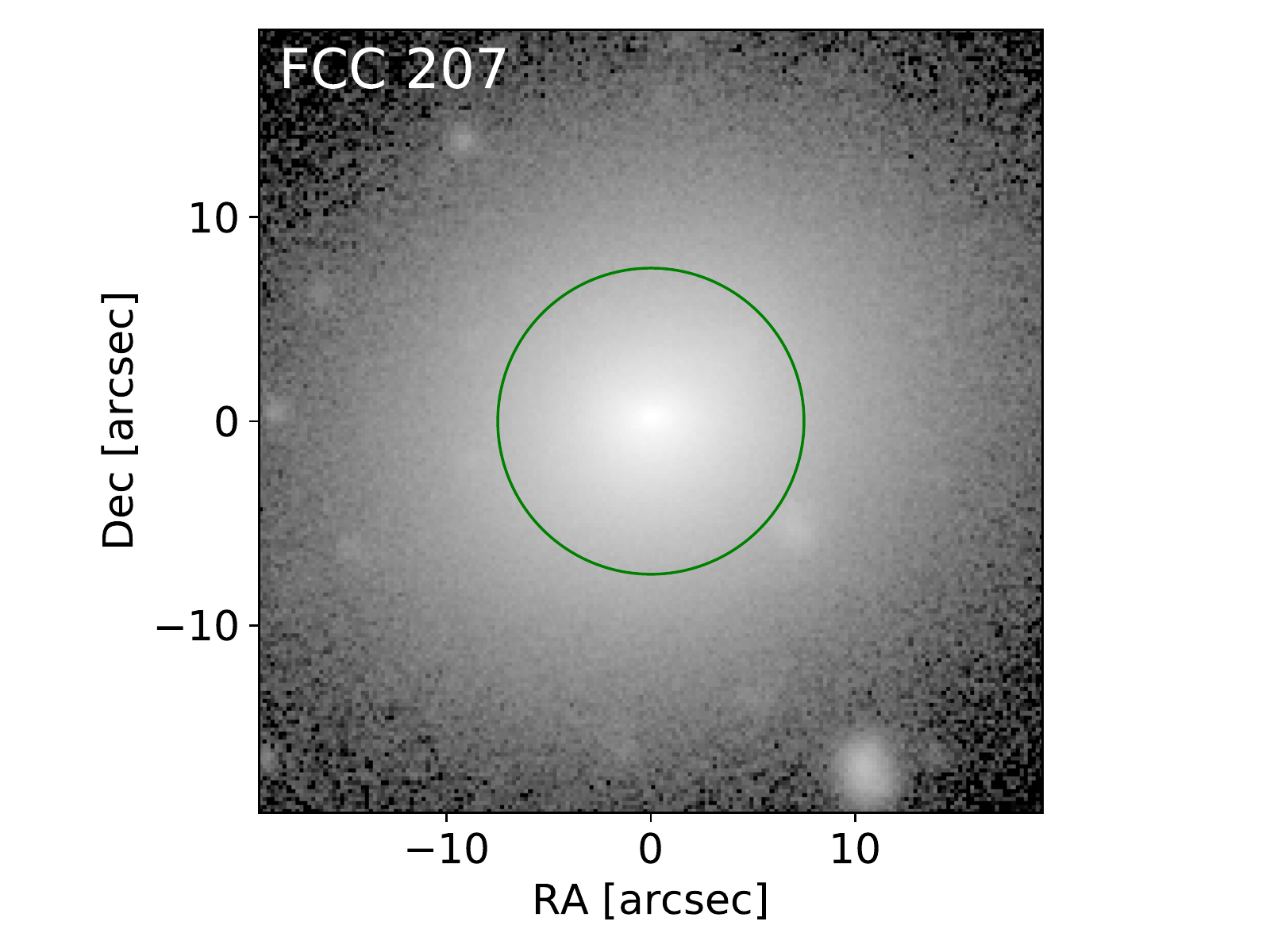}
    \includegraphics[width=2.25in,clip,trim = 20 10 30 10]{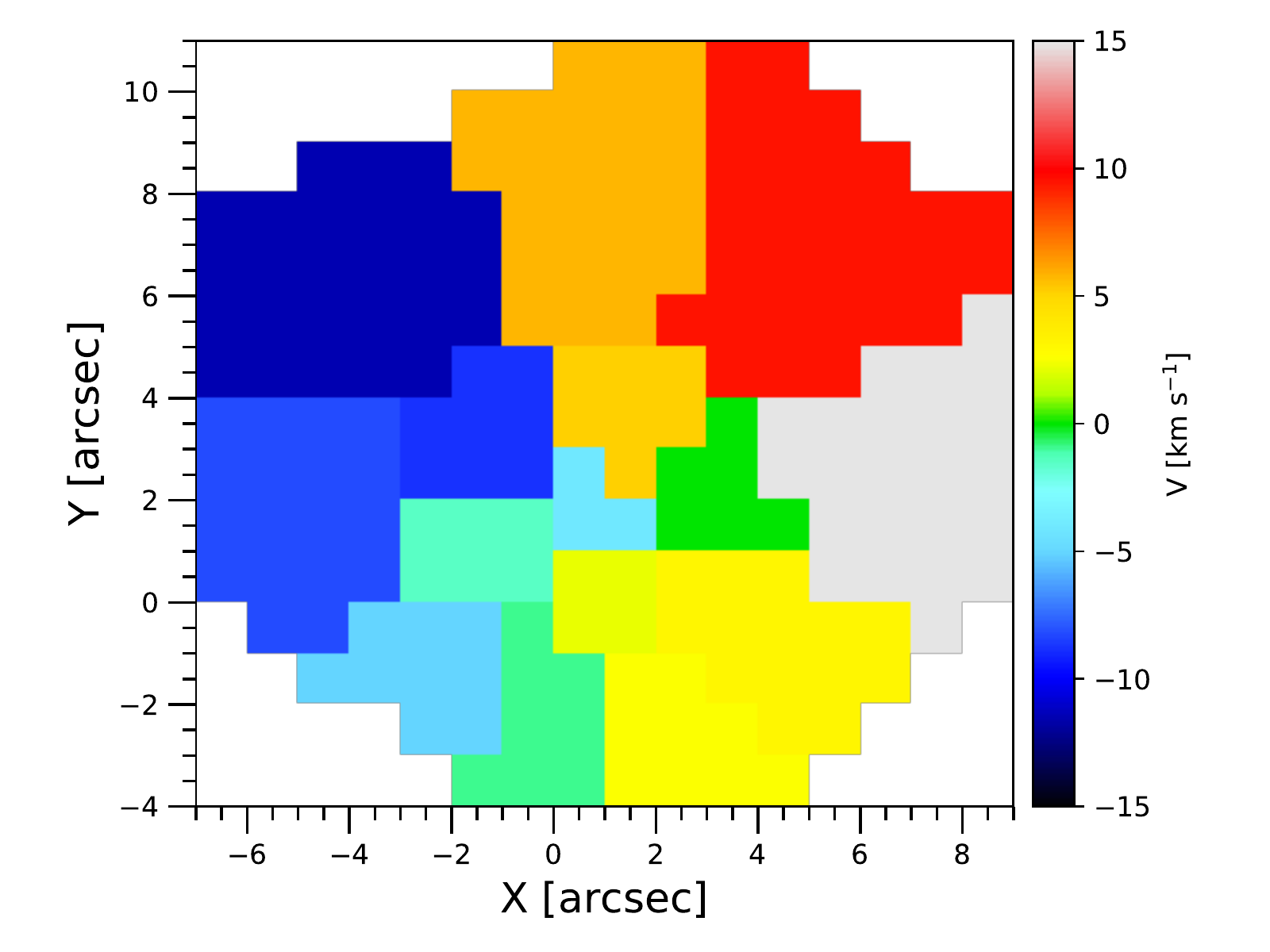}
    \includegraphics[width=2.25in,clip,trim = 20 10 30 10]{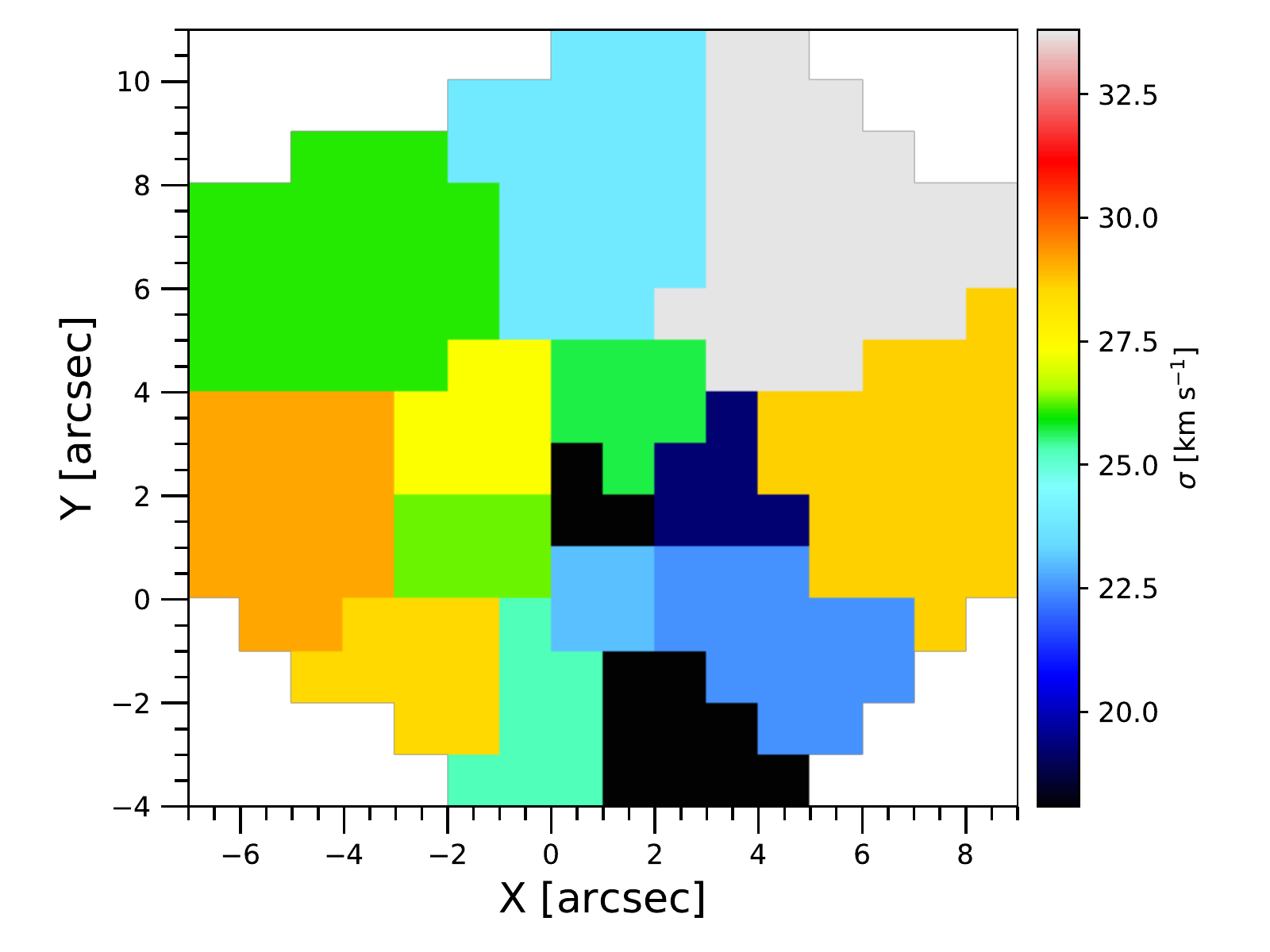}
    
    \includegraphics[width=2.in,clip,trim = 20 0 70 0]{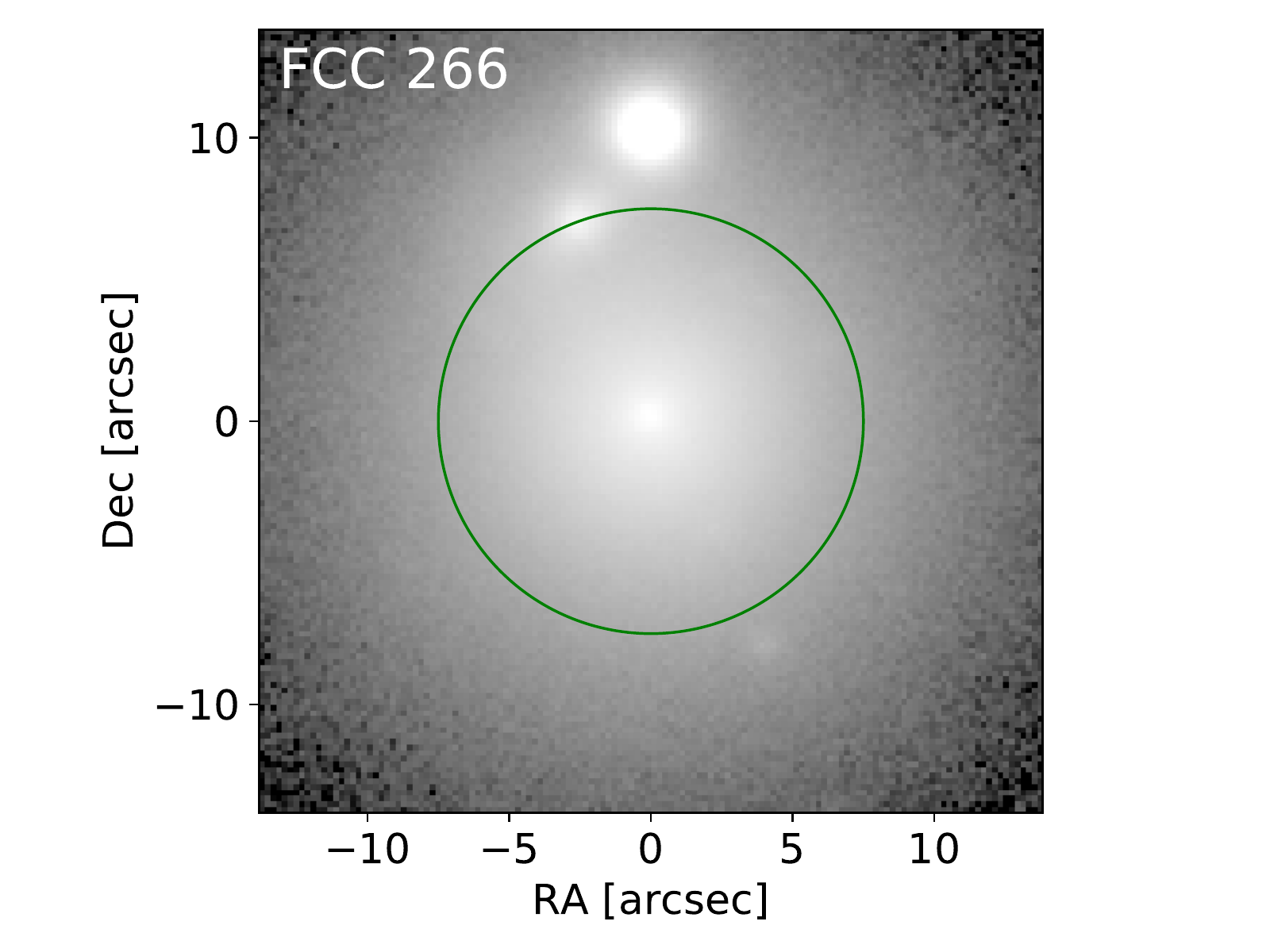}
    \includegraphics[width=2.25in,clip,trim = 20 10 30 10]{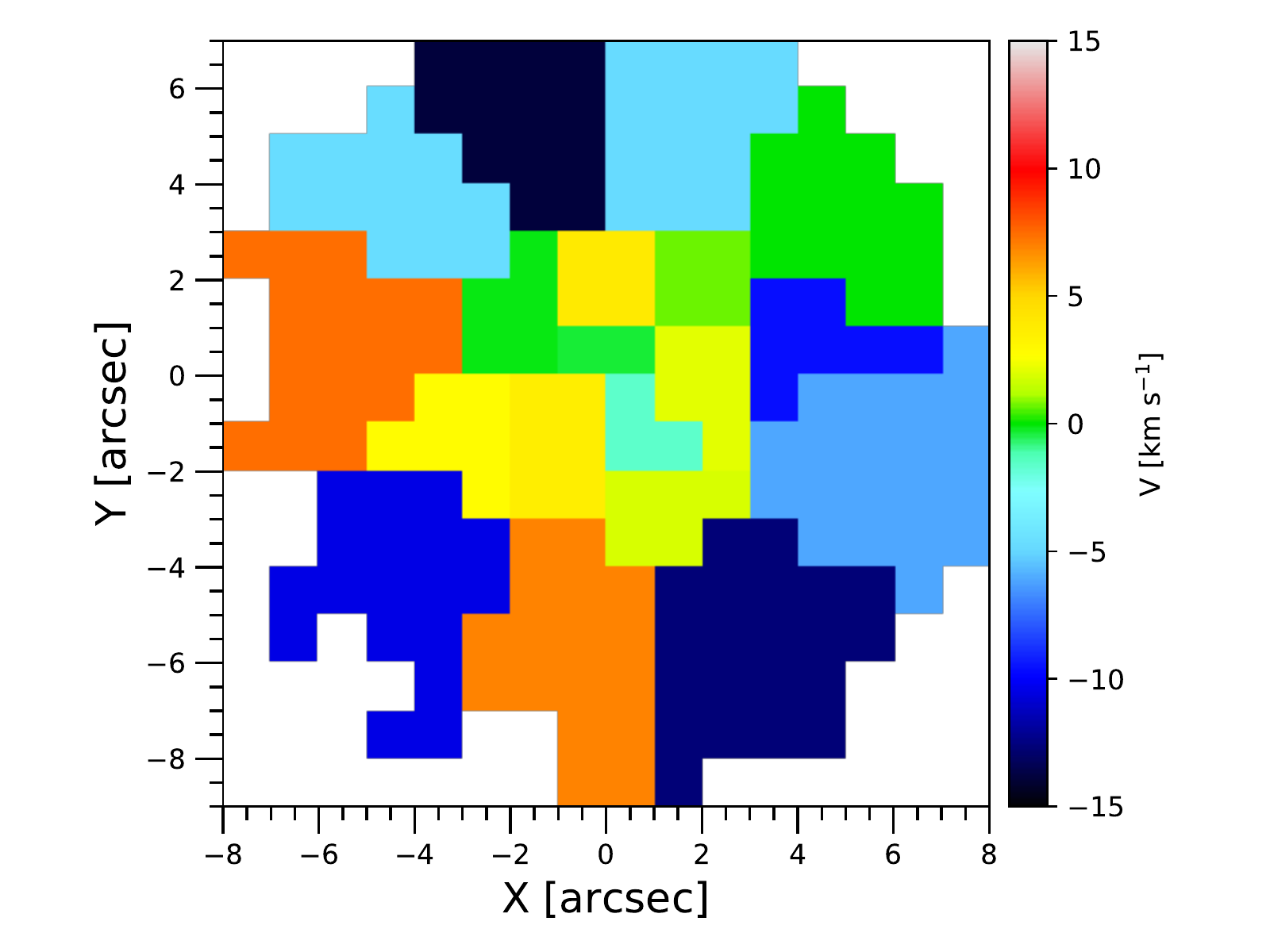}
    \includegraphics[width=2.25in,clip,trim = 20 10 30 10]{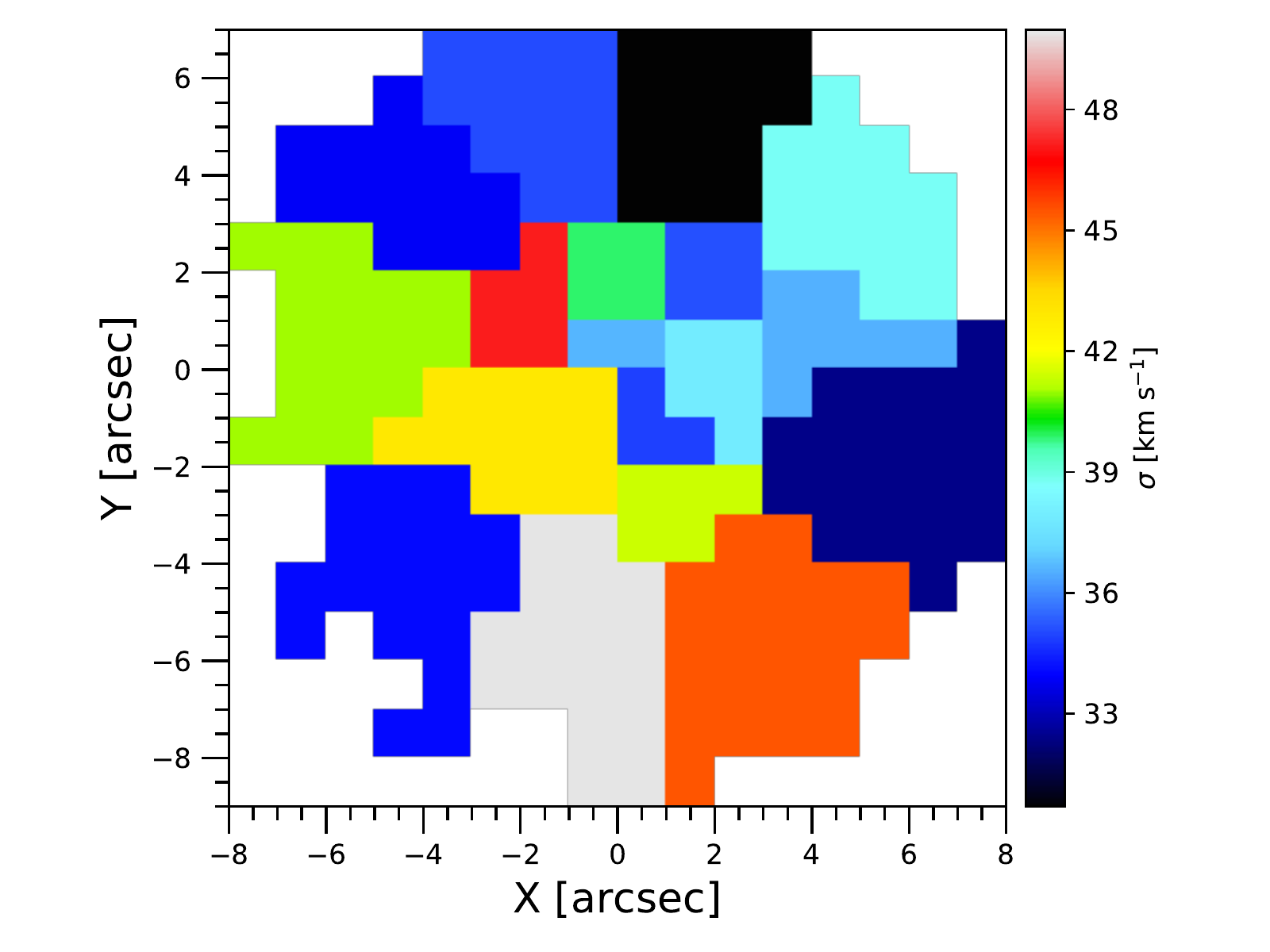}
    
    \includegraphics[width=2.in,clip,trim = 20 0 70 0]{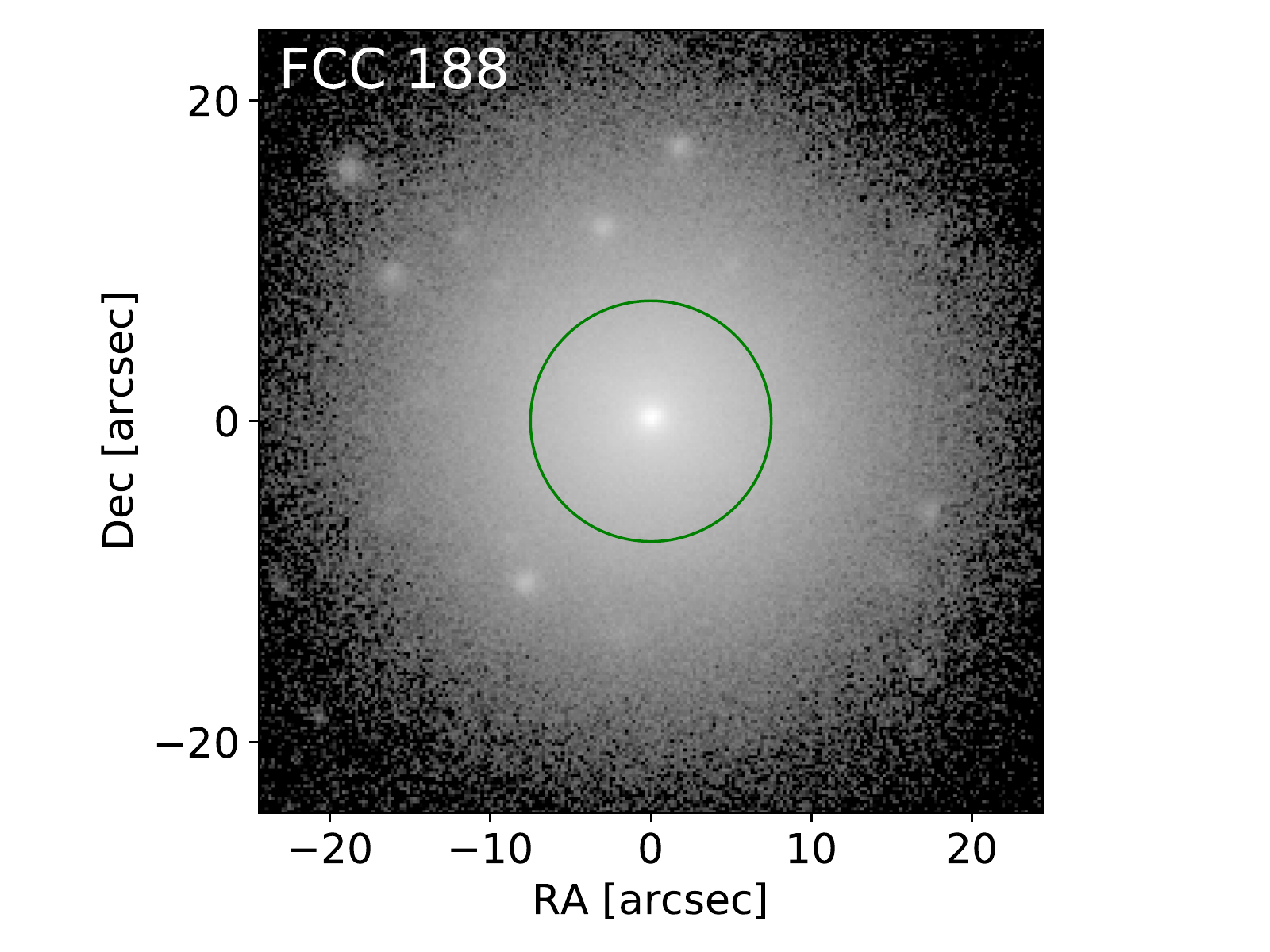}
    \includegraphics[width=2.25in,clip,trim = 10 10 20 10]{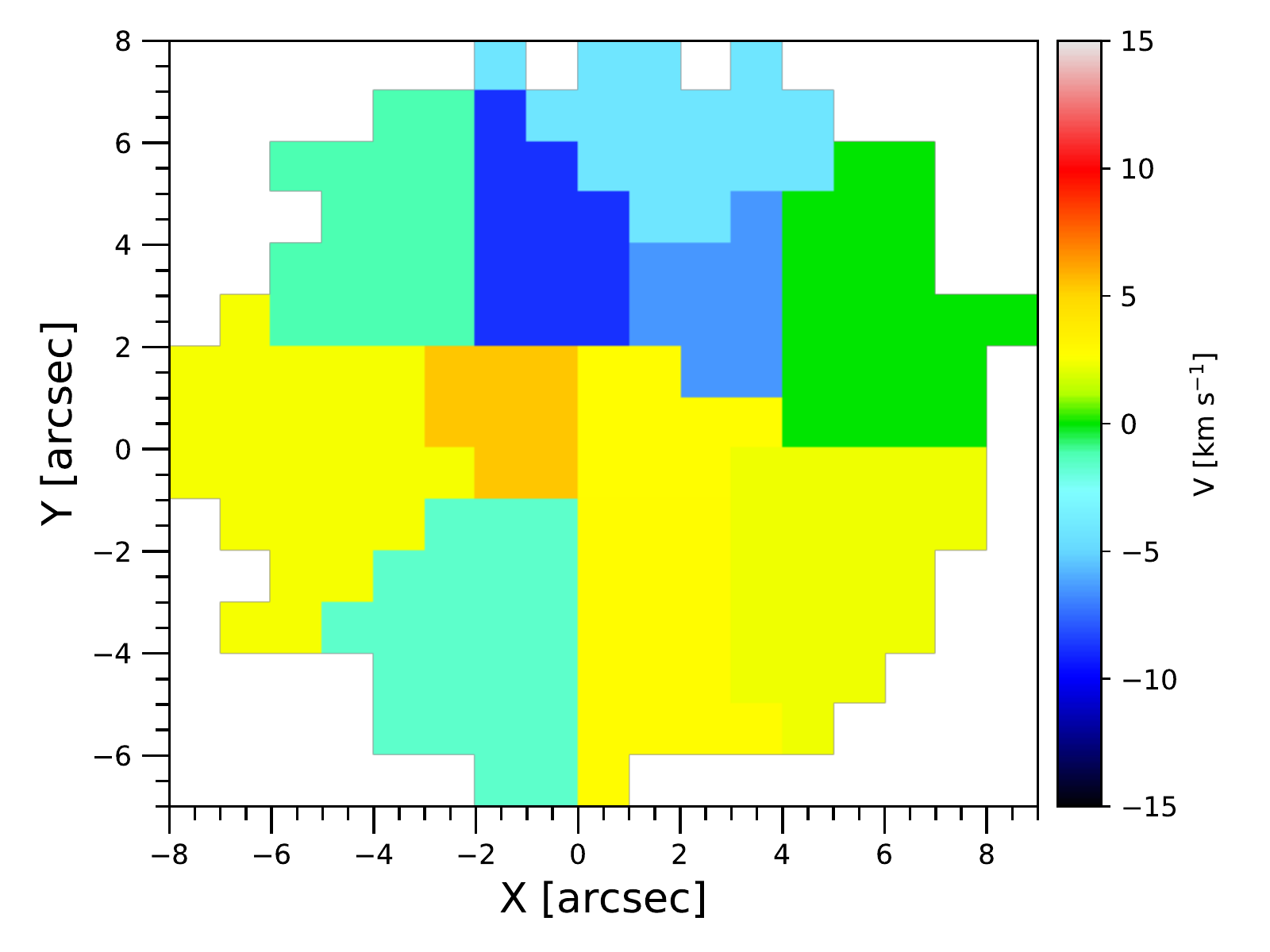}
    \includegraphics[width=2.25in,clip,trim = 10 10 20 10]{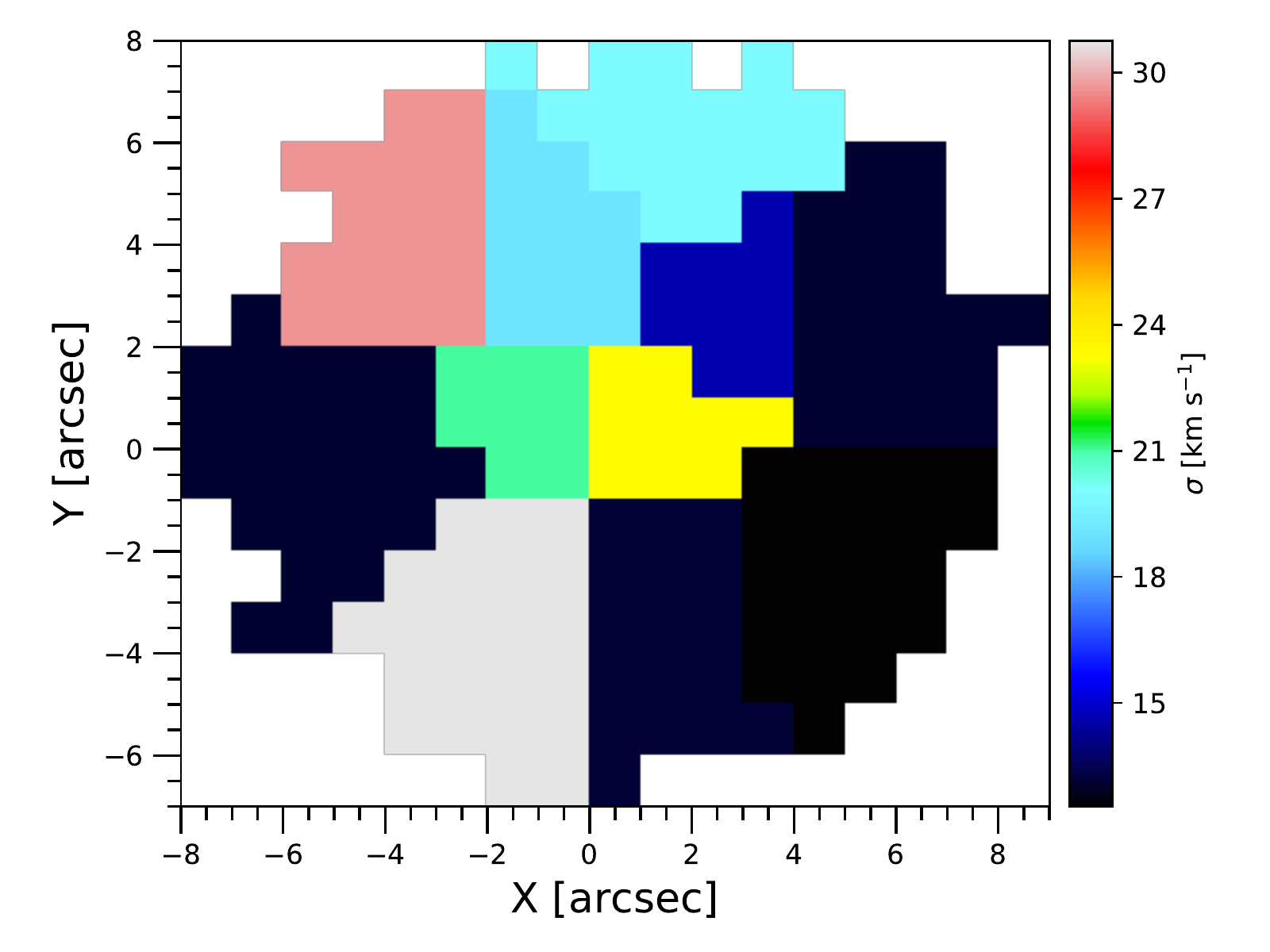}
    
    \includegraphics[width=2.in,clip,trim = 20 0 70 0]{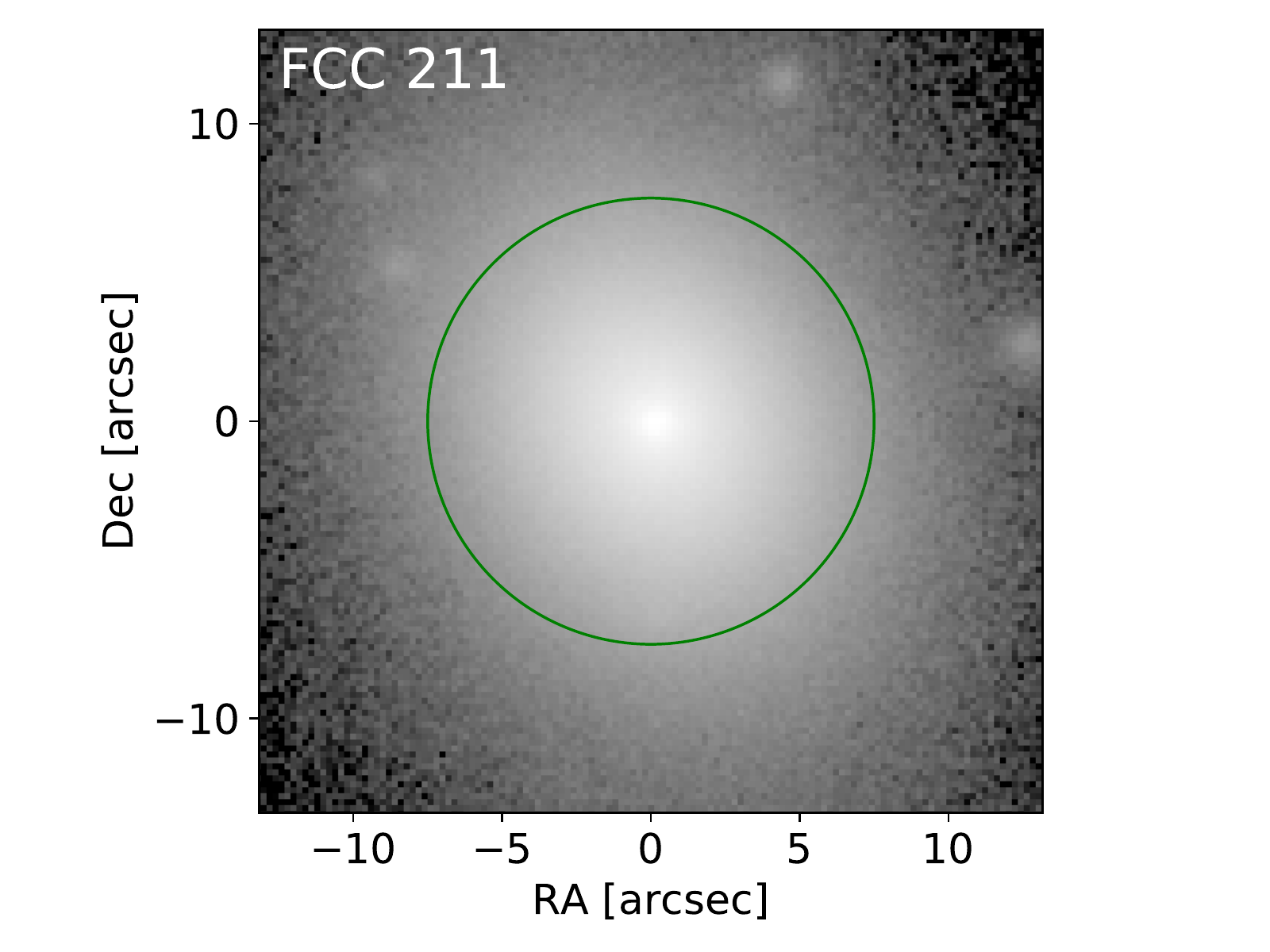}
    \includegraphics[width=2.25in,clip,trim = 20 10 30 10]{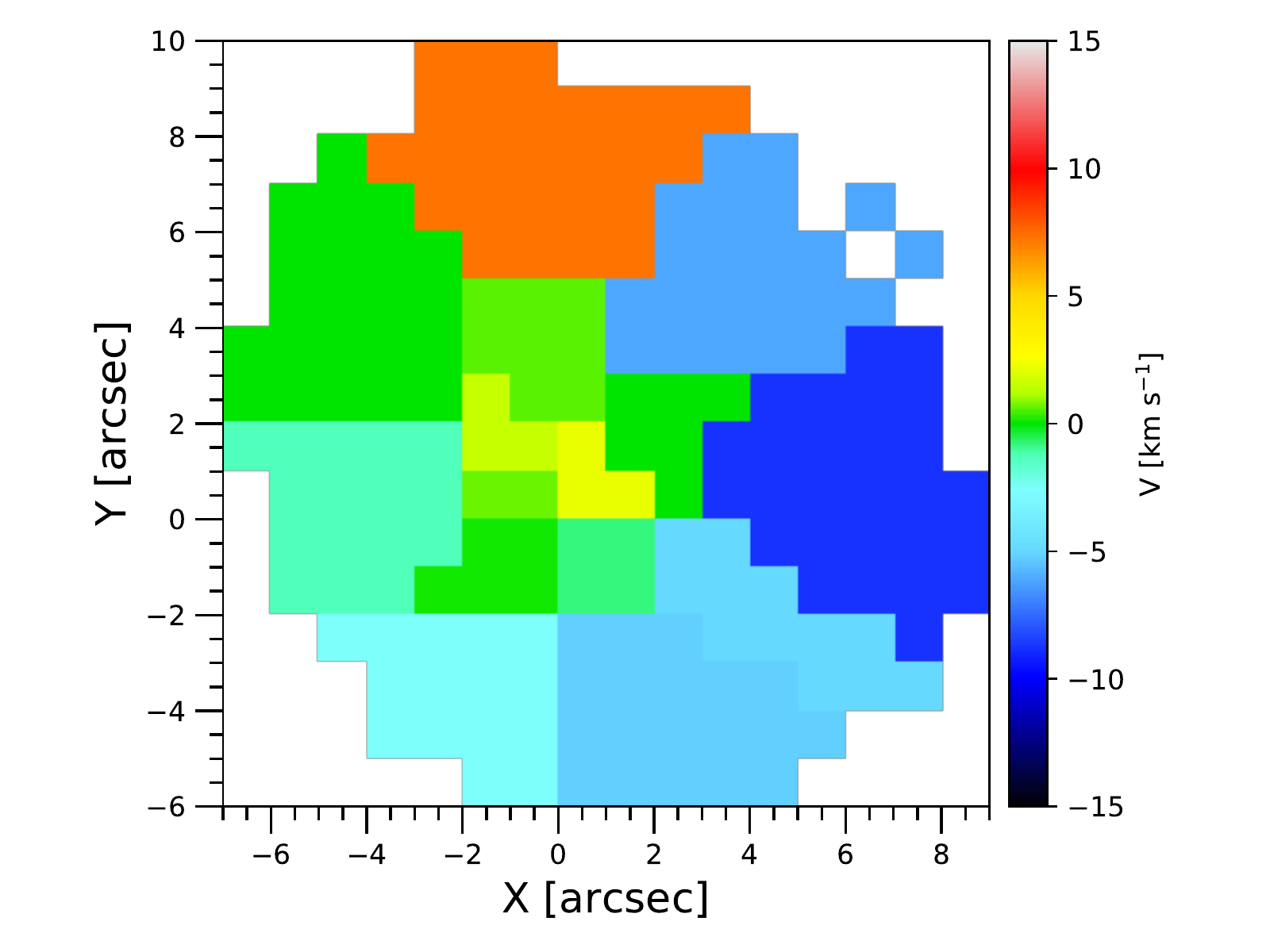}
    \includegraphics[width=2.25in,clip,trim = 20 10 30 10]{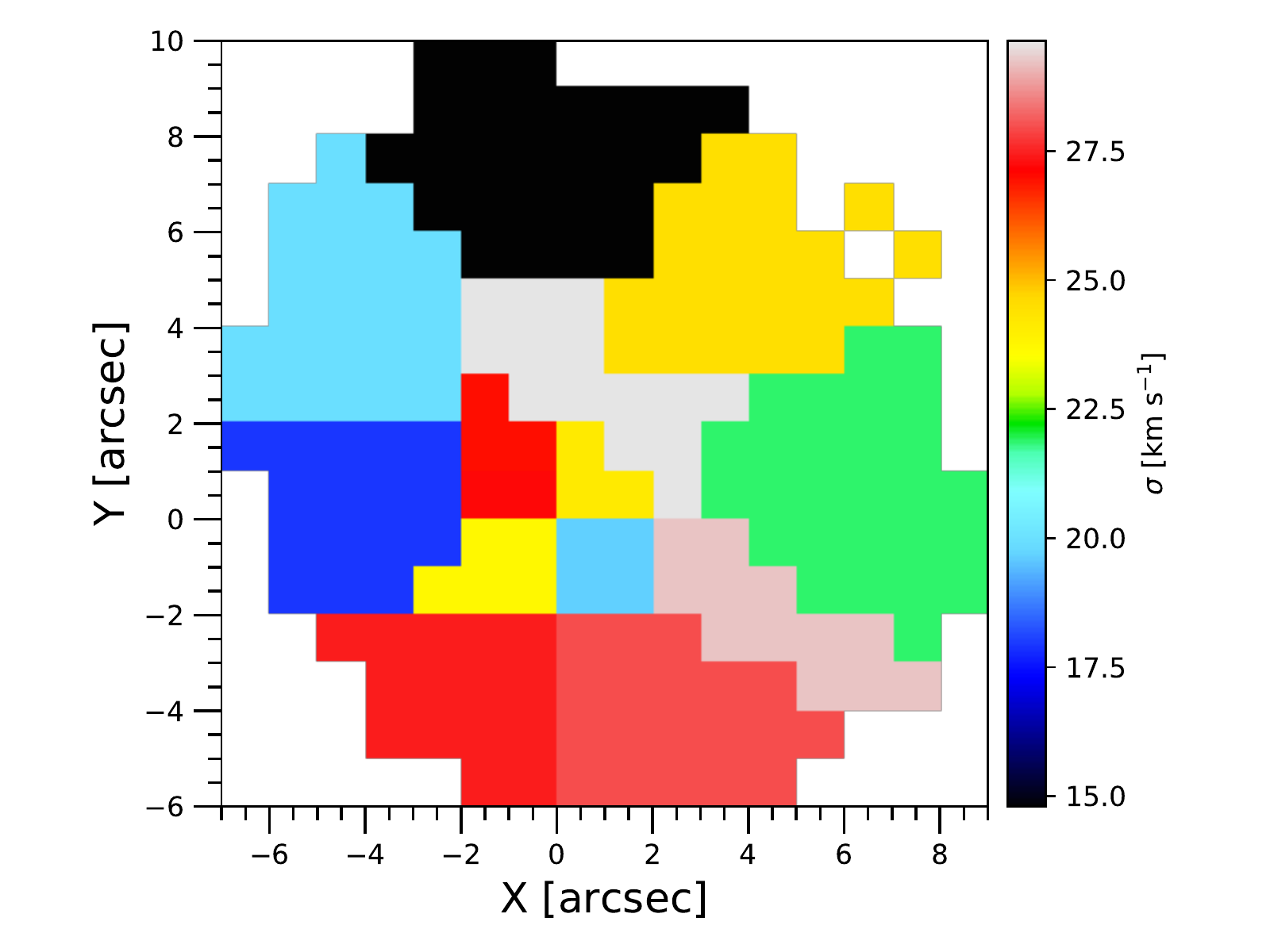}
    
    \includegraphics[width=2.in,clip,trim = 20 0 70 0]{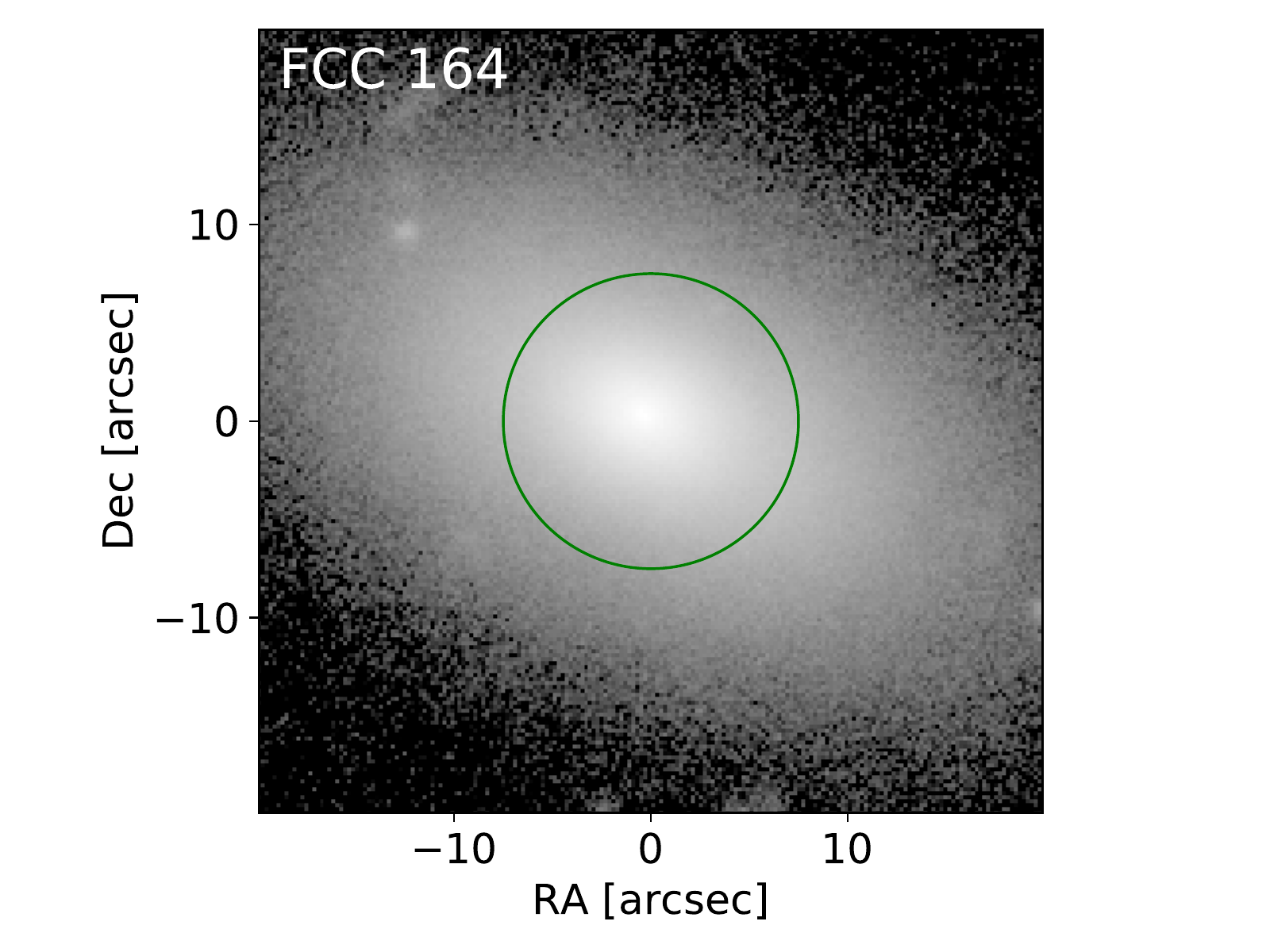}
    \includegraphics[width=2.25in,clip,trim = 20 10 30 10]{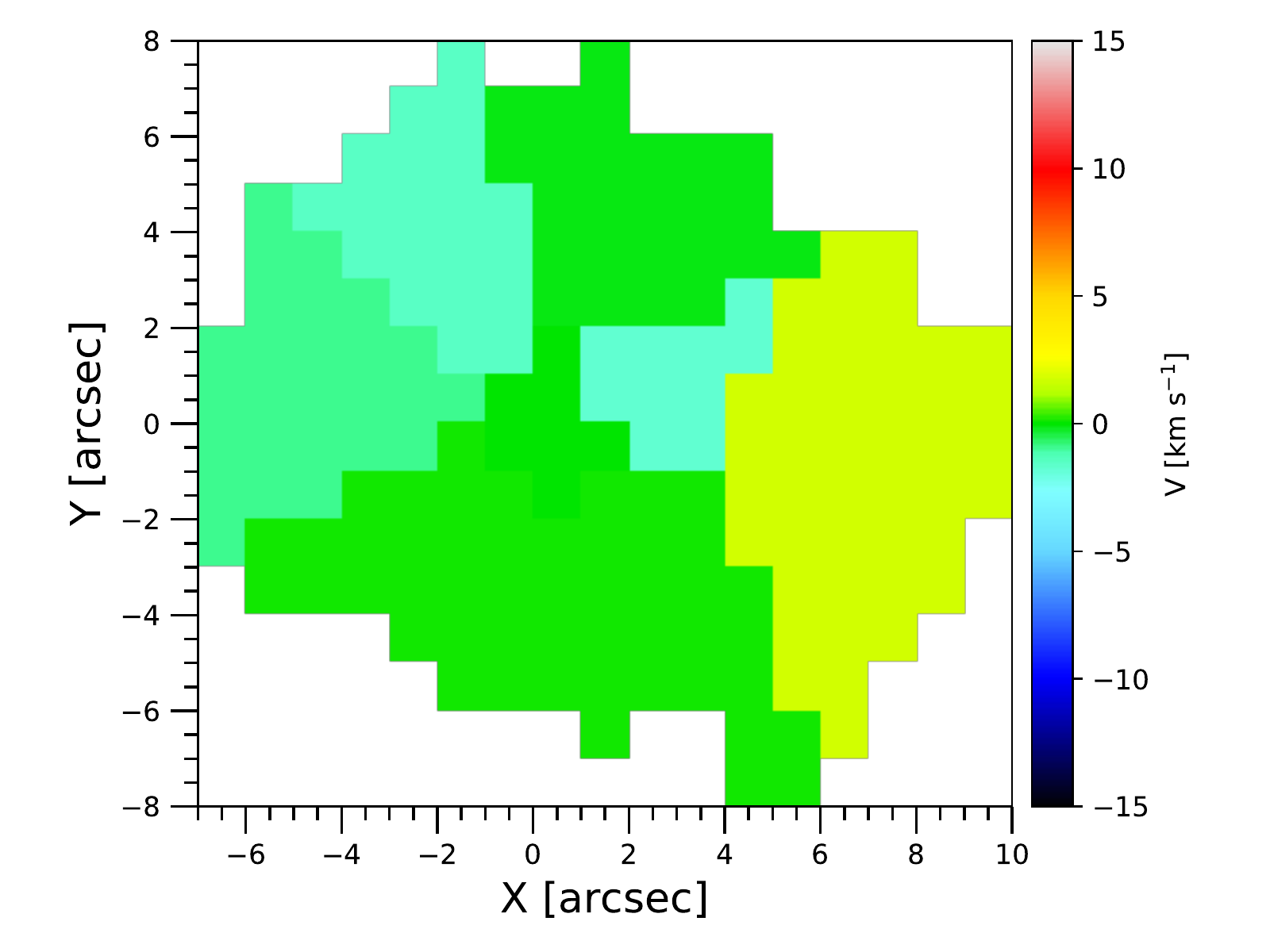}
    \includegraphics[width=2.25in,clip,trim = 20 10 30 10]{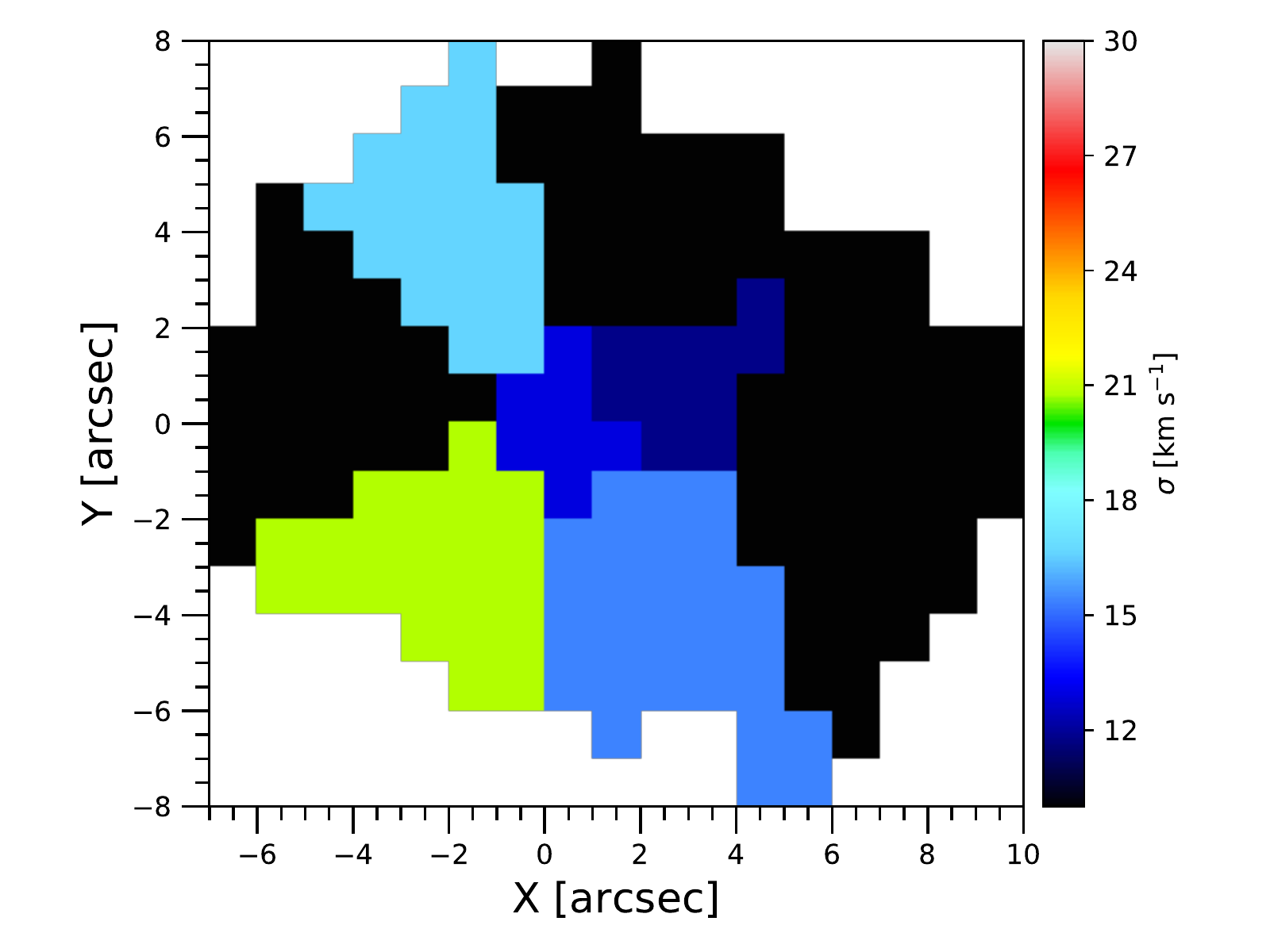}
    \vspace{3mm}
    {\bf Figure \ref{fig:primary_maps_app}.} continued
\end{figure*}

\begin{figure*}
    \centering
    \includegraphics[width=2.in,clip,trim = 20 0 70 0]{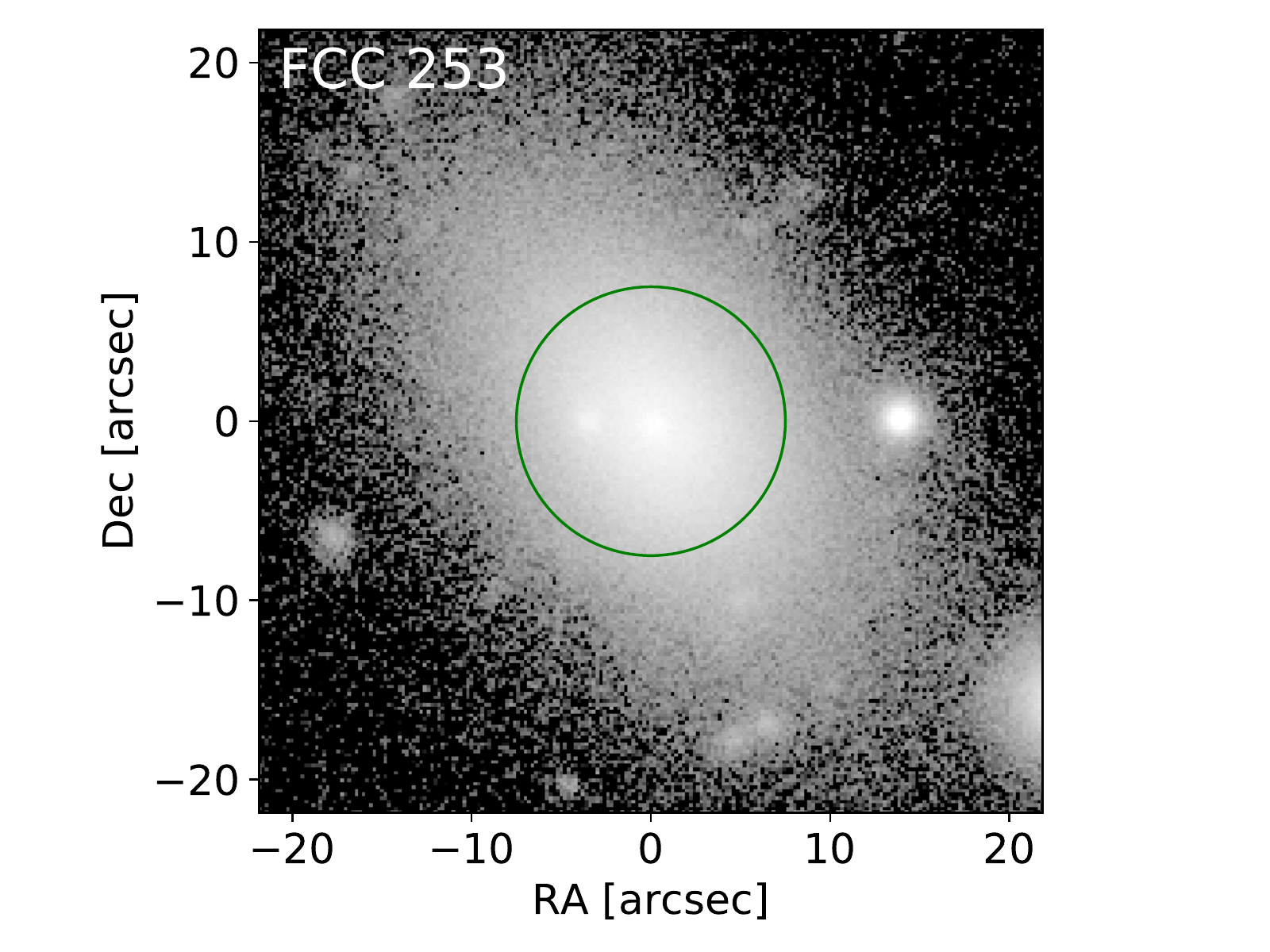}
    \includegraphics[width=2.25in,clip,trim = 20 10 40 10]{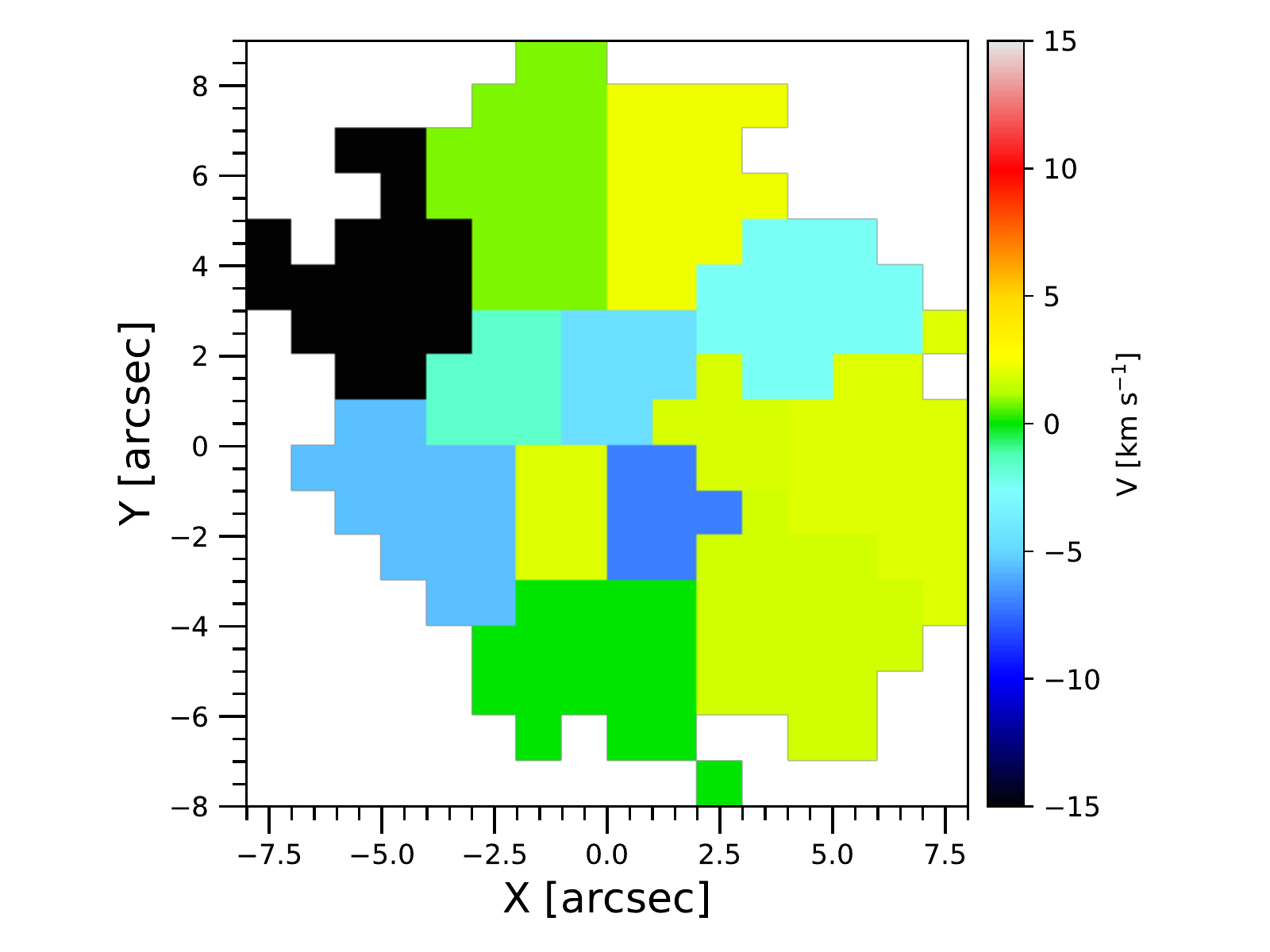}
    \includegraphics[width=2.25in,clip,trim = 20 10 40 10]{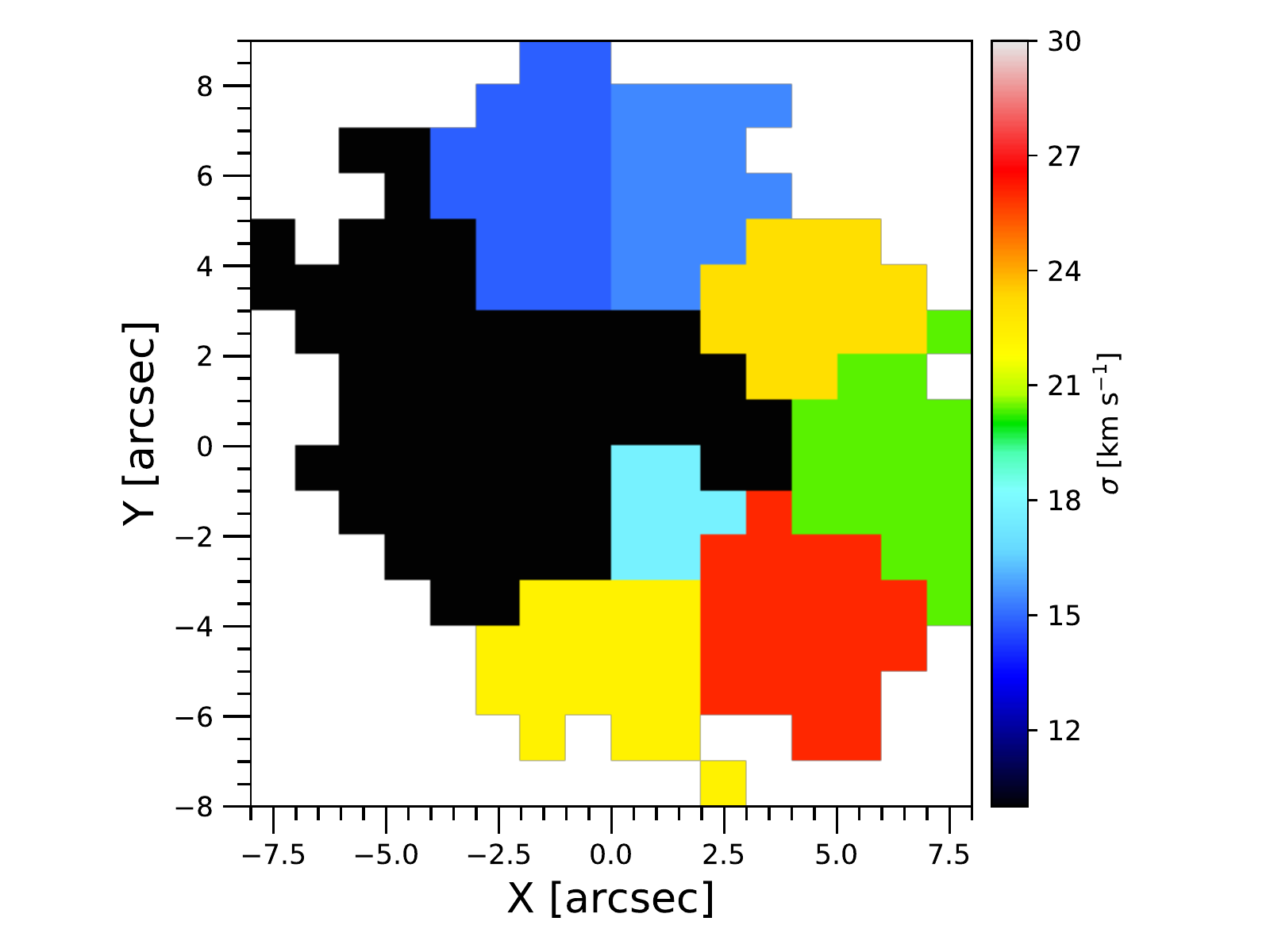}
    
    \includegraphics[width=2.in,clip,trim = 20 0 70 0]{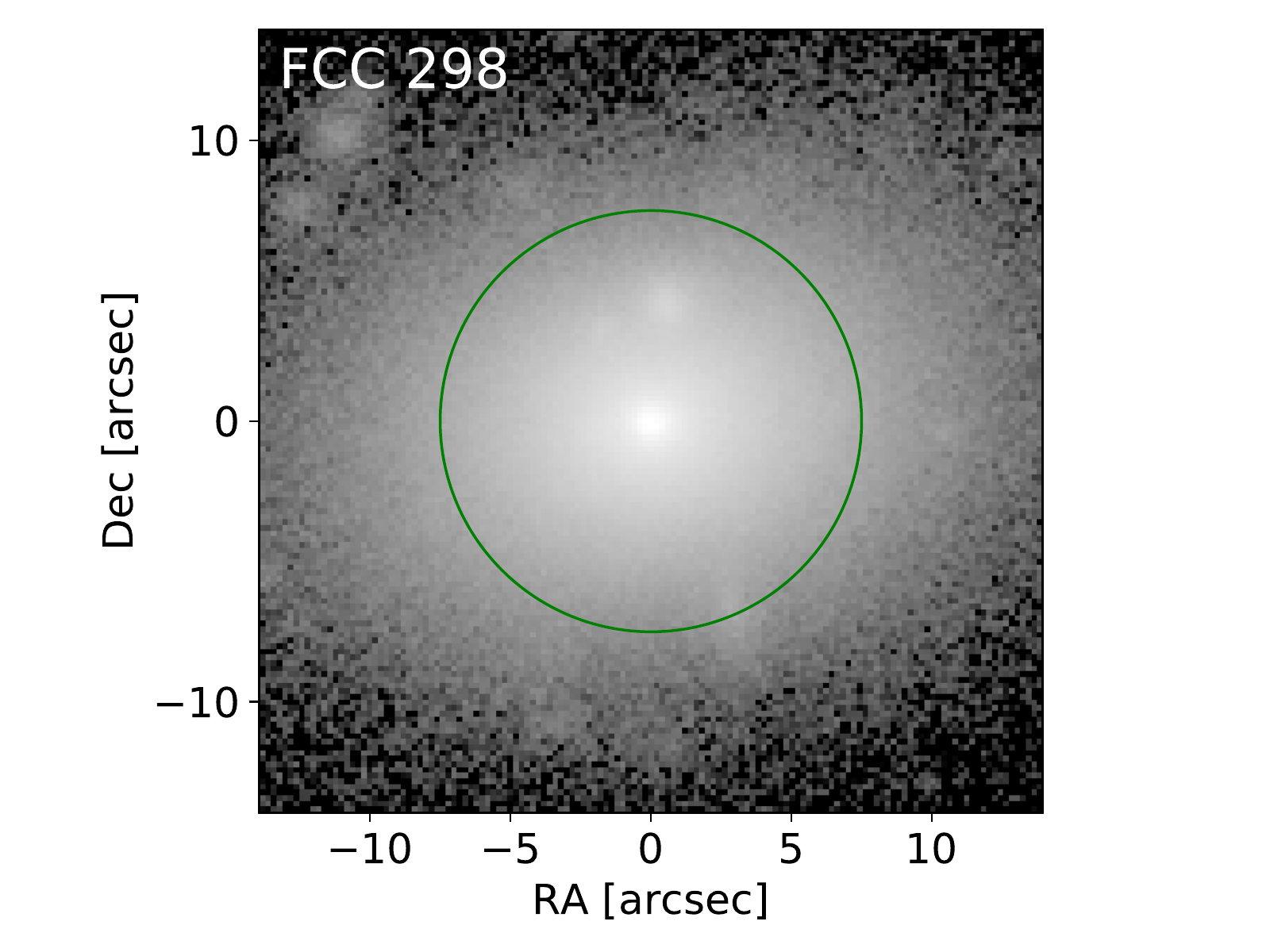}
    \includegraphics[width=2.25in,clip,trim = 10 10 10 10]{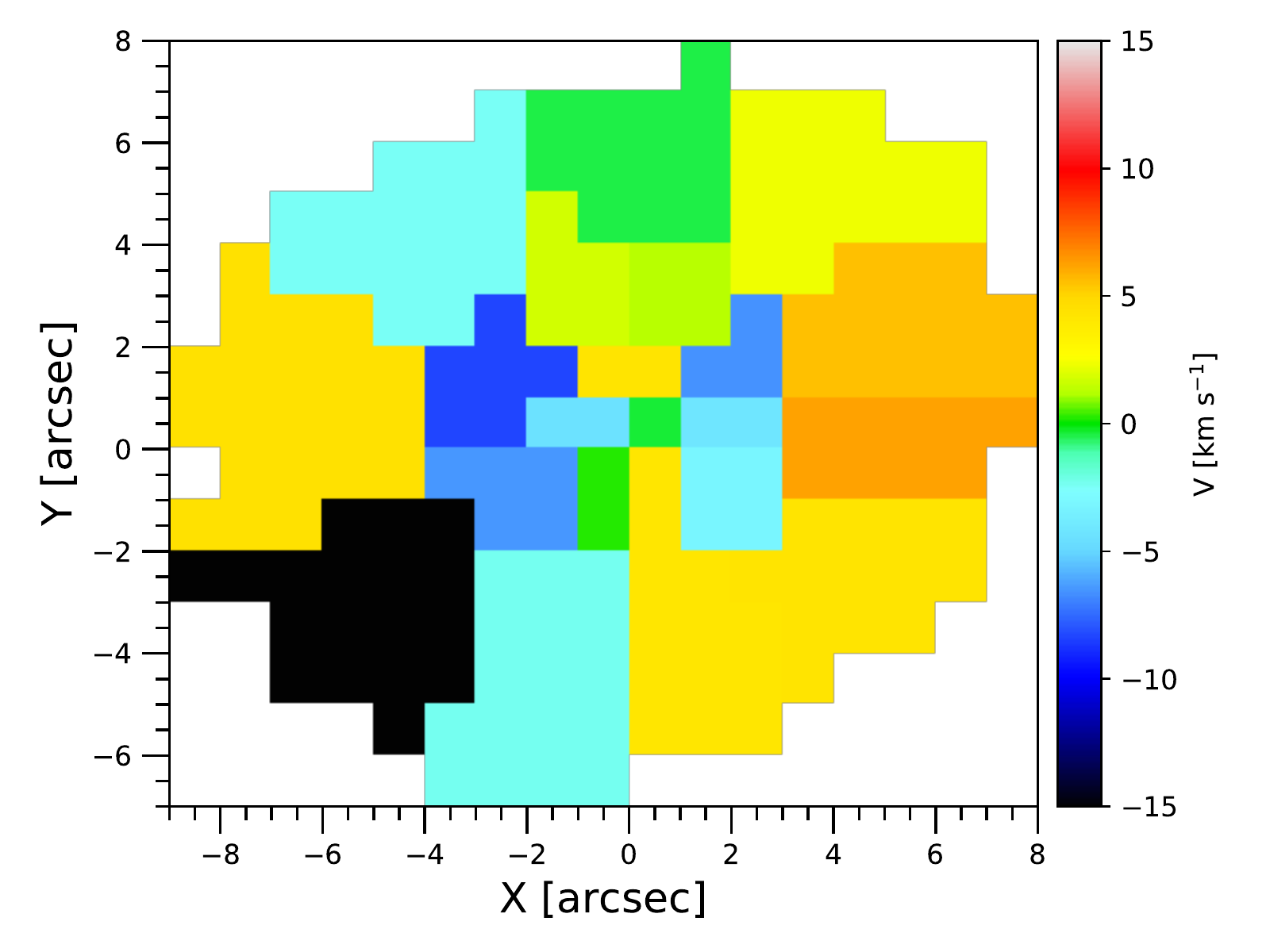}
    \includegraphics[width=2.25in,clip,trim = 20 10 10 10]{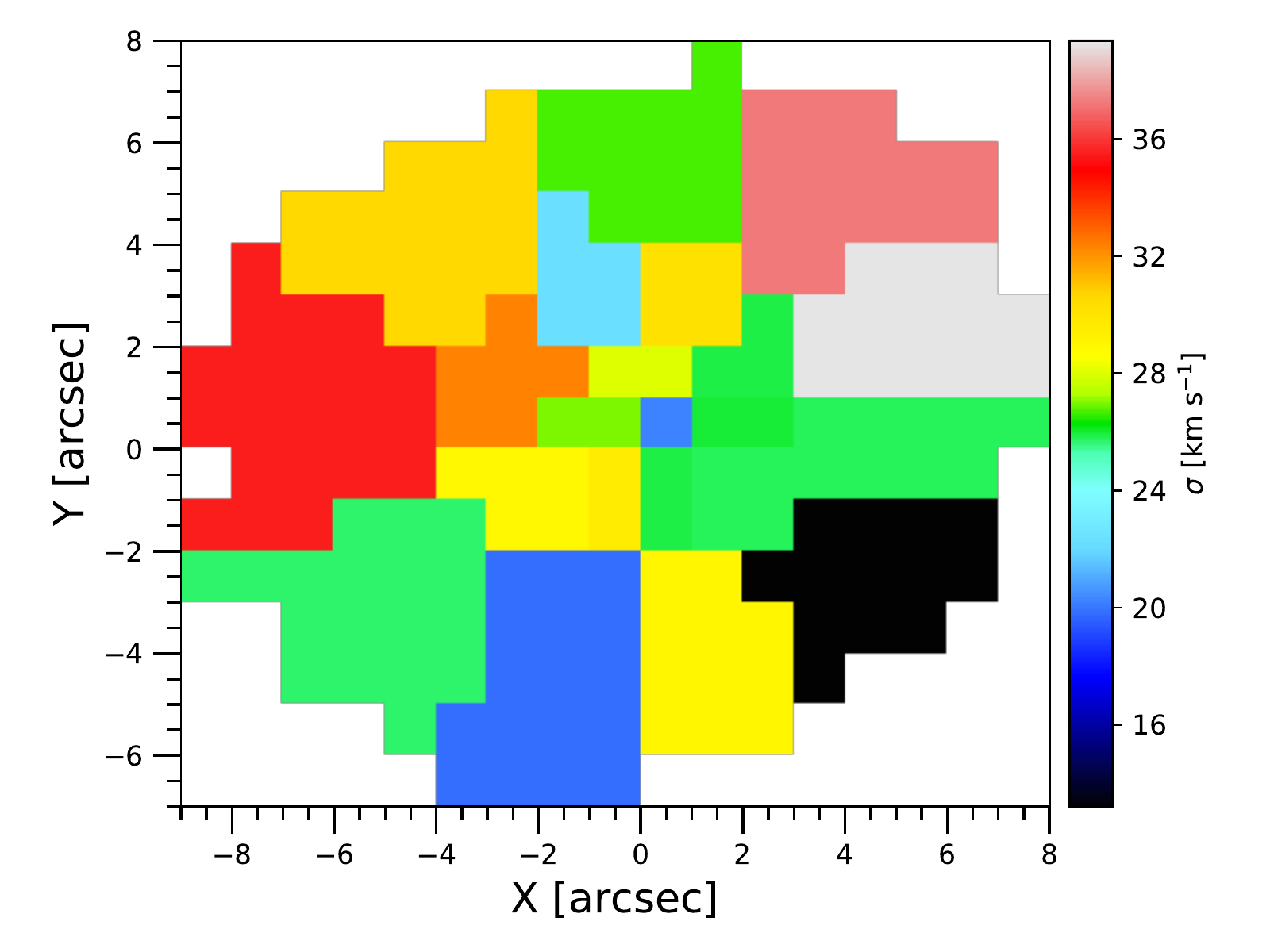}
    \vspace{3mm}
    {\bf Figure \ref{fig:primary_maps_app}.} continued
\end{figure*}

\begin{figure*}
    \centering
    \includegraphics[width=2.in,clip,trim = 20 0 70 0]{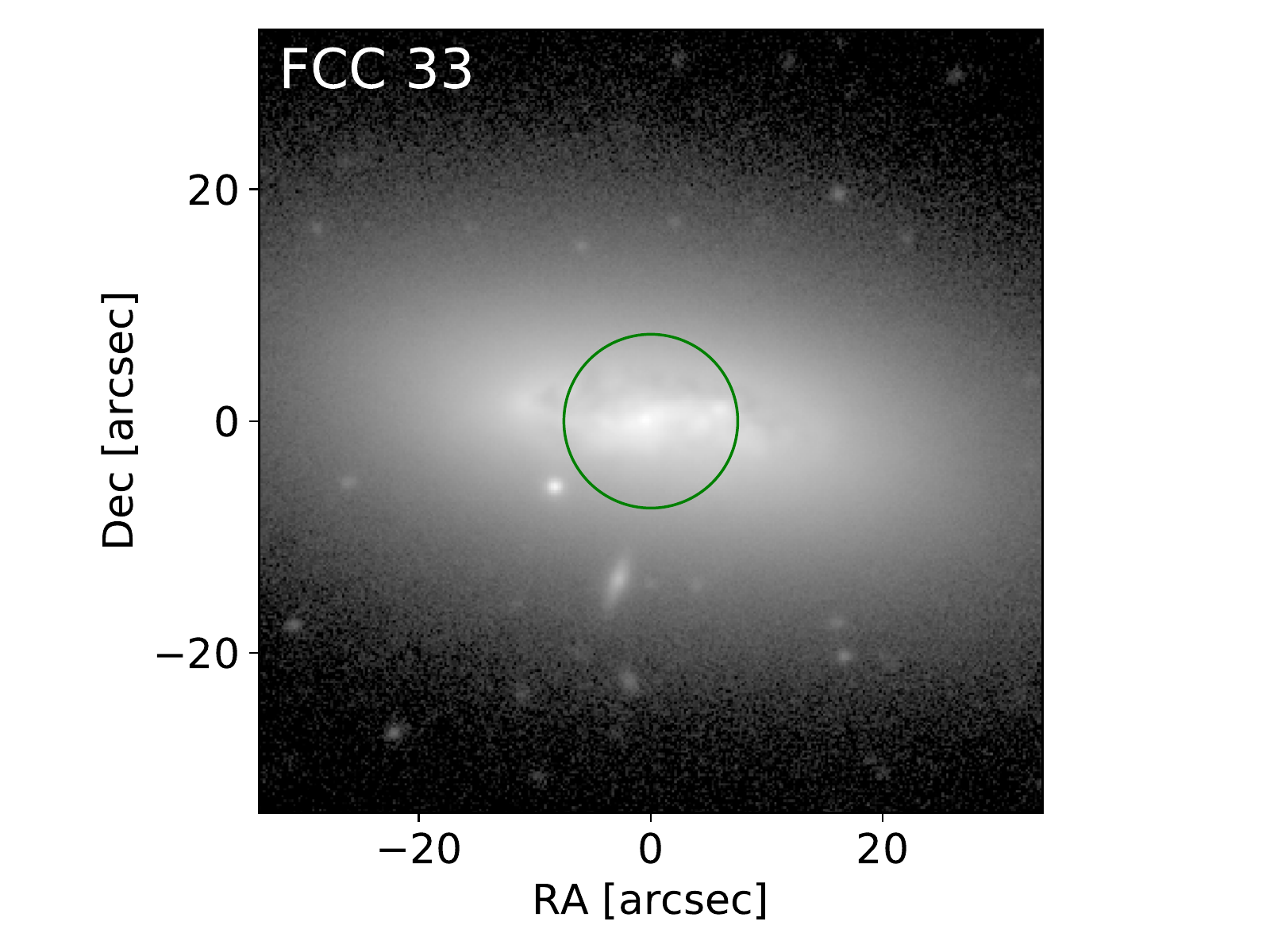}
    \includegraphics[width=2.25in,clip,trim = 20 10 10 10]{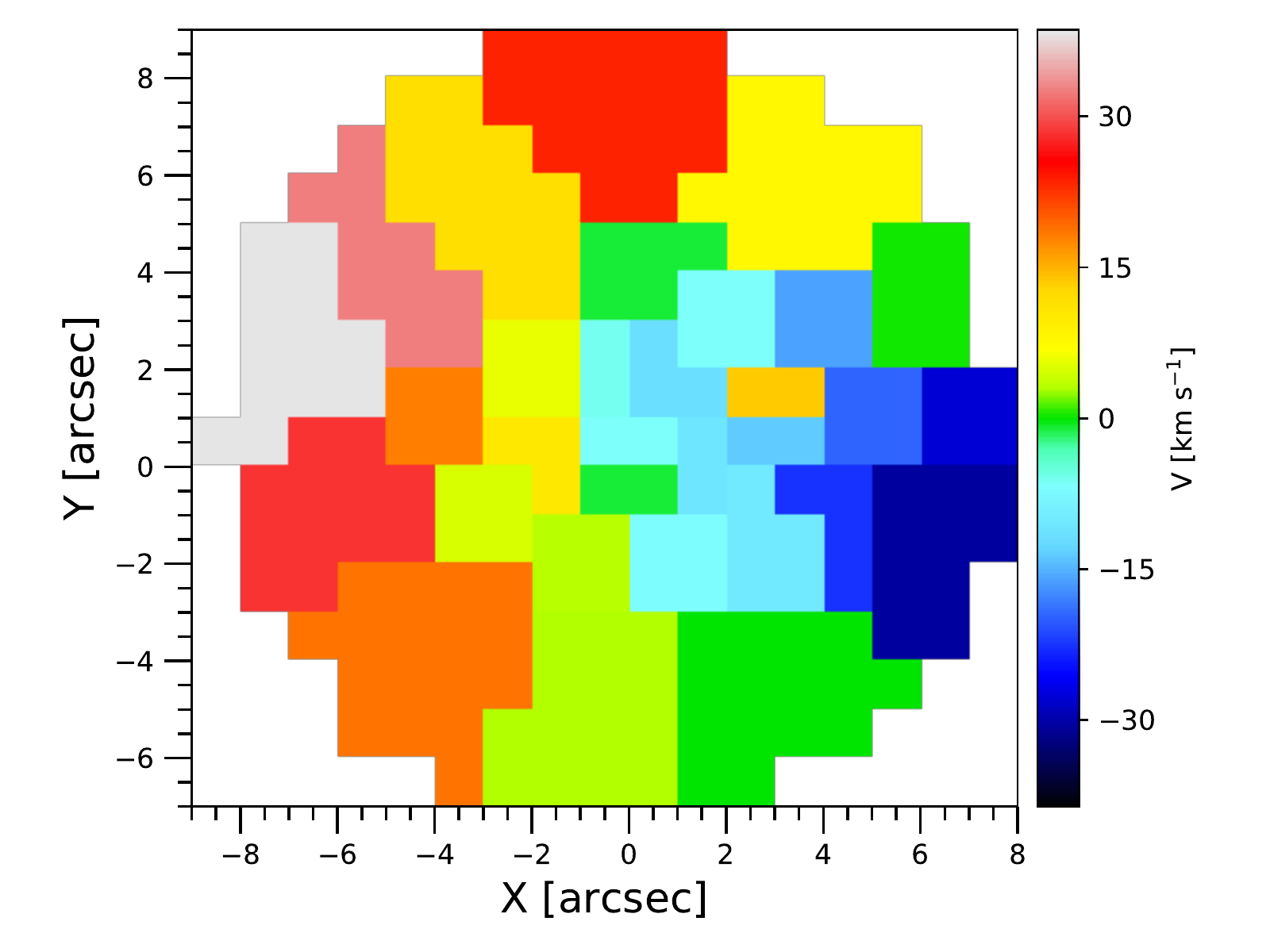}
    \includegraphics[width=2.25in,clip,trim = 20 10 10 10]{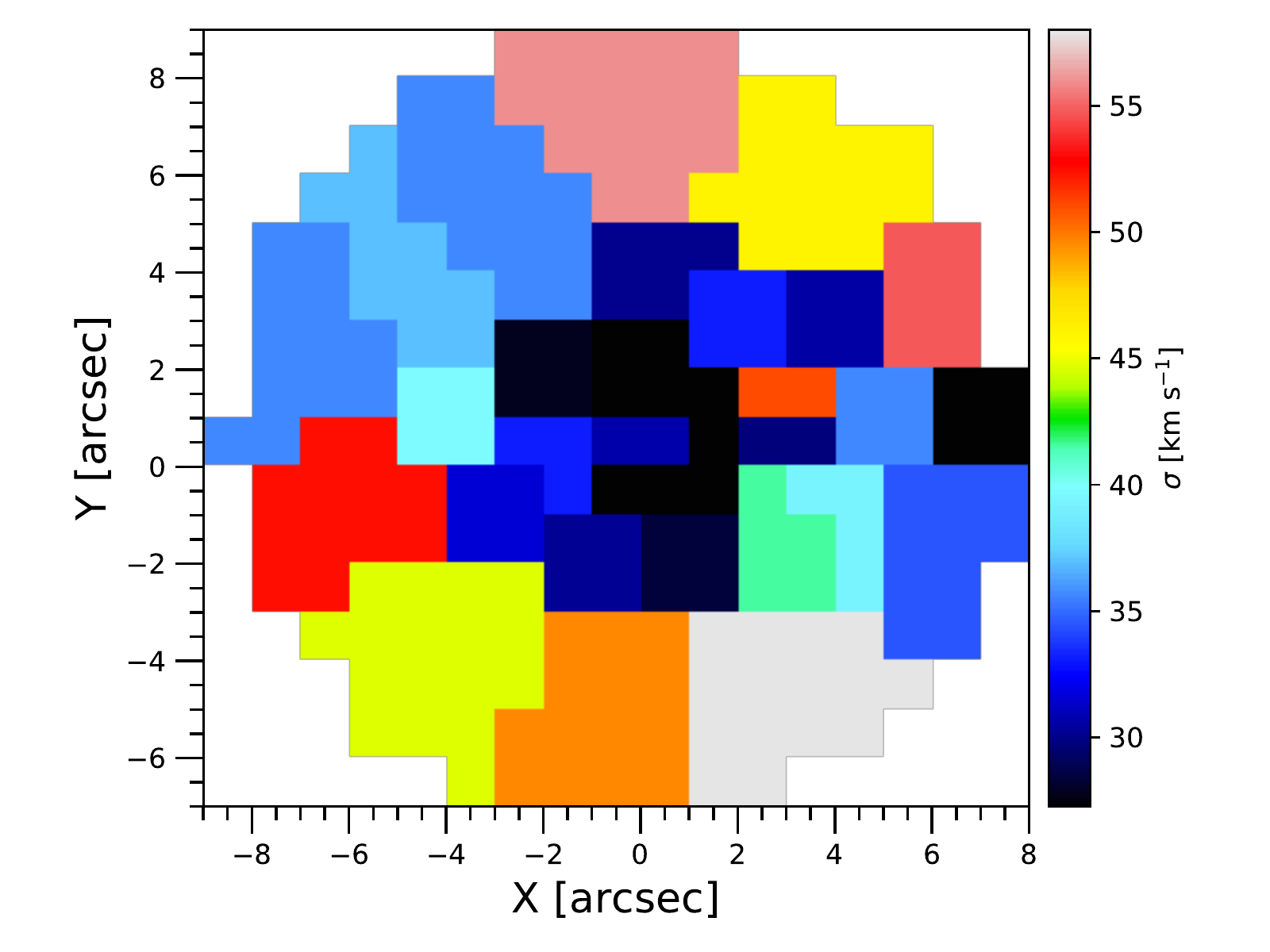}

    \includegraphics[width=2.in,clip,trim = 20 0 70 0]{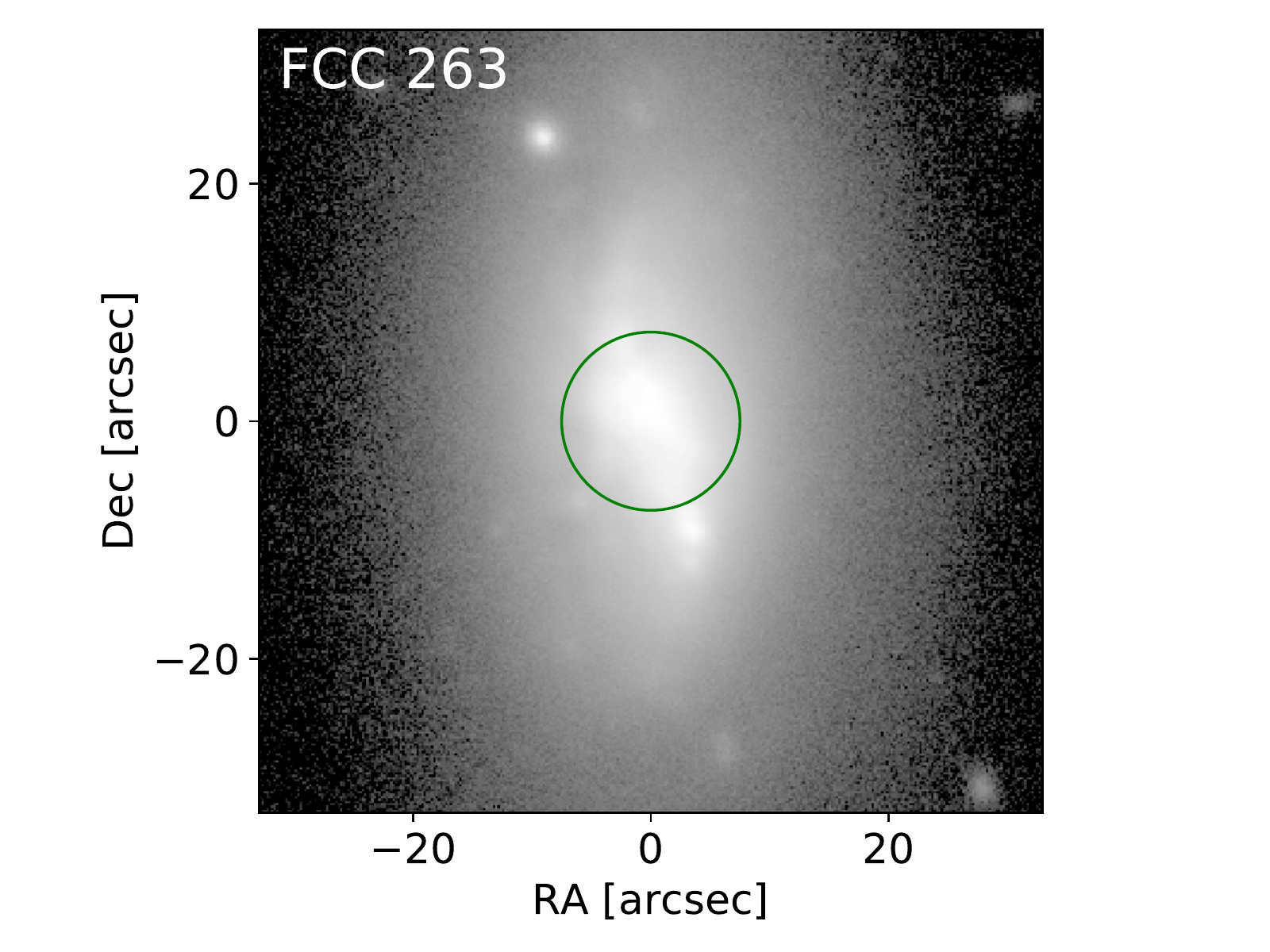}
    \includegraphics[width=2.25in,clip,trim = 20 10 10 10]{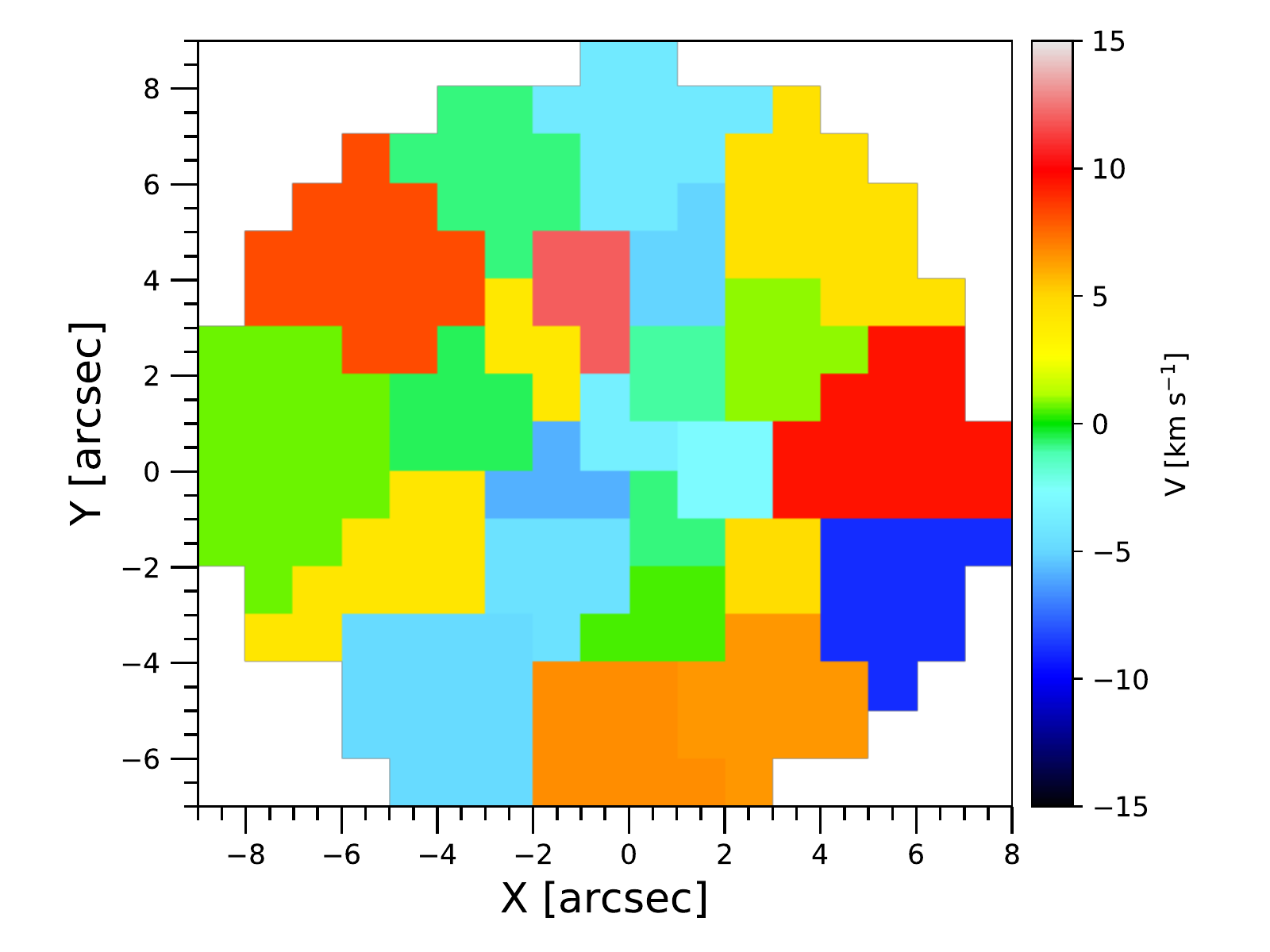}
    \includegraphics[width=2.25in,clip,trim = 20 10 10 10]{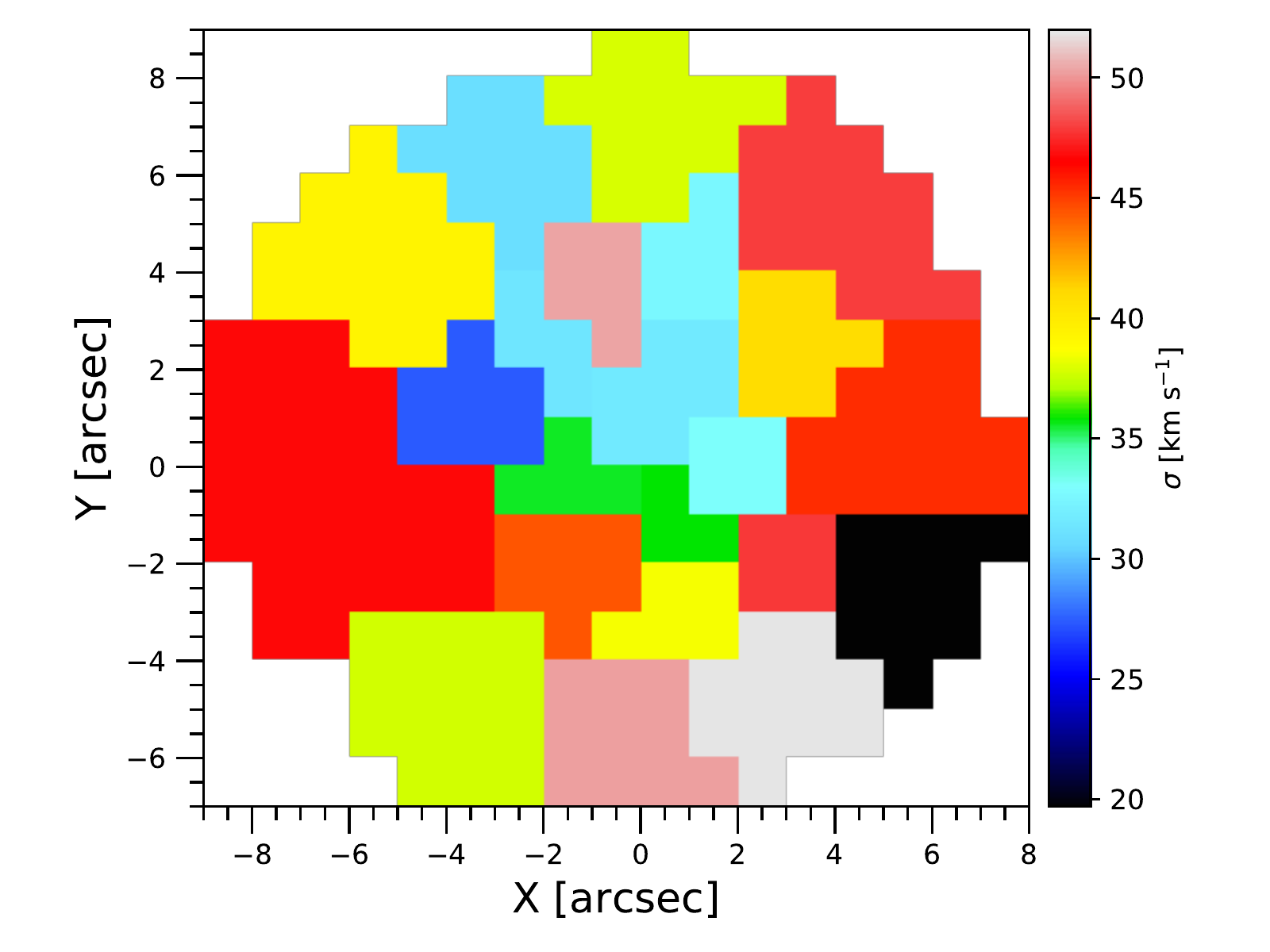}
    
    \includegraphics[width=2.in,clip,trim = 20 0 70 0]{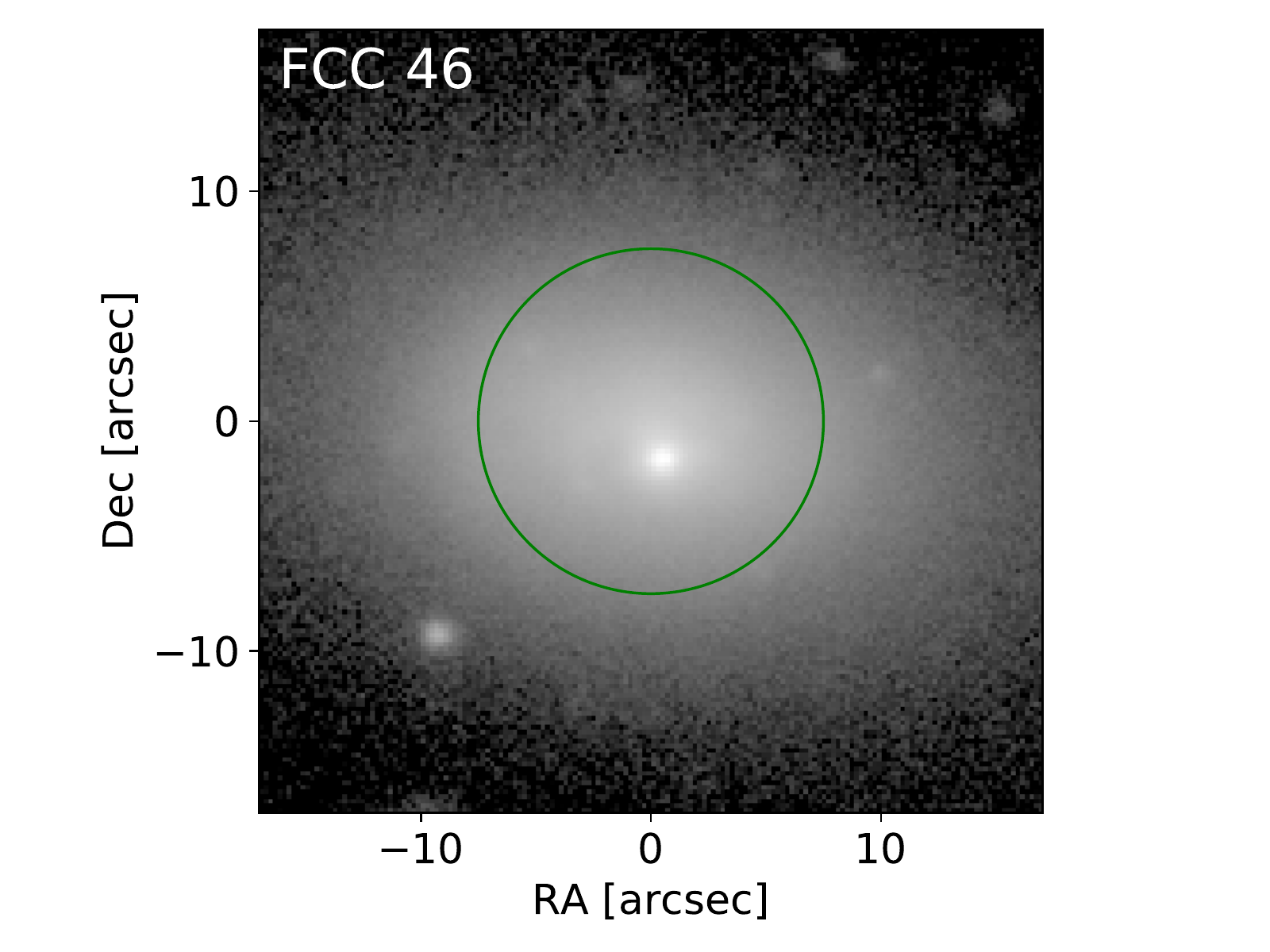}
    \includegraphics[width=2.25in,clip,trim = 5 10 10 10]{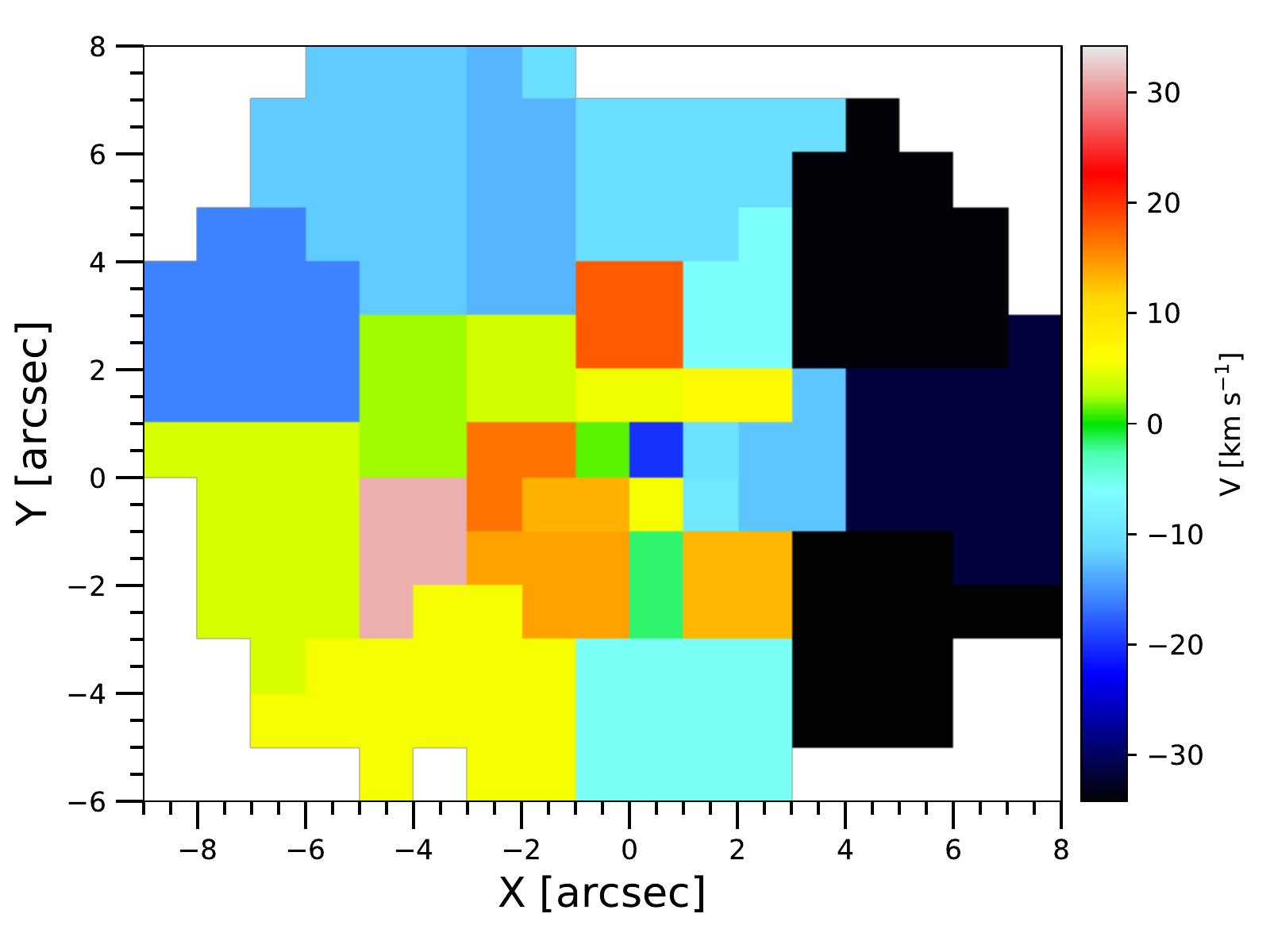}
    \includegraphics[width=2.25in,clip,trim = 20 10 10 10]{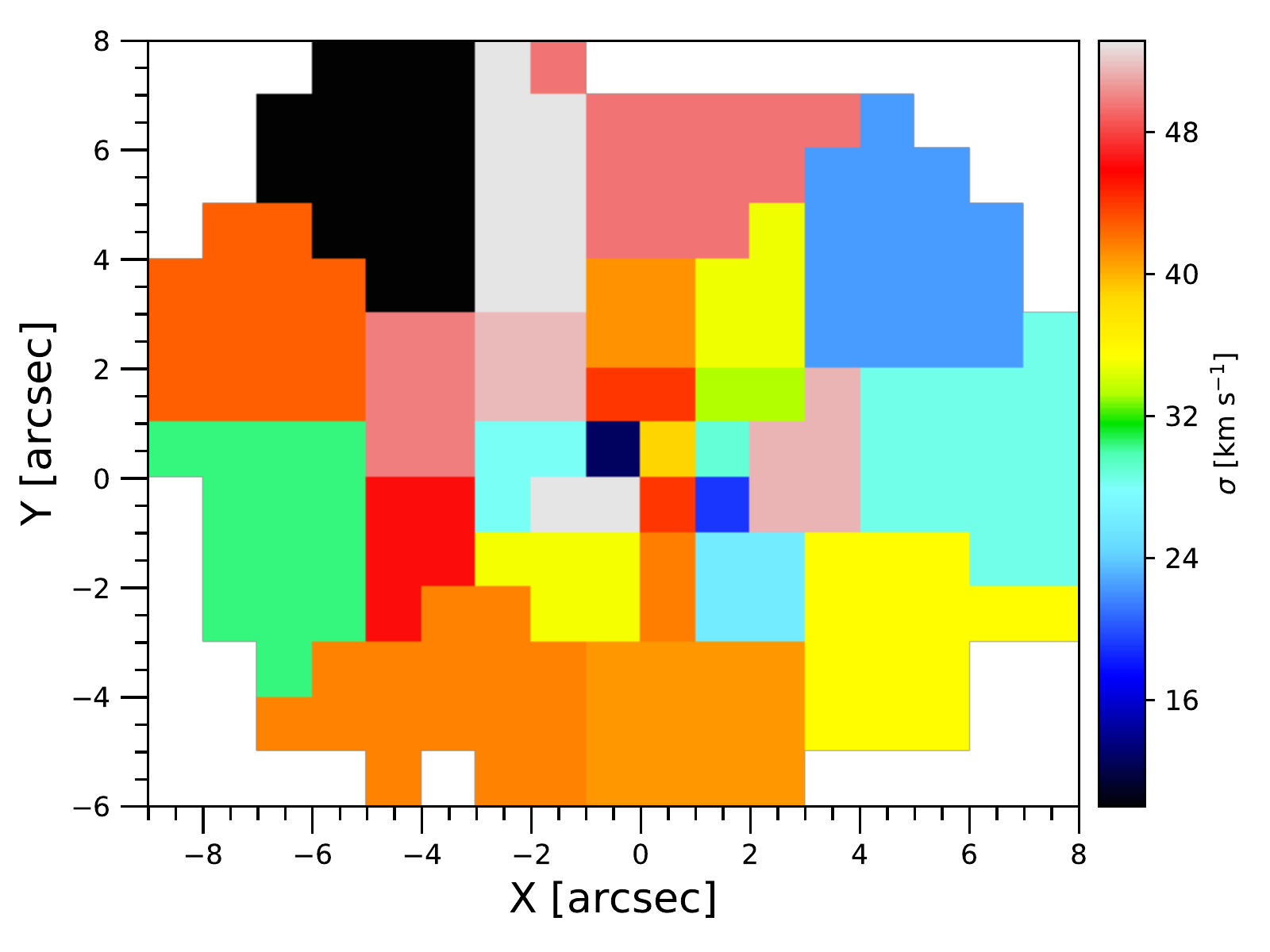}
    
    \includegraphics[width=2.in,clip,trim = 20 0 70 0]{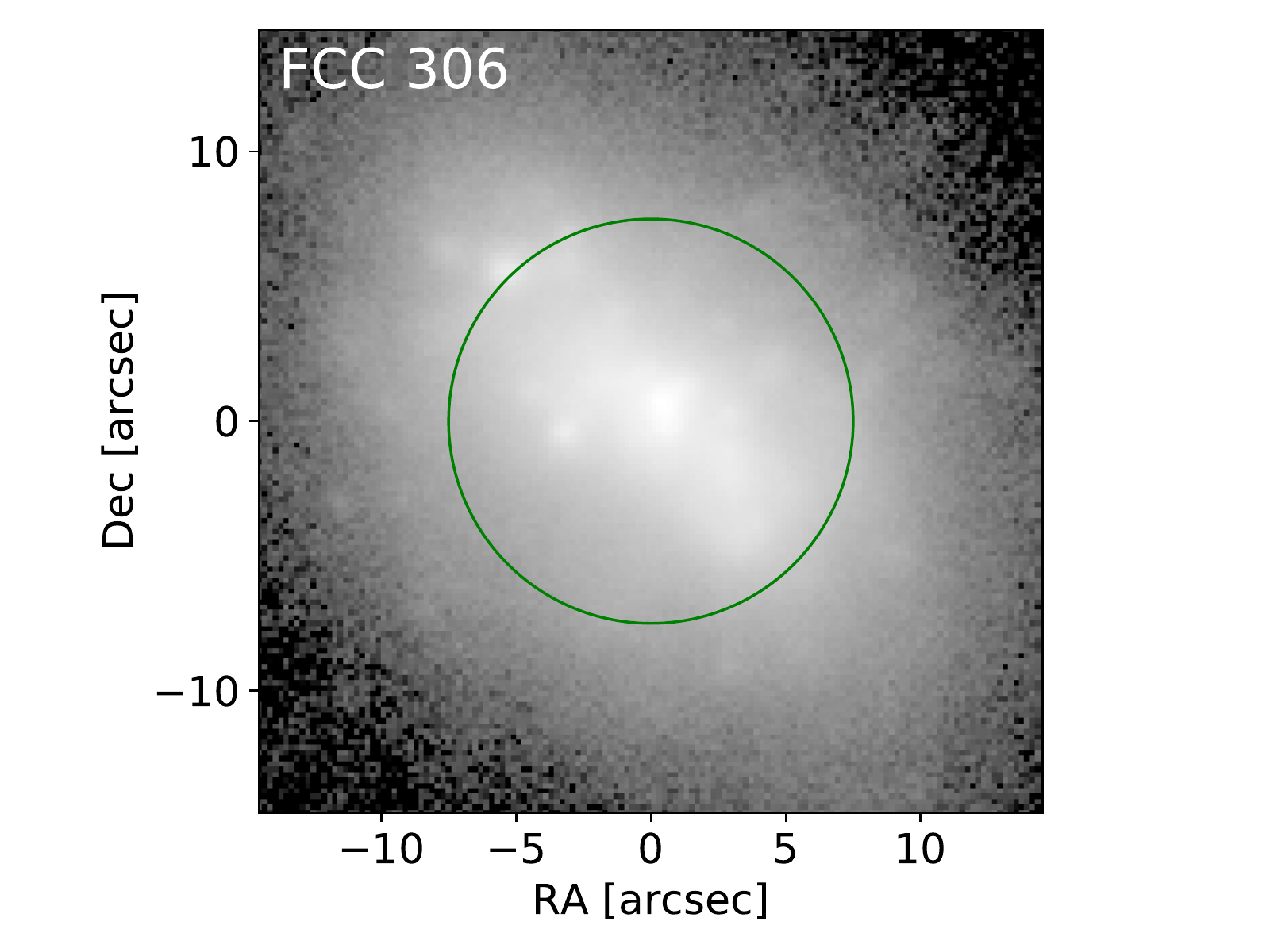}
    \includegraphics[width=2.25in,clip,trim = 20 10 40 10]{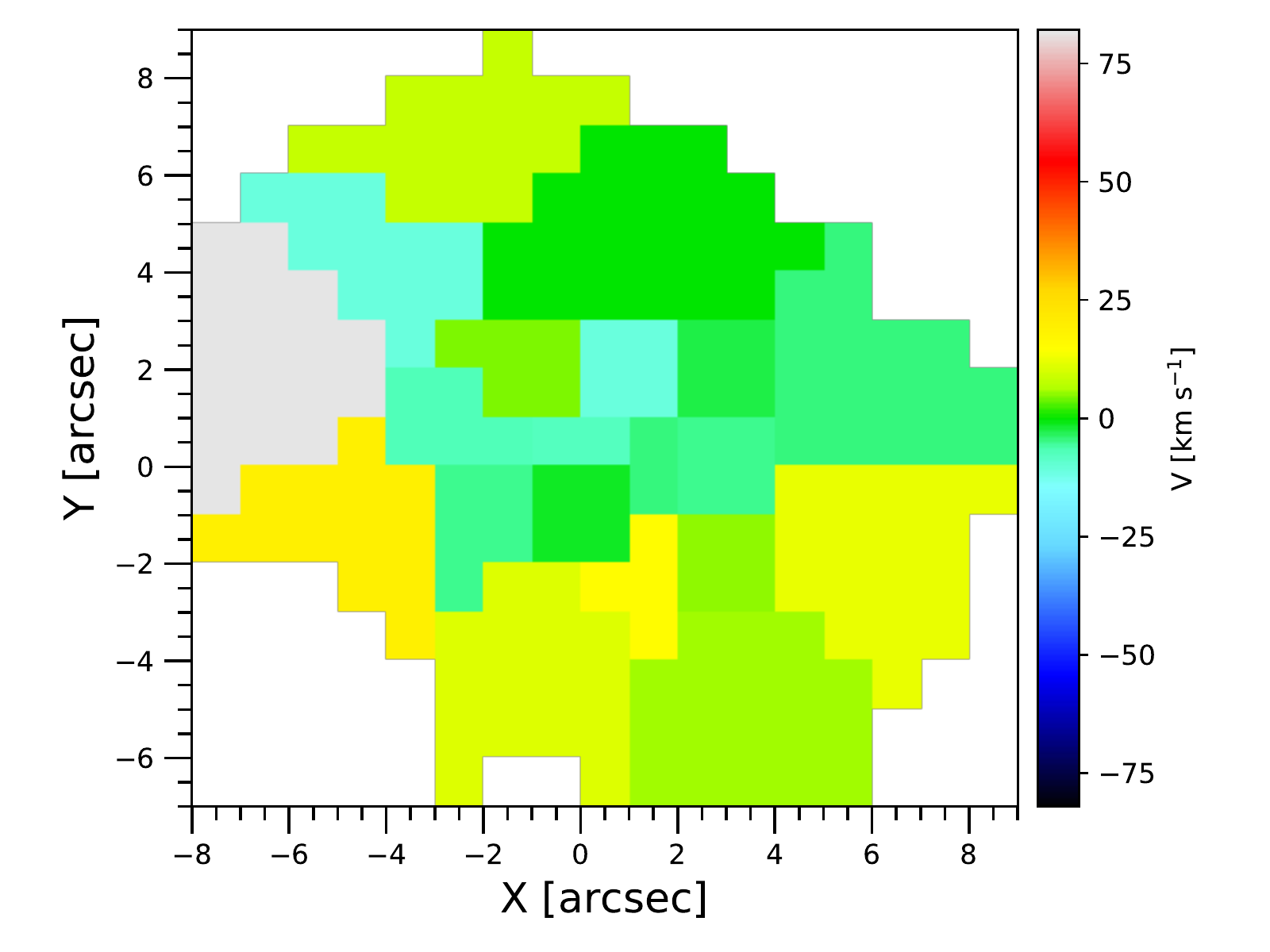}
    \includegraphics[width=2.25in,clip,trim = 20 10 40 10]{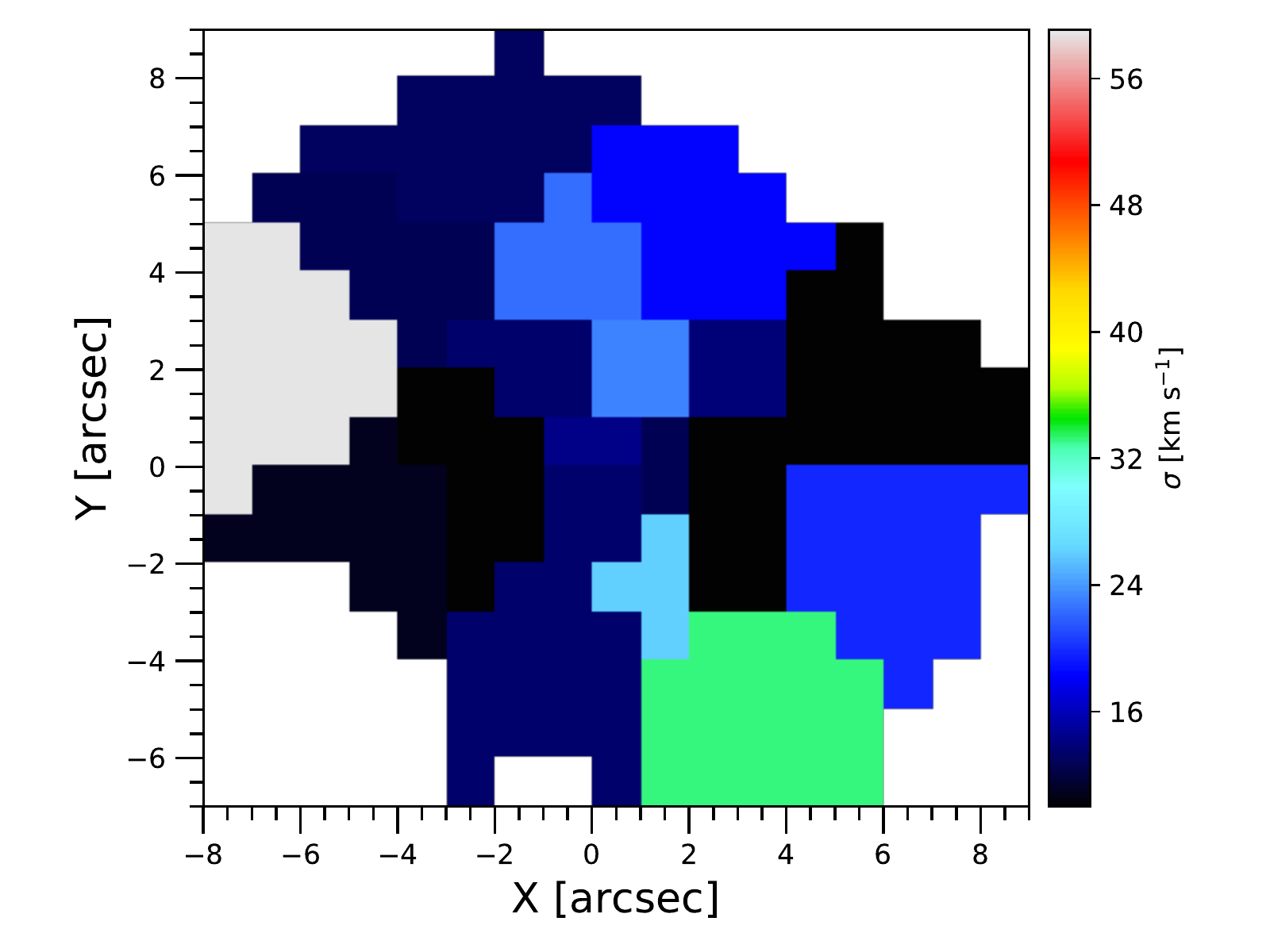}
    \caption{As Fig \ref{fig:primary_maps_app}, but for dwarf late-type galaxies.}
    \label{fig:latetype_maps}
\end{figure*}

\section{Secondary targets}
Here we provide details and stellar kinematics for the secondary targets observed as part of the survey. As with the primary targets, several of the secondary targets exhibit unusual kinematics and we discuss these objects in detail below:

{\it FCC 29:} This galaxy shows features of non-regular rotation in its velocity map, with a distinct twist seen along the kinematic axis. The dispersion map shows a central peak surrounded by a low-dispersion ring, with a significant increase outside this ring. The {\it r-}band image shows two prominent ring features and a stellar bar, though our kinematic measurements are restricted to the bulge region.

{\it FCC 153:} This galaxy shows ordered rotation and a narrow low-dispersion structure along the major axis, consistent with a cold disk dominating the light.

{\it FCC 167:} The IFU placement for this galaxy was offset from the centre, however the observed kinematics are fully consistent with regular rotation.

{\it FCC 177:} On large scales this galaxy exhibits regular rotation, but in the central $\sim 4$ arcsec we see evidence for a distinct kinematic component with increased dispersion and a non-aligned kinematic axis. \citet{Pinna:2019} examined this galaxy in detail, but did not identify the central kinematic anomaly we find, likely due to the lower spectral resolution of their data. They do however identify a region of young stellar age and enhanced stellar metallicity coincident with our kinematic anomaly. Using high spatial resolution SINFONI observations, \citet{Lyubenova:2019} report no clear rotation in the central region of this galaxy, however their kinematic map does show weak rotation consistent with that reported here.

{\it FCC 190:} As with FCC 303, this galaxy has ordered rotation and a strong central dip in velocity dispersion \citep[consistent with][]{Iodice:2019}. The elongated low-dispersion structures suggest the presence of a central cold component,likely a small inner disk, in both galaxies.

{\it FCC 193:} This galaxy shows regular rotation, but an unusual hourglass-shaped structure in it's dispersion map, with low dispersion along the major axis. This feature is also seen in \citet{Iodice:2019}, whose larger-scale measurements show this is due to a small-scale cold, inner disk that dominates the SAMI kinematic measurements.

{\it FCC 219:} This galaxy exhibits a so-called $2-\sigma$ structure, with two distinct peaks in its velocity dispersion map and evidence for a small counter-rotating component in its central regions. This is consistent with the kinematic measurements of \citet{Scott:2014} and \citet{Iodice:2019}.

{\it FCC 249:} This galaxy shows a clear kinematically decoupled core embedded in an outer region with net zero rotation, consistent with the findings of \citet{Iodice:2019}.

\begin{table*}
    \caption{SAMI --- Fornax Dwarf Survey secondary targets}
    \label{tab:secondary_sample}
    \centering
\begin{tabular}{ccccccl}
FDS & FCC & RA & DEC & $\lambda_R$ & Max Rad & Notes \\
\hline
 & & (deg) & (deg) & & (R$_e$) & \\
 \hline
 \hline
21\_D055a & 51 & 51.72132 & -36.79869 & -- & -- &  Faint early-type dwarf \\
11\_D330 & 197 & 54.4204166 & -35.29472 & -- & -- & Faint early-type dwarf \\
13\_D044 & 201 & 54.48148 & -37.85596 & -- & -- &  Faint early-type dwarf \\
11\_D047 & 214 & 54.6520833 & -35.8325 & -- & -- &  Faint early-type dwarf \\
11\_D203 & 227 & 54.95875 & -35.52111 & -- & -- &  Faint early-type dwarf \\
11\_D311 & 228 & 54.96375 & -35.32055 & -- & -- &  Faint early-type dwarf \\
11\_D134 & 229 & 54.98 & -35.66083 & -- & -- &  Faint early-type dwarf \\
13\_D230 & 239 & 55.07741 & -37.49941 & -- & -- &  Faint early-type dwarf \\
6\_D374 & 269 & 55.48875 & -35.29083 & -- & -- &  Faint early-type dwarf \\
6\_D352 & 284 & 55.7283333 & -35.34194 & -- & -- &  Faint early-type dwarf \\
16\_D062 & 132 & 53.57628885 & -35.79456749 & -- & -- &  Faint dwarf \\
-- & 218 & 54.68875 & -35.26444 & -- & -- &  Faint dwarf \\
25\_D008 & 35 & 51.26734 & -36.92768 & -- & -- &  Late-type dwarf \\
11\_D110 & 247 & 55.17634 & -35.66098 & -- & -- &  Late-type dwarf \\
6\_D176 & 302 & 56.30116 & -35.57086 & -- & -- &  Late-type dwarf \\
13\_D058 & B1436 & 55.12618 & -37.828 & -- & -- &  Late-type dwarf \\
25\_D000 & 29 & 50.98484 & -36.46444 & 0.47 & 0.3 &  Giant \\
19\_D000 & 83 & 52.64565 & -34.85394 & 0.44 & 0.3 &  Giant \\
16\_D001 & 147 & 53.8191283 & -35.22618122 & 0.20 & 0.4 &  Giant \\
15\_D002 & 153 & 53.879288 & -34.447097 & 0.52 & 0.6 &  Giant \\
11\_D006 & 167 & 54.11509563 & -34.97343539 & 0.04 & 0.1 &  Giant \\
10\_D000 & 177 & 54.19778633 & -34.73982557 & 0.26 & 0.45 &  Giant \\
12\_D003 & 179 & 54.192658 & -35.999275 & 0.46 & 0.25 &  Giant \\
11\_D001 & 184 & 54.23755131 & -35.50661218 & 0.22 & 0.2 &  Giant \\
11\_D005 & 190 & 54.28722412 & -35.1949903 & 0.25 & 0.65 &  Giant \\
11\_D000 & 193 & 54.29891392 & -35.74612523 & 0.33 & 0.8 &  Giant \\
11\_D003 & 213 & 54.621179 & -35.450742 & 0.04 & 0.15 &  Giant \\
11\_D166 & 219 & 54.716321 & -35.594392 & 0.09 & 0.45 &  Giant \\
13\_D000 & 249 & 55.17543 & -37.51077 & 0.11 & 1.35 &  Giant \\
6\_D001 & 276 & 55.58096 & -35.39253 & 0.10 & 0.15 &  Giant \\
6\_D000 & 290 & 55.90451 & -35.85309 & 0.34 & 0.3 & Giant \\
6\_D657 & 272 & 55.5454166 & -35.44222 & -- & -- & likely background  \\
19\_D300 & B664 & 52.66547 & -34.4118 & -- & -- & likely background\\
13\_D010 & -- & 55.13828 & -37.93255 & -- & -- & likely background\\
13\_D050 & B1378 & 54.94089 & -37.84422 & -- & -- & background at z $\sim$ 0.083 \\
13\_D191 & B1244 & 54.56846 & -37.55945 & -- & -- & background at z $\sim$ 0.046 \\
15\_D217 & B746 & 52.96215 & -34.59884 & -- & -- & background at z $\sim$ 0.064 \\
15\_D370 & B878 & 53.43658 & -34.34742 & -- & -- & likely background \\
15\_D404 & B750 & 52.98334 & -34.26382 & -- & -- & background at z=0.094\\
19\_D175 & B676 & 52.7069 & -34.6462 & -- & -- & background at z $\sim$ 0.095 \\
25\_D011a & B192 & 50.91508 & -36.90845 & -- & -- & background at z $\sim$ 0.065 \\
25\_D216 & B329 & 51.43603 & -36.41562 & -- & -- & background at z $\sim$ 0.095 \\
6\_D062 & B1814 & 56.34685 & -35.7656 & -- & -- & background at z $\sim$ 0.1 \\
6\_D166 & B1714 & 56.00279 & -35.59143 & -- & -- & background at z $\sim$ 0.06 \\
10\_D186? & B1106 & 54.19807014 & -34.54157219 & -- & -- & background at z$\sim$ 0.06 \\
11\_D151 & B1116 & 54.22483981 & -35.59791335 & -- & -- & background at z $\sim$ 0.07 \\
11\_D061 & B1195 & 54.46386272 & -35.79329419 & -- & -- & background at z $\sim$ 0.11 \\
-- & -- & 54.96887 & -35.07328 & -- & -- & UCD \\
-- & -- & 55.10383 & -35.11031 & -- & -- & UCD \\
-- & -- & 54.83542 & -35.32058 & -- & -- & UCD \\
-- & -- & 55.08917 & -35.40756 & -- & -- & UCD \\
-- & -- & 54.93129 & -35.44969 & -- & -- & UCD \\
\hline
\end{tabular}
\end{table*}

\begin{figure*}
    \centering
    \includegraphics[width=2.in,clip,trim = 20 0 70 0]{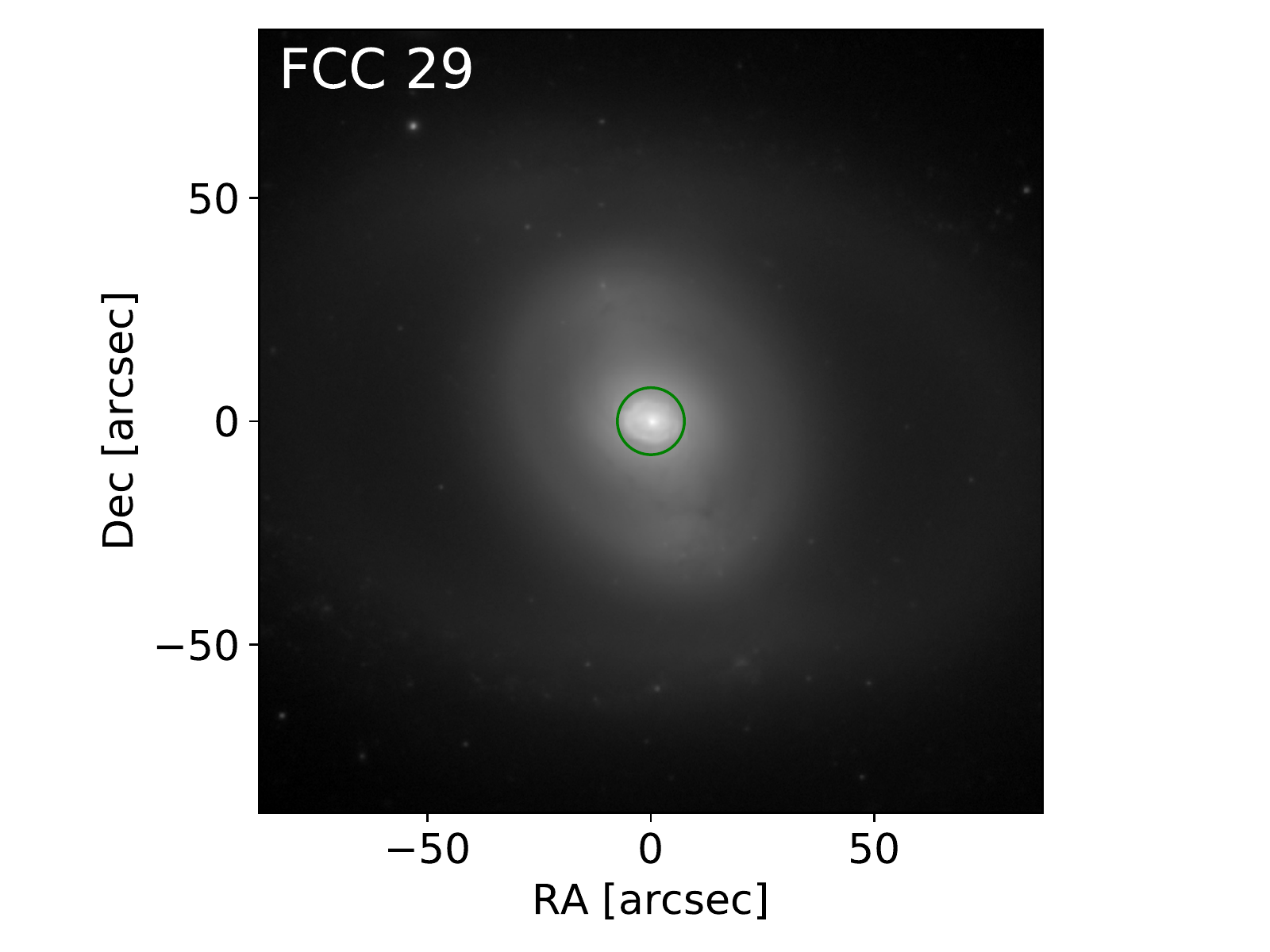}
    \includegraphics[width=2.25in,clip,trim = 20 10 10 10]{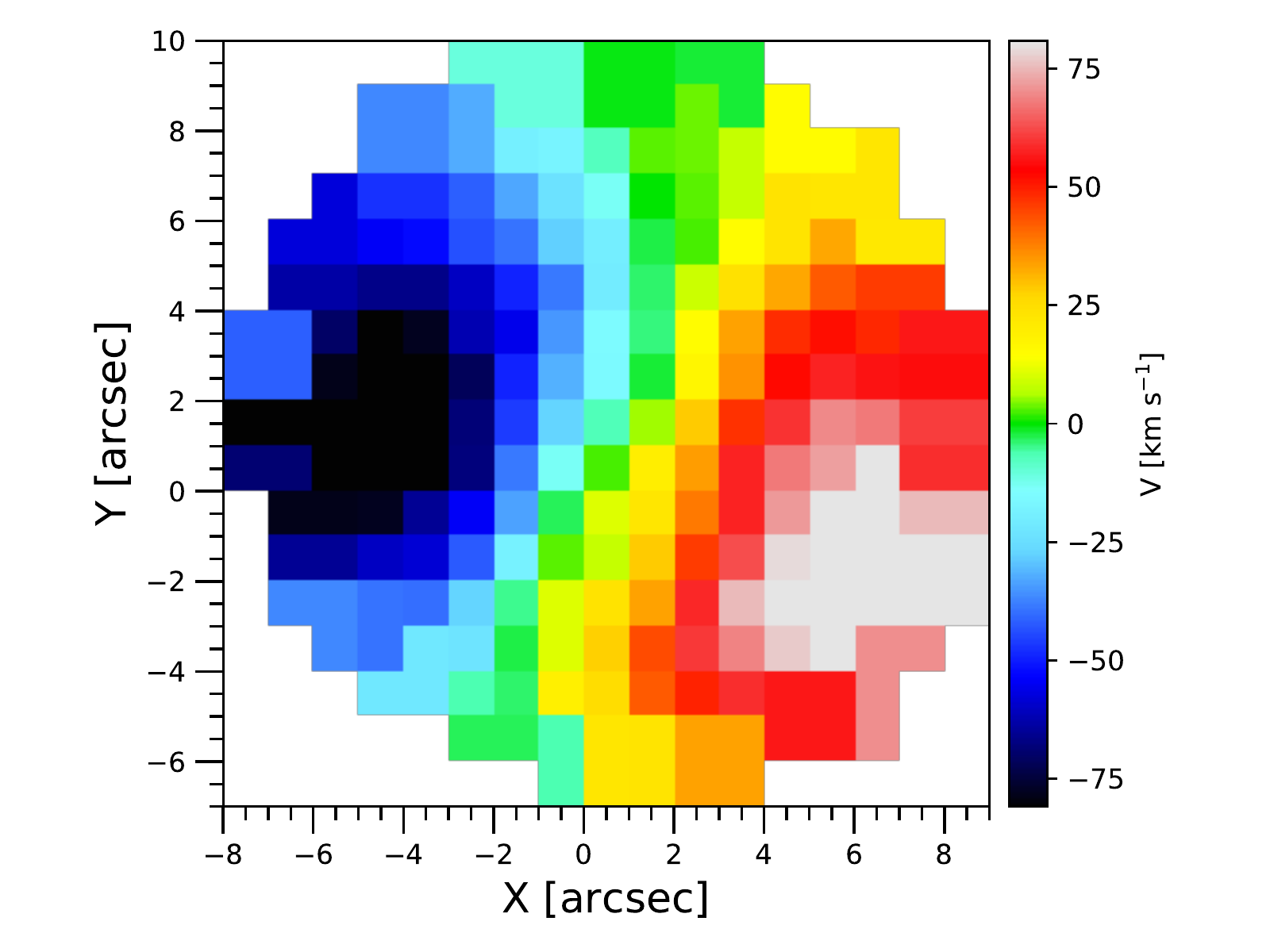}
    \includegraphics[width=2.25in,clip,trim = 20 10 10 10]{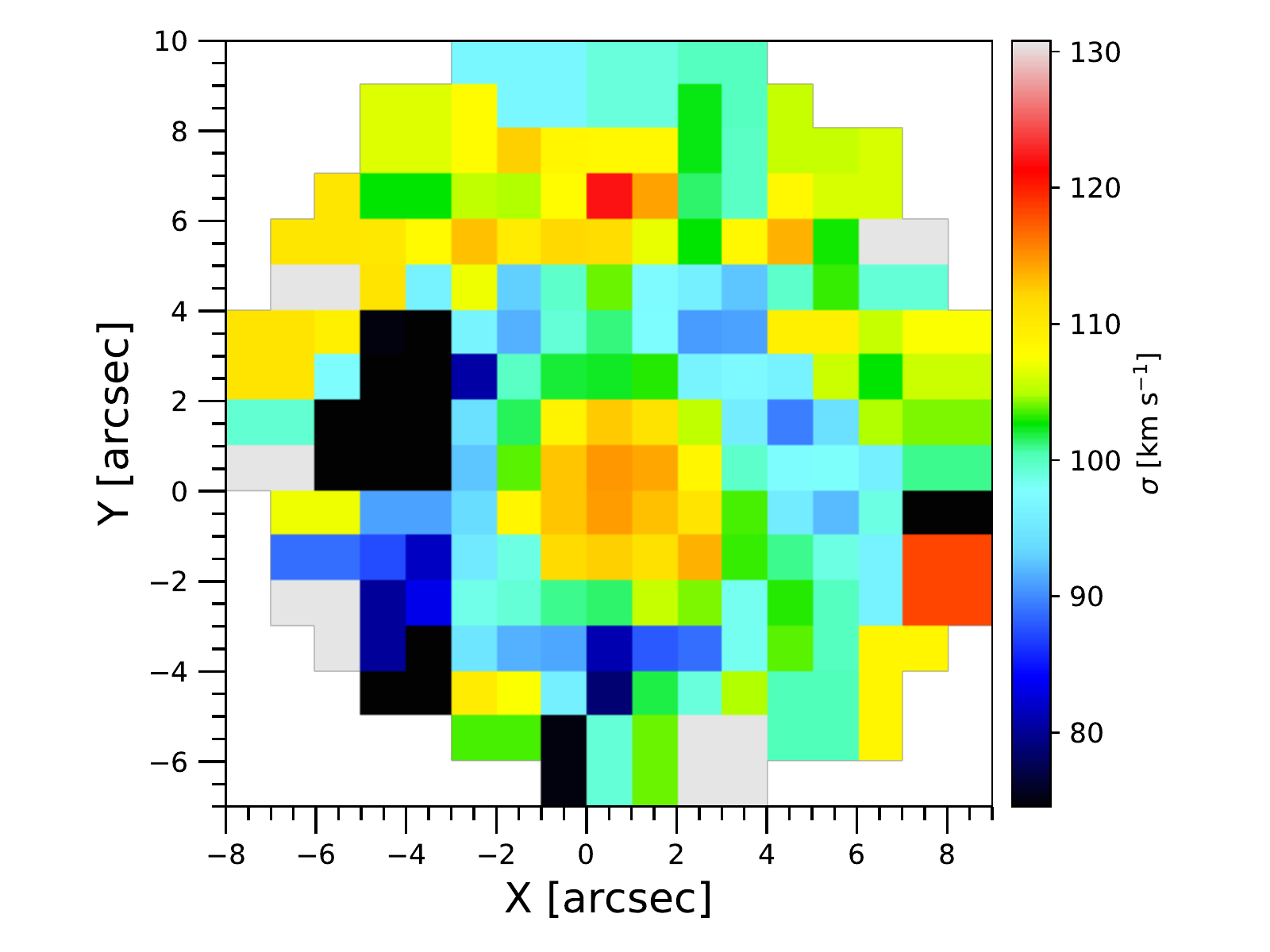}
    
    \includegraphics[width=2.in,clip,trim = 20 0 70 0]{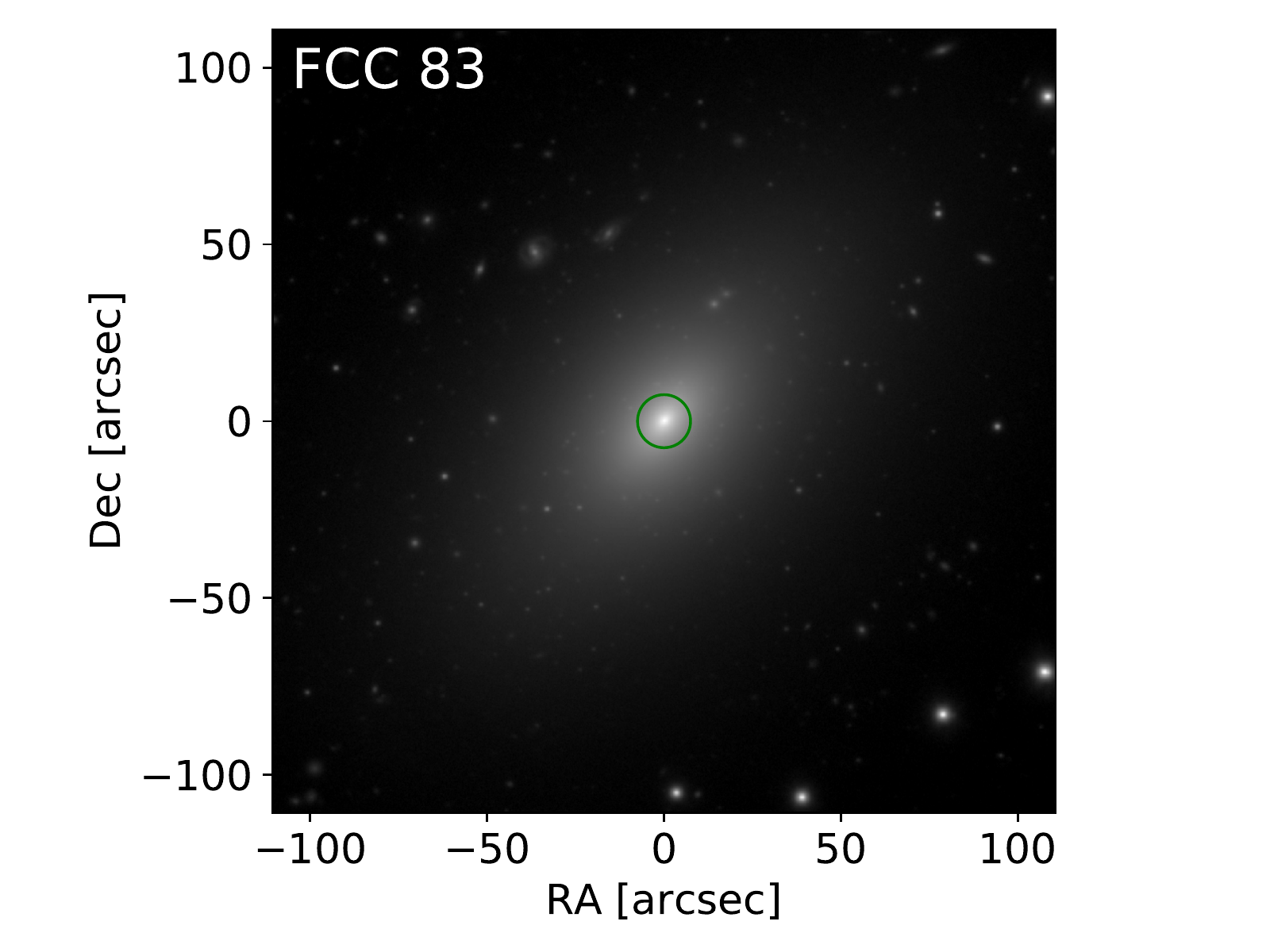}
    \includegraphics[width=2.25in,clip,trim = 20 10 40 10]{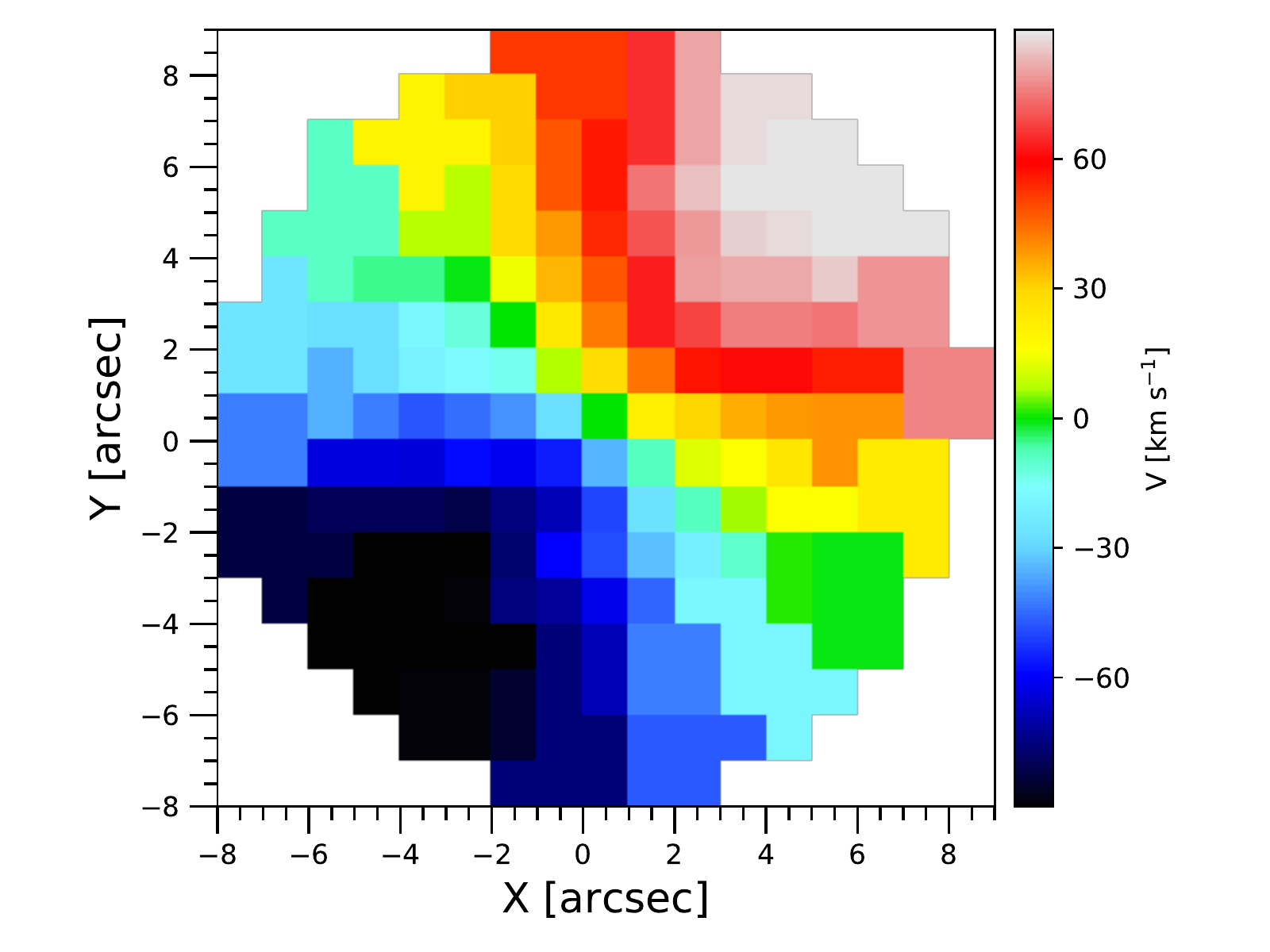}
    \includegraphics[width=2.25in,clip,trim = 20 10 40 10]{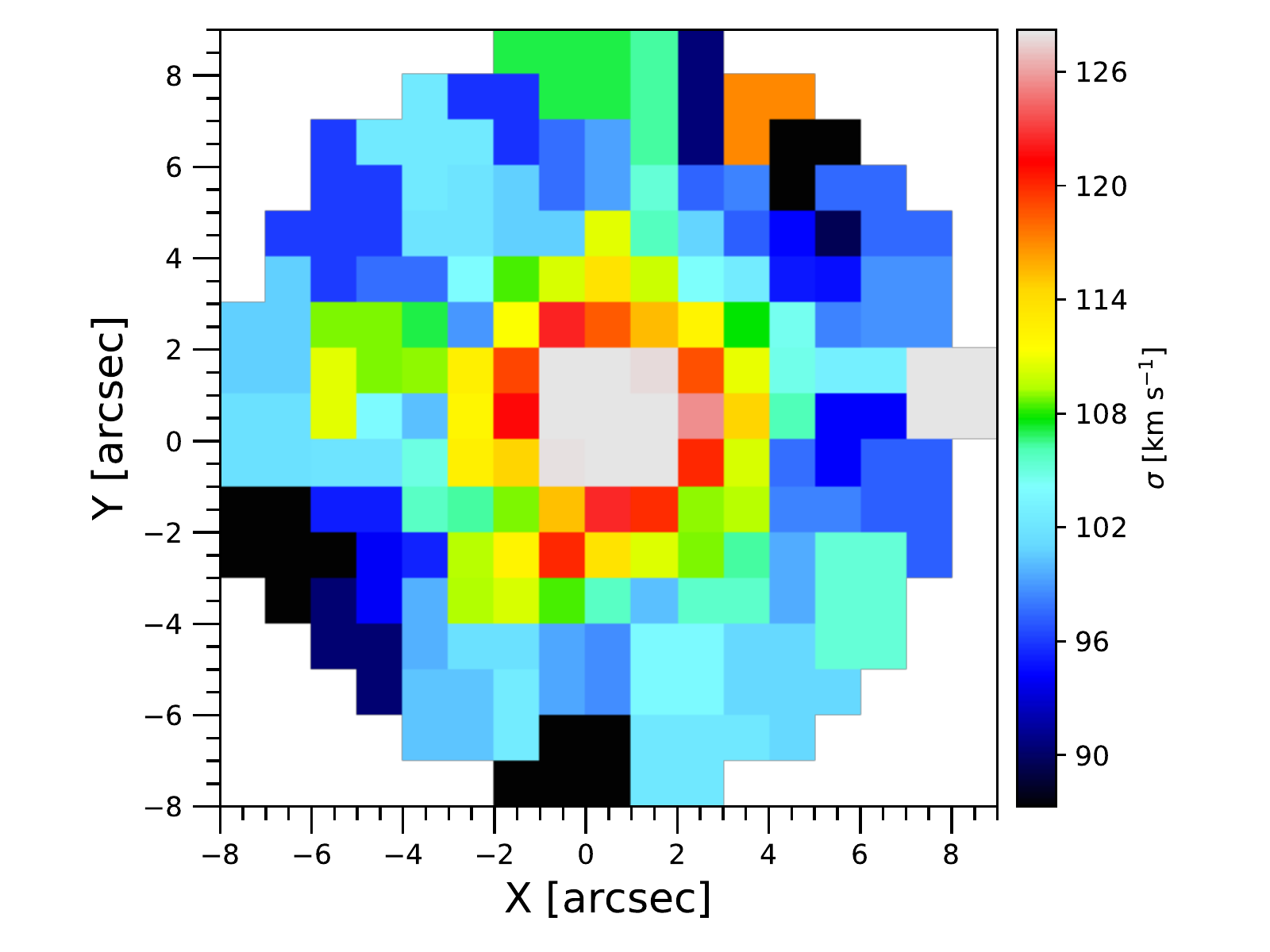}

    \includegraphics[width=2.in,clip,trim = 20 0 70 0]{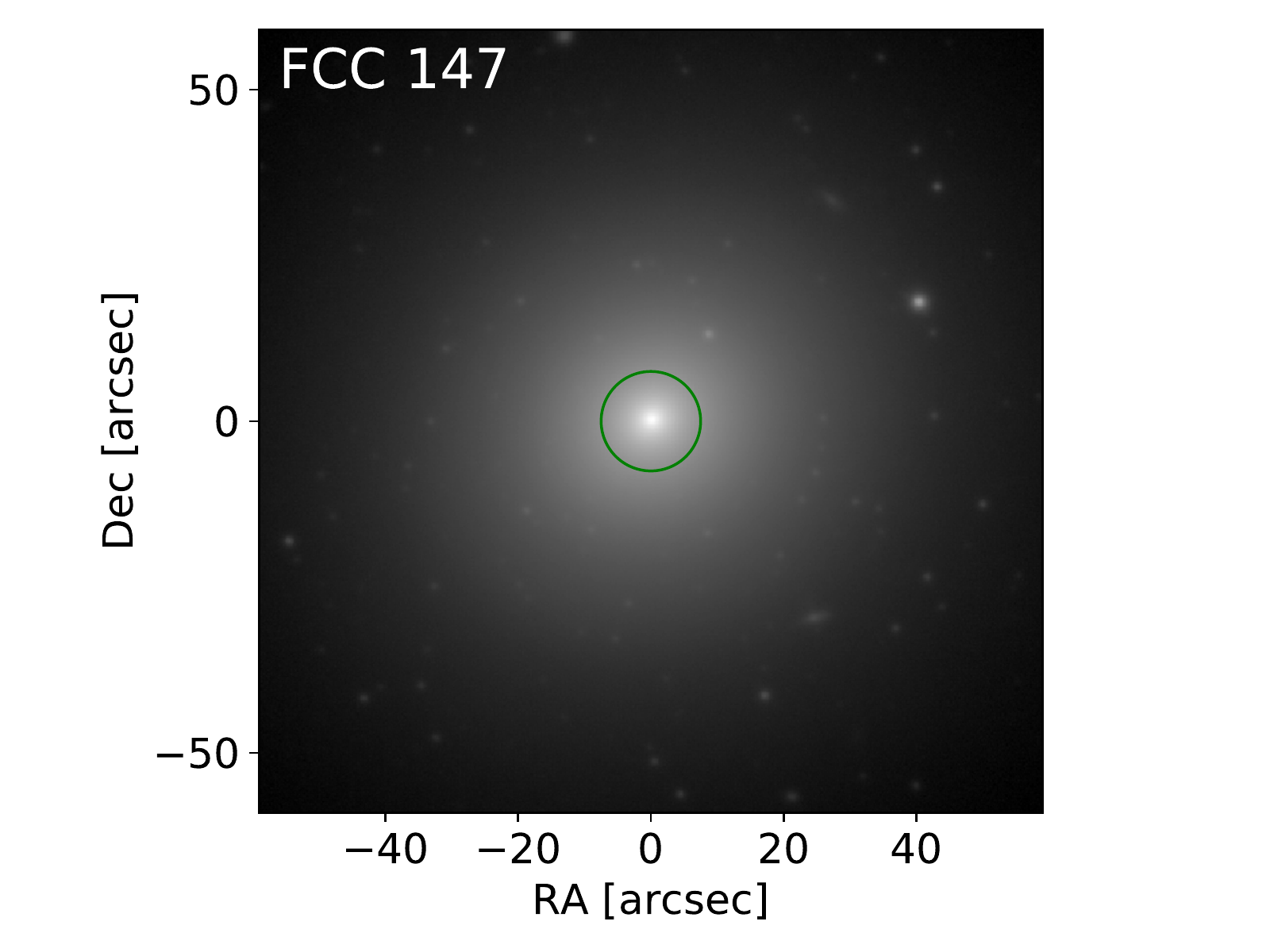}
    \includegraphics[width=2.25in,clip,trim = 20 10 10 10]{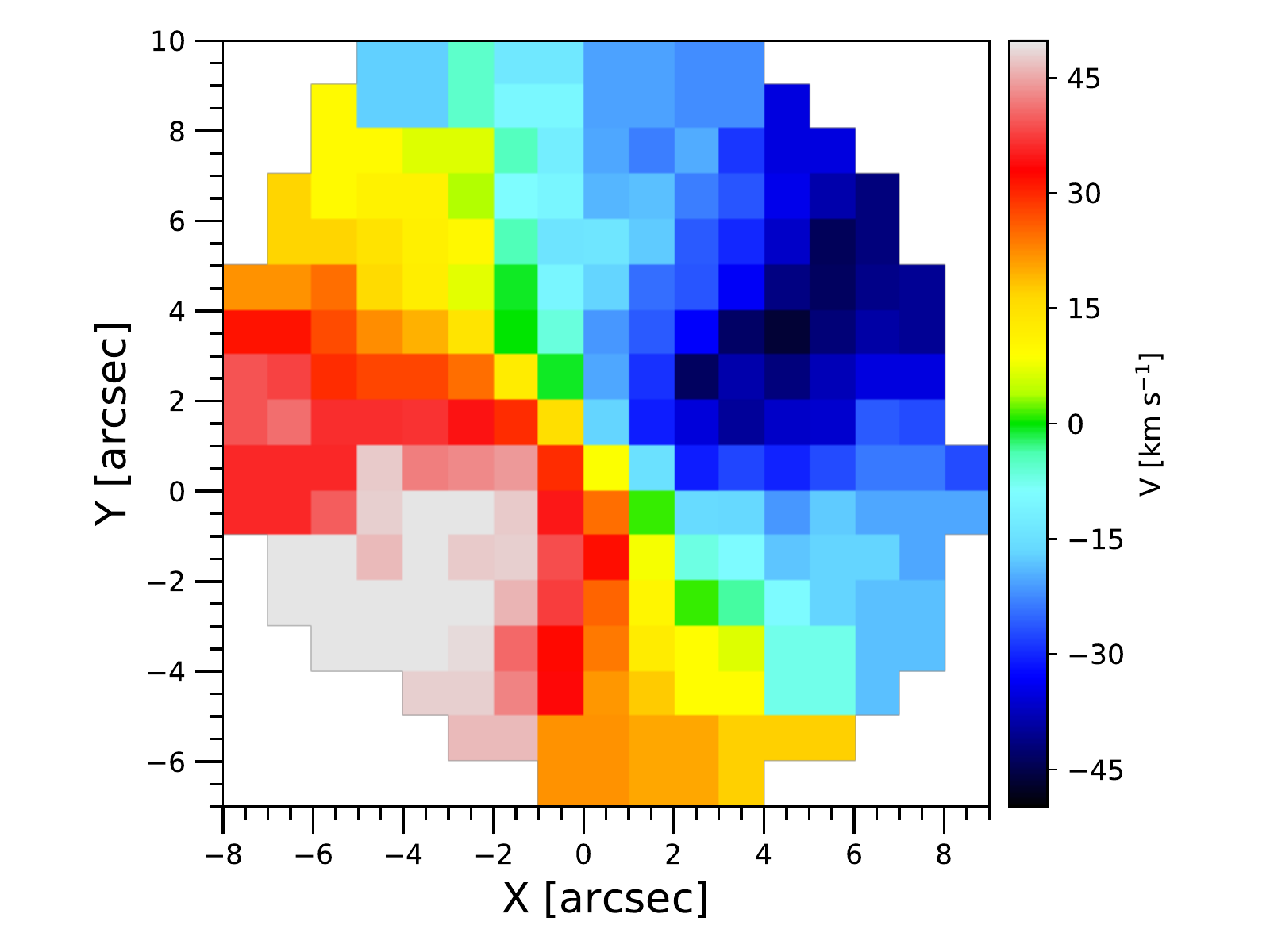}
    \includegraphics[width=2.25in,clip,trim = 20 10 10 10]{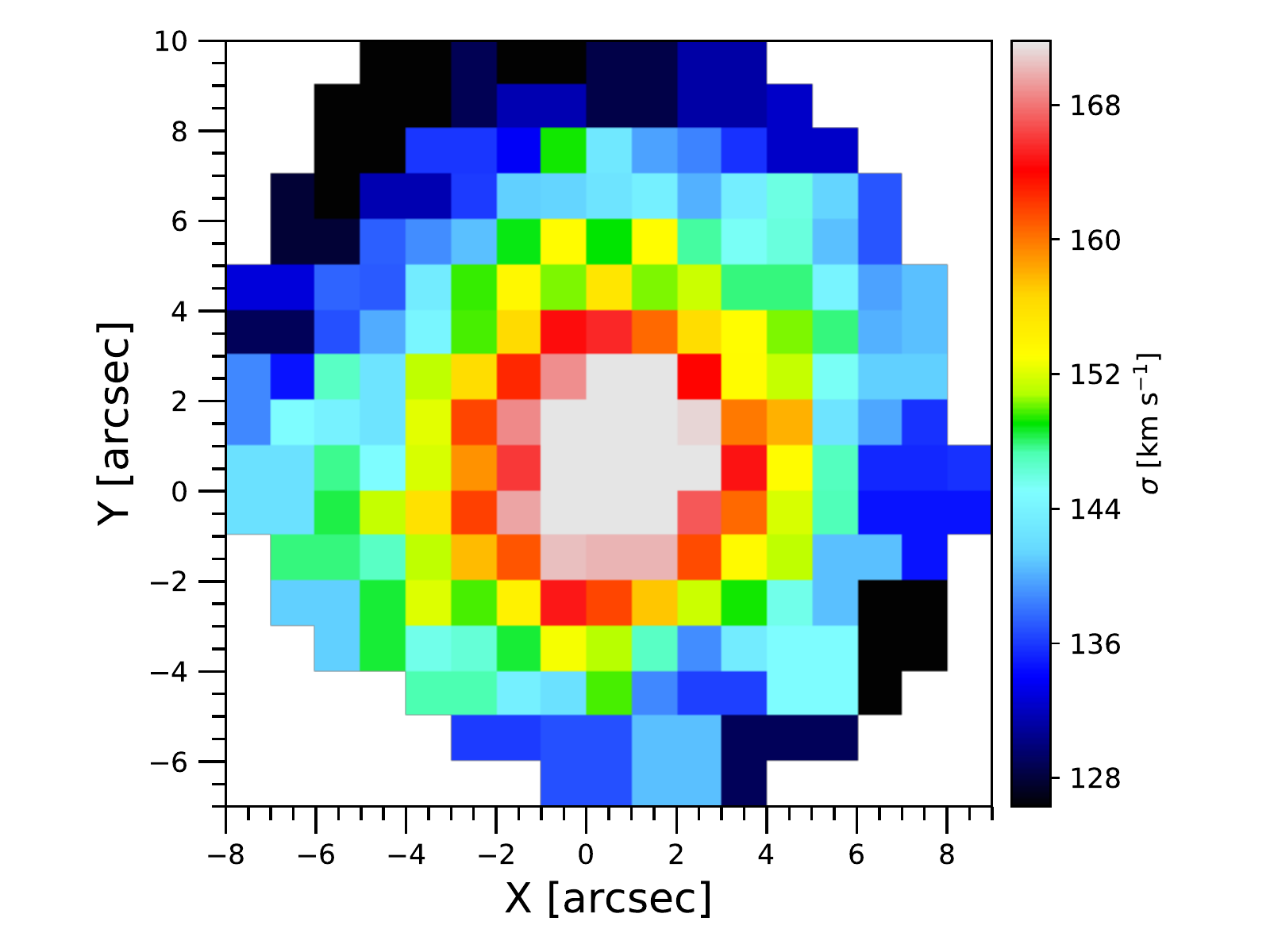}

    \includegraphics[width=2.in,clip,trim = 20 0 70 0]{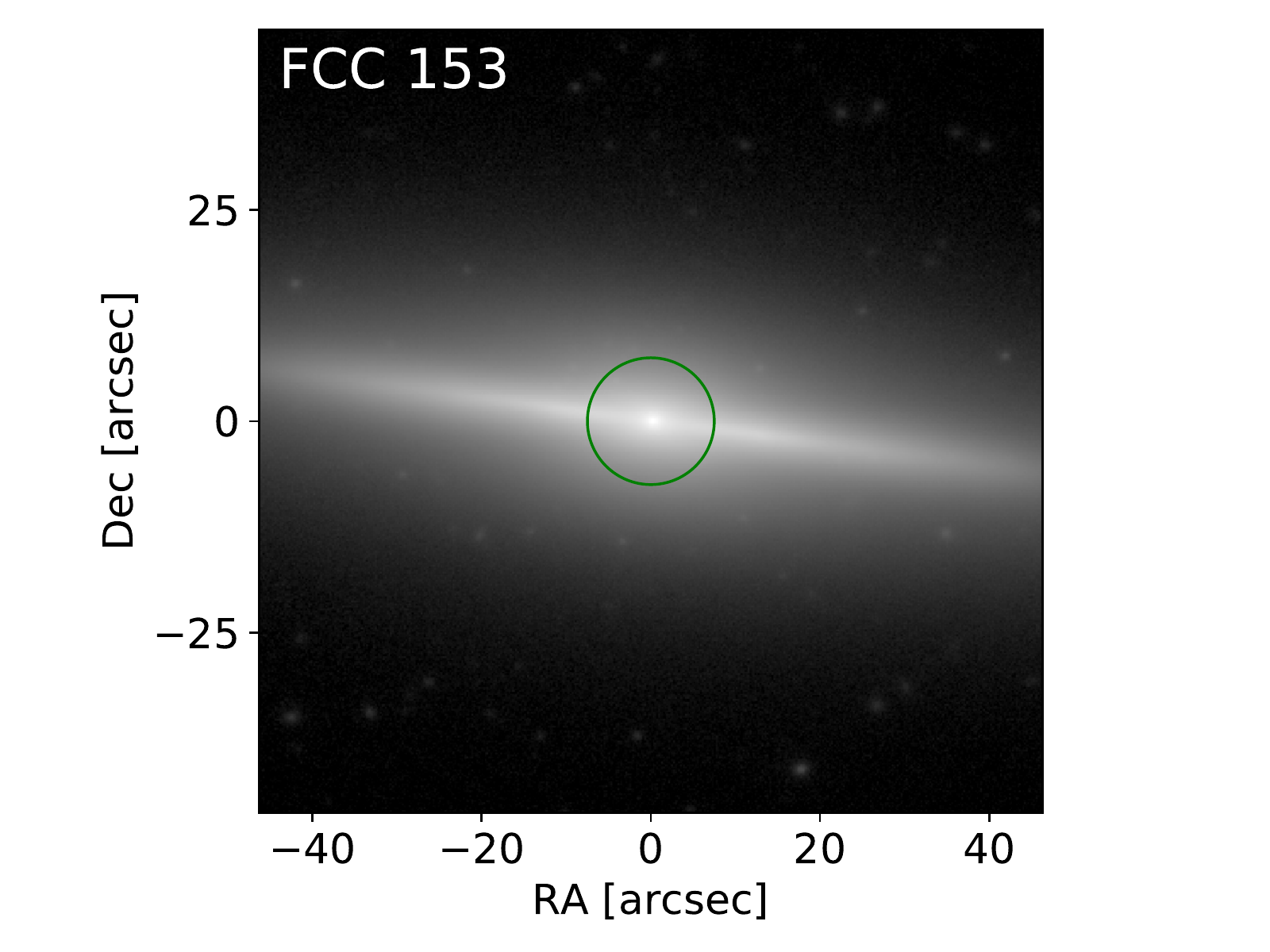}
    \includegraphics[width=2.25in,clip,trim = 20 10 10 10]{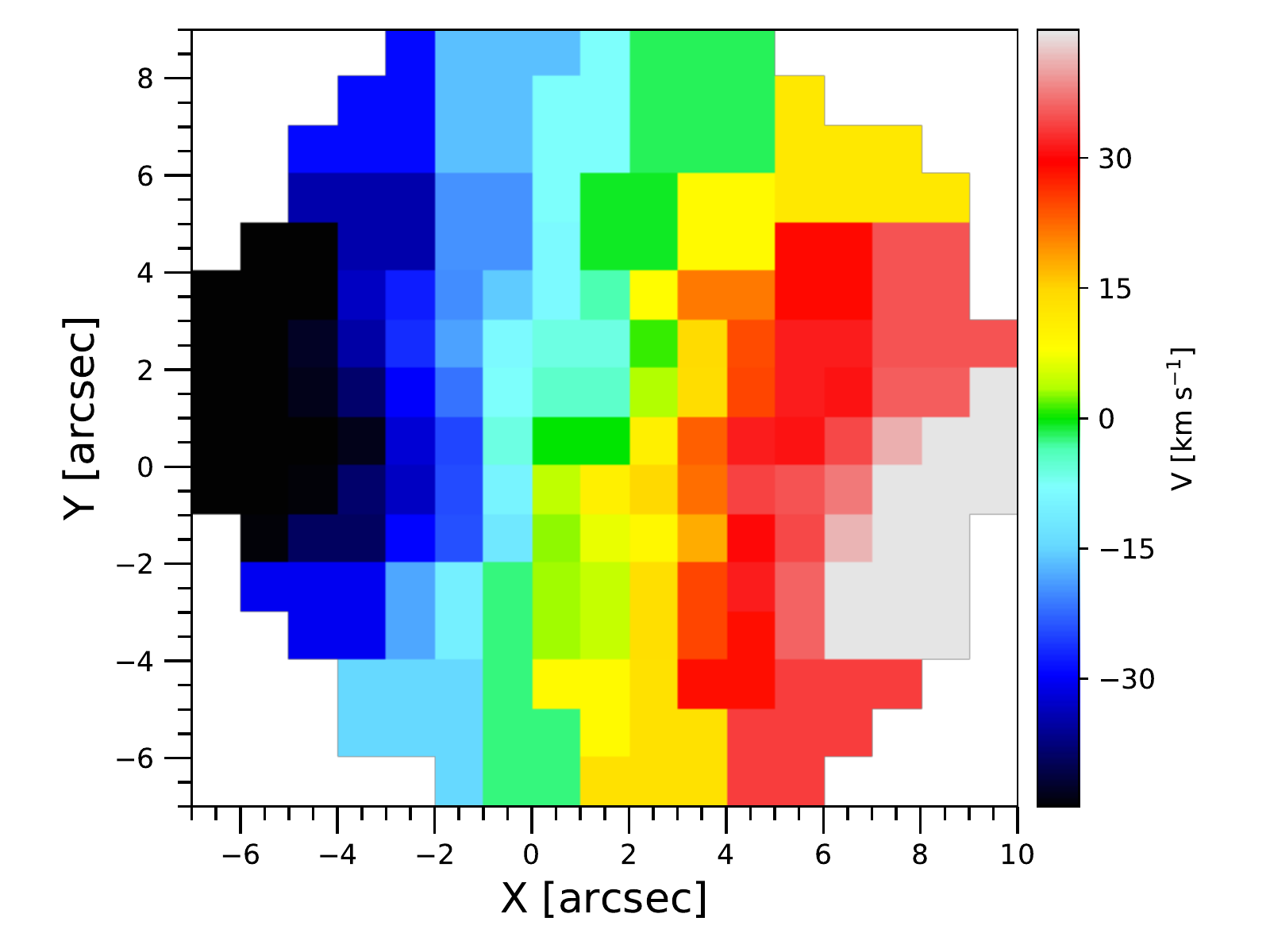}
    \includegraphics[width=2.25in,clip,trim = 20 10 10 10]{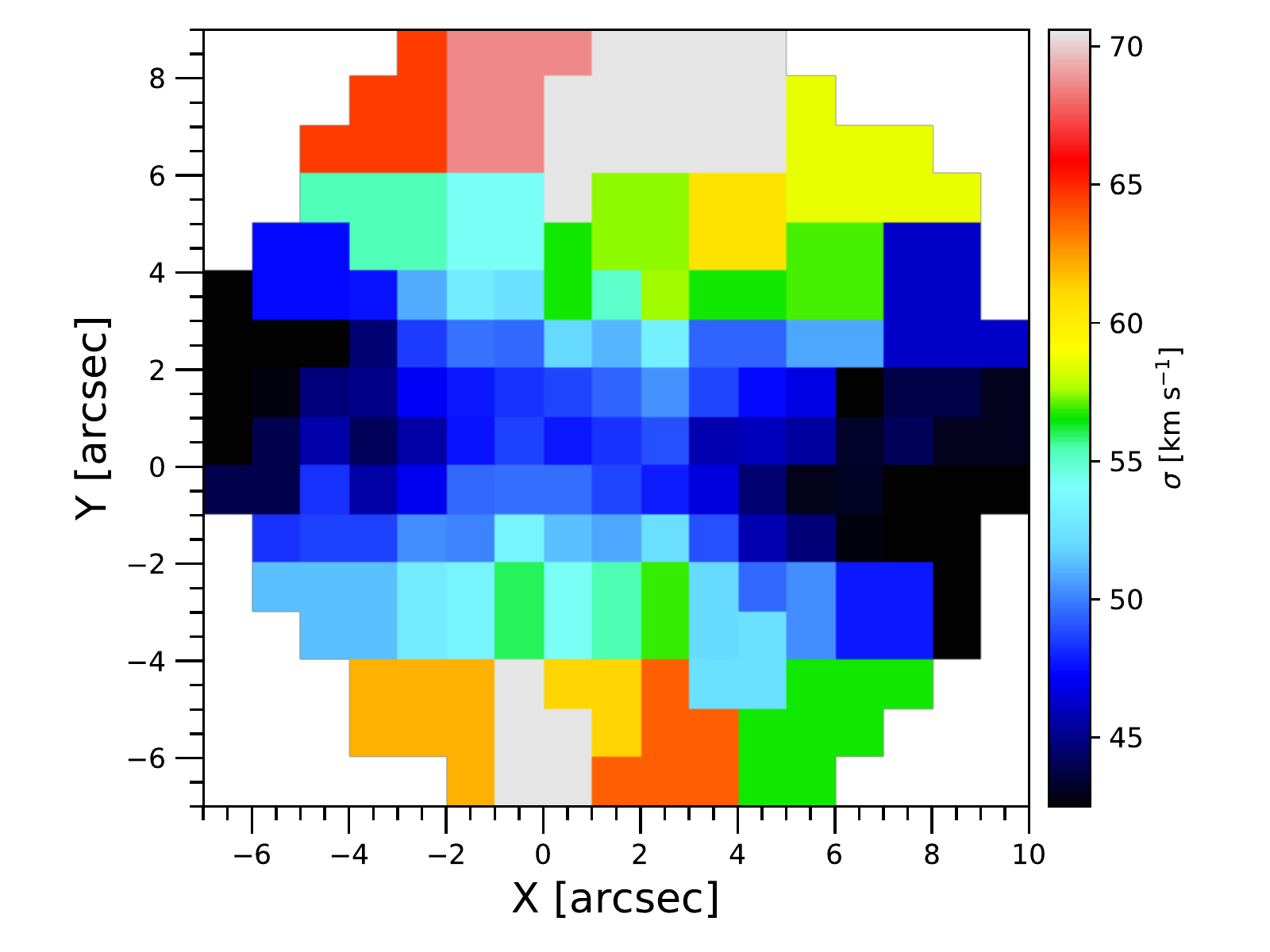}
    
    \includegraphics[width=2.in,clip,trim = 20 0 70 0]{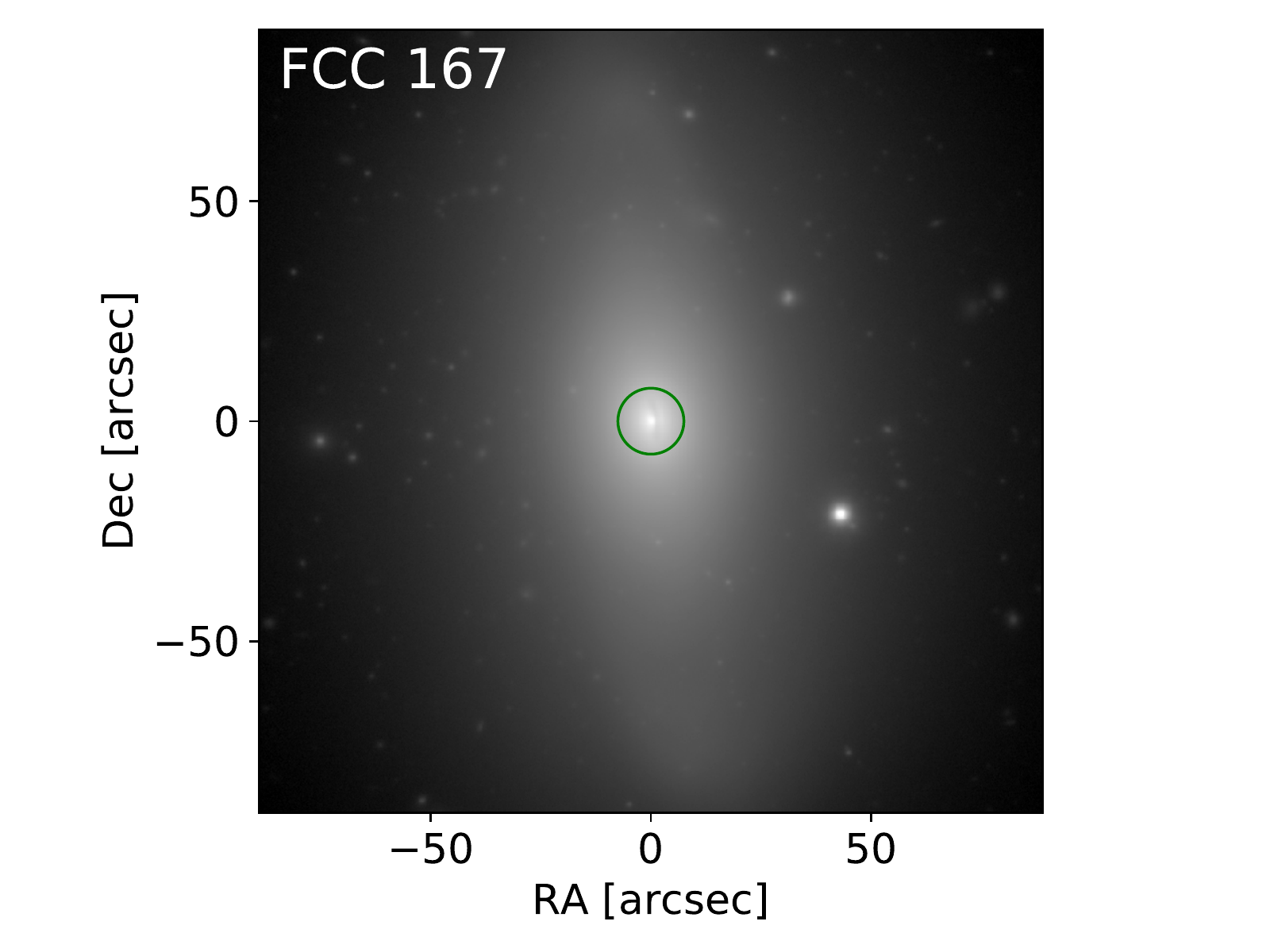}
    \includegraphics[width=2.25in,clip,trim = 5 10 10 10]{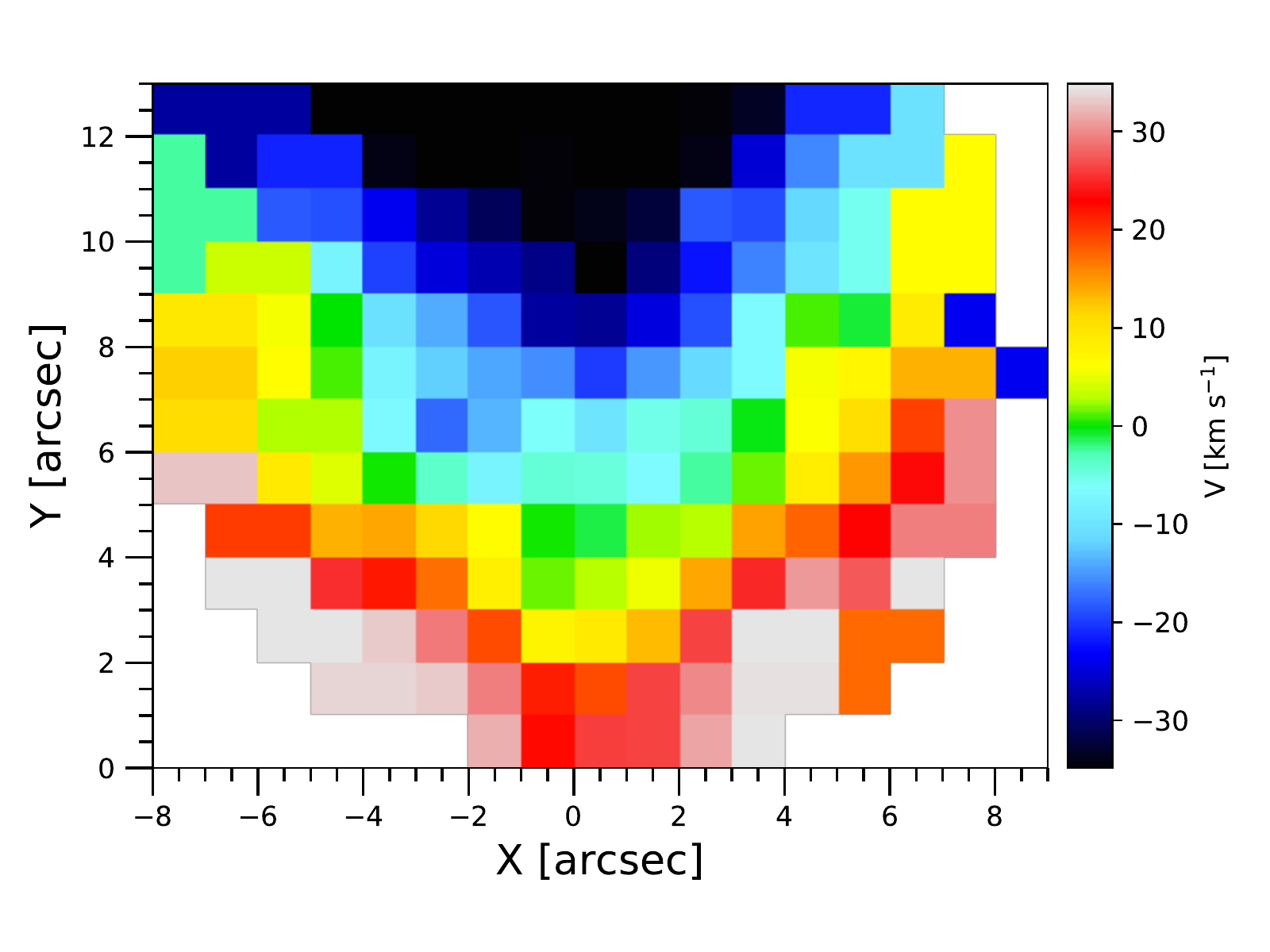}
    \includegraphics[width=2.25in,clip,trim = 20 10 10 10]{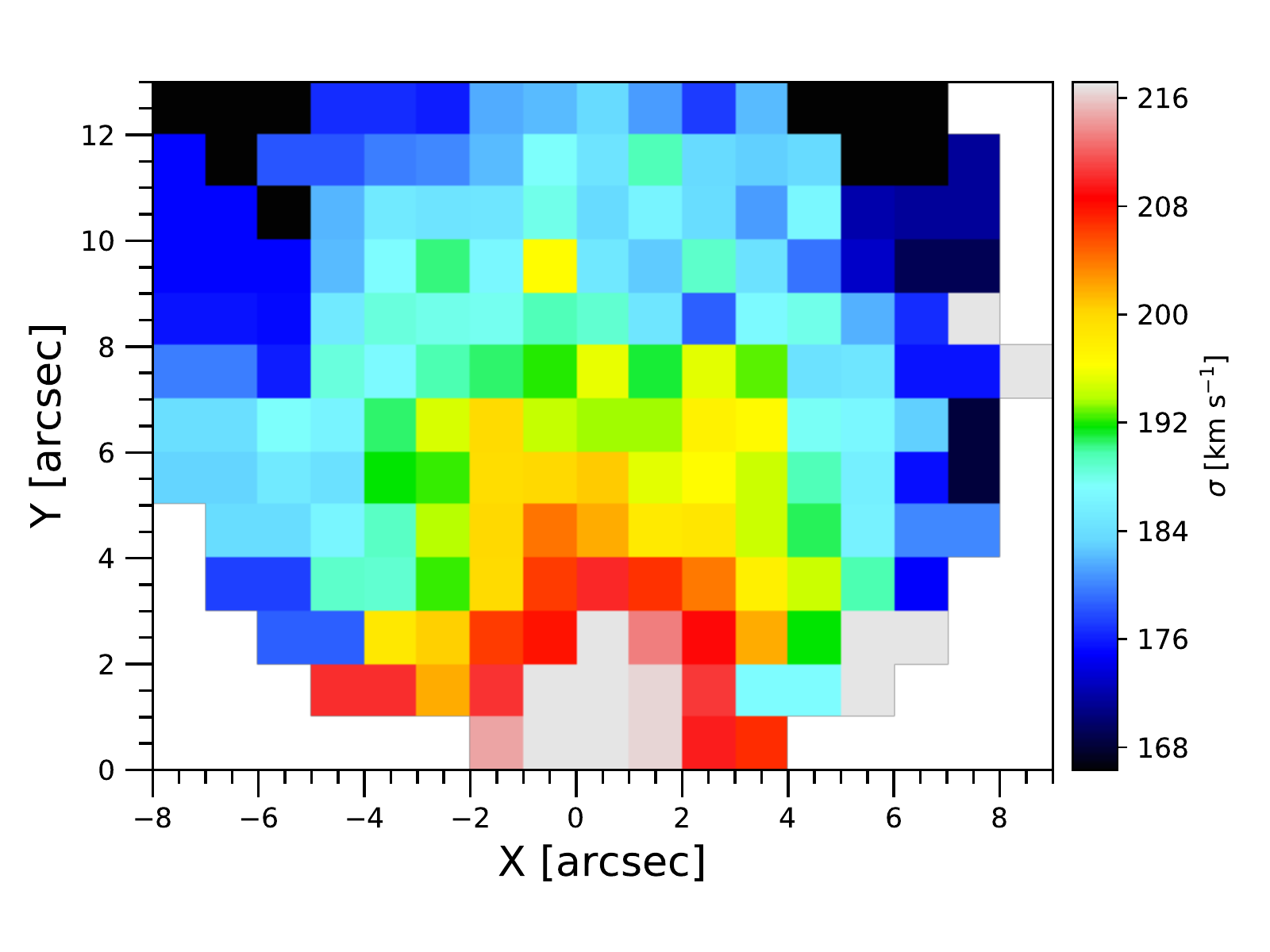}

    \caption{As Fig \ref{fig:primary_maps_app}, but for giant galaxies, ordered by FCC number.}
    \label{fig:secondary_maps}
\end{figure*}

\begin{figure*}
    \centering
    \includegraphics[width=2.in,clip,trim = 20 0 70 0]{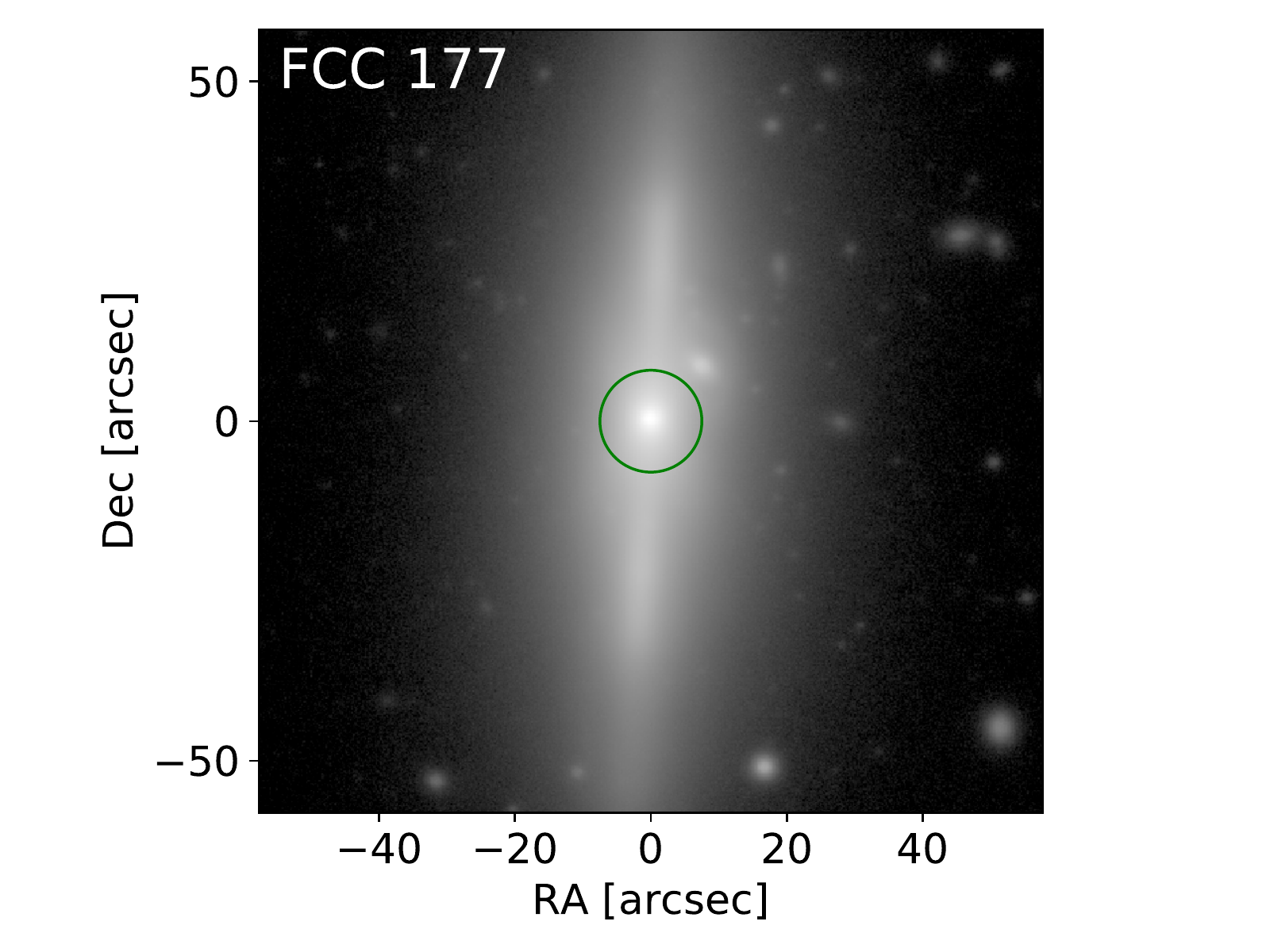}
    \includegraphics[width=2.25in,clip,trim = 20 10 40 10]{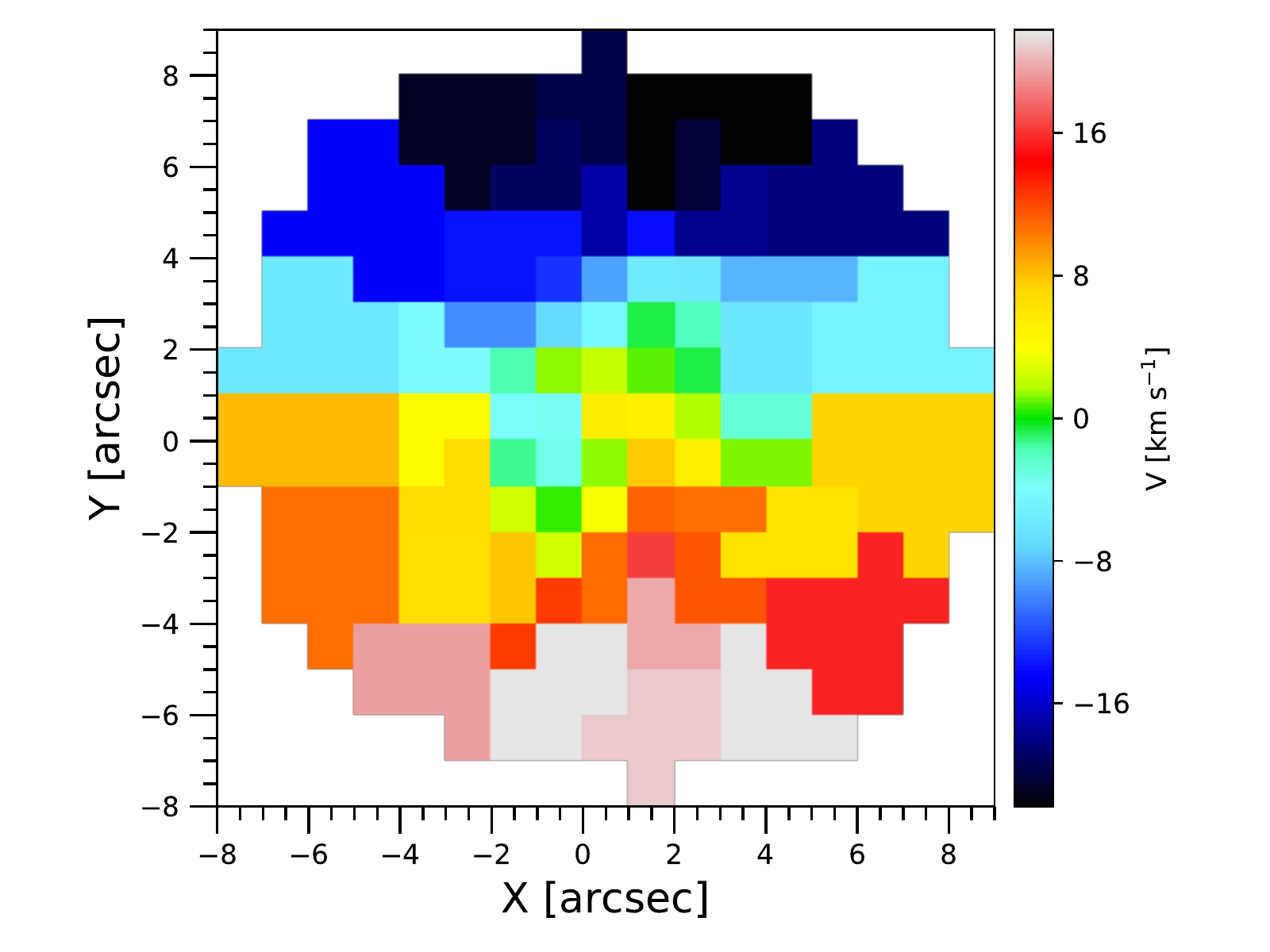}
    \includegraphics[width=2.25in,clip,trim = 20 10 40 10]{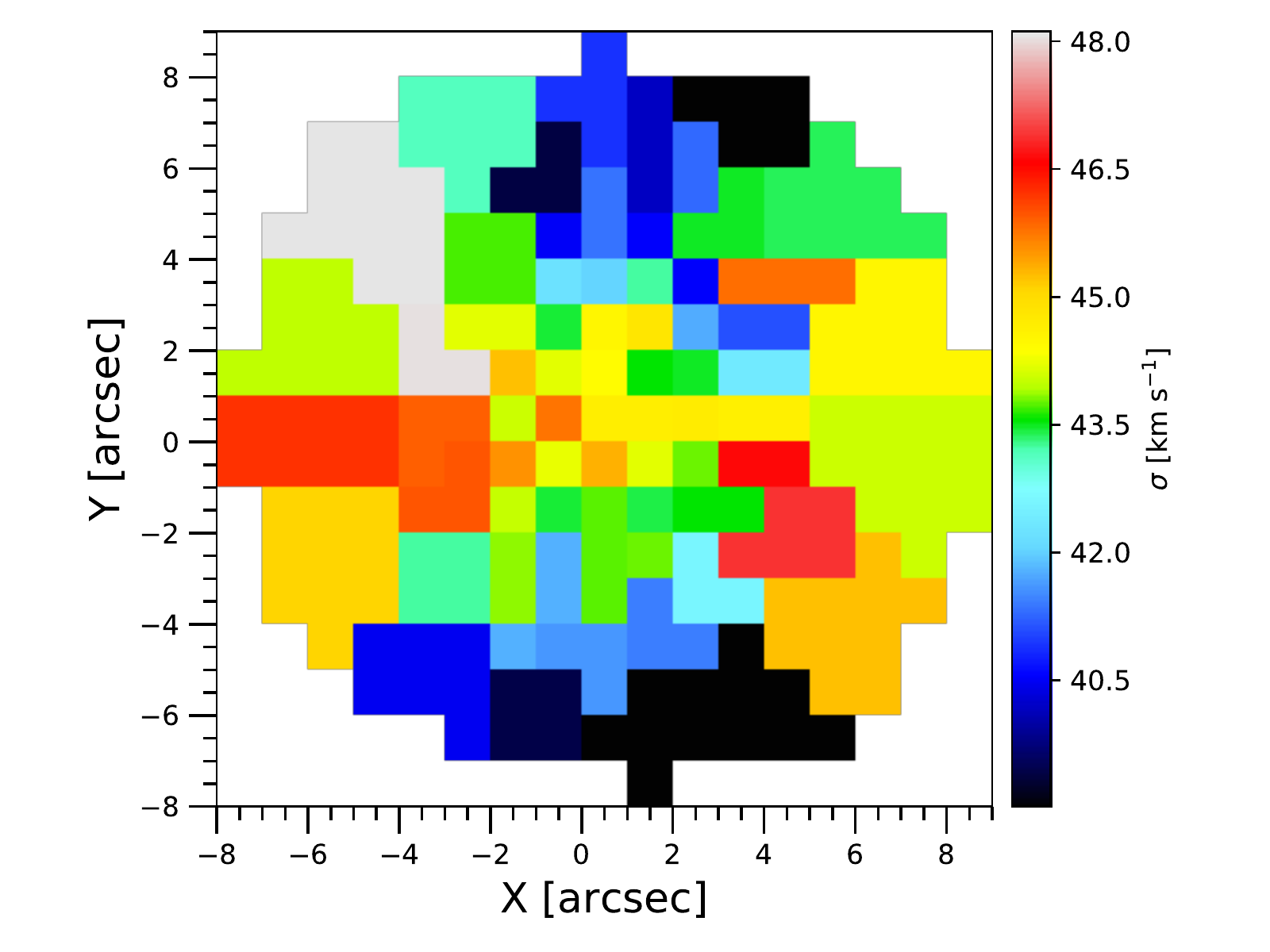}
    
    \includegraphics[width=2.in,clip,trim = 20 0 70 0]{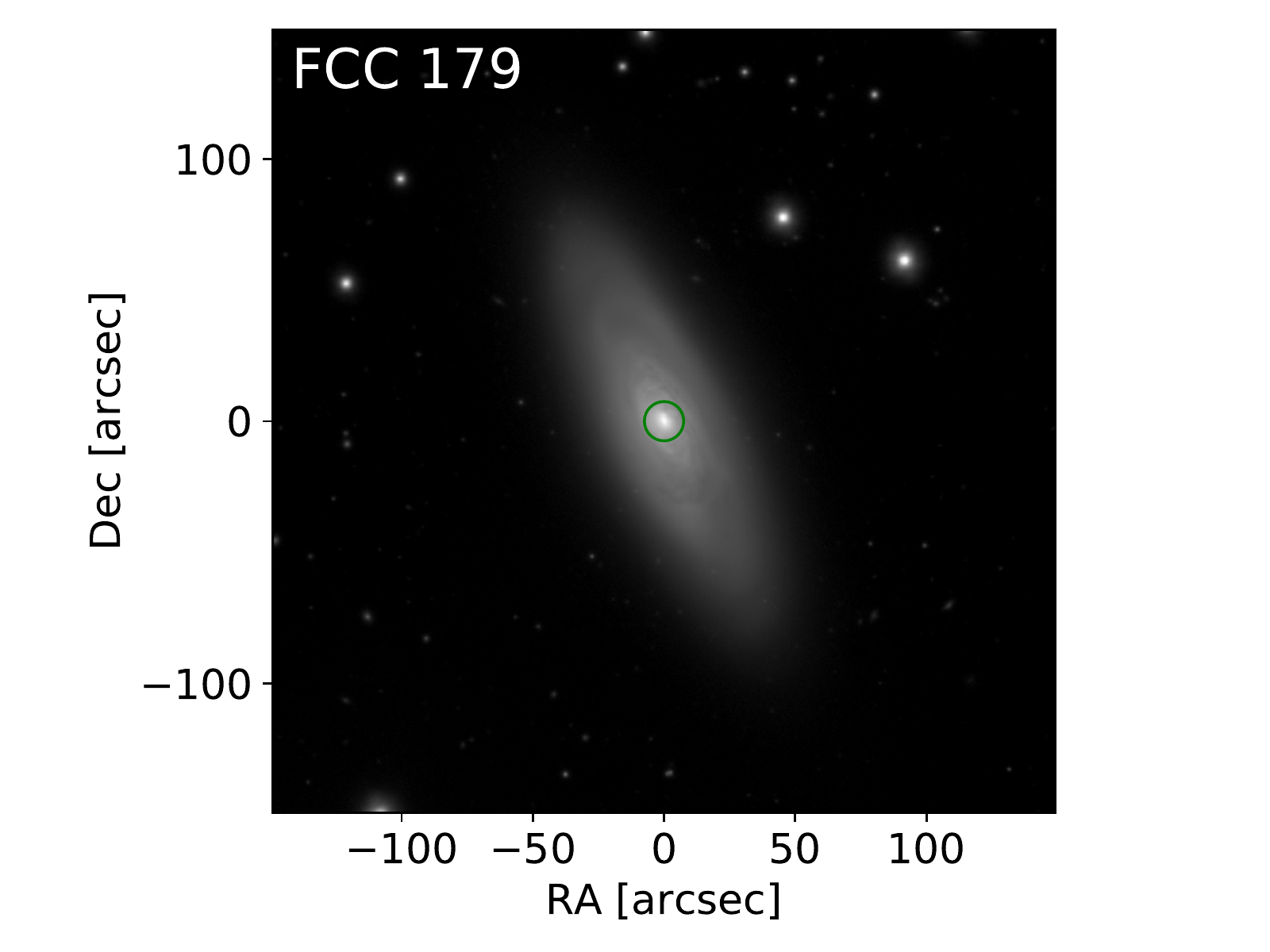}
    \includegraphics[width=2.25in,clip,trim = 20 10 10 10]{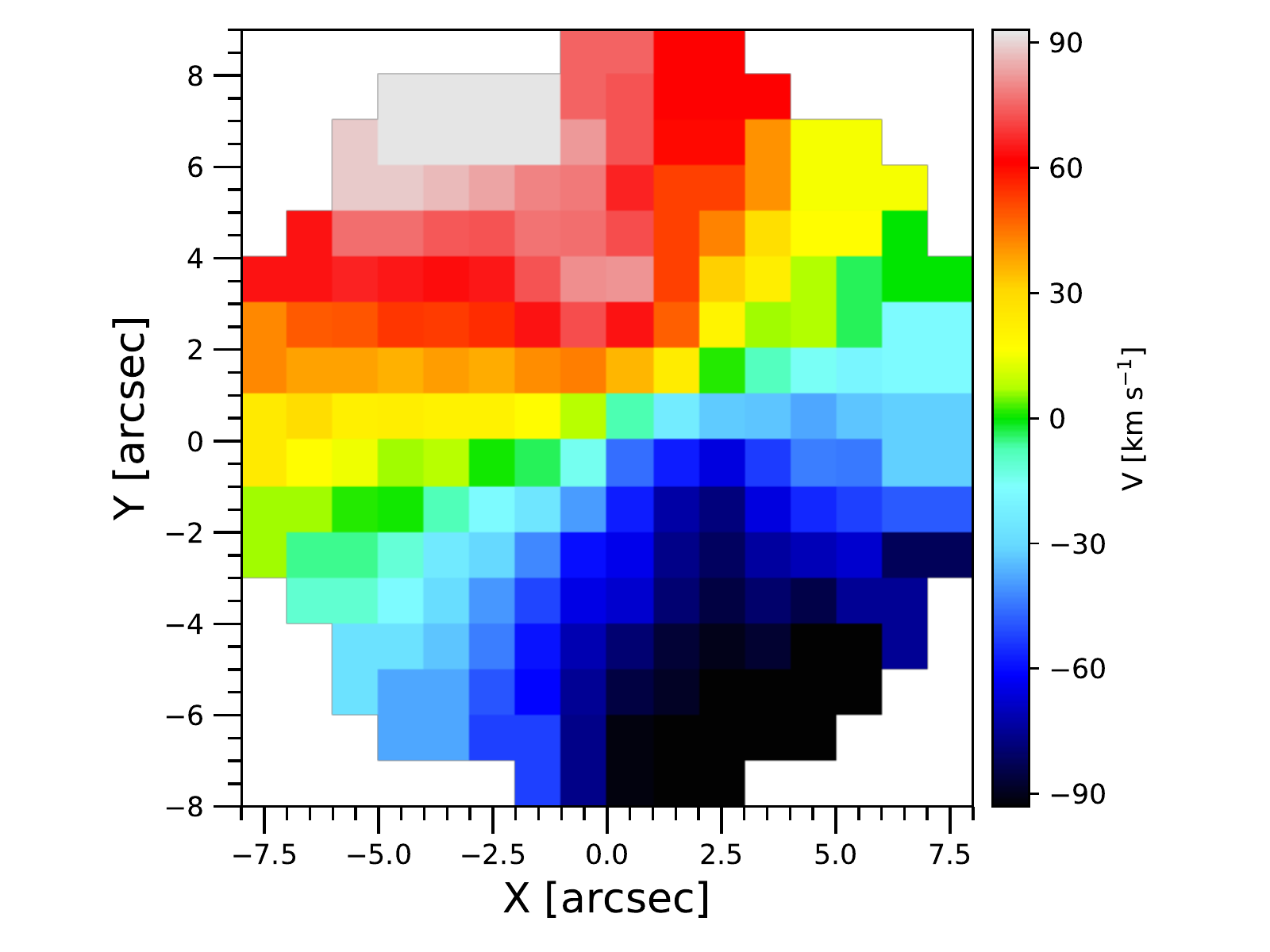}
    \includegraphics[width=2.25in,clip,trim = 20 10 10 10]{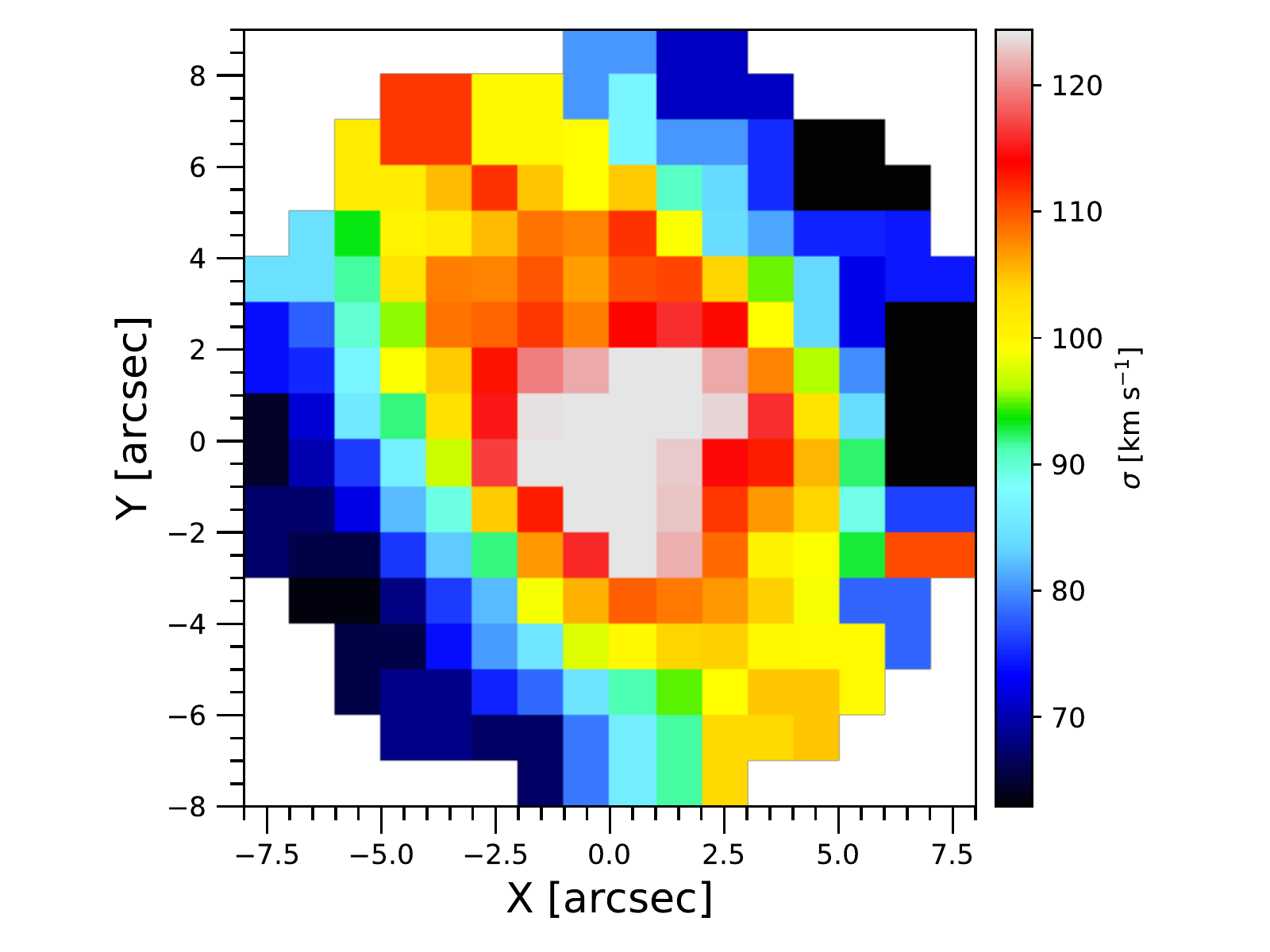}
    
    \includegraphics[width=2.in,clip,trim = 20 0 70 0]{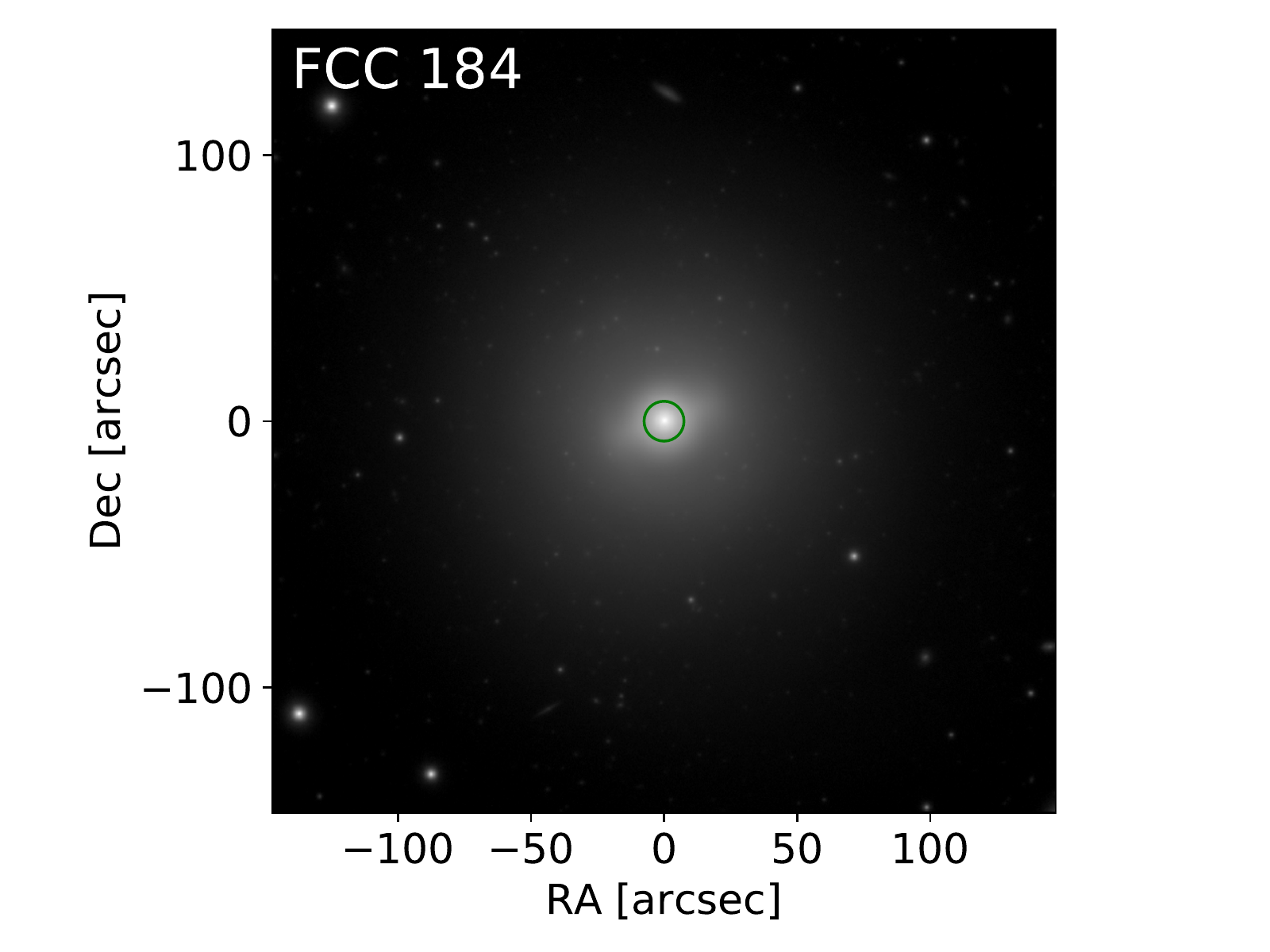}
    \includegraphics[width=2.25in,clip,trim = 20 10 30 10]{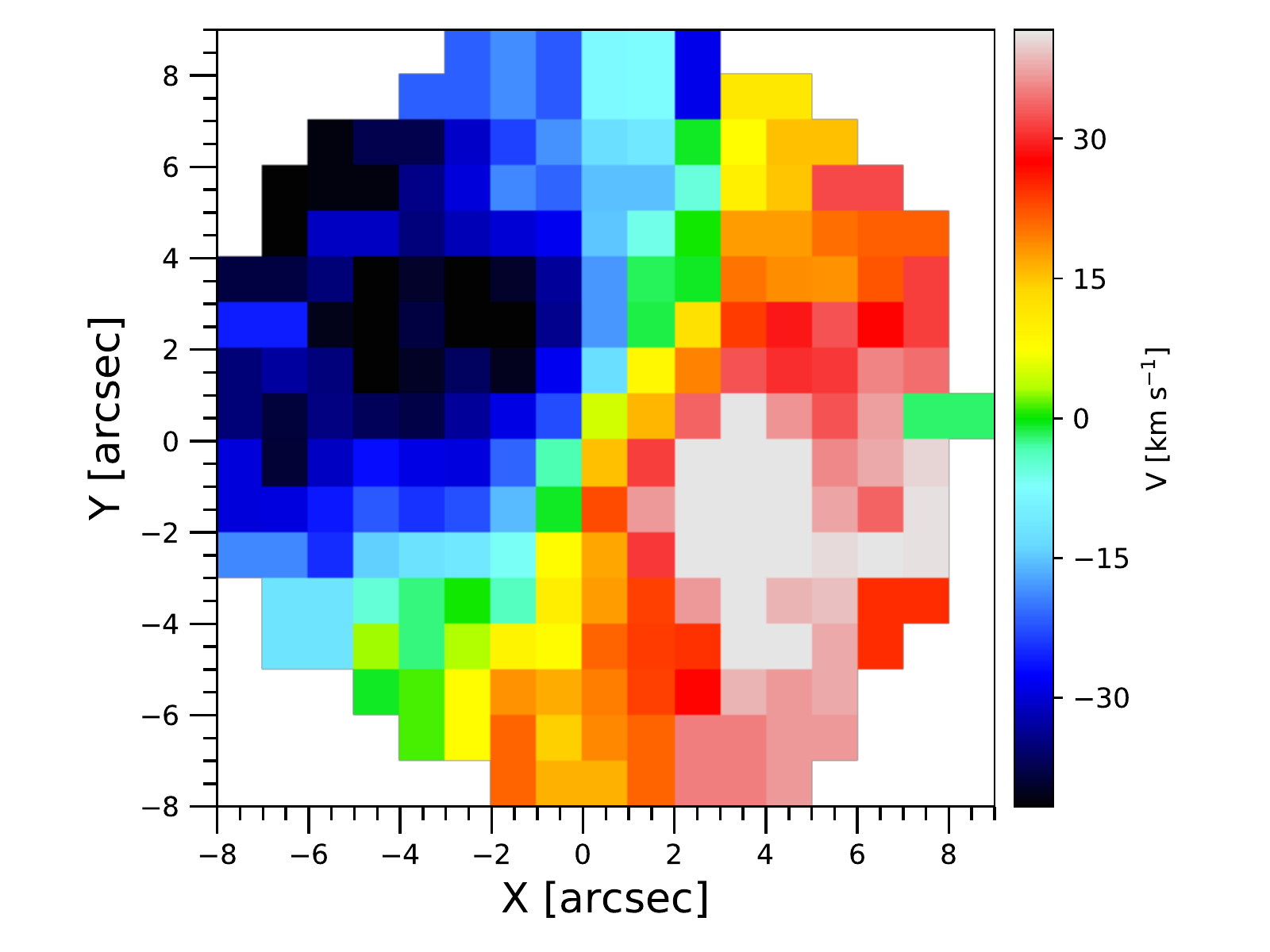}
    \includegraphics[width=2.25in,clip,trim = 20 10 30 10]{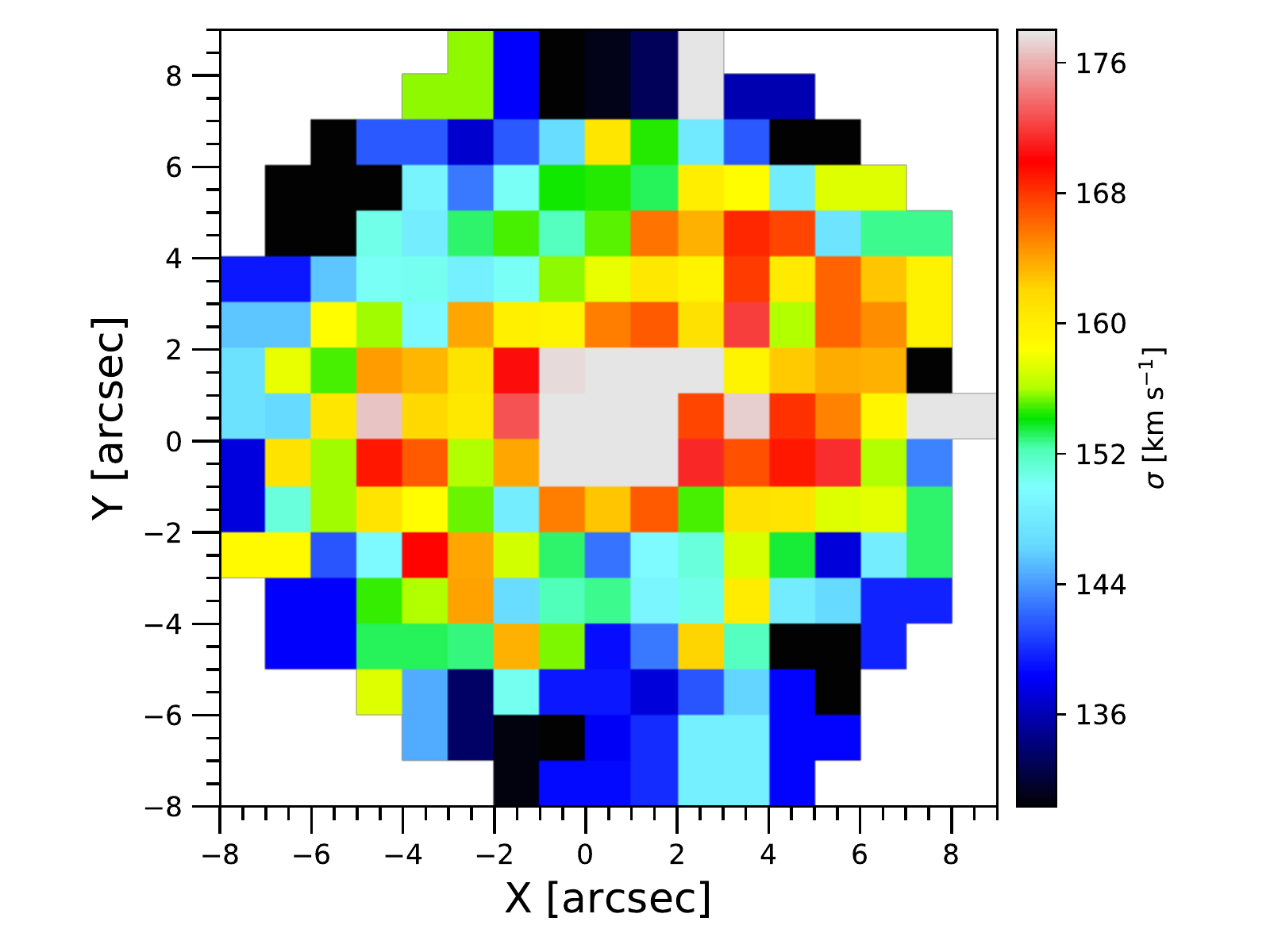}
    
    \includegraphics[width=2.in,clip,trim = 20 0 70 0]{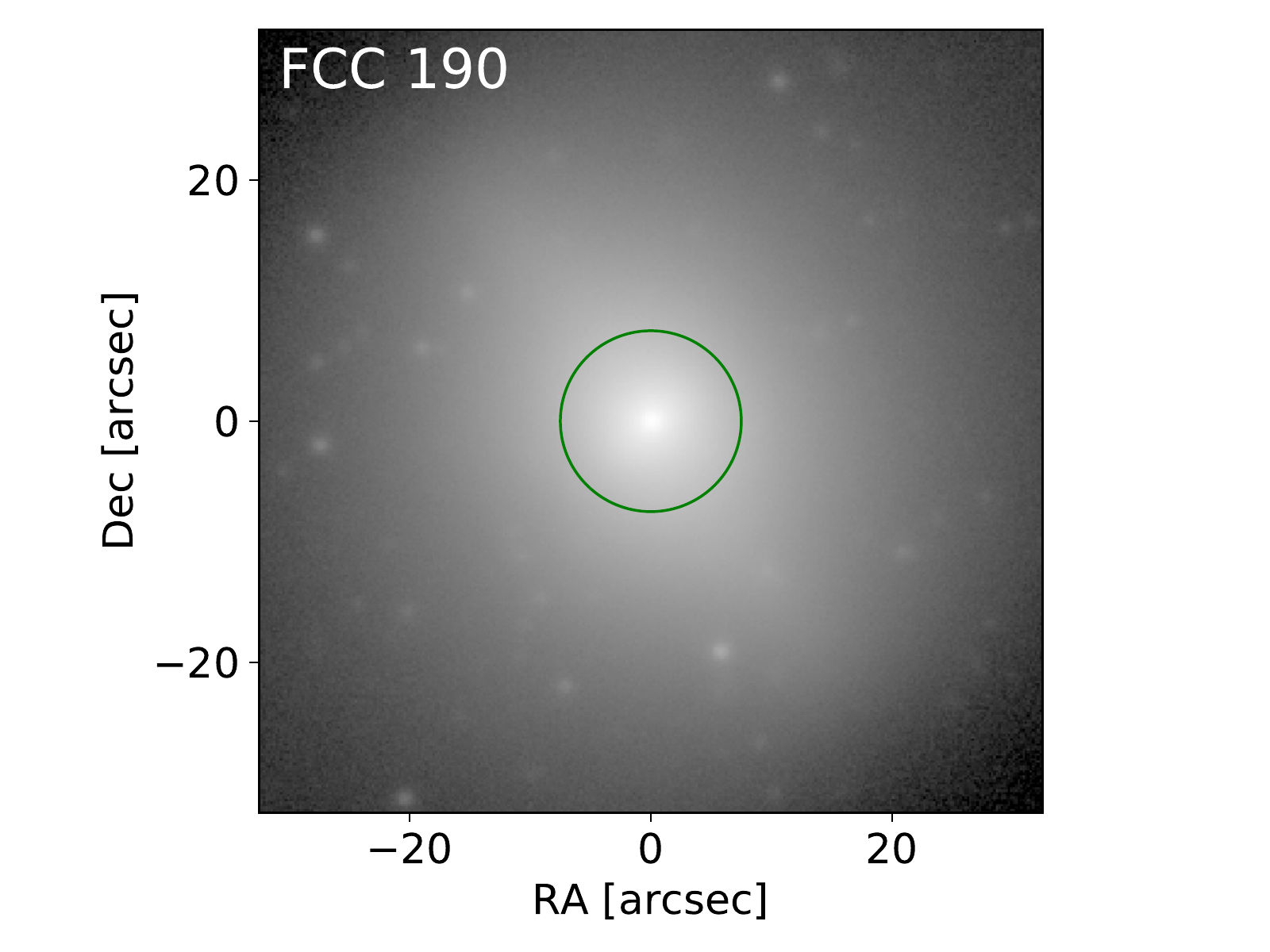}
    \includegraphics[width=2.25in,clip,trim = 20 10 40 10]{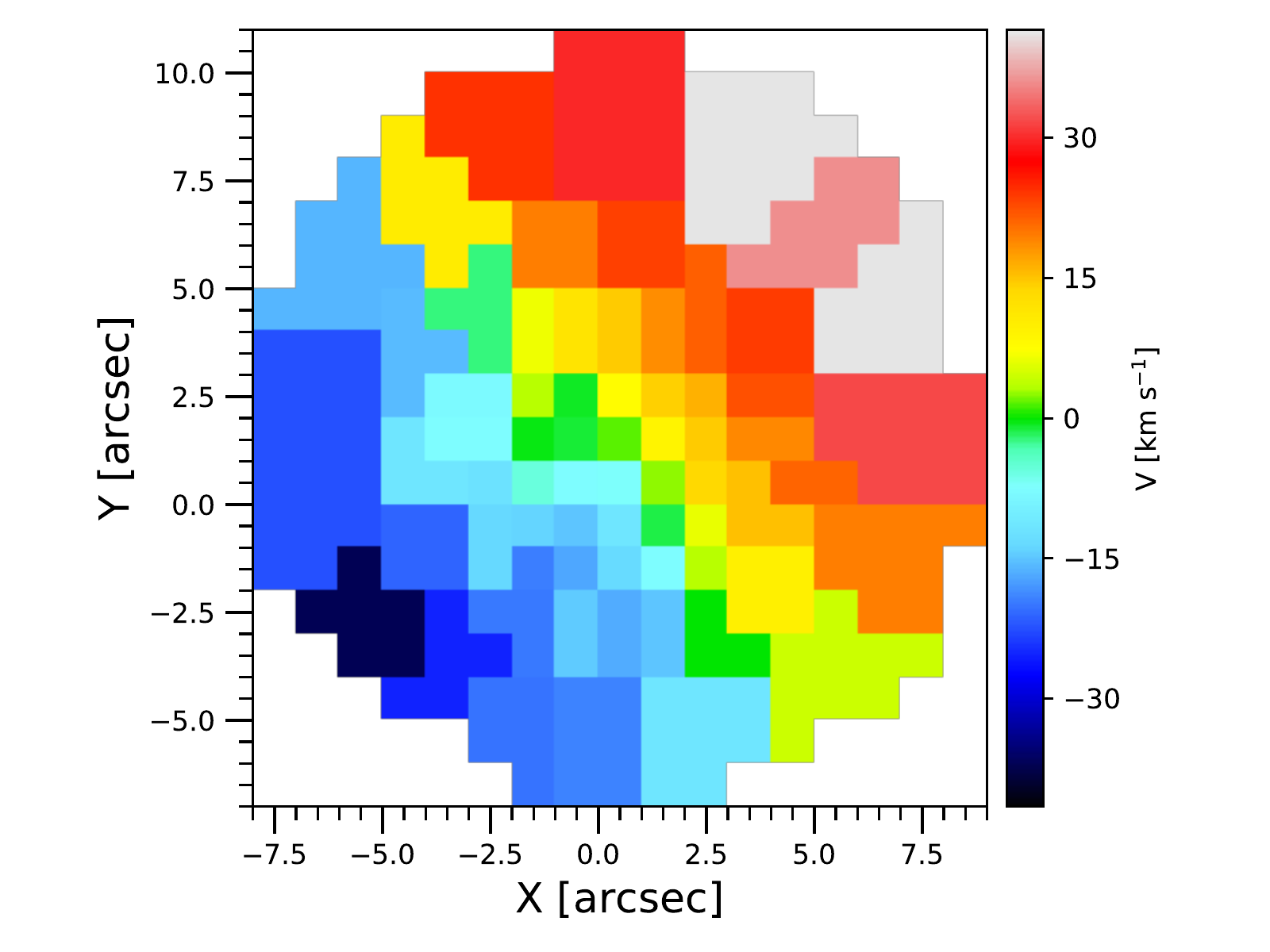}
    \includegraphics[width=2.25in,clip,trim = 20 10 40 10]{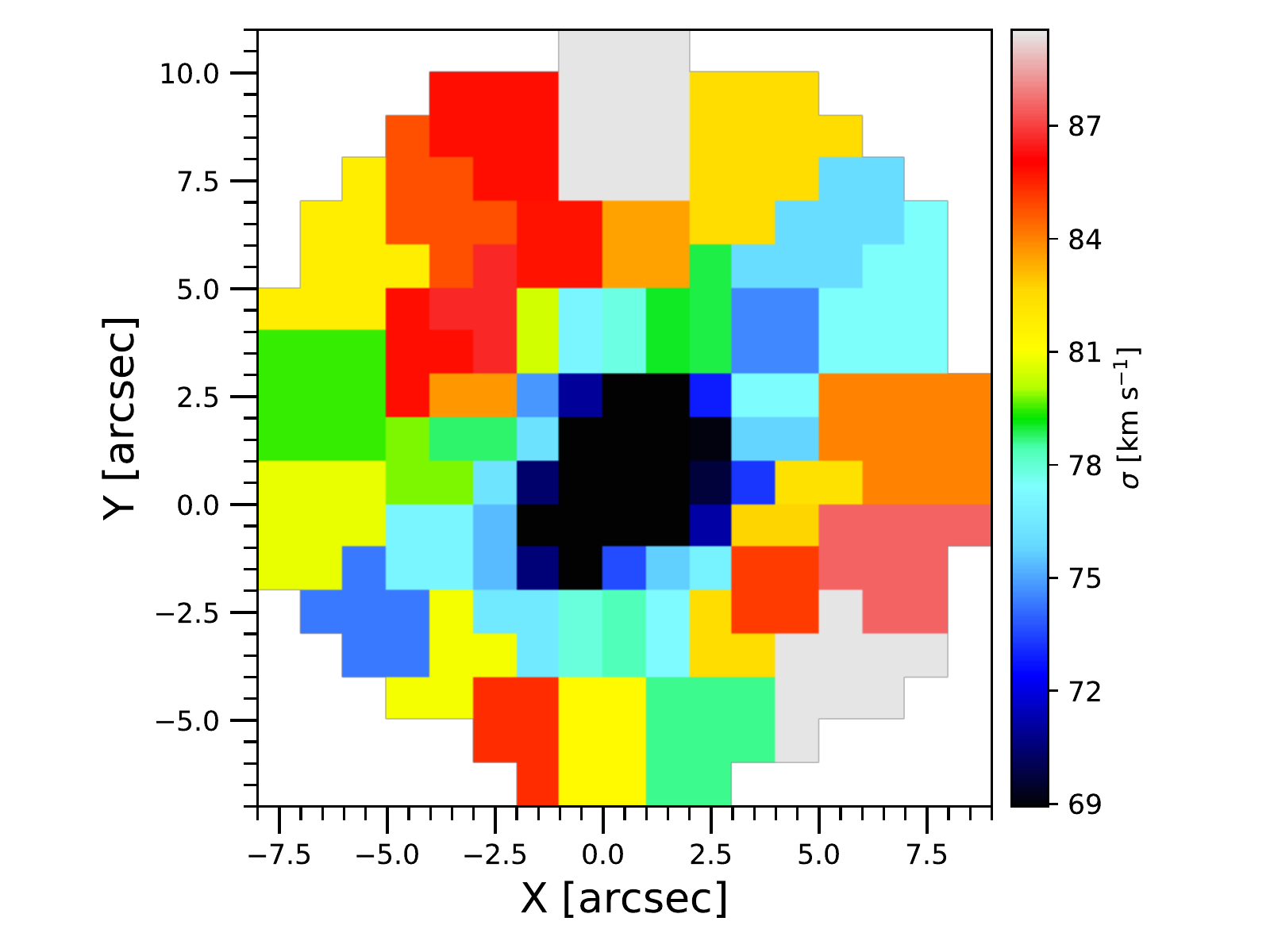}    

    \includegraphics[width=2.in,clip,trim = 20 0 70 0]{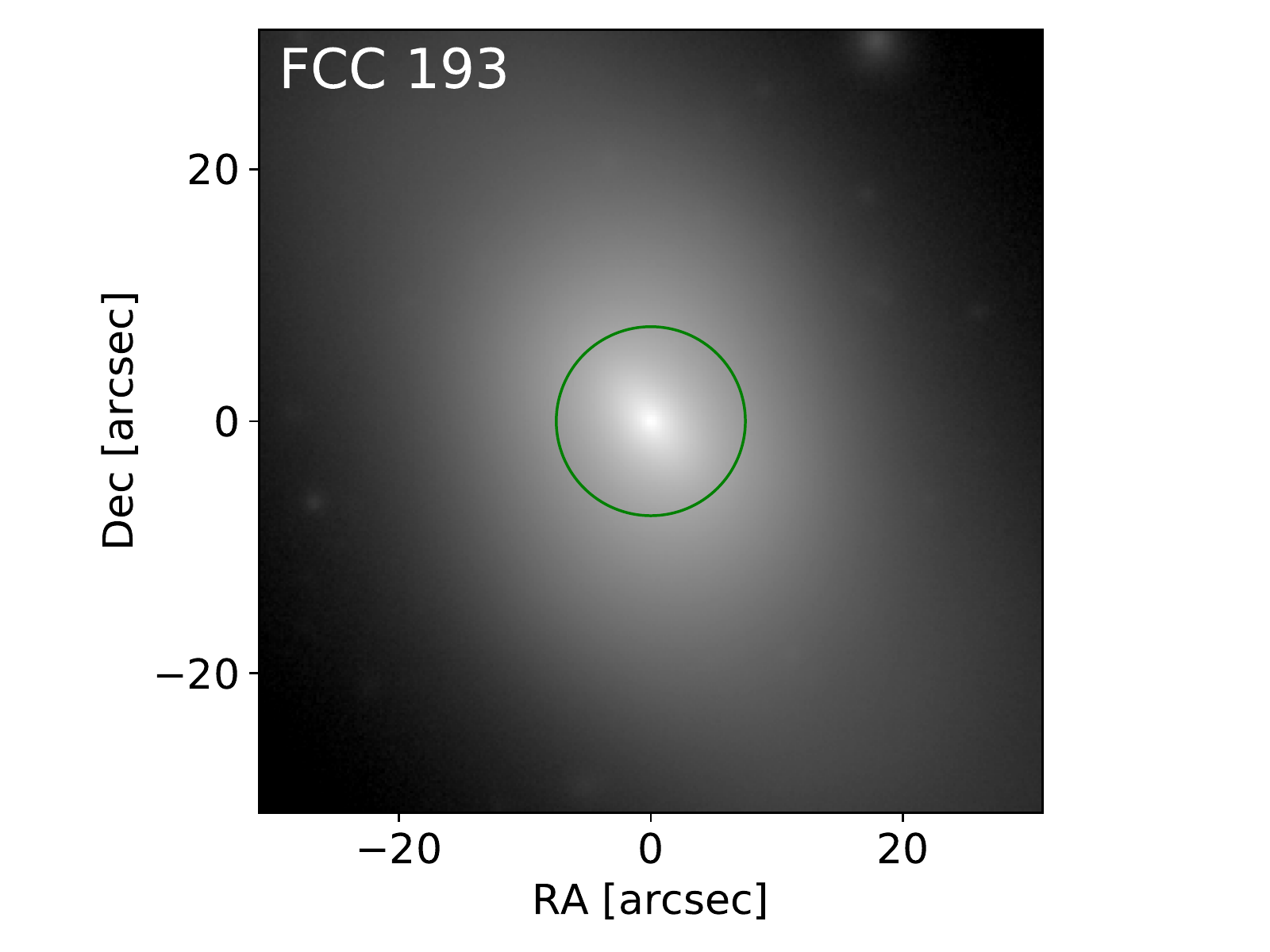}
    \includegraphics[width=2.25in,clip,trim = 20 10 30 10]{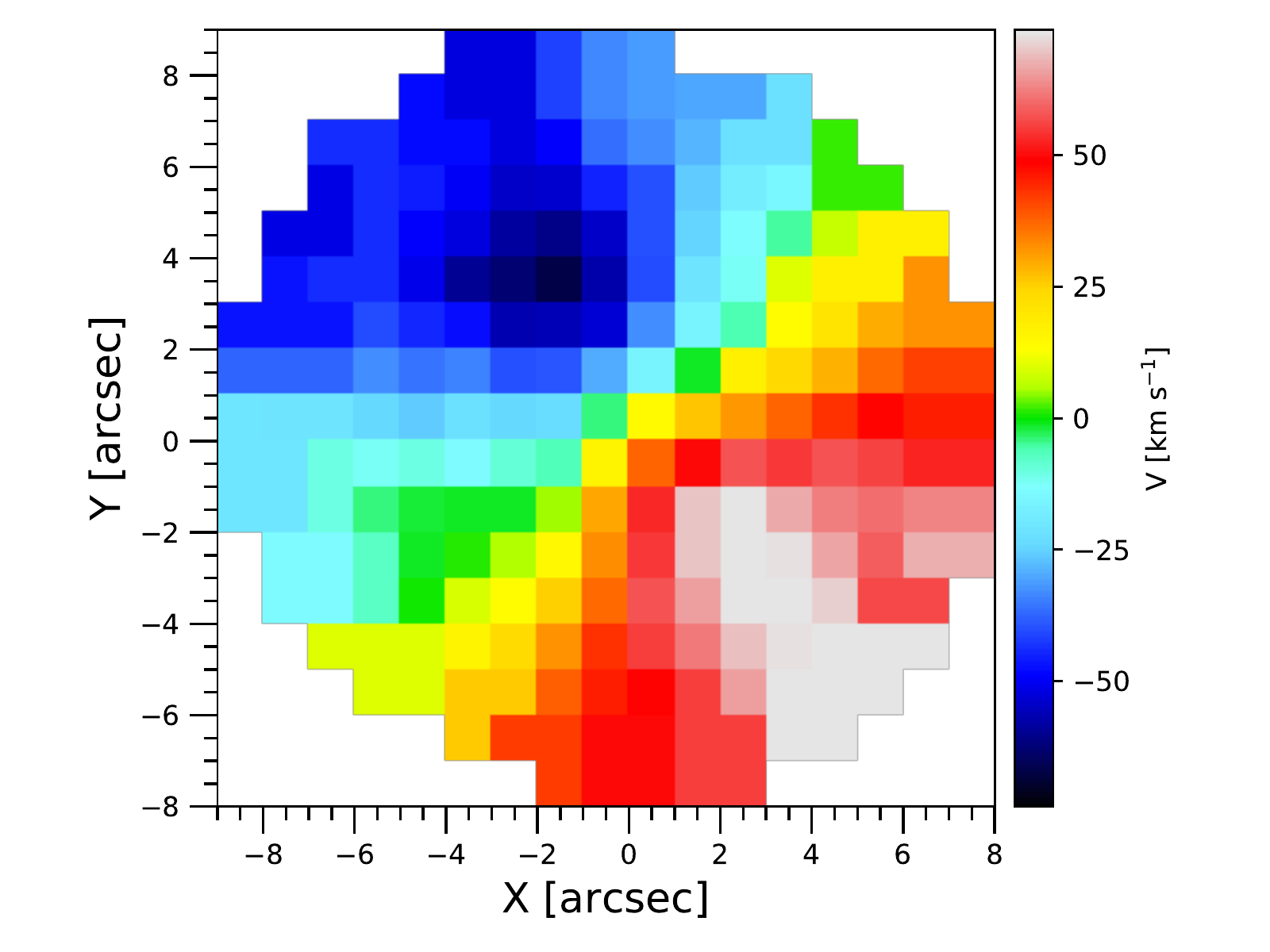}
    \includegraphics[width=2.25in,clip,trim = 20 10 30 10]{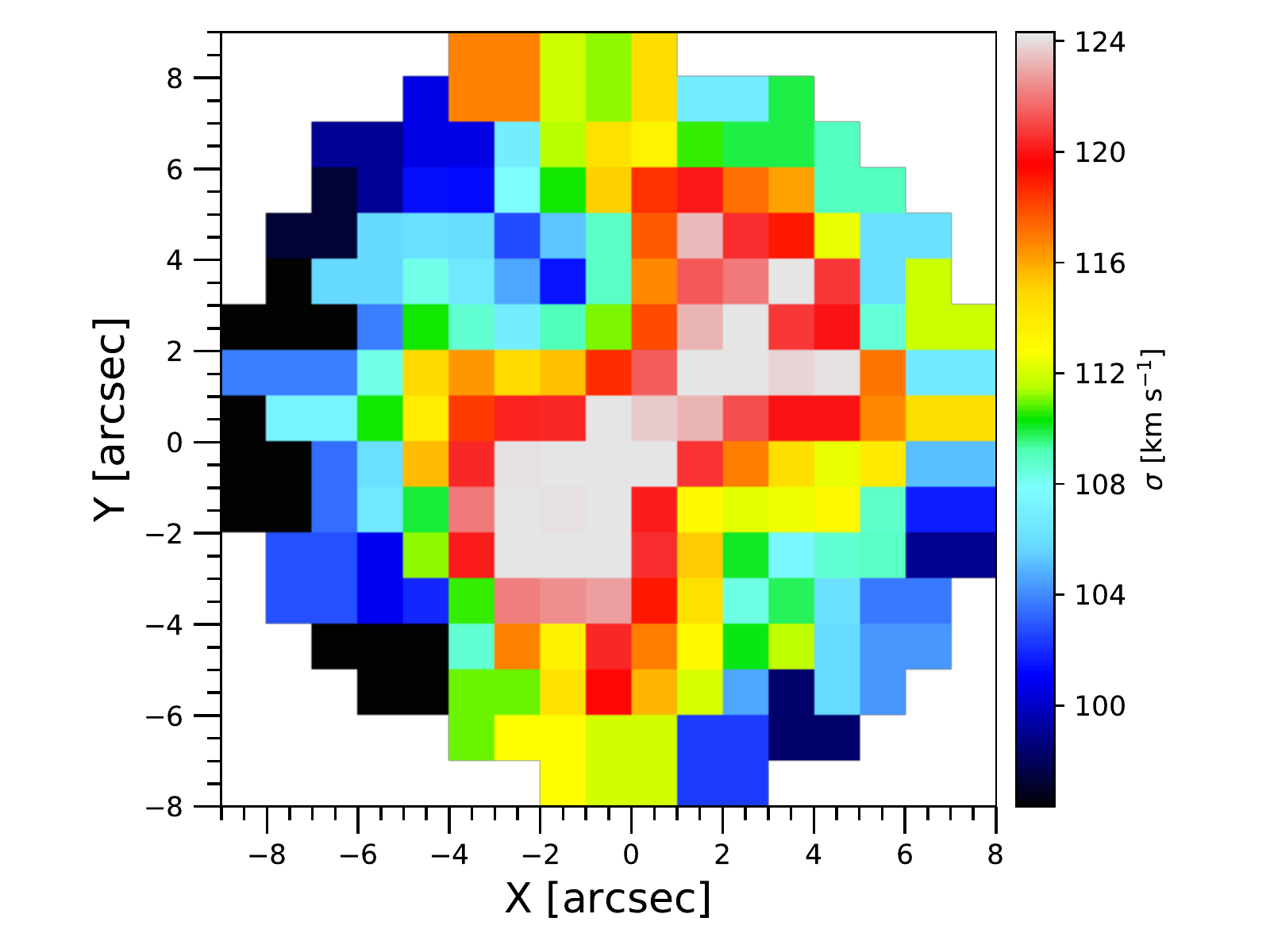}
    \vspace{3mm}
    {\bf Figure \ref{fig:secondary_maps}} continued
\end{figure*}

\begin{figure*}
    \centering
    \includegraphics[width=2.in,clip,trim = 20 0 70 0]{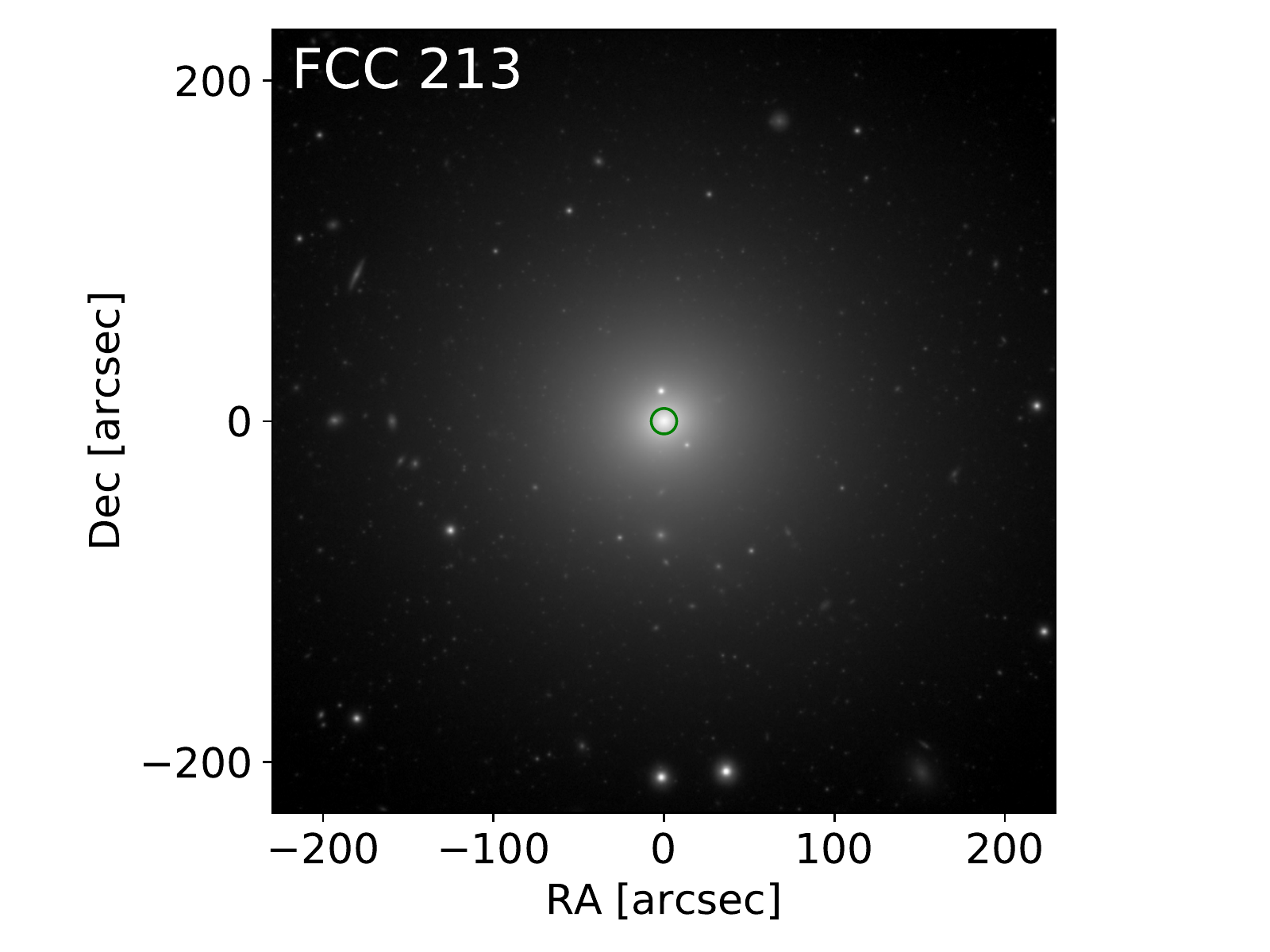}
    \includegraphics[width=2.25in,clip,trim = 20 10 10 10]{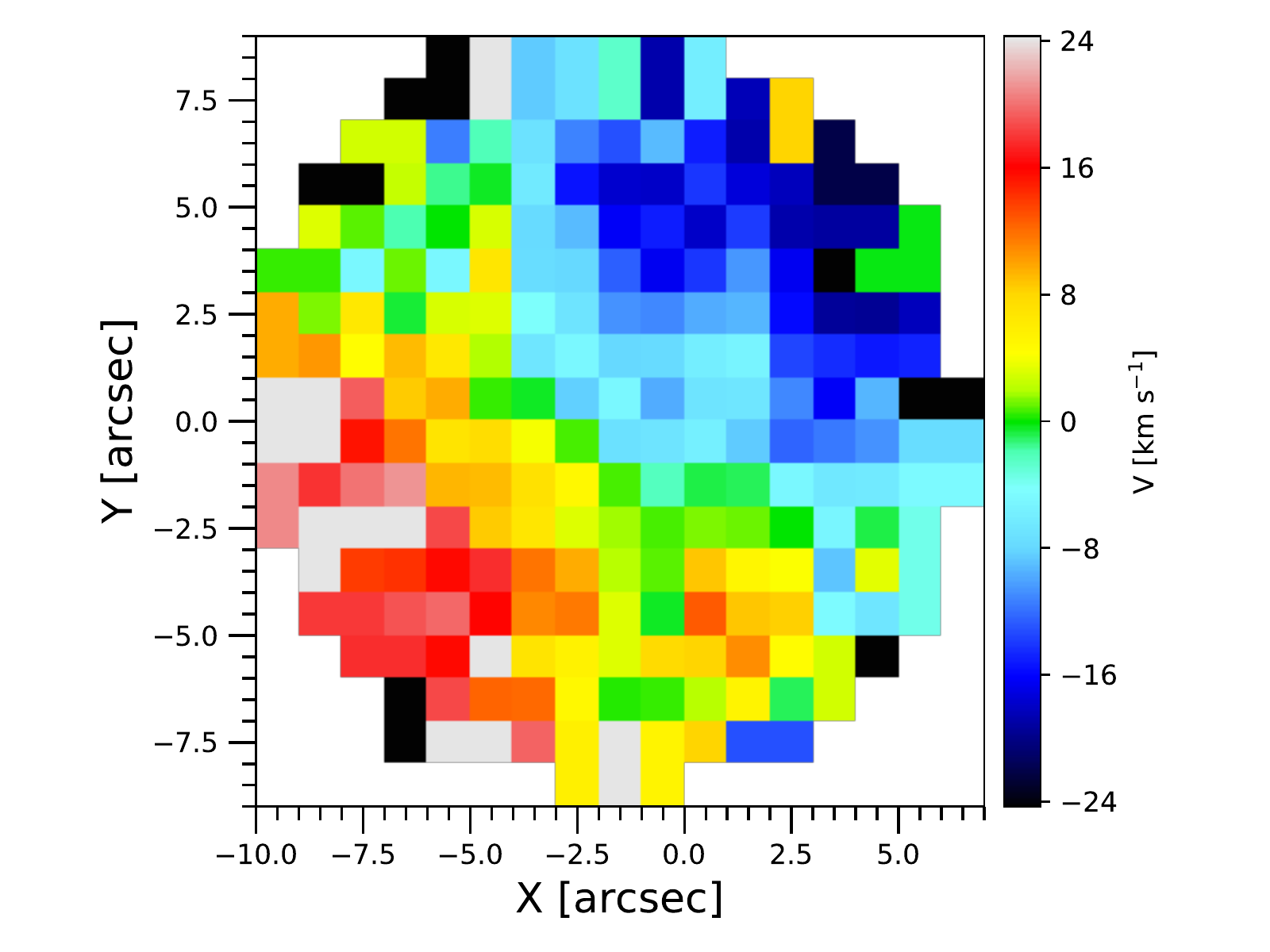}
    \includegraphics[width=2.25in,clip,trim = 20 10 10 10]{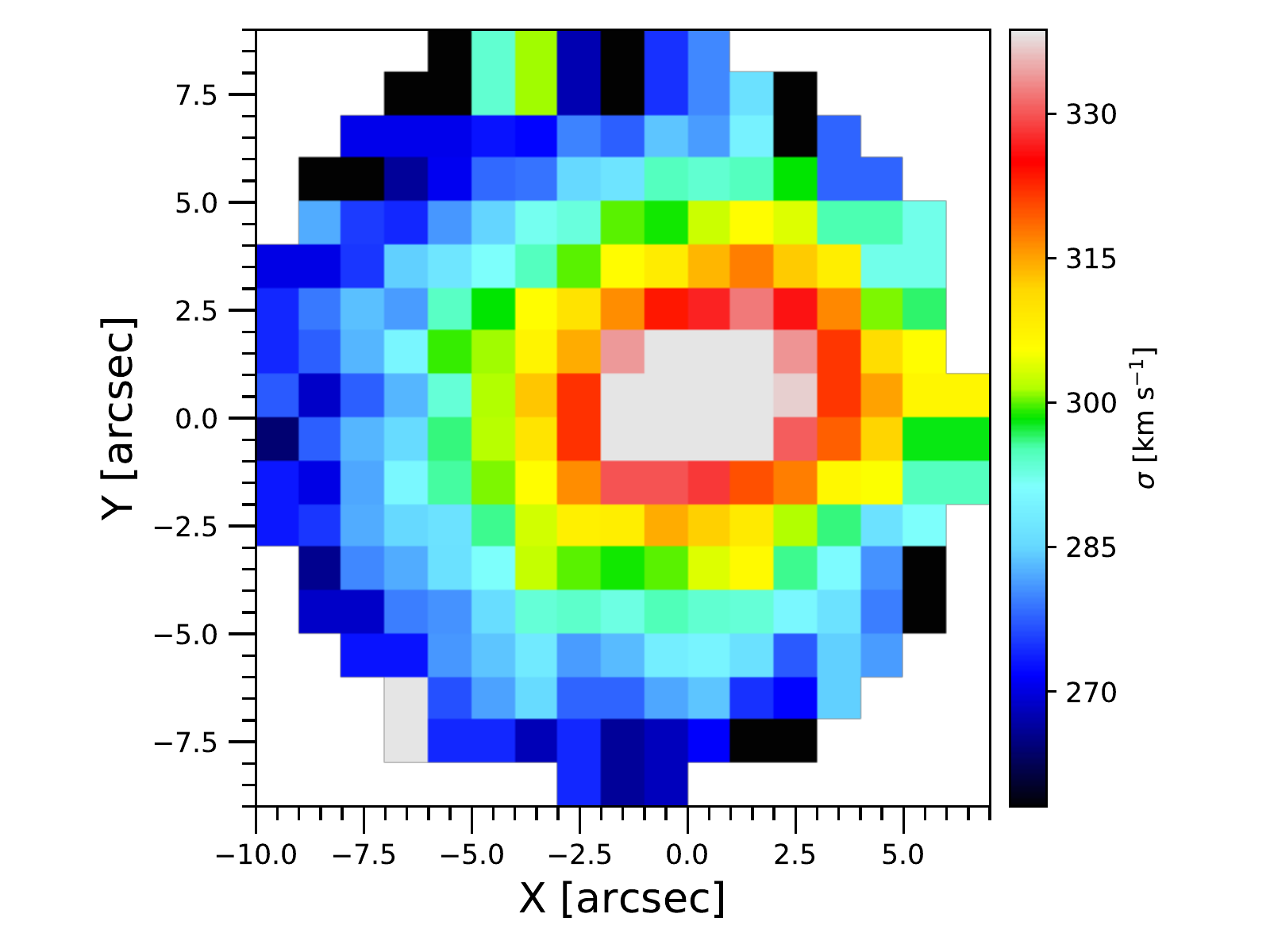}

    \includegraphics[width=2.in,clip,trim = 20 0 70 0]{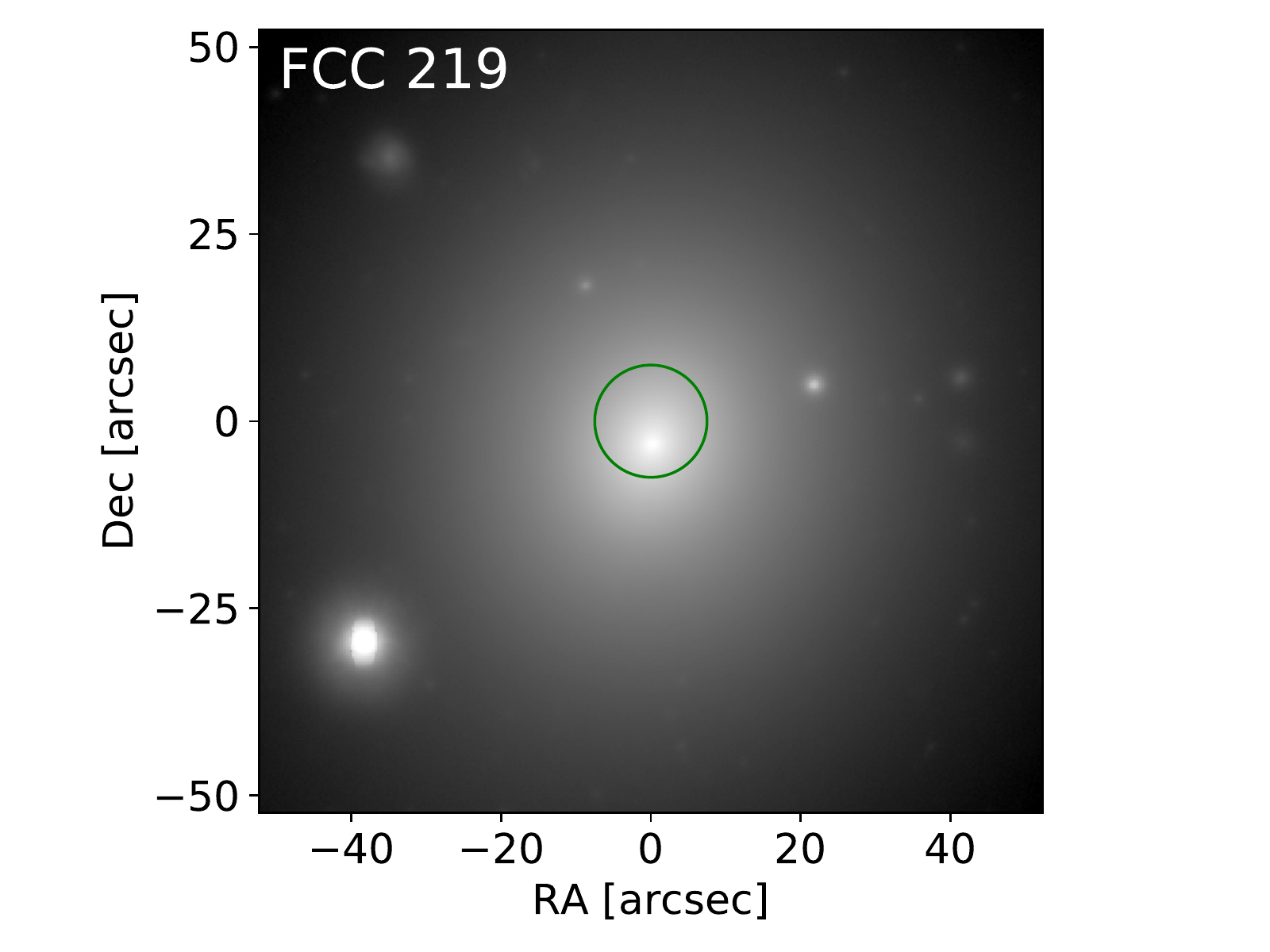}
    \includegraphics[width=2.25in,clip,trim = 15 10 10 10]{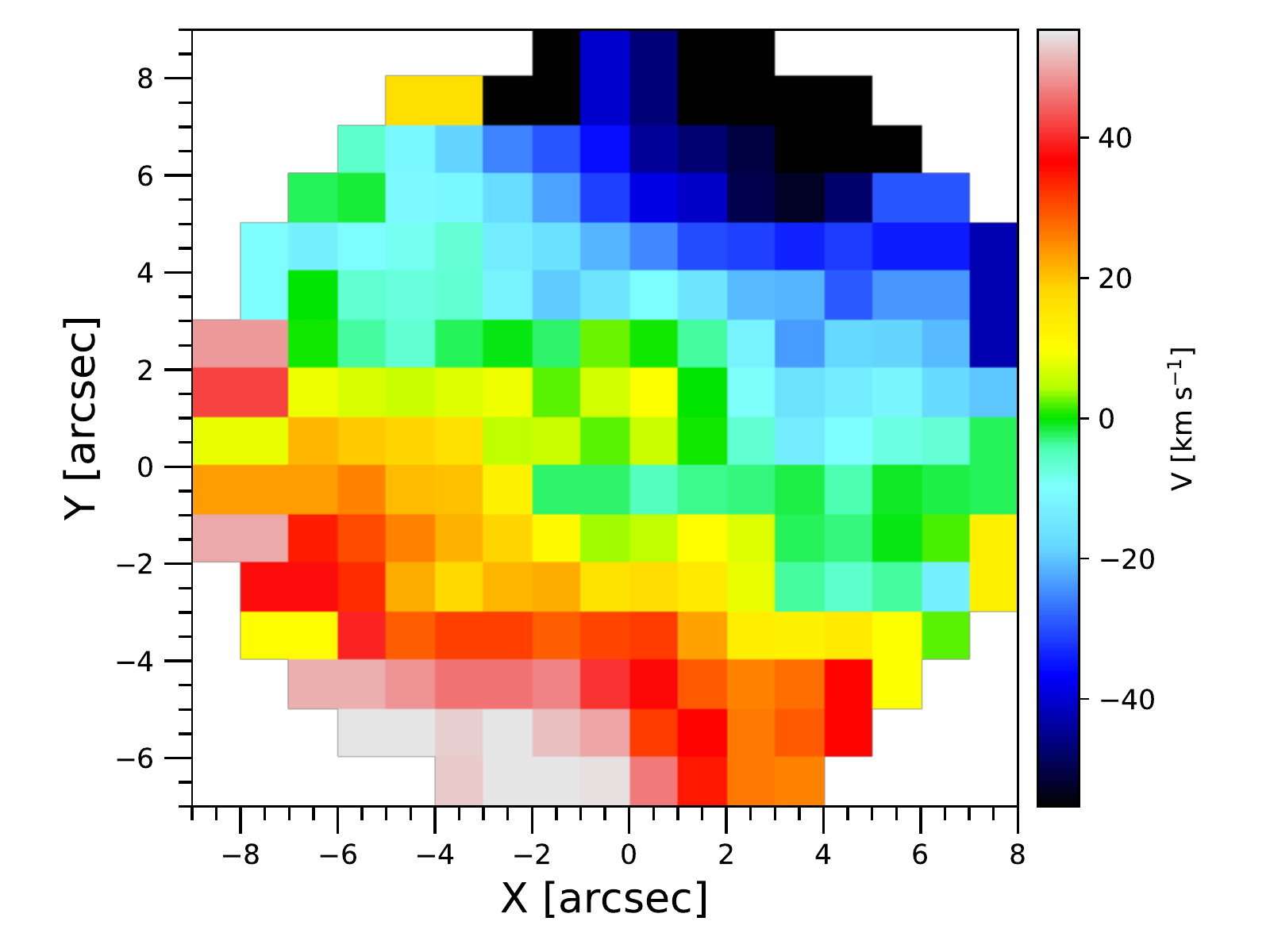}
    \includegraphics[width=2.25in,clip,trim = 20 10 10 10]{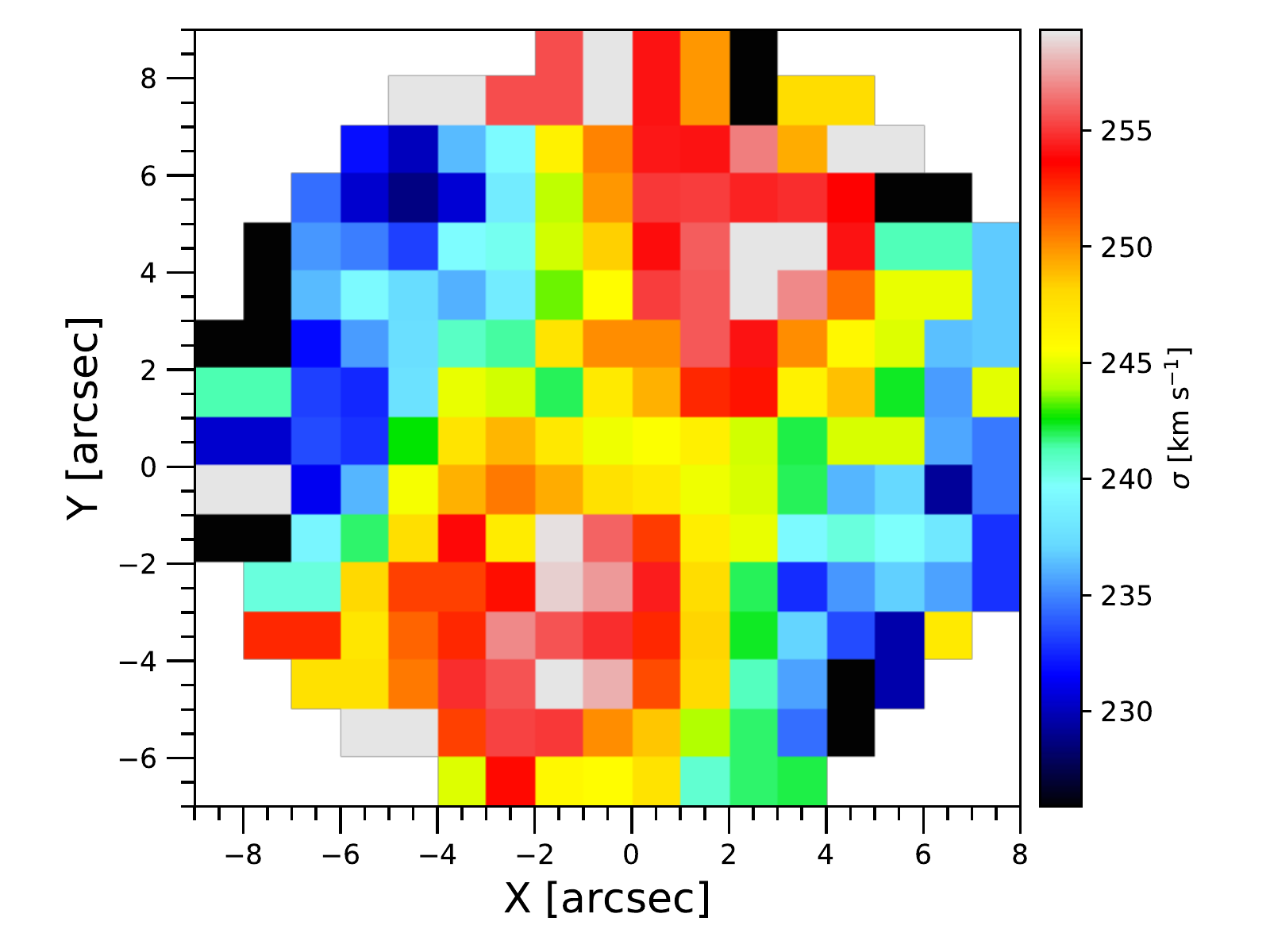}
    
    \includegraphics[width=2.in,clip,trim = 20 0 70 0]{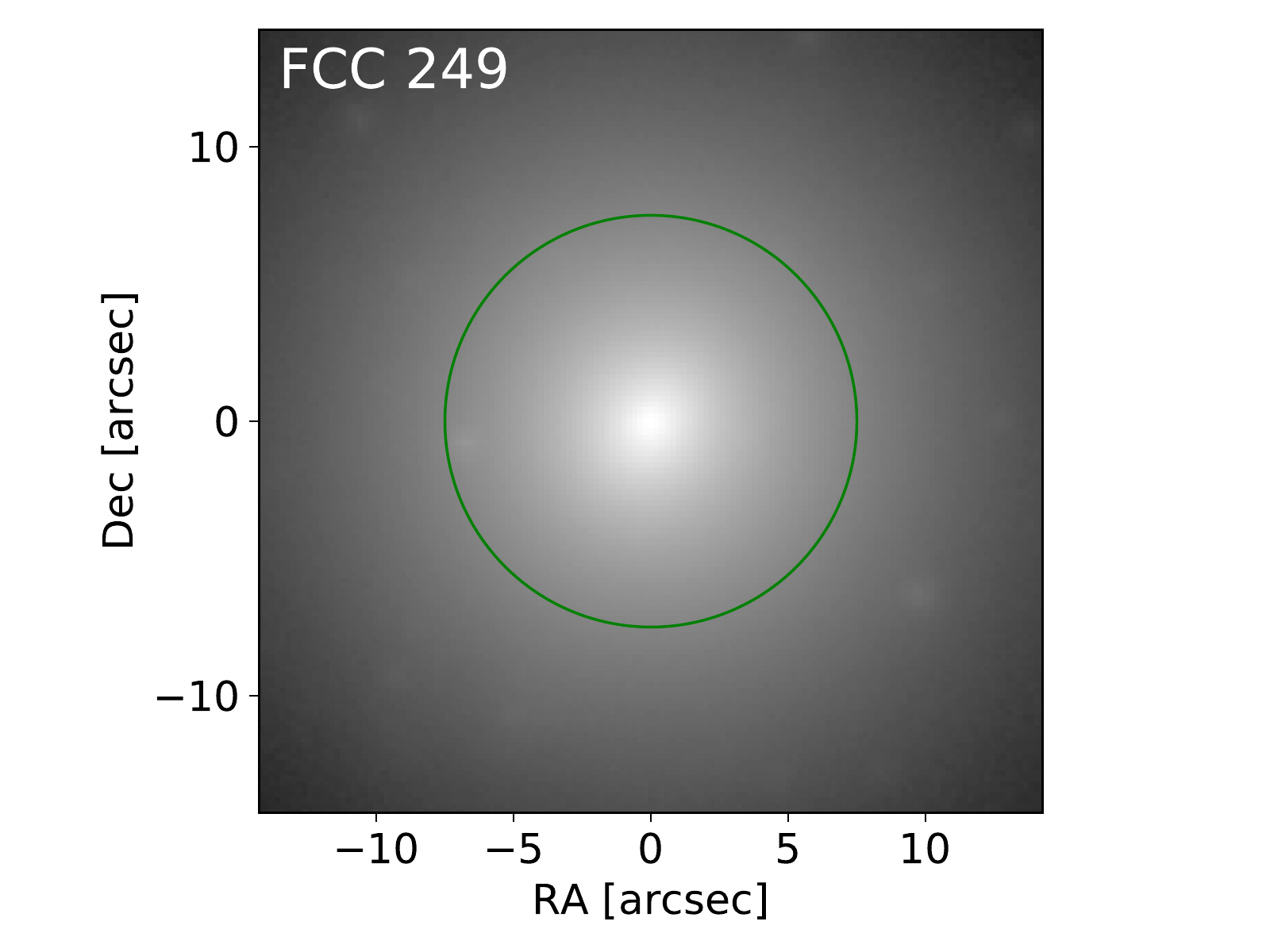}
    \includegraphics[width=2.25in,clip,trim = 20 10 40 10]{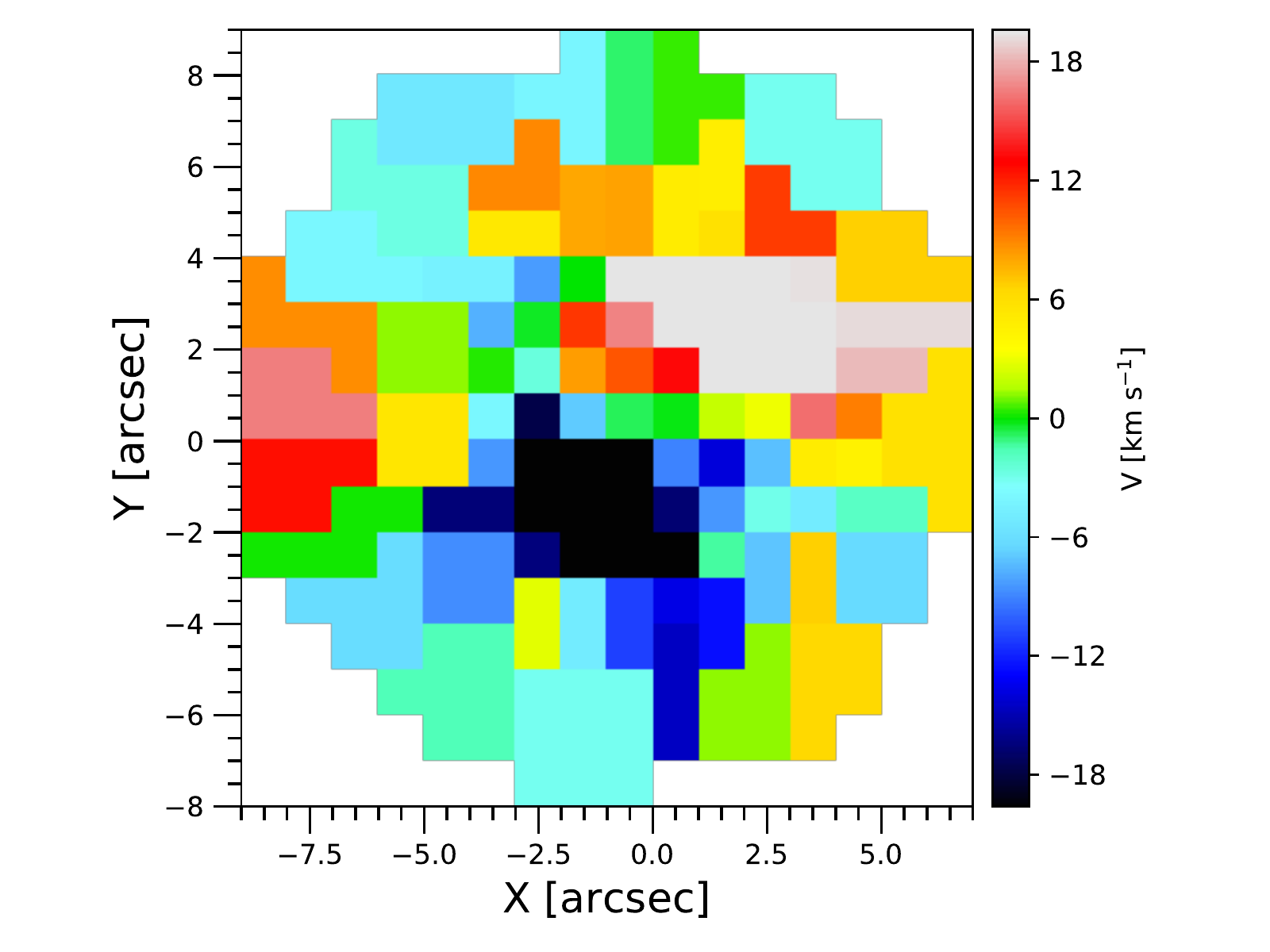}
    \includegraphics[width=2.25in,clip,trim = 20 10 40 10]{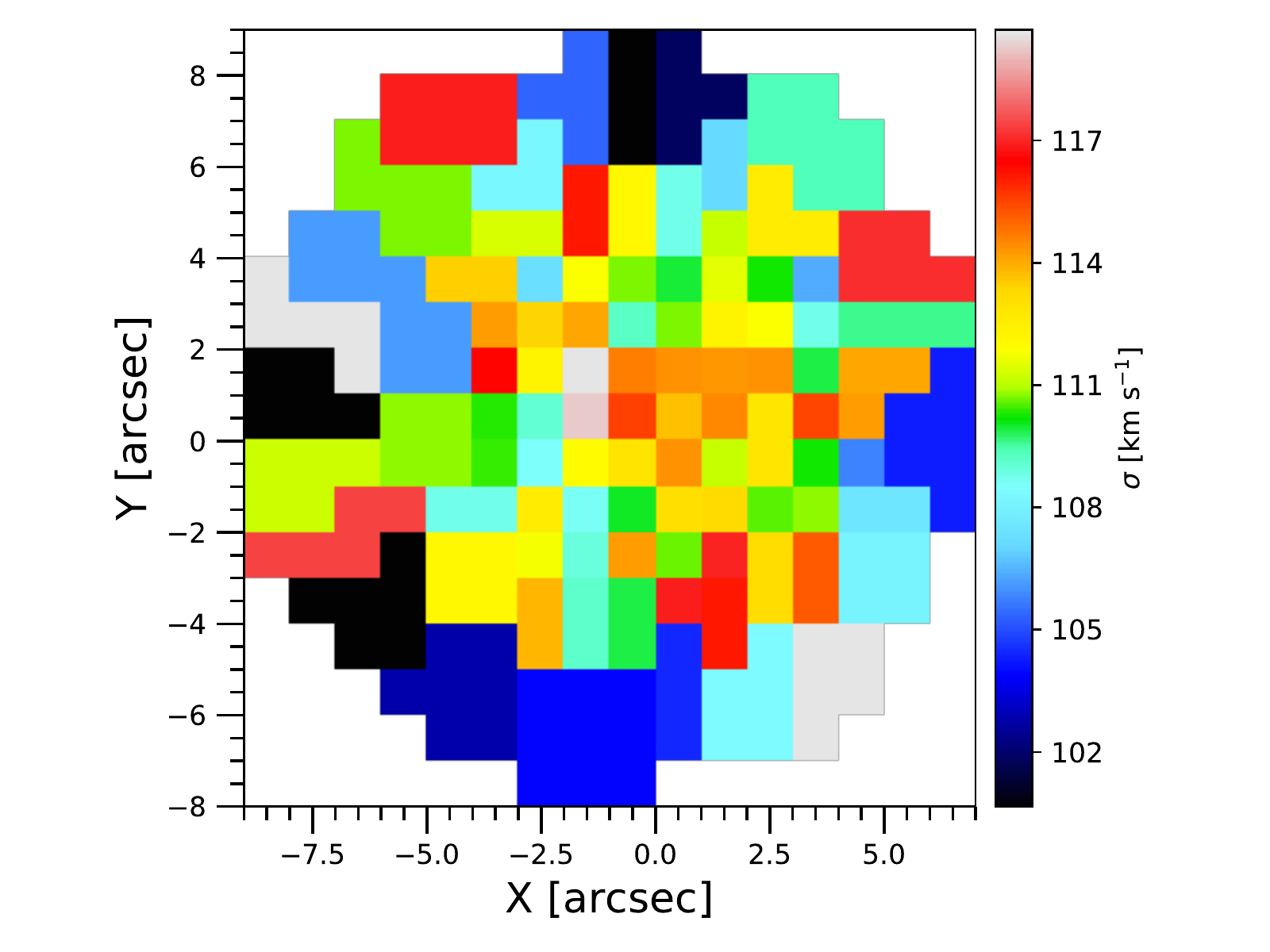}
    
    \includegraphics[width=2.in,clip,trim = 20 0 70 0]{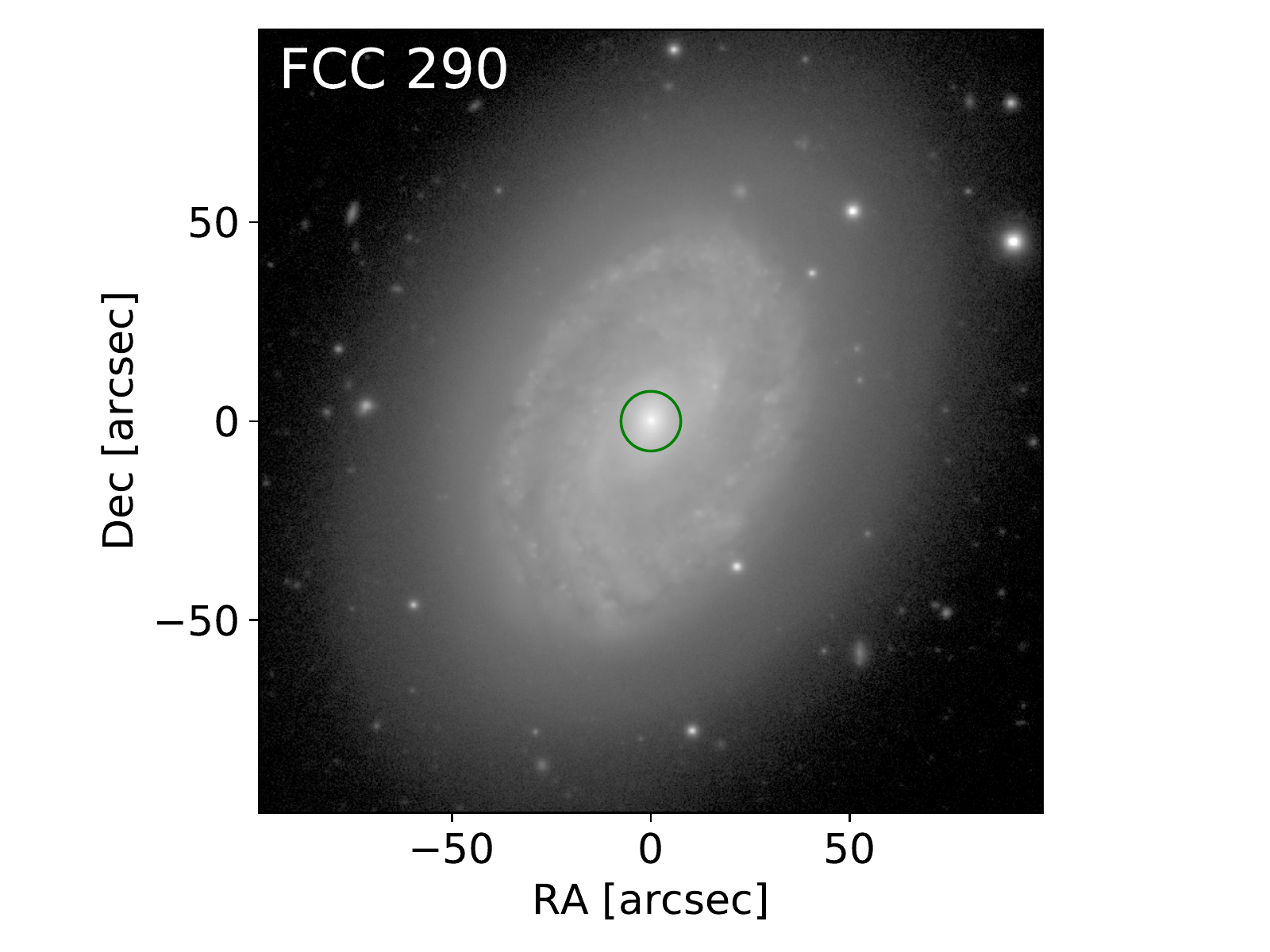}
    \includegraphics[width=2.25in,clip,trim = 20 10 10 10]{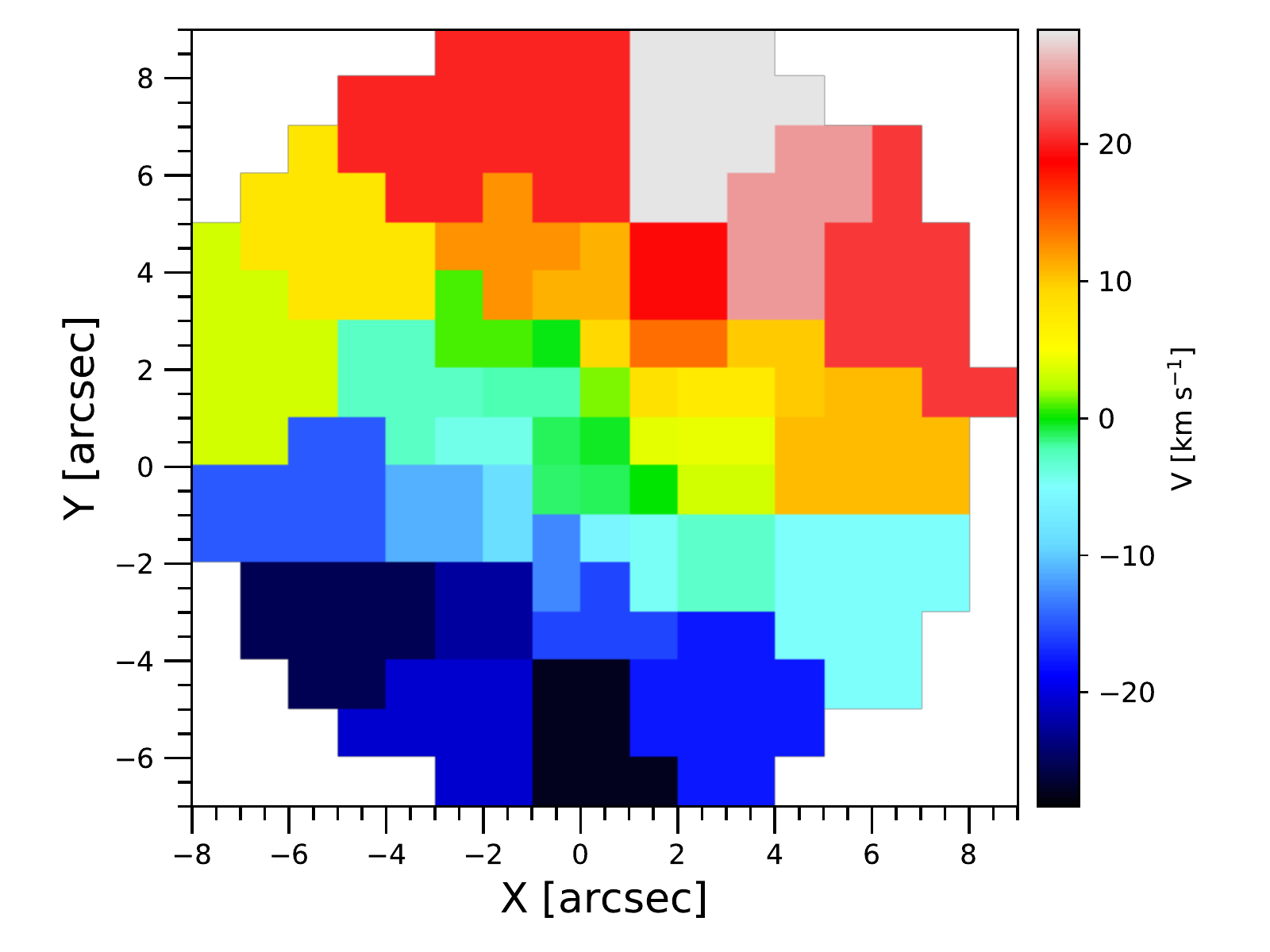}
    \includegraphics[width=2.25in,clip,trim = 20 10 10 10]{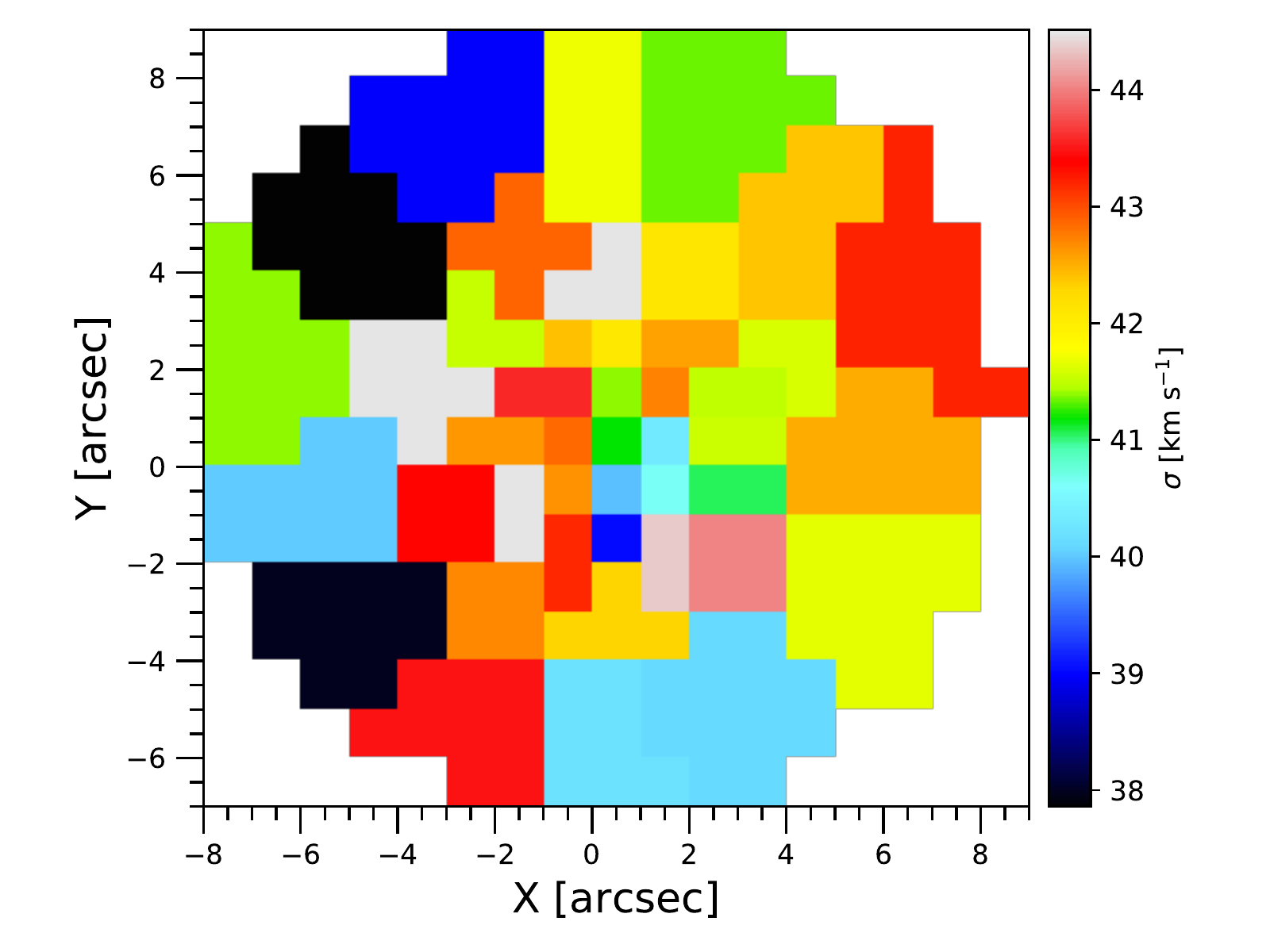}
    
    \includegraphics[width=2.in,clip,trim = 20 0 70 0]{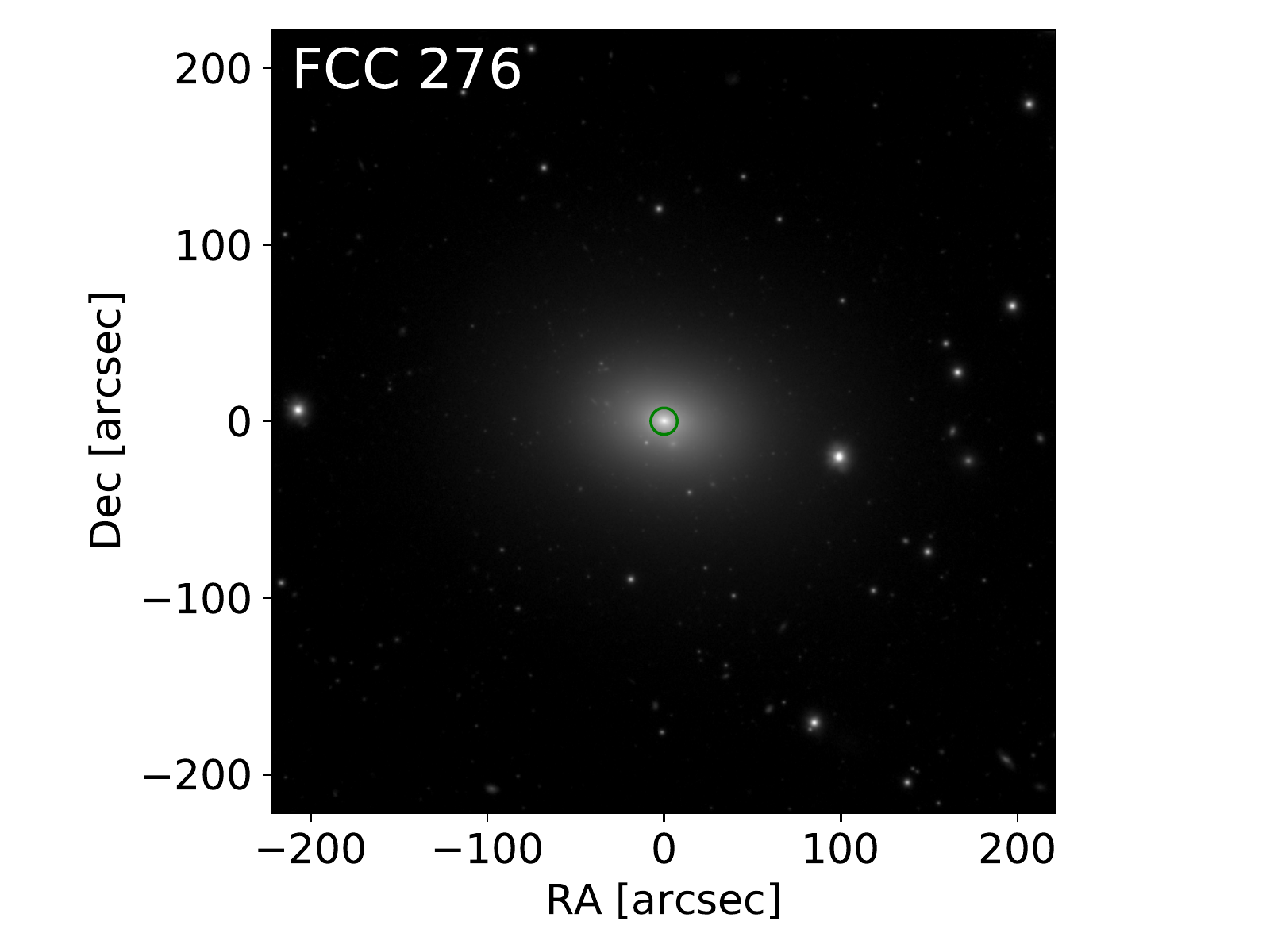}
    \includegraphics[width=2.25in,clip,trim = 20 10 30 10]{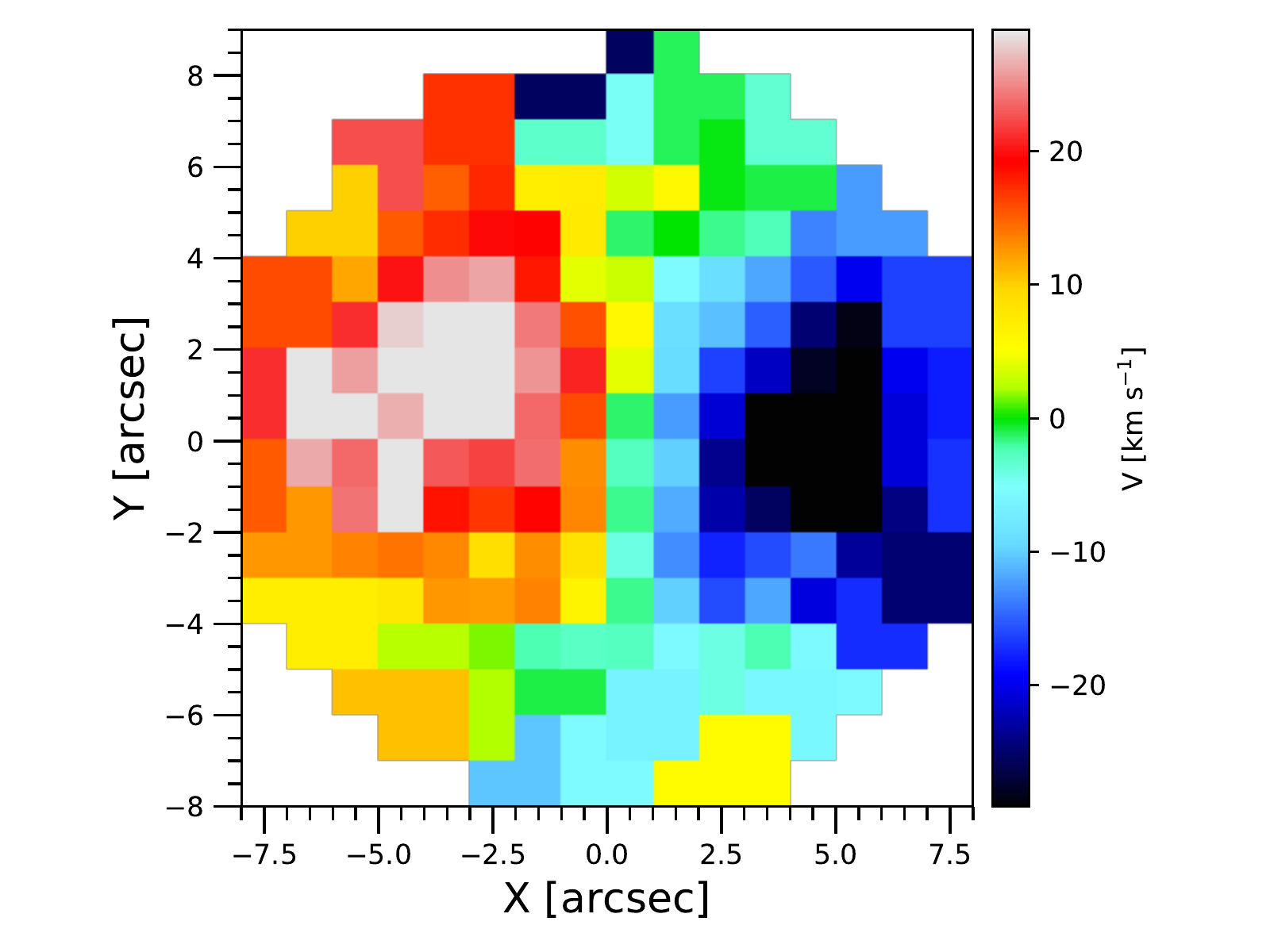}
    \includegraphics[width=2.25in,clip,trim = 20 10 30 10]{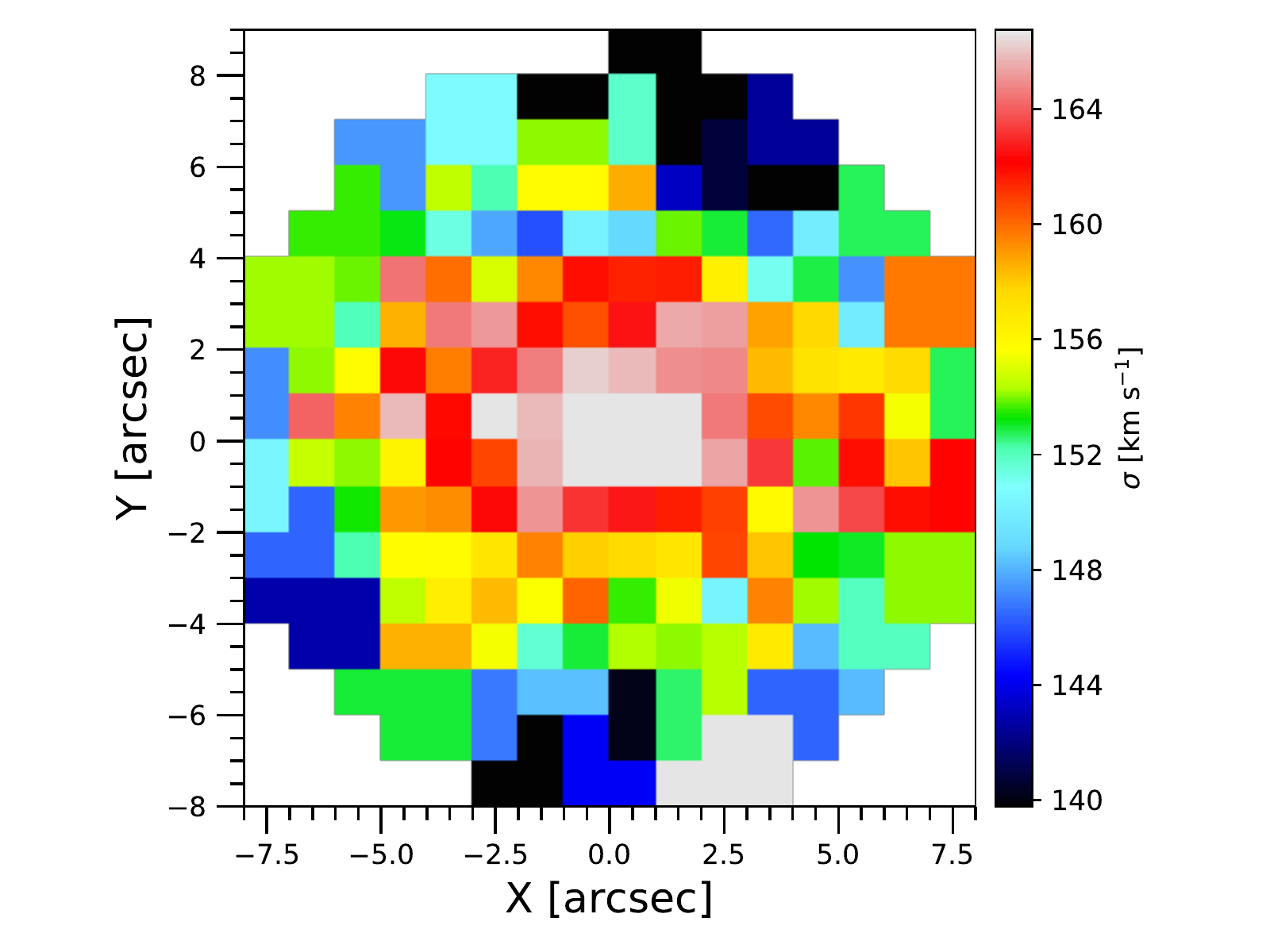}    
    \vspace{3mm}
    {\bf Figure \ref{fig:secondary_maps}} continued
\end{figure*}


\bsp	
\label{lastpage}
\end{document}